\documentclass[a4paper,11pt]{book}
\usepackage{stmaryrd}
\usepackage{amsmath}
\usepackage{revsymb}
\usepackage{mathrsfs}
\usepackage{graphicx}
\usepackage{amssymb}
\usepackage{euscript}
\usepackage[section]{placeins}
\usepackage{fancyhdr}
\allowdisplaybreaks
\DeclareMathAlphabet\Euler{U}{eur}{m}{n}
\SetMathAlphabet\Euler{bold}{U}{eur}{b}{n}
\begin{document}
\numberwithin{equation}{section}
\pagestyle{empty}
\begin{center}
\large

\textbf{UNIVERSIT\`A DEGLI STUDI DI TRIESTE}\\

\normalsize

\vspace{0.5cm}

\textbf{Facolt\`a di Scienze Matematiche, Fisiche e Naturali}\\

\vspace{0.5cm}

\textbf{Dipartimento di Fisica Teorica}\\

\vspace{1.5cm}

XVI CICLO DEL\\
DOTTORATO DI RICERCA IN FISICA\\

\vspace{1.5cm}

\huge

\textbf{Particle Production in a Strong,}\\
\textbf{Slowly-Varying Magnetic Field}\\
\textbf{With an Application to Astrophysics}\\

\normalsize

\vspace{1.5cm}

\begin{tabular}{cc}
DOTTORANDO & COORDINATORE DEL COLLEGIO DEI DOCENTI \\
Antonino Di Piazza & CHIAR.MO PROF. Gaetano Senatore\\
& Dipartimento di Fisica Teorica della\\
& Universit\`a degli Studi di Trieste\\
& \\
& \\
& FIRMA. . . . . . . . . . . . . . . . . . . . . . . . . . . . .\\
& \\
& \\
& \\
& \\
& \\
& \\
& TUTORE e RELATORE\\
& CHIAR.MO PROF. Giorgio Calucci\\
& Dipartimento di Fisica Teorica della\\
& Universit\`a degli Studi di Trieste\\
& \\
& \\
\end{tabular}
\end{center}
\frontmatter
\pagestyle{fancy}
\fancyhf{}
\fancyhead[LE,RO]{\thepage}
\fancyhead[LO,RE]{Contents}
\tableofcontents
\chapter{Notation}
\fancyhead[LE,RO]{\thepage}
\fancyhead[LO,RE]{Notation}
\begin{itemize}
\item[--] Starting from Chap. 2 natural units with $\hbar=1$ and $c=1$ are used.
\item[--] The Minkowski metric signature is $(1,-1,-1,-1)$.
\item[--] The Greek indices are spacetime indices running from $0$ to $3$. The ones at the beginning of the Greek alphabet like $\alpha$, $\beta$, $\gamma$ or $\delta$ label quantities referring to Minkowski spacetime while the others in the middle of the Greek alphabet like $\lambda$, $\mu$, $\nu$ or $\rho$ label quantities referring to a generic curved spacetime.
\item[--] The Latin indices at the beginning of the Latin alphabet like $a$, $b$ or $c$ are spinor indices running from $1$ to $4$. The Latin index $i$ is a space index running from $1$ to $3$. Finally, the Latin indices at the end of the Latin alphabet like $u$ or $v$ are coordinate indices standing for $x$, $y$ or $z$.
\item[--] The index $j$ embodies all the quantum numbers of a relativistic electron (positron) in the presence of a constant and uniform magnetic field in Minkowski spacetime. The index $J$ embodies the corresponding quantum numbers but in curved spacetime. Instead, the index $\jmath$ indicates a set of quantum numbers to be determined.
\item[--] The sans serif letters like $\mathsf{x}$ or $\mathsf{y}$ embody the four general coordinates of a spacetime fourpoint.
\item[--] The subscript ``$\shortuparrow$'' refers to quantities concerning the physical situation in which the magnetic field is always directed along the $z$ axis.
\item[--] The subscript ``$\nnearrow$'' refers to quantities concerning the physical situation in which the magnetic field lies in the $y\text{--}z$ plane with a nonzero $y$ component. 
\item[--] The subscripts ``$\perp$'' and ``$\parallel$'' refer to quantities perpendicular and parallel to the magnetic field respectively. 
\item[--] The superscripts ``(ann)'', ``(syn)'' and ``(dir)'' refer to photon production mechanisms and stand for ``annihilation'', ``synchrotron'' and ``direct'' respectively.
\item[--] The superscripts ``lin'' and ``exp'' refer to the magnetic field time variation and stand for ``linear'' and ``exponential'' respectively.
\item[--] The superscript ``(R)'' refers to quantities calculated in Rindler spacetime.
\item[--] The primed classical and quantum fields, the primed propagators and the primed $S$-matrix elements refer to the physical situation in which the magnetic field is always directed along the $z$ axis.
\item[--] The capital calligraphic letters like $\mathcal{P}$ or $\mathcal{H}$ refer to one-particle electron and positron quantum operators.
\item[--] The variant capital calligraphic letters like $\mathscr{L}$ or $\mathscr{H}$ refer to field density quantities.
\item[--] The ``Euler'' font letters like $\Euler{A}$, $\Euler{F}$ or $\Euler{e}$ refer to quantities concerning the photon radiation field.
\end{itemize}
\chapter{Introduction and outline}
\fancyhead[LE,RO]{\thepage}
\fancyhead[LO,RE]{Introduction and outline}
During the years Quantum Electrodynamics (QED) has received a large amount 
of experimental confirmations and today it is considered the 
most reliable microscopic physical theory we have. It is enough to think about the excellent agreement between the predicted \cite{Kinoshita} and the measured \cite{Mohr} anomalous magnetic moment of the electron. The main subject of this thesis concerns 
a very interesting and fascinating branch of QED: 
Quantum Electrodynamics in the presence of External Fields (QEDEF) \cite{Greiner,Fradkin}. As the name itself suggests, QEDEF concerns those quantum electromagnetic processes happening when a classical (nonquantized) field is present. The 
external fields are typically electromagnetic fields and they have two main general characteristics:
\begin{enumerate}
\item they are produced by sources not belonging to the system under 
study and their spatiotemporal evolution is assigned;
\item they are so intense that:
\begin{enumerate}
\item[2a.] quantizing them would be useless;
\item[2b.] trying to compute their effects by means of 
standard perturbative techniques would be in most cases conceptually wrong.
\end{enumerate}
\end{enumerate}

Now, many processes studied in pure QED can also be analyzed in the presence of external fields such as particle scattering and so on. But there is a process that is typical of QEDEF: the production of particles \emph{from vacuum}. The reason is understandable because pure QED describes a closed system whose energy is conserved and then particles can not be created from vacuum. Instead, in QEDEF just the external fields can supply the energy necessary to create electron-positron pairs or photons directly from vacuum. The seminal work about this subject is that by Schwinger \cite{Schwinger} in which he shows that in the presence of a constant and uniform electric field, real pairs\footnote{From now on, when it is not specified, it is understood that ``pair'' (``pairs'') stands for ``electron-positron pair'' (``electron-positron pairs'').} can be produced directly from vacuum. The production probability results significantly different from zero only for electric field strengths much larger than $E_{cr}=m^2c^3/(\hbar e)=1.3\times 10^{16}\;\text{V/cm}$ where $m$ and $-e<0$ are the mass and the charge of the electron respectively. This value of $E_{cr}$ represents a sort of benchmark over which the effects of an external electric field become important. Nevertheless, we will see in the next Chapter that $E_{cr}$ is much larger than the electric fields that today can be produced in terrestrial laboratories and the experimental confirmations concerning QEDEF are less numerous than those concerning pure QED. 

The expression of $E_{cr}$ can be seen qualitatively as that of the electric field strength whose energy in a volume with typical length of the order of the Compton length $\lambdabar=\hbar/(mc)$ is large enough to produce an electron and a positron at rest. By using the same energetic argument it can be seen that a magnetic field strength stronger than $B_{cr}=m^2c^3/(\hbar e)=4.4\times 10^{13}\;\text{gauss}$ is capable to ``break'' the vacuum and create a pair.\footnote{It must be pointed out that, although from an \emph{energetic} point of view pairs can be created in the presence of a constant and uniform magnetic field with strength larger than $B_{cr}$, it is impossible, as Schwinger also showed in \cite{Schwinger}, from a \emph{dynamical} point of view essentially because the Lorentz force does not do any work.} The value of $B_{cr}$ is also much larger than that of the strongest, steady magnetic field ever produced in a terrestrial laboratory which is of the order of $10^5\;\text{gauss}$ \cite{NHMFL}. Instead, we will see in the next Chapter that there are various indirect evidences that 
around astrophysical compact objects (neutron stars, black holes) magnetic fields 
much stronger than $B_{cr}$ are present. Actually, this is still not a 
good enough reason to start studying QED in the presence 
of such strong magnetic fields if an (at least) indirect experimental 
effect of their presence is not at hand. Now, one of the most intriguing and 
mysterious astrophysical phenomena seems to be originated just around 
neutron stars or black holes surrounded by an accretion disk. I am referring to the so-called 
Gamma-Ray Bursts (GRBs) that are huge pulses of soft gamma-rays that 
our satellites register on average once a day. The exact GRBs production mechanism is still not completely well understood. Actually, following the widely accepted \emph{fireball} model of GRBs, the photons making a GRB are produced, through standard electromagnetic mechanisms such as synchrotron emission, by a ``fireball'' in turn made mostly of electrons, positrons and photons themselves. Nevertheless, it is not still clear \emph{how} the fireball itself is generated. On the other hand, as we will see in the next Chapter, it seems almost sure that the fireball is produced near a neutron star or a black hole surrounded by an accretion disk and that the huge magnetic fields that are there play a fundamental role in this process. In this respect, trying to find a microscopic underlying mechanism responsible of the formation of the fireball represents for a theoretical physicist a very challenging reason to 
study those electromagnetic processes that can be primed in the presence of such strong magnetic fields. In fact, in this way, some general experimental features of GRBs could be explained and interpreted.

In this thesis I will focus in particular on particle production processes in the presence of strong (\textit{i.e.}, much larger than $B_{cr}$), uniform and slowly-varying magnetic fields. The theoretical justifications of these assumptions about the structure of the magnetic fields considered in the present work will be given in the following Chapters that are organized as follows. Chap. 1 is divided into two independent parts: in the first one I briefly review some developments and theoretical predictions about the production of pairs in the presence of external electromagnetic fields; instead, in the second one I give a brief introduction about the phenomenological characteristics of GRBs and about the fireball model of GRBs. In Chap. 2 I just outline the following three ``textbook'' subjects: the motion of a relativistic charged particle in the presence of a constant and uniform magnetic field, the adiabatic perturbation theory in quantum mechanics and the quantum field theory in curved spacetime. Even if these subjects (at least the first two) are well-known I treat them for the sake of completeness because they represent the theoretical tools I have used to derive my own results. These last are presented in Chaps. 3-5. In particular, in Chap. 3 I study the production of an electron-positron pair from vacuum in the presence of a strong, uniform and slowly-varying magnetic field. In particular two different magnetic field time evolutions are considered. Since, as I have said, a possible application of these calculations concerns the mechanisms accounting for the production of a GRB, in Chap. 4 I study some processes through that the electrons and the positrons already created in a strong, uniform and slowly-rotating magnetic field can produce photons. In particular, I calculate the energy spectra of photons produced through pair annihilation and as synchrotron radiation. Also, by using the so-called \emph{effective Lagrangean method} I obtain the energy spectrum of the photons produced directly from vacuum in the presence of a strong, uniform and slowly-rotating magnetic field. Finally, in the astrophysical scenario I have just sketched, it is interesting to examine what the role of the compact object's gravitational field is. I devote the last Chapter to this subject by analyzing, in the context of quantum field theory in curved spacetime, how the presence of a weak and of a strong gravitational field modifies some results on the production of electron-positron pairs already obtained in Chap. 3.

\mainmatter
\chapter{Physical background and motivations}
\fancyhf{}
\fancyhead[LE,RO]{\thepage}
\fancyhead[LO]{Chapter \thechapter}
\fancyhead[RE]{Section \thesection}
As I have sketched in the Introduction, despite the possible application to the study of GRBs, this work is principally theoretical. As a consequence, in this Chapter I want to firstly describe the scientific ``landscape'' where this work is included with particular reference to pair production in the presence of strong electromagnetic fields (Sect. \ref{QEDPEF}). Also, in order to make the cited astrophysical application clear I give a brief description of the general experimental features of GRBs, of the physical scenario where they are supposed to be originated and of the widely accepted fireball model of GRBs (Sect. \ref{GRB}).
%
%
\section{Pair production in the presence of strong electromagnetic fields: a short review}
\label{QEDPEF}
This work will concern mainly the theoretical study of the production of pairs in the presence of a strong, uniform and time-varying magnetic field in the framework of QEDEF. Then, it is worth giving a brief review about, in general, the production of pairs in the presence of external electromagnetic fields. Actually, books \cite{Greiner,Fradkin} and conferences \cite{Zavattini} have been devoted to this subject and then my review will be unavoidably incomplete. 

\begin{figure}[ht]
\begin{center}
\includegraphics[width=\textwidth]{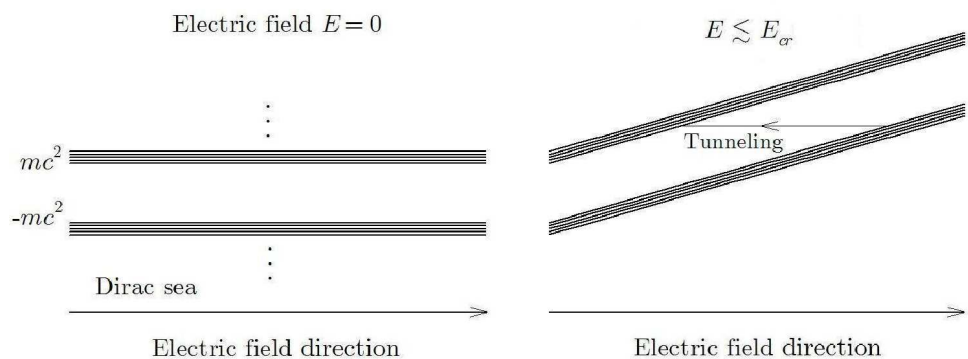}
\end{center}
\caption{Tunneling of an electron from a negative to a positive energy level in the presence of a strong electric field $E$.}
\label{Pair_prod}
\end{figure}

As I have said in the Introduction, the possibility that real pairs can be produced from vacuum was first investigated by Schwinger in \cite{Schwinger}. In particular, he discussed the pair production process in the presence of an external constant and uniform electric field. The calculations are rather difficult but the production mechanism, sometimes called \emph{Schwinger mechanism}, is qualitatively easy to be understood. In fact, it is known that virtual pairs are spontaneously produced and then annihilated as ``vacuum fluctuations''. The electron and the positron making a virtual pair ``live'' at a very short distance between each other of the order of the Compton length $\lambdabar= 3.9\times 10^{-11}\;\text{cm}$. Nevertheless, if the external electric field strength is larger than $E_{cr}=1.3\times 10^{16}\;\text{V/cm}$, then the electron and the positron forming the virtual pair can be separated becoming \emph{real} particles. In terms of the Dirac picture of the quantum vacuum, in the presence of an electric field $E$ with strength of order of $E_{cr}$ the electron levels are so distorted that an electron with negative energy can ``tunnel'' to the positive energy levels leaving a positively charged ``hole'' (see Fig. \ref{Pair_prod}).
The interpretation of the pair production mechanism in terms of the tunnel effect is confirmed by the \emph{typically nonperturbative} expression of the production probability. In fact, if the external field strength $E$ is much lower than $E_{cr}$, it results proportional to $\exp(-\pi E_{cr}/E)$ \cite{Schwinger}. The same previous qualitative interpretation but also rigorous calculations show that the production of pairs from vacuum is impossible in the presence of a pure, constant and uniform magnetic field. In fact, unlike the electric force, the Lorentz force can not do the work necessary to ``separate'' the electron and the positron forming a virtual pair and to make them real particles.

The extraordinarily high value of $E_{cr}$ makes impossible to test in a terrestrial laboratory the pair production Schwinger mechanism. A further difficulty is represented by the fact that in the Schwinger formalism the electric field is assumed to be constant and uniform while from an experimental point of view it is easier to produce strong alternating electric fields (lasers). A wide literature has been devoted to the theoretical study of pair production in the presence of uniform, alternating fields. The first work was that by Brezin and Itzykson \cite{Brezin}. In that paper it is found that if $\Omega_0$ is the electric field rotational frequency then the pair production is significantly different from zero if the electric field strength $E$ is such that $E> m\Omega_0 c/[e\sinh[\hbar\Omega_0/(4mc^2)]]$. Even using the X-ray lasers actually at our disposal, the effect is still too tiny to be observed. In fact, the peak value of the electric field strength at the focus of an X-ray laser necessary to measure some effects of pair creation is of the order of $10^{15}\;\text{V/cm}$, that is five orders of magnitude larger than those of the now available X-ray lasers \cite{Melissinos,Ringwald}.

Due to this unavoidable ``inadequacy'' of the experimental resources, various electromagnetic field configurations have been analyzed by theoretical physicists in order to find the most efficient one from the point of view of particle production. For example, in \cite{Narozhnyi,Dunne} the production of pairs in the presence of a uniform and time-dependent (but nonoscillating) electric field is considered. The authors use the \emph{effective Lagrangean method} [similar to that used by Schwinger in \cite{Schwinger}] to calculate approximatively the number of pairs per unit volume and unit time created from vacuum as twice the imaginary part of the effective Lagrangean density of the system \cite{Dittrich1,Dittrich2}. Instead, in \cite{Hounkonnou} the same technique is used to estimate numerically the production of electrons and positrons in the presence of electric and magnetic fields coupled in various configurations: an alternating electric field superimposed to a uniform magnetic field or an alternating electric field superimposed to an also alternating magnetic field. Finally, in another class of papers the possible multiple pair production in the electromagnetic field of two colliding heavy ions is studied \cite{Baur,Best} but till now the experimental evidence of this mechanism of particle production is inconclusive and controversial.

Finally, a different theoretical method has been used to predict, in general, particle production in the presence of external time-dependent fields: the \emph{Bogoliubov transformation}. In order to describe the physical content of the Bogoliubov transformation,  I assume, as usual in quantum field theory, to work in the Heisenberg picture. Now, when a time-dependent external field is present the second quantized number operators associated with a generic quantum field change nontrivially with time. It may happen that even if the \emph{time-independent} state of the quantum field system is the vacuum state of the number operators calculated at $t\to-\infty$ it is not the vacuum state of the number operators at $t\to\infty$.\footnote{It can be shown that these two classes of operators are connected by means of a linear transformation called \emph{Bogoliubov transformation}.} In this case, one just concludes that while in the far past no particles were present, they are there in the far future and then that the external time-dependent field induced their production directly from vacuum. As it is evident from this qualitative description, the use of the Bogoliubov transformation technique is not limited to the study of \emph{electron-positron} pairs production in the presence of time-dependent \emph{electromagnetic} fields. In fact, it has also been used, for example, in \cite{Parker2,Parker1,Zeldovich,Frieman,Parker3,Mak} to calculate the production rate of \emph{different kinds of particles} in the presence of time-dependent cosmological \emph{gravitational} fields.
%
%
\section{GRBs: a brief introduction}
\label{GRB}
In this Section I want to give a brief introduction concerning GRBs, their general features and so on. The literature about GRBs is really endless but the present introduction results mostly from my reading of some nonspecialistic papers \cite{Fishman,Musser,Gehrels,Lamb} and some review papers \cite{Piran_r,Meszaros_r,Piran,Waxman_r}. In the following I will not quote the previous references anymore but only other ones where I found specific information.
%
%
\subsection{General experimental characteristics of GRBs}
\label{GRB_char}
\begin{figure}[ht]
\begin{center}
\includegraphics[angle=-90,width=\textwidth]{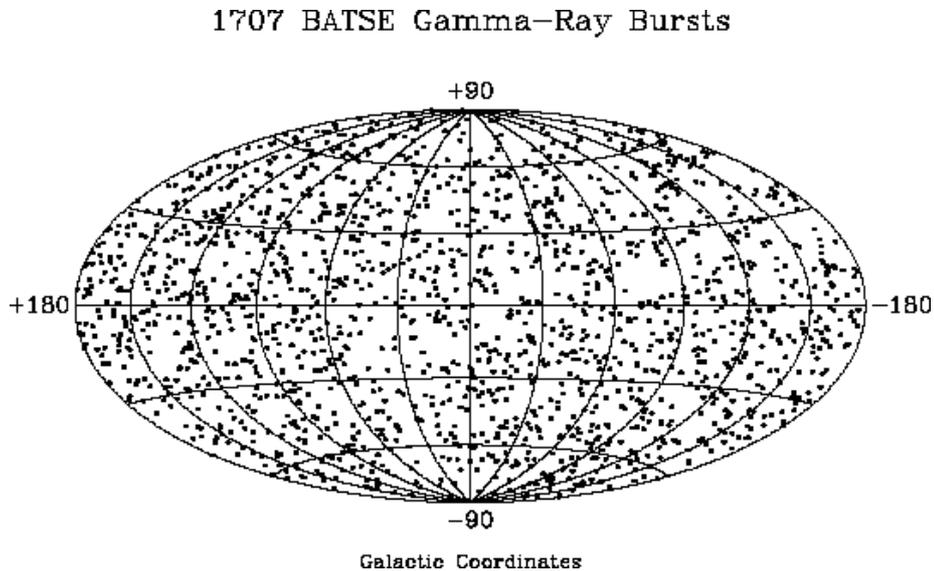}
\end{center}
\caption{Spatial distribution in the sky of $1707$ GRBs detected by BATSE.}
\label{GRB_distr}
\end{figure}

The first GRB was detected by chance on July 2, 1967 by a military satellite watching for nuclear tests in space. The experimental data concerning that burst were published much later and, actually, the scientific community started studying systematically GRBs only after the launch of the Compton Gamma-Ray Observatory (CGRO) and its detector BATSE (Burst and Transient Experiment). BATSE made a crucial contribution in establishing the \emph{distance scale} of GRBs. In fact, it detected GRBs from all directions in the sky with a completely isotropic distribution (see Fig. \ref{GRB_distr}) strongly supporting the idea that GRBs are originated at cosmological ($\sim 10^{28}\;\text{cm}$) distances from us.\footnote{By contrast, the isotropic distribution of GRBs rules out the possibility that, as it was firstly believed, GRBs are produced in our galaxy because it is not spherical and we do not occupy a privileged position in it.}
The cosmological origin of GRBs has tremendous consequences about the total energy carried by them. In fact, the detected GRBs fluences range typically from $10^{-7}\; \text{erg/cm$^2$}$ to $10^{-4}\;\text{erg/cm$^2$}$ \cite{Piran}, then, assuming an isotropic emission, the total energy carried by a GRB is of the order of $10^{50}\;\text{erg}\text{--}10^{53}\;\text{erg}$. For this reason, GRBs have been characterized as \emph{the brightest explosions in the Universe after Big Bang}. 

\begin{figure}[ht]
\begin{center}
\includegraphics[angle=90,width=\textwidth]{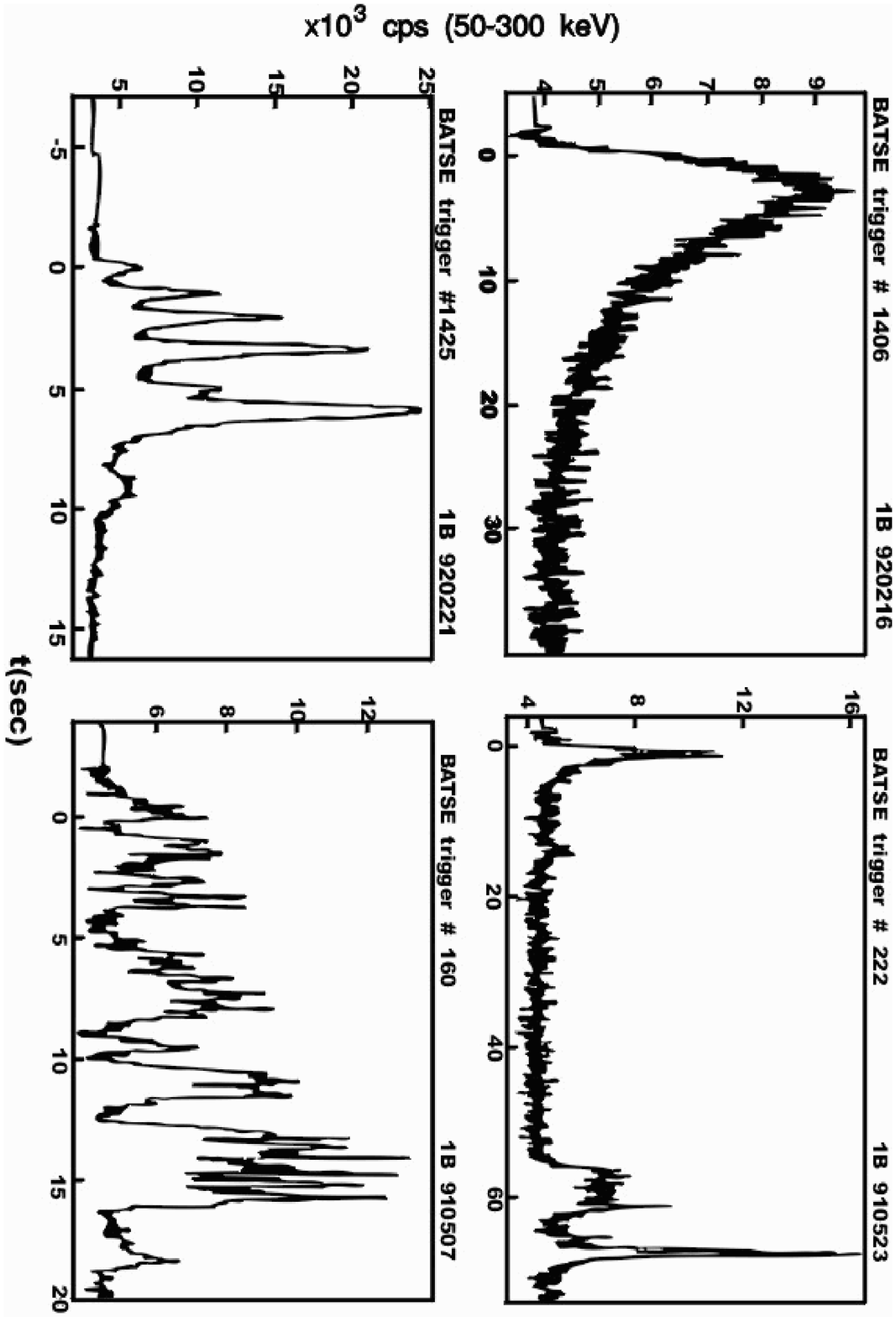}
\end{center}
\caption{Various GRBs time profiles.}
\label{GRB_t_p}
\end{figure}

Now, thousands of GRBs have been detected by high-energy astrophysics satellites and, as we will see in the next Paragraph, a feature that makes them hard to be understood is just their wide variety. For example, Fig. \ref{GRB_t_p} shows very different GRBs time profiles, \textit{i. e.} the number of photons detected per unit time as a function of time. 
In this respect, GRBs typically last from $0.01\;\text{s}$ to $100\;\text{s}$ and are generally classified into two large categories: \emph{short} bursts lasting less than $2\;\text{s}$ and \emph{long} bursts the others. We know very little about short bursts. In general, their energy spectra are harder than those of the long GRBs and they are supposed to be originated nearer with respect to long bursts \cite{Cline}. Instead, our knowledge about long bursts has been made deeper thanks to the discovery of their X-ray \emph{afterglows} by means of the Italian-Dutch satellite Beppo-SAX. Actually, the data from Beppo-SAX showed that long GRBs are followed by a multiwavelength ``afterglow'' made of X-ray photons, ultraviolet photons down to radio photons. On the one hand, the \emph{optical} observations confirmed the cosmological origin of GRBs. On the other hand, the presence of iron lines in the afterglows was the first indication of a connection, that has become stronger and stronger, between GRBs and supernovae explosions. In fact, it is known that during these explosions large amount of iron atoms are synthesized and ejected.

\begin{figure}[ht]
\includegraphics[angle=90,width=\textwidth]{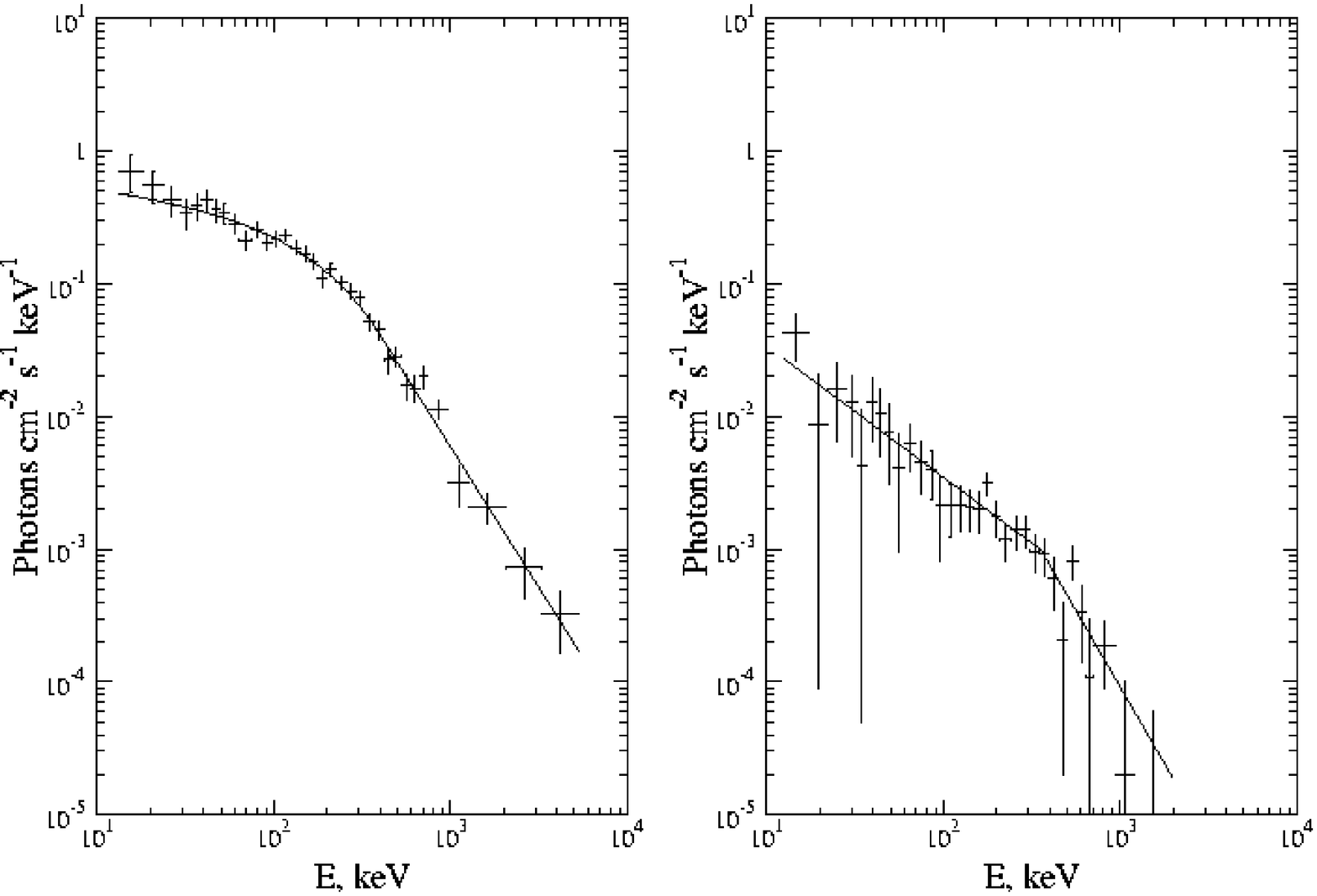}
\caption{Two typical GRBs energy spectra.}
\label{GRB_en_sp}
\end{figure}

I want to conclude this Paragraph by quoting three other general features of GRBs that give quite striking conditions mostly on the mechanism(s) responsible of their production: 
\begin{itemize}
\item the energy spectrum of GRBs is not thermal;
\item the time variability of the GRBs signals is observed down to time scales of less than $10\;\text{ms}$ (see Fig. \ref{GRB_t_p});
\item the prompt gamma-ray emission of many GRBs shows a high linear polarization degree \cite{Coburn}. 
\end{itemize}
Concerning the first point, Fig. \ref{GRB_en_sp} shows two typical GRBs energy spectra.
In general, called $n(\omega)$ the number of photons per unit area and unit time as a function of the photon energy $\omega$ (indicated as E in Fig. \ref{GRB_en_sp}), a typical experimental GRB energy spectrum is well fitted by the following piecewise function
\begin{equation}
\label{GRB_spectr_eq}
\frac{dn(\omega)}{d\omega}\propto
\begin{cases}
\omega^{-\alpha}\exp\left[-\dfrac{(\beta-\alpha)\omega}{\omega_b}\right] & \text{if $\omega \le \omega_b$}\\
\omega_b^{-\alpha}\left(\frac{\omega}{\omega_b}\right)^{-\beta}\exp\left[-(\beta-\alpha)\right] & \text{if $\omega>\omega_b$}
\end{cases}
\end{equation}
with $\alpha\simeq 1$ and $\beta\simeq 2\text{--}3$ and with $\omega_b$ a \emph{break} energy. Eq. (\ref{GRB_spectr_eq}) represents essentially a double power-law smoothly joined at $\omega=\omega_b$ by a decreasing exponential. The break energies $\omega_b$ typically lie between $0.1\;\text{MeV}$ and $0.3\;\text{MeV}$, even if bursts with $\omega_b>1\;\text{MeV}$ have been detected \cite{Preece,Briggs}. 
%
%
\subsection{The fireball model of GRBs}
\label{Fireball_model}
In this Paragraph I want to show how the previous experimental evidences led astrophysicists to work out the so-called \emph{fireball} model \cite{Cavallo,Paczynski_f} that is the most widely accepted model of \emph{long} GRBs.\footnote{Actually, due to the great variety of GRBs, many models of GRBs have been proposed during the years describing GRBs as jets from pulsars \cite{Blackman_p} or as bursts emitted by \emph{cannonballs} in turn originated during supernovae explosions \cite{Dado} etc....}

In the fireball model the photons forming a GRB are thought to be emitted by a ``fireball'', a plasma made essentially of electrons, positrons and photons, that expands high relativistically \cite{Paczynski}.\footnote{One of the strength points of the fireball model is its almost complete independence from the nature of the \emph{central engine} that produces the fireball (see Sect. \ref{Centr_eng}) and from the mechanism(s) giving rise to the fireball itself.} In general, a \emph{direct} emission of the photons forming a GRB would be inconsistent with the experimental evidence about the nonthermal spectrum of GRBs. In fact, as we have seen, during a burst $10^{50}\;\text{erg}\text{--}10^{53}\;\text{erg}$ are released on average in less than $100\;\text{s}$. Also, the fact that GRBs show in the soft-gamma region variabilities at time scales of the order of $1\;\text{ms}$ implies, by using the causality limit, that their source should have a linear length of the order of $1\;\text{ms}\times c=3.0\times 10^7\;\text{cm}$. In these extreme conditions, even if initially only photons were produced then their energy ($\lesssim 1\;\text{MeV}$) and their density would be high enough to create electron-positron pairs through the reaction $\gamma\gamma\to e^-e^+$. In turn, the strong electromagnetic interactions among the electrons, the positrons and the photons would allow these last to escape only after so much time that they would thermalize.

\begin{figure}[ht]
\begin{center}
\includegraphics[angle=-90,width=\textwidth]{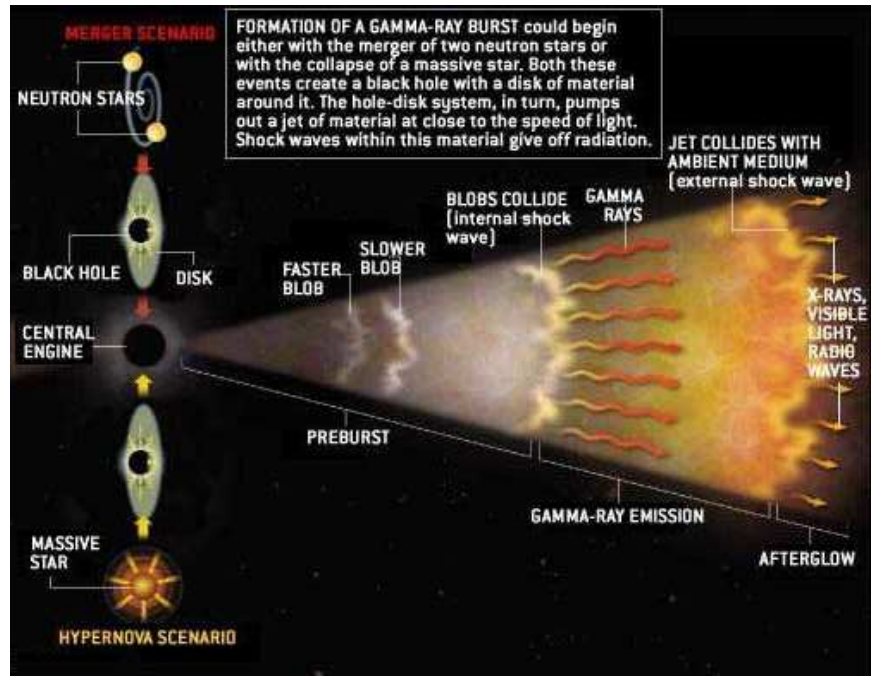}
\end{center}
\caption{Schematic representation of the mechanisms producing a long GRB and its afterglow in the framework of the fireball model. The left part of the Figure and the inset refer to two possible progenitors of the fireball that are discussed in Par. \ref{Centr_eng}.}
\label{fireball}
\end{figure}

As I have only mentioned at the beginning of this Paragraph, the fireball model is intended to describe only long GRBs. In fact, the feature that mostly has contributed to its success is just the prediction of the existence of the afterglow. Actually, the fireball model gives quite naturally a mechanism for producing the true burst and the afterglow (see Fig. \ref{fireball}). 
In general, in this model a GRB and its afterglow are produced at the expense of the kinetic energy of the fireball. In particular, the true burst is produced as a consequence of the \emph{internal} shocks among different layers of the fireball travelling at different speeds \cite{Rees_2}. Of course, although these shocks may last hours, the resulting burst can last only few seconds because the fireball expands high relativistically with a Lorentz factor also more than $100$. Also, the fact that the electrons and the positrons in the fireball are very energetic explains why during a burst photons can be emitted with energies of the order of $1\;\text{GeV}$ or more and why the time profile of a burst can also show structures at the millisecond scale. Instead, the afterglow of a GRB is produced as a consequence of the shocks between the \emph{external} layer of the fireball and the surrounding medium \cite{Mezsaros_3}. The two mechanisms producing the true burst and the afterglow are independent from each other but, in general, the internal shocks and the expansion slow down the fireball before it encounters the interstellar medium. In this way the photons forming the afterglow result, in general, to be less energetic than those forming the true burst and the afterglow itself lasts much longer than the true burst.\footnote{Nevertheless, some GRBs have been detected having an afterglow ``harder'' than the true burst.}

If we want to go a bit more deeply into the electromagnetic processes responsible of the production of the photons forming a GRB and its afterglow we have to take into account the indications suggested by the experimental data (see the previous Paragraph). Firstly, as I have just mentioned before, the photons forming a GRB can not be those present in the fireball because their energetic spectrum would be thermal. Moreover, all the experimental indications support the idea that the main mechanisms responsible of the production of the photons forming a GRB are the \emph{inverse Compton effect} and, mostly, the \emph{synchrotron emission}. The inverse Compton effect, that is the scattering between an energetic electron (positron) and a soft photon resulting in a less energetic electron (positron) and a harder photon, is important in explaining the high-energy (of the order of $\text{GeV}$) part of the photon spectrum of a GRB \cite{Mezsaros_4}. Instead, the photons forming the other less energetic parts of GRBs photon spectra are thought to be produced as synchrotron radiation by the shock-accelerated electrons and positrons forming the fireball. Firstly, this photon production mechanism explains the presence of the break energy $\omega_b$ in the energy spectra (see Fig. \ref{GRB_en_sp}) and the power-law behaviour below $\omega_b$ [see Eq. (\ref{GRB_spectr_eq})]. Most important, the recent discovery of the highly linear polarization degree [$\Pi=(80\pm 20)\;\%$] of the gamma-ray emission of the GRB detected on December, 6 2002 \cite{Coburn} strongly supports the idea that those photons are emitted through synchrotron radiation just because synchrotron radiation is theoretically known to be high linearly polarized \cite{Waxman,Granot}. Obviously, in order that synchrotron radiation can be emitted, a magnetic field must be present in the emission region. The high polarization degree itself suggests that the magnetic field in the emission region should be nearly uniform and strong enough to ``order'' the motion of the emitting particles in the fireball \cite{Coburn,Lyutikov}. This is in turn a clear indication that the true GRB is generated in regions where the magnetic field is that produced by a macroscopic object that, as we will see in the next Paragraph, is identified with the central engine that powers the fireball \cite{Zhang}. By contrast, the polarization degree of the afterglow photons has been measured to be of the order of $10\;\%$. This much lower value indicates that the afterglow is produced in regions where the magnetic field can also have a ``turbulent'' spatial distribution and can be produced by the shock-accelerated electrons and positrons themselves in the fireball \cite{Medvedev,Gruzinov,Greiner_J}.
%
%
\subsection{Central engine}
\label{Centr_eng}
As I have said before, the fireball model is quite independent from the nature of the source, the central engine that primes the formation of the fireball. Obviously, there is no direct evidence of what the central engine is and, for this reason, many candidates have been proposed \cite{Woosley}. In any case, the fact that during a GRB a so large amount of energy is ejected in a so relatively short time leads to believe that their sources should be systems storing very much energy and that a sudden, explosive event should be at the origin of the burst. Two kinds of astrophysical systems have been proposed as the possible progenitors powering GRBs (see Fig. \ref{fireball}): on the one hand, binary merging systems such as neutron star/neutron star systems or neutron star/black hole systems \cite{Goodman,Paczynski} and, on the other hand, very massive (more than $20$ solar masses) stars with a collapsing core, called \emph{failed supernovae} or \emph{collapsars} \cite{Woosley_2}.\footnote{The adjective ``failed'' refers to the fact that initially it was supposed that such supernovae explosions did not eject outside the envelope of the star.} The result of the merging in the first case and of the collapse in the second case is the same: a massive rotating black hole surrounded by an accretion disk.\footnote{A variant of the collapsar model has been proposed in \cite{Vietri}. In this model, called \emph{supranova} model, the supernova explosion gives rise to a supermassive, rapidly rotating neutron star that, in turn, collapses producing a black hole and the fireball. With respect to the collapsar model, in the supranova model the fireball is generated much later than the supernova explosion (approximatively a week later) in a ``cleaner'' environment poor of heavy baryons. In this way, the energy transfer to the GRB results much more efficient.} Clearly, the observation that many GRBs have been localized in star forming regions is an evidence against the first kind of progenitors. In fact, the merging of two compact objects comes even billions of years later than their formation, then the composite system has enough time to drift far from its birth location. Also, the fact that iron (but also silicon, sulfur and so on) lines have been detected in GRBs afterglows corroborates the hypothesis of the supernova explosion as the ``dramatic'' event producing the fireball because during supernovae explosions such kinds of atoms are synthesized and ejected into the interstellar medium. Finally, there are now also many \emph{direct} evidences that GRBs are detected where also a supernova is ``seen'' \cite{Galama,Bloom,Matheson,Kawabata}. 
It is worth noting that all the previous indications against the binary-merging model are obtained from afterglow measurements. For this reason, since no afterglows have been so far detected for short GRBs, the hypothesis that the binary-merging model can describe the formation of short GRBs has been carried out \cite{Katz,Popham} (since, from now on, I will not deal with short GRBs anymore but only with the long ones, I will omit the adjective ``long'' that, actually, will be understood).

It is clear that a massive rotating black hole surrounded by an accretion disk is a huge container of rotational and gravitational energy. But, how this energy can be ``extracted'' from the black hole-accretion disk system to form a fireball and then a GRB? Also concerning this subject, there are no direct observations and  many extraction mechanisms have been proposed. I quote, for example, the possibility that the electric fields around \emph{charged} black holes ``break'' the vacuum producing large amounts of pairs \cite{Ruffini}. Another mechanism proposed in the framework of the collapsar model involves the annihilation into electron-positron pairs of the numerous neutrino-antineutrino pairs that are produced during the supernova explosion but quantitative estimates show this mechanism is not efficient enough. Another popular mechanism that has been invoked to explain the formation of the fireball is the \emph{Blandford-Znajek mechanism} \cite{Gibbons,Lee_GRB,Barbiellini}. This is a very complicated mechanism \cite{Blandford} but it essentially works as a dynamo that transforms rotational energy into electromagnetic energy.\footnote{In the black hole-accretion disk system also a large amount of \emph{gravitational} binding energy is converted into electromagnetic energy by means of the Blandford-Znajek mechanism.} Now, as in a dynamo, a magnetic field is necessary in order that this transformation can happen. In addition, in the Blandford-Znajek mechanism a huge, overcritic ($\gg B_{cr}=4.4\times 10^{13}\;\text{gauss}$) magnetic field is needed. In fact, the energy extraction rate is proportional to the square of the magnetic field strength and theoretical calculations estimate that in order to extract $10^{53}\;\text{erg}$ in less than $1000\;\text{s}$, magnetic field strengths of the order of $10^{15}\;\text{gauss}$ are needed \cite{Lee_GRB}. Moreover, the possible presence of such strong magnetic field around massive rotating black holes surrounded by an accretion disk is also confirmed by numerical simulations [see \cite{Popham} and Ref. [9] in \cite{Lee_GRB}]. In this respect, I want to mention two other models proposed in \cite{Usov} and in \cite{Thompson_2,Hanami} where just the magnetic energy associated with such strong magnetic fields is ``directly'' transformed to power the fireball. In these models the central engines of GRBs are ultramagnetized rapidly rotating neutron stars called \emph{magnetars} that, in fact, are able to produce dipole magnetic fields up to $10^{15}\;\text{gauss}$ \cite{Duncan,Thompson,Kouveliotou}. In particular, in \cite{Usov} the magnetic energy release to the fireball is supposed to happen during the formation of the magnetar as a consequence of the collapse of a white dwarf, while in \cite{Thompson_2,Hanami} it is supposed to happen during the collapse of the magnetar into a black hole.

As a conclusion of this Paragraph and of the previous one, \emph{I want to stress the fundamental role that the presence of strong magnetic fields around the central engine has in the production of a GRB}:
\begin{itemize}
\item it is necessary to explain the high linear polarization degree of the GRBs gamma-ray spectrum;
\item it is invoked to account for the energy ``extraction'' from the central engine to power a GRB or, even, to account for the energy itself to power a GRB. 
\end{itemize}
I have stressed the importance and the role of such strong magnetic fields because, as we will see in the following Chapters, their existence around astrophysical compact objects represents one of the fundamental hypotheses of this work and its most important experimental counterpart. In fact, I will perform all the calculations by assuming to deal with overcritic magnetic fields and having in mind the astrophysical scenario I have just described.

\chapter{Theoretical tools}
\label{II}
In this Chapter I want to resume the theoretical background I needed to obtain the main results of the thesis. The subjects can be found in many textbooks and they are also, unavoidably, very different from each other. Nevertheless, the goal of the Chapter is to give the theoretical tools to derive the final results and, also, to fix the notation that will used in the rest of the thesis.

The Chapter is divided into three Sections. In the first one I quote the main results about the motion of a charged relativistic particle in the presence of a constant and uniform magnetic field: I treat both the case of a classical and of a quantum particle. In the second Section I discuss about the effects and the transitions induced on a quantum system by an adiabatic perturbation. Finally, the third Section is devoted to a brief introduction to quantum field theory in curved spacetime with particular attention to the general covariant formalism to deal with a spinor field in the presence of a classical background gravitational field.

\section{Motion of a charged relativistic particle in the presence of a constant and uniform magnetic field}
\label{Motion}
The problem of a charged relativistic particle in the presence of a constant and uniform magnetic field can be solved exactly both at a classical and at a quantum level. Since this physical system is discussed in many textbooks I limit myself to a quotation of the results I will use in the following Chapters. A rigorous derivation of these results can be found in \cite{Jackson,Cohen} for what it concerns the classical case and in \cite{Cohen,Bagrov} for what it concerns the quantum case. 

As usual, I choose the reference system, whose coordinates are indicated as $x$, $y$ and $z$, in such a way the magnetic field lies in the positive $z$ direction that is
\begin{equation}
\label{B_p}
\mathbf{B}'=
\begin{pmatrix}
0\\
0\\
B
\end{pmatrix}
\end{equation}
with $B>0$ (I use the primed notation for later convenience). For definiteness, the charged particle is assumed to be an electron with rest mass $m$ and electric charge $-e<0$. Finally, natural units with $\hbar=1$ and $c=1$ are used throughout.
%
%
\subsection{Classical mechanics}
\begin{figure}[ht]
\begin{center}
\includegraphics[width=8.5cm]{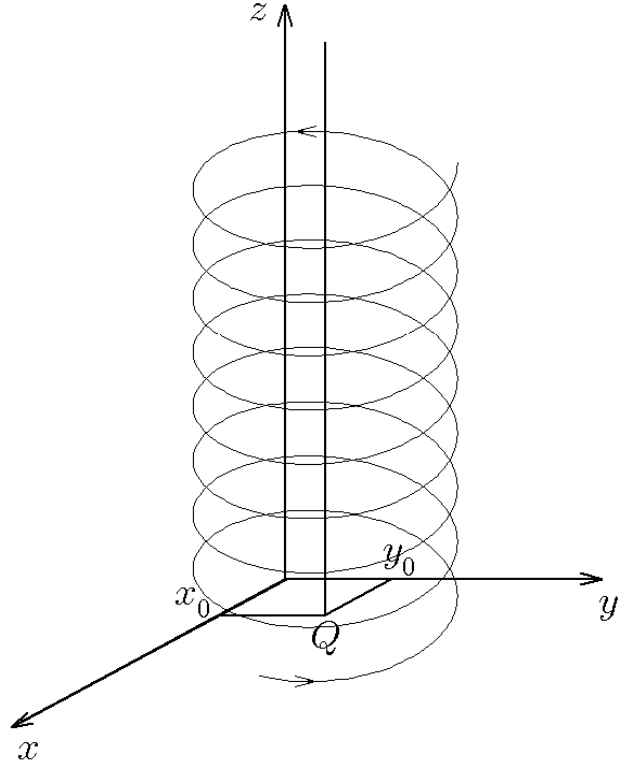}
\end{center}
\caption{Classical trajectory of an electron in the presence of a constant and uniform magnetic field directed along the positive $z$ axis.}
\label{class_mot}
\end{figure}

From a classical point of view, it is well known that the electron has a constant $z$ component $v_z$ of the velocity while it rotates uniformly and anticlockwise in the $x\text{--}y$ plane. The resulting trajectory is, in general, an helix with the axis parallel to the $z$ axis (see Fig. \ref{class_mot}). 
The axis of the helix intersects the $x\text{--}y$ plane at a point $Q$ whose nonzero coordinates have to be given as initial conditions and are indicated in Fig. \ref{class_mot} as $x_0$ and $y_0$. In this way the square distance $R^2_{xy}$ of the axis of the helix from the origin is given by $R^2_{xy}=x_0^2+y_0^2$ and it is a constant of motion. Since the Lorentz force does not do any work, the energy of the electron is also a constant of motion. Instead, the conservation of the $z$ component 
\begin{equation}
\label{ang_m}
l_z=xp_y-yp_x
\end{equation}
of the electron angular momentum depends on the gauge one chooses for the vector potential $\mathbf{A}'(\mathbf{r})$ corresponding to the magnetic field $\mathbf{B}'$ (the scalar potential can be assumed to vanish).\footnote{This is not surprising because $l_z$ is not in the present problem a true physical quantity.} In fact, if $v_x$ and $v_y$ are the $x$ and $y$ components of the electron velocity, the momenta $p_x$ and $p_y$ appearing in Eq. (\ref{ang_m}) are defined as
\begin{subequations}
\begin{align}
p_x &\equiv mv_x-eA'_x(\mathbf{r}),\\
p_y &\equiv mv_y-eA'_y(\mathbf{r})
\end{align}
\end{subequations}
and, in general, $l_z$ is not conserved. Moreover, it can easily be shown that if one chooses the so-called ``symmetric'' gauge in which
\begin{equation}
\label{A_p}
\mathbf{A}'(\mathbf{r})=-\frac{1}{2}(\mathbf{r}\times\mathbf{B}')=\frac{B}{2}
\begin{pmatrix}
-y\\
x\\
0
\end{pmatrix},
\end{equation}
then $l_z$ is also a constant of motion.
%
%
\subsection{Quantum mechanics}
\label{Motion_QM}
The quantum description of the motion of a relativistic electron in the presence of the magnetic field $\mathbf{B}'$ is very different from the classical one and it depends on the complete set of commuting observables one chooses in the Hilbert space of the system. Now, the Hamiltonian of the system is, obviously, a conserved quantity and, in the symmetric gauge, it is given by the Dirac Hamiltonian
\begin{equation}
\label{H_1p_p}
\mathcal{H}'=\boldsymbol{\alpha}\cdot [\boldsymbol{\mathcal{P}}+e\mathbf{A}'(\mathbf{r})]+\beta m
\end{equation}
where $\boldsymbol{\mathcal{P}}\equiv -i\boldsymbol{\partial}$ is the linear momentum vector operator and $\boldsymbol{\alpha}$ and $\beta$ are the $4\times 4$ Dirac matrices defined by the relations
\begin{subequations}
\label{ant_alph_bet}
\begin{align}
&\alpha_u^2=\beta^2=I && u=x,y,z,\\
&\{\alpha_u,\alpha_v\}=\{\alpha_u,\beta\}=0 && u,v=x,y,z \text{ and } u\neq v
\end{align}
\end{subequations}
where $I$ is the $4\times 4$ unit matrix and the braces indicate the anticommutator. In particular, I choose the matrices $\boldsymbol{\alpha}$ and $\beta$ in the Dirac representation
\begin{align}
\boldsymbol{\alpha}=
\begin{pmatrix}
0 & \boldsymbol{\sigma}^{(2)}\\
\boldsymbol{\sigma}^{(2)} & 0
\end{pmatrix}, && 
\beta=
\begin{pmatrix}
I^{(2)} & 0\\
0 & -I^{(2)}
\end{pmatrix}
\end{align}
where $I^{(2)}$ and $\boldsymbol{\sigma}^{(2)}$ with
\begin{align}
\sigma^{(2)}_x=
\begin{pmatrix}
0 & 1\\
1 & 0
\end{pmatrix}, &&
\sigma^{(2)}_y=
\begin{pmatrix}
0 & -i\\
i & 0
\end{pmatrix}, && 
\sigma^{(2)}_z=
\begin{pmatrix}
1 & 0\\
0 & -1
\end{pmatrix}
\end{align}
are the $2\times 2$ unit and Pauli matrices respectively.

We have seen before that in classical mechanics the $z$ component of the orbital angular momentum is a conserved quantity in the symmetric gauge. In quantum mechanics this in not true because the electron spin has to be taken into account. In fact, the $z$ component of the electron \emph{total} angular momentum defined as 
\begin{equation}
\label{J_z}
\mathcal{J}^{(1/2)}_z\equiv \mathcal{L}_z+\frac{\sigma_z}{2}\equiv x\mathcal{P}_y-y\mathcal{P}_x+\frac{\sigma_z}{2}
\end{equation}
with
\begin{equation}
\sigma_z=
\begin{pmatrix}
\sigma^{(2)}_z & 0\\
0 & \sigma^{(2)}_z
\end{pmatrix}
\end{equation}
is, actually, a constant of motion in the symmetric gauge.\footnote{In the future there will be no possibility of confusion between $2\times 2$ and $4\times 4$ matrices and I will omit the superscript ``$(2)$'' from the $2\times 2$ unit and Pauli matrices.} In order to check that $[\mathcal{H}',\mathcal{J}^{(1/2)}_z]=0$ one only needs the well-known commutator rules among the operators $\mathbf{r}$ and $\boldsymbol{\mathcal{P}}$:
\begin{subequations}
\label{comm_r_P}
\begin{align}
& [u,\mathcal{P}_v]=i\delta_{u,v} && u,v=x,y,z,\\
& [u,v]=[\mathcal{P}_u,\mathcal{P}_v]=0 && u,v=x,y,z
\end{align}
\end{subequations}
and the anticommutator rules (\ref{ant_alph_bet}).

By observing from Eqs. (\ref{H_1p_p}), (\ref{A_p}) and (\ref{J_z}) that both $\mathcal{H}'$ and $\mathcal{J}^{(1/2)}_z$ do not depend on the $z$ coordinate, one realizes that the commutators $[\mathcal{P}_z,\mathcal{H}']$ and $[\mathcal{P}_z,\mathcal{J}^{(1/2)}_z]$ also vanish. Now, since a relativistic electron has also an internal degree of freedom connected with its spin, the three operators $\mathcal{H}'$, $\mathcal{J}^{(1/2)}_z$ and $\mathcal{P}_z$ are not enough to build up a complete set of commuting observables and then to describe in a quantum complete way the motion of the electron. Instead, another operator has to be added and a good candidate is represented by the quantum operator corresponding to the quantity $R_{xy}^2=x_0^2+y_0^2$ that is, as we have seen, a constant of motion in classical mechanics. In fact, it can be shown that the operators corresponding to $x_0$ and $y_0$ (that will be indicated with the same symbols) can be written in terms of the fundamental dynamical operators $\mathbf{r}$ and $\boldsymbol{\mathcal{P}}$ as
\begin{subequations}
\label{x_0_y_0}
\begin{align}
\label{x_0}
x_0 &=\frac{x}{2}-\frac{\mathcal{P}_y}{eB},\\
\label{y_0}
y_0 &=\frac{y}{2}+\frac{\mathcal{P}_x}{eB}
\end{align}
\end{subequations}
and, by using these equations and the commutators (\ref{comm_r_P}), it is not difficult to show that the operator
\begin{equation}
\label{R_xy^2}
R_{xy}^2=\left(\frac{x}{2}-\frac{\mathcal{P}_y}{eB}\right)^2+
\left(\frac{y}{2}+\frac{\mathcal{P}_x}{eB}\right)^2
\end{equation}
commutes both with $\mathcal{H}'$ and with $\mathcal{J}^{(1/2)}_z$. Finally, since $R_{xy}^2$ does not depend on $z$ then $[\mathcal{P}_z,R_{xy}^2]=0$ and I can conclude that the operator set $\mathcal{S}'\equiv\{\mathcal{H}',\mathcal{J}^{(1/2)}_z,\mathcal{P}_z,R_{xy}^2\}$ is a complete set of commuting observables in the Hilbert space of the system ``relativistic electron in the presence of the magnetic field $\mathbf{B}'$''.

By concluding, I want to give here an alternative expression of the operator $y_0$ that will be useful in Chap. 5. In fact, from Eqs. (\ref{x_0_y_0}) and (\ref{comm_r_P}) it can easily be shown that
\begin{equation}
\label{comm_x_0}
[x_0,y_0]=\frac{i}{eB}
\end{equation}
in such a way in the representation in which $x_0$ is simply a multiplicative operator the operator $y_0$ can be written as
\begin{equation}
\label{y_0_der}
y_0=-\frac{i}{eB}\partial_{x_0}.
\end{equation}
%
%
%
\subsubsection{Electron and positron modes}
Once the complete set of commuting observables $\mathcal{S}'$ has been determined, the next task is to find the orthonormal basis of the Hilbert space built up by the common eigenstates or ``modes'' of $\mathcal{S}'$. It is well known that in doing that in the relativistic domain one has to face the problem of the appearance of the negative-energy modes. This problem is solved through the so-called \emph{second quantization} procedure that gives the possibility to interpret the pathological negative-energy modes as positive-energy antiparticle (positron) modes. The subject is well known and, for this reason, I will deal here directly with electron and positron modes without introducing the negative-energy modes, the charge-conjugation operator and so on. 

If the symbol ``$j$'' embodies all the needed quantum numbers, the electron and the positron modes can be indicated as $u'_j(\mathbf{r})$ and $v'_j(\mathbf{r})$ respectively and they are, from a mathematical point of view, fourdimensional spinors. Now, it can be shown that four quantum numbers $n_d$, $k$, $\sigma$ and $n_g$ have to be introduced and that they can assume the following values:
\begin{subequations}
\begin{align}
n_d,n_g &=0,1,\dots, \\
k &=\text{any real number},\\
\sigma &=\pm 1.
\end{align}
\end{subequations}
The physical meaning of these quantum numbers can be understood by looking at the eigenvalue equations that the spinors $u'_j(\mathbf{r})$ and $v'_j(\mathbf{r})$ satisfy:
\begin{subequations}
\label{eig_e}
\begin{align}
\label{eig_H_p_e}
\mathcal{H}'u'_j &=w_ju'_j,\\
\label{eig_P_z_e}
\mathcal{P}_zu'_j &=ku'_j,\\
\label{eig_J_z_e}
\mathcal{J}^{(1/2)}_zu'_j &=\left(n_d-n_g+\frac{\sigma}{2}\right)u'_j,\\
\label{eig_R_xy^2_e}
R_{xy}^2u'_j &=\frac{2n_g+1}{eB} u'_j
\end{align}
\end{subequations}
and
\begin{subequations}
\label{eig_p}
\begin{align}
\label{eig_H_p_p}
\mathcal{H}'v'_j &=-\tilde{w}_jv'_j,\\
\label{eig_P_z_p}
\mathcal{P}_z v'_j &=-kv'_j,\\
\label{eig_J_z_p}
\mathcal{J}^{(1/2)}_zv'_j &=-\left(n_d-n_g+\frac{\sigma}{2}\right)v'_j,\\
\label{eig_R_xy^2_p}
R_{xy}^2v'_j &=\frac{2n_d+1}{eB}v'_j.
\end{align}
\end{subequations}
In these equations I introduced the energies
\begin{subequations}
\label{L_l}
\begin{align}
\label{L_l_e}
w_j &=\sqrt{m^2+k^2+eB(2n_d+1+\sigma)},\\
\label{L_l_p}
\tilde{w}_j &=\sqrt{m^2+k^2+eB(2n_g+1-\sigma)}
\end{align}
\end{subequations}
that are called electron and positron \emph{Landau levels} respectively.\footnote{I mention the fact that even if I will work in the strong magnetic field regime in which $B/B_{cr}\gg 1$, it can be shown that the radiative corrections to the electron and positron Landau levels are logarithmic in the ratio $B/B_{cr}$ \cite{Tsai} and then that they can be safely neglected.} I point out that the electron (positron) Landau levels do not depend on $n_g$ ($n_d$).\footnote{The electron (positron) Landau levels have a further kind of degeneration, in fact, if $j_+=\{n_d,k,+1,n_g\}$ and $j_-=\{n_d+1,k,-1,n_g\}$ ($\tilde{j}_+=\{n_d,k,+1,n_g+1\}$ and $\tilde{j}_-=\{n_d,k,-1,n_g\}$) then $w_{j_+}=w_{j_-}$ ($\tilde{w}_{\tilde{j}_+}=\tilde{w}_{\tilde{j}_-}$). I will have to take into account this further degeneration in Chap. 5.} Now, from Eqs. (\ref{eig_P_z_e}), (\ref{eig_J_z_e}), (\ref{eig_P_z_p}) and (\ref{eig_J_z_p}) one realizes that $k$ ($-k$) and $\sigma$ ($-\sigma$) are the linear momentum and the polarization of the electron (positron) along the magnetic field or, in other words, the \emph{longitudinal} linear momentum and the \emph{longitudinal} polarization of the electron (positron). Instead, Eqs. (\ref{eig_J_z_e}), (\ref{eig_R_xy^2_e}), (\ref{eig_J_z_p}) and (\ref{eig_R_xy^2_p}) show that the quantum numbers $n_d$ and $n_g$ are connected with the \emph{transverse} motion of the electron and of the positron that is with the motion of the electron and of the positron in the plane perpendicular to $\mathbf{B}'$. It is worth giving here some details on how the quantum numbers $n_d$ and $n_g$ appear in the electron eigenvalue equations (\ref{eig_H_p_e}), (\ref{eig_J_z_e}) and (\ref{eig_R_xy^2_e}) [their appearance in the corresponding positron eigenvalue equations (\ref{eig_H_p_p}), (\ref{eig_J_z_p}) and (\ref{eig_R_xy^2_p}) can be understood in a completely analogous way]. The essential point is that in order to solve these equations one introduces the operators $a_d$, $a_d^{\dag}$ and $a_g$, $a_g^{\dag}$ defined as
\begin{subequations}
\label{a_a_dag}
\begin{align}
\label{a_d}
a_d &=\frac{1}{2}\left[\sqrt{\frac{eB}{2}}(x-iy)+\sqrt{\frac{2}{eB}}
\left(\mathcal{P}_y+i\mathcal{P}_x\right)\right],\\
\label{a_d_c}
a_d^{\dag} &=\frac{1}{2}\left[\sqrt{\frac{eB}{2}}(x+iy)+\sqrt{\frac{2}{eB}}
\left(\mathcal{P}_y-i\mathcal{P}_x\right)\right],\\
\label{a_g}
a_g &=\frac{1}{2}\left[\sqrt{\frac{eB}{2}}(x+iy)-\sqrt{\frac{2}{eB}}
\left(\mathcal{P}_y-i\mathcal{P}_x\right)\right],\\
\label{a_g_c}
a_g^{\dag} &=\frac{1}{2}\left[\sqrt{\frac{eB}{2}}(x-iy)-\sqrt{\frac{2}{eB}}
\left(\mathcal{P}_y+i\mathcal{P}_x\right)\right].
\end{align}
\end{subequations}
By means of these definitions the transverse position and momentum operators $x$, $y$, $\mathcal{P}_x$ and $\mathcal{P}_y$ can be written as
\begin{subequations}
\label{x_y_P_x_P_y}
\begin{align}
\label{x}
x &=\frac{1}{2}\sqrt{\frac{2}{eB}}(a_g+a_g^{\dag}+a_d+a_d^{\dag}),\\
\mathcal{P}_x &=\frac{1}{2i}\sqrt{\frac{eB}{2}}
(a_g-a_g^{\dag}+a_d-a_d^{\dag}),\\
\label{y}
y &=\frac{1}{2i}\sqrt{\frac{2}{eB}}(a_g-a_g^{\dag}-a_d+a_d^{\dag}),\\
\label{p_y}
\mathcal{P}_y &=-\frac{1}{2}\sqrt{\frac{eB}{2}}
(a_g+a_g^{\dag}-a_d-a_d^{\dag}).
\end{align}
\end{subequations}
By performing these substitutions on the operators $\mathcal{H}'$, $\mathcal{J}^{(1/2)}_z$ and $R_{xy}^2$, one sees that $\mathcal{H}'$, $\mathcal{J}^{(1/2)}_z$ and $R_{xy}^2$ themselves depend on $a_d$, $a_g$, $a^{\dag}_d$ and $a^{\dag}_g$ only through the quadratic operators $\mathcal{N}_d=a_d^{\dag}a_d$ and $\mathcal{N}_g=a_g^{\dag}a_g$. Now, starting from Eqs. (\ref{a_a_dag}) and from the commutators (\ref{comm_r_P}) it can be seen that
\begin{subequations}
\begin{align}
\label{comm_a}
[a_g,a_g^{\dag}] &=[a_d,a_d^{\dag}]=1,\\
[a_g,a_d] &=[a_g,a_d^{\dag}]=0
\end{align}
\end{subequations}
and then that $a_d$, $a_d^{\dag}$ and $a_g$, $a_g^{\dag}$ can be interpreted as two independent pairs of ladder operators. The corresponding operators $\mathcal{N}_d$ and $\mathcal{N}_g$ can be interpreted consequently as number operators and their eigenvalues are just the previously introduced nonnegative integers numbers $n_d$ and $n_g$.

I want to discuss now the structure of the electron and positron modes $u'_j(\mathbf{r})$ and $v'_j(\mathbf{r})$ and their orthonormalization relations. By solving step by step Eqs. (\ref{eig_e}) and (\ref{eig_p}) one finds that $u'_j(\mathbf{r})$ and $v'_j(\mathbf{r})$ are given by
\begin{subequations}
\label{u_p_v_p}
\begin{align}
\label{u_p}
u'_j(\mathbf{r}) &=\sqrt{\frac{w_j+m}{2w_j}}
\begin{pmatrix}
      \varphi'_j(\mathbf{r})\\ 
\dfrac{\mathcal{V}'}{w_j+m}\varphi'_j(\mathbf{r})
      \end{pmatrix}, \\
\label{v_p}
v'_j(\mathbf{r}) &=\sigma\sqrt{\frac{\tilde{w}_j+m}{2\tilde{w}_j}}
\begin{pmatrix}
      -\dfrac{\mathcal{V}'}{\tilde{w}_j+m}\chi'_j(\mathbf{r})\\
      \chi'_j(\mathbf{r})
      \end{pmatrix}.
\end{align}
\end{subequations}
In these expressions I introduced the operator $\mathcal{V}'$ and the two twodimensional
spinors $\varphi'_j(\mathbf{r})$ and $\chi'_j(\mathbf{r})$. On the one hand, the operator $\mathcal{V}'$ is defined as
\begin{equation}
\label{V_p_d}
\mathcal{V}'\equiv\boldsymbol{\sigma}\cdot
\left[\boldsymbol{\mathcal{P}}+e\mathbf{A}'(\mathbf{r})\right]=\boldsymbol{\sigma}\cdot
\left[\boldsymbol{\mathcal{P}}-\frac{e}{2}\left(\mathbf{r}\times\mathbf{B}'\right)\right]
\end{equation}
and, by substituting the expressions (\ref{x_y_P_x_P_y}), it can written as
\begin{equation}
\label{V_p}
\mathcal{V}'=i\sqrt{2eB}(a_d^{\dag}\sigma_--
a_d\sigma_+)+\sigma_z\mathcal{P}_z
\end{equation}
with $\sigma_{\pm}=(\sigma_x\pm i\sigma_y)/2$. On the other hand, the twodimensional spinors $\varphi'_j(\mathbf{r})$ and $\chi'_j(\mathbf{r})$ are given by
\begin{subequations}
\label{phi_p_chi_p}
\begin{align}
\label{phi_p}
\varphi'_j(\mathbf{r})
&=f'_{\sigma}\theta'_{n_d,n_g}(x,y)\frac{\exp(ikz)}{\sqrt{L_z}},\\
\label{chi_p}
\chi'_j(\mathbf{r})&=f'_{-\sigma}\theta'_{n_g,n_d}(x,y)\frac{\exp(-ikz)}{\sqrt{L_z}}
\end{align}
\end{subequations}
where
\begin{align}
f'_{+1}=\begin{pmatrix}
        1\\
        0
        \end{pmatrix}, &&
f'_{-1}=\begin{pmatrix}
        0\\
        1
        \end{pmatrix}
\end{align}
and where the scalar functions
\begin{equation}
\label{theta_p}
\begin{split}
\theta'_{l_1,l_2}(x,y)&=\sqrt{\frac{eB}{2\pi}\frac{1}{l_1!}\frac{1}{l_2!}}
(a_d^{\dag})^{l_1}(a_g^{\dag})^{l_2}\exp\left[-\frac{eB(x^2+y^2)}{4}\right]\\
&=\frac{(a_d^{\dag})^{l_1}}{\sqrt{l_1!}}
\sqrt{\frac{1}{\pi l_2!}\left(\frac{eB}{2}\right)^{l_2+1}}(x-iy)^{l_2}
\exp\left[-\frac{eB(x^2+y^2)}{4}\right]
\end{split}
\end{equation}
depend only on the coordinates $x$ and $y$ [in this equation the operators $a_d^{\dag}$ and $a_g^{\dag}$ are supposed to be expressed as in
Eqs. (\ref{a_d_c}) and (\ref{a_g_c})].

In Eqs. (\ref{phi_p_chi_p}) I also introduced the symbol $L_z$ which is the length of the quantization volume in the $z$ direction. In fact, in the course of the calculations it is easier to deal with normalizable electron and positron modes. To do this, I assume the whole space to be confined between the two planes $z=-L_z/2$ and $z=L_z/2$ and the modes to satisfy the periodic boundary conditions
\begin{subequations}
\label{bound_cond}
\begin{align}
u'_j(x,y,-L_z/2) &=u'_j(x,y,L_z/2),\\
v'_j(x,y,-L_z/2) &=v'_j(x,y,L_z/2).
\end{align}
\end{subequations}
In this way, the longitudinal momentum $k$ assumes only the discrete values\footnote{For notational simplicity, I do not indicate the dependence of $k$ on $\ell$.}
\begin{align}
\label{k_discr}
k=\pm\frac{2\pi\ell}{L_z} && \ell=0,1\dots
\end{align}
and only in the final results the continuum limit $L_z\to\infty$ is performed. By using the boundary conditions (\ref{bound_cond}) it can be shown that the twodimensional spinors $\varphi'_j(\mathbf{r})$ and $\chi'_j(\mathbf{r})$ satisfy the following orthonormalization relations
\begin{equation}
\label{ort_phi_chi_1}
\int d\mathbf{r}\varphi_j^{\prime\dag}(\mathbf{r})\varphi'_{j'}(\mathbf{r})=\int d\mathbf{r}\chi_j^{\prime\dag}(\mathbf{r})\chi'_{j'}(\mathbf{r})=\delta_{j,j'}
\end{equation}
where $\delta_{j,j'}\equiv\delta_{n_d,n'_d}\delta_{k,k'}\delta_{\sigma,\sigma'}
\delta_{n_g,n'_g}$ and where the integrals on $z$ are intended to be performed from $-L_z/2$ to $L_z/2$. Analogously, if we calculate the square of the operator $\mathcal{V}'$ as given in Eq. (\ref{V_p}) we have
\begin{equation}
\label{V_p_2}
\mathcal{V}^{\prime 2}=2eB\left[N_d(\sigma_-\sigma_++
\sigma_+\sigma_-)+\sigma_+\sigma_-\right]+\mathcal{P}_z^2=eB\left(2N_d+I+\sigma_z\right)+\mathcal{P}_z^2,
\end{equation}
and then, from the expressions (\ref{L_l}) of the electron and positron Landau levels, we obtain
\begin{subequations}
\label{ort_phi_chi_2}
\begin{align}
\label{ort_phi_2}
\int d\mathbf{r}\varphi_j^{\prime\dag}(\mathbf{r})
\frac{\mathcal{V}'}{w_j+m}\frac{\mathcal{V}'}
{w_{j'}+m}\varphi'_{j'}(\mathbf{r}) &=\frac{w_j-m}{w_j+m}\delta_{j,j'},\\
\label{ort_chi_2}
\int d\mathbf{r}\chi_j^{\prime\dag}(\mathbf{r})
\frac{\mathcal{V}'}{\tilde{w}_j+m}\frac{\mathcal{V}'}
{\tilde{w}_{j'}+m}\chi'_{j'}(\mathbf{r}) &=
\frac{\tilde{w}_j-m}{\tilde{w}_j+m}\delta_{j,j'}.
\end{align}
\end{subequations}
Finally, from these equations and from Eqs. (\ref{ort_phi_chi_1}) and (\ref{u_p_v_p}) the orthonormalization relations 
\begin{subequations}
\begin{align}
&\int d\mathbf{r}u^{\prime\dag}_j(\mathbf{r})u'_{j'}(\mathbf{r})=\int d\mathbf{r}v^{\prime\dag}_j(\mathbf{r})v'_{j'}(\mathbf{r})=\delta_{j,j'},\\
&\int d\mathbf{r}u^{\prime\dag}_j(\mathbf{r})v'_{j'}(\mathbf{r})=0
\end{align}
\end{subequations}
can be immediately found.
%
%
\subsubsection{Transverse ground states}
\label{TGSs}
By looking at the Landau levels (\ref{L_l}), one sees that there is a class of electron (positron) states characterized by the quantum numbers $n_d=0$ and $\sigma=-1$ ($n_g=0$ and $\sigma=+1$) whose energies do not depend on the magnetic field strength $B$. From a physical point of view, when the electron or the positron is in one of these states the energy associated with the interaction of the particle spin with the magnetic field compensates for the energy associated to the particle transverse rotational motion. These states will play a fundamental role in this work and they will be called \emph{Transverse Ground States} (TGSs) to distinguish them from the other ones that will be generically indicated as \emph{excited} Landau levels. In fact, as I have said at the end of the previous Chapter, I will deal mostly with pair production in the presence of strong magnetic fields such that $B/B_{cr}\gg 1$. In this regime a TGS characterized by a given value $k$ of the longitudinal linear momentum has an energy much smaller than that of the other excited Landau levels with the same $k$. In this respect, it is intuitively understandable that if there is no any other dynamical constraint the electrons and the positrons are more likely to be produced in TGSs than in other states. For this reason, it is worth giving here the explicit expression of the electron and positron TGSs. In order to simplify the notation, I will label them only by two indices $n$ and $k$ (since there is no possibility of confusion I omit the indices ``$d$'' and ``$g$'' on $n_d$ and $n_g$) that is [see Eqs. (\ref{u_p_v_p}) and (\ref{phi_p_chi_p})]
\begin{subequations}
\label{u_v_p_g}
\begin{align}
\label{u_p_g}
u'_{n,k}(\mathbf{r}) &\equiv u'_{0,k,-1,n}(\mathbf{r})=\sqrt{\frac{\varepsilon_k+m}{2\varepsilon_k}}
\begin{pmatrix}
0\\
1\\
0\\
-\dfrac{k}{\varepsilon_k+m}
\end{pmatrix}\theta'_n (x,y)\frac{\exp(ikz)}{\sqrt{L_z}},\\
\label{v_p_g}
v'_{n,k}(\mathbf{r}) &\equiv v'_{n,k,+1,0}(\mathbf{r})=\sqrt{\frac{\varepsilon_k+m}{2\varepsilon_k}}
\begin{pmatrix}
0\\
-\dfrac{k}{\varepsilon_k+m}\\
0\\
1
\end{pmatrix}\theta'_n (x,y)\frac{\exp(-ikz)}{\sqrt{L_z}}
\end{align}
\end{subequations}
where [see the second equality in Eq. (\ref{theta_p})]
\begin{equation}
\label{theta_p_g}
\theta'_n (x,y)\equiv\theta'_{n,0}(x,y)=\sqrt{\frac{1}{\pi n!}
\left(\frac{eB}{2}\right)^{n+1}}(x-iy)^n\exp\left[-\frac{eB}{4}(x^2+y^2)\right]
\end{equation}
and where
\begin{equation}
\label{w_g}
\varepsilon_k\equiv w_{0,k,-1,n}=\tilde{w}_{n,k,+1,0}=\sqrt{m^2+k^2}
\end{equation}
are the energies of the TGSs that have the same expression for the electrons and the positrons. Another feature of the TGSs that it is worth stressing is that \emph{they are all eigenstates of the operator $\sigma_z$ with the same eigenvalue $-1$}. I will show in the next Chapter how this feature will give the possibility to state some very important selection rules concerning the pair production process from vacuum in the presence of a strong, uniform and slowly-varying magnetic field.
%
%
\subsubsection{Rotated magnetic field}
All the previous results and equations are valid when the magnetic field is given by Eq. (\ref{B_p}) that is when it lies on the positive $z$ direction. Actually, I will also deal with a magnetic field $\mathbf{B}$ that, more generally, lies on the $y\text{--}z$ plane that is
\begin{equation}
\label{B}
\mathbf{B}=
\begin{pmatrix}
0\\
B_y\\
B_z
\end{pmatrix}=B\begin{pmatrix}
0\\
\sin\vartheta\\
\cos\vartheta
\end{pmatrix}
\end{equation}
with
\begin{subequations}
\begin{align}
\label{B_mod}
B &\equiv\sqrt{B_y^2+B_z^2},\\
\label{theta}
\tan\vartheta &\equiv\frac{B_y}{B_z}.
\end{align}
\end{subequations}
The magnetic fields $\mathbf{B}$ and $\mathbf{B}'$ have been assumed to have the same strength in order to exploit the results obtained in the previous Paragraphs. In fact, the magnetic field $\mathbf{B}$ can be obtained from $\mathbf{B}'$ by rotating the sources of $\mathbf{B}'$ clockwise around the $x$ axis by an angle $\vartheta$. For this reason, by introducing the rotation unitary operator
\begin{equation}
\mathcal{R}^{(1/2)}_x(\vartheta)=\exp(-i\vartheta \mathcal{J}^{(1/2)}_x)
\end{equation}
with $\mathcal{J}^{(1/2)}_x=y\mathcal{P}_z-z\mathcal{P}_y+\sigma_x/2$ the $x$ component of the electron total angular momentum operator, some assertions can be immediately stated:
\begin{itemize}
\item a complete set of commuting observables $\mathcal{S}$ in the Hilbert space of the system ``relativistic electron in the presence of the magnetic field $\mathbf{B}$'' is built up by 
\begin{enumerate}
\item the Hamiltonian
\begin{equation}
\label{H_1p}
\mathcal{H}=\mathcal{R}^{(1/2)\dag}_x(\vartheta)\mathcal{H}'\mathcal{R}^{(1/2)}_x(\vartheta)=
\boldsymbol{\alpha}\cdot \left[\boldsymbol{\mathcal{P}}+e\mathbf{A}(\mathbf{r})\right]+\beta m
\end{equation}
with
\begin{equation}
\label{A}
\mathbf{A}(\mathbf{r})=-\frac{1}{2}(\mathbf{r}\times\mathbf{B})
\end{equation}
the vector potential in the symmetric gauge corresponding to the magnetic field $\mathbf{B}$,
\item the longitudinal (with respect to $\mathbf{B}$) linear momentum
\begin{equation}
\mathcal{P}_{\parallel}=\mathcal{R}^{(1/2)\dag}_x(\vartheta)\mathcal{P}_z\mathcal{R}^{(1/2)}_x(\vartheta),
\end{equation}
\item the longitudinal total angular momentum
\begin{equation}
\mathcal{J}^{(1/2)}_{\parallel}=\mathcal{R}^{(1/2)\dag}_x(\vartheta)\mathcal{J}^{(1/2)}_z\mathcal{R}^{(1/2)}_x(\vartheta),
\end{equation}
\item the square transverse (with respect to $\mathbf{B}$) distance
\begin{equation}
\label{R_perp^2}
R_{\perp}^2=\mathcal{R}^{(1/2)\dag}_x(\vartheta)R_{xy}^2\mathcal{R}^{(1/2)}_x(\vartheta);
\end{equation}
\end{enumerate}
\item the spinors
\begin{subequations}
\label{u_v}
\begin{align}
\label{u}
u_j(\mathbf{r}) &=\mathcal{R}^{(1/2)\dag}_x(\vartheta)u'_j(\mathbf{r}),\\
\label{v}
v_j(\mathbf{r}) &=\mathcal{R}^{(1/2)\dag}_x(\vartheta)v'_j(\mathbf{r})
\end{align}
\end{subequations}
build up the common eigenstates of the operators in $\mathcal{S}$ and are an orthonormal basis of the Hilbert space of the system;
\item the spinors $u_j(\mathbf{r})$ and $v_j(\mathbf{r})$ satisfy the eigenvalue equations
\begin{subequations}
\begin{align}
\mathcal{H}u_j &=w_ju_j,\\
\mathcal{P}_{\parallel}u_j &= ku_j,\\
\mathcal{J}^{(1/2)}_{\parallel}u_j &=\left(n_d-n_g+\frac{\sigma}{2}\right)u_j,\\
R_{\perp}^2u_j &=\frac{2n_g+1}{eB}u_j,
\end{align}
\end{subequations}
and
\begin{subequations}
\begin{align}
\mathcal{H}v_j &=-\tilde{w}_jv_j,\\
\mathcal{P}_{\parallel}u_j &=-kv_j,\\
\mathcal{J}^{(1/2)}_{\parallel}v_j &=-\left(n_d-n_g+\frac{\sigma}{2}\right)v_j,\\
R_{\perp}^2v_j &=\frac{2n_d+1}{eB}v_j,
\end{align}
\end{subequations}
and the orthonormalization relations
\begin{subequations}
\label{ort_uv}
\begin{align}
\label{ort_uv_1}
&\int d\mathbf{r}u^{\dag}_j(\mathbf{r})u_{j'}(\mathbf{r})=\int d\mathbf{r}v^{\dag}_j(\mathbf{r})v_{j'}(\mathbf{r})=\delta_{j,j'},\\
\label{ort_uv_2}
&\int d\mathbf{r}u^{\dag}_j(\mathbf{r})v_{j'}(\mathbf{r})=0.
\end{align}
\end{subequations}
\end{itemize}
%
%
\section{Adiabatic perturbation theory}
\label{Ad_pert_th}
The adiabatic perturbation theory concerns the effects induced on a physical system by an external perturbation whose typical time evolution is much larger than the typical free time evolution of the system itself. I am interested here in adiabatic perturbations acting on a quantum system whose time evolution is described by the Schroedinger equation \cite{Bates,Migdal}. I also suppose that 
\begin{enumerate}
\item the system under study is characterized by a Hamiltonian $H(\xi(t))$ that depends on time through a parameter $\xi(t)$ representing the external perturbation;
\item for any fixed value $\xi$ the eigenvalue equation
\begin{equation}
\label{eig}
H(\xi)|n,\xi\rangle=\epsilon_n(\xi)|n,\xi\rangle
\end{equation}
can be solved exactly and all the resulting quantum numbers embodied in the symbol $n$ are discrete. 
\end{enumerate}
Now, the evolution of the system when $\xi(t)$ varies with time is determined provided the Schroedinger equation
\begin{equation}
\label{Sch}
i\frac{d|t\rangle}{dt}=H(\xi(t))|t\rangle
\end{equation}
is solved. Since, at any time $t$ the states $|n,\xi(t)\rangle$ build up an orthonormal basis of the Hilbert space of the system, I can write the state $|t\rangle$ in the form
\begin{equation}
\label{t}
|t\rangle=\sum_na_n(t)\exp\left[-i\int_{t_0}^tdt'\epsilon_n(\xi(t'))\right]|n,\xi(t)\rangle
\end{equation}
where it has been assumed that the perturbation starts changing with time at $t_0$ and where the complex coefficients $a_n(t)$ have to be determined. By substituting Eq. (\ref{t}) in Eq. (\ref{Sch}) and by projecting the resulting equation on the state $|m,\xi(t)\rangle$, I obtain the following differential equation for the coefficient $a_m(t)$:
\begin{equation}
\label{da_m_dt}
\frac{da_m}{dt}+\sum_n\langle m, \xi(t)|\partial_{\xi}|n,\xi(t)\rangle\dot{\xi}(t)a_n(t)\exp\left[i\int_{t_0}^tdt'\Delta\epsilon_{mn}(\xi(t'))\right]=0,
\end{equation}
where
\begin{equation}
\Delta\epsilon_{mn}(\xi(t))\equiv\epsilon_m(\xi(t))-\epsilon_n(\xi(t)).
\end{equation}

Obviously, Eq. (\ref{da_m_dt}) is equivalent to the Schroedinger equation (\ref{Sch}). Nevertheless, since the parameter $\xi(t)$ changes slowly with time, its time derivative $\dot{\xi}(t)$ is a small quantity (in a sense that will be specified at the end of this Section) and then a perturbative solution of Eq. (\ref{da_m_dt}) in powers of $\dot{\xi}(t)$ can be built up. Up to zero order in $\dot{\xi}(t)$ one simply obtains $a_m^{(0)}(t)=a_m(t_0)$. Also, by assuming that the system is at time $t_0$ in the state labeled by the index $i$ and that $a_i(t_0)=1$, then $a_m^{(0)}(t)=\delta_{m,i}$ and the first-order coefficients $a^{(1)}_m(t)$ are given by
\begin{equation}
\label{a_m_1}
a^{(1)}_m(t)=-\int_{t_0}^tdt'\langle m, \xi(t')|\partial_{\xi}|i,\xi(t')\rangle\dot{\xi}(t')\exp\left[i\int_{t_0}^{t'}dt''\Delta\epsilon_{mi}(\xi(t''))\right].
\end{equation}
This expression can be put in a more useful form. In fact, by deriving Eq. (\ref{eig}) with respect to the parameter $\xi$ and by projecting on $|m,\xi\rangle$, one obtains
\begin{equation}
\langle m,\xi|[\partial_{\xi}H(\xi)]|n,\xi\rangle=[\partial_{\xi}\epsilon_n(\xi)]\delta_{n,m}-\Delta\epsilon_{mn}(\xi)\langle m,\xi|\partial_{\xi}|n,\xi\rangle.
\end{equation}
By assuming that the eigenstates $|n,\xi\rangle$ are nondegenerate the previous equation is equivalent to the following ones:
\begin{subequations}
\begin{align}
\langle n,\xi|\partial_{\xi}|n,\xi\rangle &=\frac{1}{2}\partial_{\xi}\langle n,\xi|n,\xi\rangle=0,\\
\label{m_d_n}
\langle m,\xi|\partial_{\xi}|n,\xi\rangle &=-\frac{\langle m,\xi|[\partial_{\xi} H(\xi)]|n,\xi\rangle}{\Delta\epsilon_{mn}(\xi)} && \text{if $m\neq n$}.
\end{align}
\end{subequations}
I point out that in writing the first equation one tacitly exploits the following facts:
\begin{enumerate}
\item being the states $|n,\xi\rangle$ nondegenerate, they can be chosen to be real; 
\item being the indices $n$ discrete, the states $|n,\xi\rangle$ are normalizable.
\end{enumerate}
In conclusion, by substituting the previous equations in Eq. (\ref{a_m_1}) one obtains
\begin{subequations}
\begin{align}
\label{a_i_1}
a^{(1)}_i(t) &=0,\\
\label{a_m_1_f}
a^{(1)}_m(t)&=\int_{t_0}^tdt'\frac{\dot{H}_{mi}(t')}{\Delta\epsilon_{mi}(\xi(t'))}\exp\left[i\int_{t_0}^{t'}dt''\Delta\epsilon_{mi}(\xi(t''))\right] && \text{if $m\ne i$}
\end{align}
\end{subequations}
with
\begin{equation}
\label{dot_H_mi}
\dot{H}_{mi}(t')\equiv\langle m,\xi(t')|\dot{H}(\xi(t'))|i,\xi(t')\rangle=\langle m,\xi(t')|[\partial_{\xi}H(\xi(t'))]|i,\xi(t')\rangle\dot{\xi}(t').
\end{equation}
In particular, Eq. (\ref{a_i_1}) implies that the coefficient $a_i(t)$ is up to first order in $\dot{\xi}(t)$ equal to one. In other words, in the previous approximations the depletion of the initial state is at least a second-order effect in $\dot{\xi}(t)$.

Finally, it must be pointed out that the first-order perturbative approximation can be safely used only if the variation of the coefficients $a^{(1)}_m(t)$ with $m\neq i$ is much smaller than one in the typical time given by the inverse of the Bohr frequency $\Delta\epsilon_{mi}(\xi(t))$, that is only if
\begin{equation}
\label{ad_cond}
\left\vert\frac{\dot{H}_{mi}(t)}{[\Delta\epsilon_{mi}(\xi(t))]^2}\right\vert\ll 1
\end{equation}
at any $t\ge t_0$ and for any $m\neq i$.
%
%
\section{Quantum field theory in curved spacetime}
\label{QFTCS}
It is well known that, despite many efforts, a completely satisfying quantum theory of the gravitational field is still missing. Consequently, the interaction of a matter field and of the gravitational field has not been described in a completely quantum way. A less ambitious attempt to take into account the effects of the gravitational field on a matter field consists in \cite{Birrell,Fulling,Wald} 
\begin{enumerate}
\item treating classically the gravitational field itself as a modification of the metric properties of the spacetime;
\item quantizing the matter field in the resulting \emph{curved} spacetime.
\end{enumerate} 
I do not want to go into the conceptual difficulties that \emph{quantum field theory in curved spacetime} involves and the exact physical conditions that have to be satisfied in order that it can be safely applied [see in particular \cite{Birrell}]. In general, it can be said that if the typical lengths involved in the problem under study are much larger than the Planck scale $L_P\equiv\sqrt{G}=1.6\times 10^{-33}\;\text{cm}$ with $G$ the gravitational constant, then the quantum effects of the gravitational field can be neglected and it can be treated classically. In this dynamical regime the gravitational field is described mathematically by means of the metric tensor $g_{\mu\nu}(\mathsf{x})$ of the spacetime where $\mathsf{x}=(x^0,\ldots,x^3)$ characterizes a generic point of the spacetime itself. When no gravitational field is present the spacetime is the Minkowski spacetime and, by definition, it is possible to choose the coordinates in such a way the metric tensor has the simple structure $\eta_{\alpha\beta}=\text{diag}(1,-1,-1,-1)$.\footnote{For future notational convenience, the tensor indices of quantities referring to Minkowski spacetime are indicated with the first Greek letters $\alpha$, $\beta$, $\gamma$ or $\delta$, while those of quantities referring to a generic curved spacetime are indicated with the Greek letters in the middle of the Greek alphabet such as $\lambda$, $\mu$, $\nu$ or $\rho$.} The form itself of the metric tensor $\eta_{\alpha\beta}$ suggests a natural way to distinguish a time coordinate from the other three space coordinates. For this reason I will indicate the four coordinates of a generic \emph{event} in Minkowski spacetime by using the noncovariant notation $(t,\mathbf{r})$.

Now, in this work I will deal essentially with electrons and positrons then I will be interested only in spinor matter fields. The treatment of a spinor field in general relativity is more complicated than that of a scalar field or, in general, of a tensor field and it requires a detailed analysis which is independent of the quantization procedure. For this reason in the next Paragraph I will discuss the ``pedagogical'' case of the quantization of a free, real scalar field in curved spacetime and then in Par. \ref{II_III_II} I will outline the formalism to describe a spinor field in a general covariant way.
%
%
\subsection{Quantization of a free, real scalar field in curved spacetime}
The study of quantum field theory in curved spacetime gives the possibility to understand that the quantization of a matter field in Minkowski spacetime and the consequent interpretation in terms of identical particles can be carried out in a coherent way only because of the particular metric features of this spacetime. Actually, in a completely generic spacetime it is impossible to characterize what a quantum particle is and how to detect it \cite{Birrell}. In the present Paragraph I firstly review the main steps one follows in quantizing a free, real scalar field in Minkowski spacetime and then I show which features a curved spacetime has to share in order that the same real scalar field can also be quantized coherently in it. In this respect, I observe that a scalar field in curved spacetime is intended as a \emph{general scalar} that is a scalar quantity under \emph{general} coordinate transformations, while a scalar field in Minkowski spacetime is intended as a \emph{Lorentz scalar} that is a scalar quantity under \emph{Lorentz} transformations. Nevertheless, with an abuse of notation I will indicate both these fields and their related quantities with the same symbol. 

If I call $\phi(t,\mathbf{r})$ a free, real and Lorentz scalar field, then its Lagrangean density $\mathscr{L}$ is given by
\begin{equation}
\label{L_sc}
\mathscr{L}=\eta^{\alpha\beta}(\partial_{\alpha}\phi)(\partial_{\beta}\phi)-m_0^2\phi^2
\end{equation}
where, after the quantization of the field itself, $m_0$ will be interpreted as the mass of the resulting identical particles. The Lagrangean density $\mathscr{L}$ is assumed to be a Lorentz scalar in such a way the action
\begin{equation}
S=\int dtd\mathbf{r}\mathscr{L}
\end{equation}
is also a Lorentz scalar and the equation of motion of the field
\begin{equation}
\label{eq_mot_sc}
\left(\partial_t^2-\boldsymbol{\partial}^2+m_0^2\right)\phi=0,
\end{equation}
obtained from the stationary condition $\delta S=0$, is covariant. The next step in the quantization procedure consists in finding the general solution of Eq. (\ref{eq_mot_sc}) as a sum of \emph{normal modes} that build up, in fact, an orthonormal basis of the space of the solutions of Eq. (\ref{eq_mot_sc}) itself. As it follows from the general theory of partial differential equations, the normal modes can be unambiguously determined only after imposing some boundary conditions at a fixed time $t=t_0$. If one chooses to impose \emph{periodic} boundary conditions in a finite cubic volume $V=L^3$, the modes $\phi_{\mathbf{k}}(t,\mathbf{r})$ are characterized by the three discrete momenta $\mathbf{k}=\pm 2\boldsymbol{\ell}\pi/L$ with $\boldsymbol{\ell}$ a vector of natural numbers. In particular they can be written as
\begin{equation}
\phi_{\mathbf{k}}(t,\mathbf{r})=\frac{\exp[-i(\omega_kt-\mathbf{k}\cdot\mathbf{r})]}{\sqrt{2V\omega_k}}
\end{equation}
with $\omega_k=\sqrt{m_0^2+k^2}$ and they are orthonormal in the sense that they satisfy the relations
\begin{equation}
\label{orth_norm_phi}
(\phi_{\mathbf{k}},\phi_{\mathbf{k}'})=\delta_{\mathbf{k},\mathbf{k}'}
\end{equation}
with the scalar product between two scalar functions $\phi_1(t,\mathbf{r})$ and $\phi_2(t,\mathbf{r})$ defined as
\begin{equation}
\label{sc_prod_sc}
(\phi_1,\phi_2)\equiv-i\int_Vd\mathbf{r}[\phi_1(t,\mathbf{r})\partial_t \phi_2^*(t,\mathbf{r})-\phi_2^*(t,\mathbf{r})\partial_t \phi_1(t,\mathbf{r})].
\end{equation}
By using the modes $\phi_{\mathbf{k}}(t,\mathbf{r})$, the general solution of Eq. (\ref{eq_mot_sc}) can be written as
\begin{equation}
\label{phi_exp}
\phi(t,\mathbf{r})=\sum_{\mathbf{k}}\left[a_{\mathbf{k}}\phi_{\mathbf{k}}(t,\mathbf{r})+a^*_{\mathbf{k}}\phi^*_{\mathbf{k}}(t,\mathbf{r})\right]
\end{equation}
and the quantization of the field consists in transforming the complex numbers $a_{\mathbf{k}}$ into operators and in imposing the quantization rules
\begin{subequations}
\label{comm_a_a_d}
\begin{align}
[a_{\mathbf{k}},a^{\dag}_{\mathbf{k}'}] &=\delta_{\mathbf{k},\mathbf{k}'},\\
[a_{\mathbf{k}},a_{\mathbf{k}'}] &=0.
\end{align}
\end{subequations}
Finally, the particle interpretation of the quantized field is achieved by introducing the total linear momentum $\mathbf{P}$ of the field:
\begin{equation}
\label{P_tot_M}
\mathbf{P}\equiv -\int_Vd\mathbf{r}\pi\boldsymbol{\partial}\phi
\end{equation}
and the total Hamiltonian $H$ of the field:
\begin{equation}
\label{H_tot_M}
H\equiv\int_Vd\mathbf{r}(\pi\partial_t\phi-\mathscr{L})
\end{equation}
where
\begin{equation}
\pi(t,\mathbf{r})\equiv\frac{\partial \mathscr{L}}{\partial(\partial_t\phi)}=\partial_t\phi(t,\mathbf{r})
\end{equation}
is the momentum field conjugated to $\phi(t,\mathbf{r})$. In fact, by substituting the expansion (\ref{phi_exp}) in Eqs. (\ref{P_tot_M}) and (\ref{H_tot_M}) and by exploiting the orthonormalization relations (\ref{orth_norm_phi}) and the quantization rules (\ref{comm_a_a_d}), one easily finds that
\begin{subequations}
\begin{align}
\mathbf{P} &=\sum_{\mathbf{k}}\mathbf{k}(a_{\mathbf{k}}a^{\dag}_{\mathbf{k}}+a^{\dag}_{\mathbf{k}}a_{\mathbf{k}})=\sum_{\mathbf{k}}\mathbf{k}N_{\mathbf{k}},\\
H &=\frac{1}{2}\sum_{\mathbf{k}}\omega_k(a_{\mathbf{k}}a^{\dag}_{\mathbf{k}}+a^{\dag}_{\mathbf{k}}a_{\mathbf{k}})=\sum_{\mathbf{k}}\omega_kN_{\mathbf{k}}+W_0
\end{align}
\end{subequations}
with $N_{\mathbf{k}}\equiv a^{\dag}_{\mathbf{k}}a_{\mathbf{k}}$ and with $W_0\equiv 1/2\sum_{\mathbf{k}}\omega_k$ an \emph{infinite} constant that will be discussed below. Now, by using again Eqs. (\ref{comm_a_a_d}) one sees that the operators $N_{\mathbf{k}}$ commute among them and that they have nonnegative integer eigenvalues $n_{\mathbf{k}}$. Their generic common eigenstate $|n_{\mathbf{k}}\rangle$ is such that
\begin{subequations}
\begin{align}
\mathbf{P}|n_{\mathbf{k}}\rangle &=\left(\sum_{\mathbf{k}}\mathbf{k}n_{\mathbf{k}}\right)|n_{\mathbf{k}}\rangle,\\
\label{E_f_g}
H|n_{\mathbf{k}}\rangle &=\left(\sum_{\mathbf{k}}\omega_kn_{\mathbf{k}}+W_0\right)|n_{\mathbf{k}}\rangle
\end{align}
\end{subequations}
and it can be interpreted as \emph{the state in which $n_{\mathbf{k}}$ relativistic particles with mass $m_0$, i. e. particles with fourmomentum $(\omega_k,\mathbf{k})=(\sqrt{m_0^2+k^2},\mathbf{k})$ are present}. In this picture the infinite energy $W_0$ is interpreted as the energy of the \emph{vacuum} state $|0\rangle$ characterized by $n_{\mathbf{k}}=0$ for all $\mathbf{k}$. In Minkowski spacetime and, in general, when the time evolution of the spacetime metric is assigned, the presence of $W_0$ in Eq. (\ref{E_f_g}) does not play any role because it is just a constant zero-point energy and it is neglected.\footnote{The presence of $W_0$ must be taken into account if one wants to compute the backreaction effects that the matter field can have on the time evolution of the spacetime metric \cite{Birrell}.}

In what follows, I want to do the analogous steps I have already done in Minkowski spacetime but in a spacetime with general coordinates $\mathsf{x}=(x^0,\dots,x^3)$ and with metric tensor $g_{\mu\nu}(\mathsf{x})$. Firstly, I generalize the Lagrangean density (\ref{L_sc}) to transform it into a \emph{general} scalar Lagrangean density. The easiest way to do this is to replace the Minkowski metric tensor $\eta^{\alpha\beta}$ with $g^{\mu\nu}(\mathsf{x})$ and to multiply the resulting Lagrangean density by $\sqrt{-g(\mathsf{x})}$ with $g(\mathsf{x})\equiv\det(g_{\mu\nu}(\mathsf{x}))$: 
\begin{equation}
\label{L_sc_g}
\mathscr{L}=\sqrt{-g(\mathsf{x})}\left[g^{\mu\nu}(\mathsf{x})(\partial_{\mu}\phi)(\partial_{\nu}\phi)-m_0^2\phi^2\right].
\end{equation}
Actually, other terms proportional, for example, to the scalar curvature of the spacetime could be added to this Lagrangean density but I am not interested in them here. Analogously, the action is defined here as
\begin{equation}
S=\int d^4\mathsf{x}\mathscr{L}
\end{equation}
in order that it is also a general scalar and the equation of motion
\begin{equation}
\label{eq_mot_sc_g}
\frac{1}{\sqrt{-g(\mathsf{x})}}\partial_{\mu}\big(\sqrt{-g(\mathsf{x})}\partial^{\mu}\phi\big)+m_0^2\phi=0
\end{equation}
is generally covariant. Now, analogously to what I have done in Minkowski spacetime, I have to determine the normal modes building up an orthonormal basis of the solutions of the partial differential equation (\ref{eq_mot_sc_g}). Preliminarily, I also have to define a threedimensional hypersurface, corresponding to the ordinary threedimensional space in Minkowski spacetime, where
\begin{enumerate}
\item the scalar product between two general scalar functions can be defined;
\item some boundary conditions can be imposed in such a way a set of normal modes can be unambiguously determined.
\end{enumerate}
Now, a rigorous treatment of these subjects can be found in \cite{Hawking_b} and here only some stated results will be quoted. In particular, it can be shown that a \emph{Cauchy hypersurface}, that is a spacelike hypersurface such that any timelike or lightlike curve intersects it only once, shares both the previous features \cite{Hawking_b,Fulling}. A generic spacetime does not admit the existence of a Cauchy hypersurface. Instead, it can be shown that the so-called \emph{global hyperbolic} spacetimes not only admit a Cauchy hypersurface but they can also be ``foliated'' by means of Cauchy hypersurfaces. By ignoring here the exact mathematical definition of a global hyperbolic spacetime [it can be found, for example, in \cite{Hawking_b}], one can conclude that the topological structure of a globally hyperbolic fourdimensional spacetime is $\mathbb{R}\times \mathbb{M}^3$ where $\mathbb{R}$ will act as the time axis and $\mathbb{M}^3$ as the threedimensional manifold representing the ordinary space. If one also assumes that each ``leaf'' of the foliation can be represented by an equation like $x^0=\text{const}.$, one can consider $x^0$ as the time coordinate and $(x^1,x^2,x^3)$ as the three space coordinates. In this way, to work in a globally hyperbolic spacetime allows to
\begin{enumerate}
\item define, analogously to Eq. (\ref{sc_prod_sc}), the scalar product between two general scalar functions $\phi_1(\mathsf{x})$ and $\phi_2(\mathsf{x})$ as
\begin{equation}
\label{sc_prod_sc_g}
(\phi_1,\phi_2)\equiv-i\int_{\Sigma}d\Sigma n^{\mu}(\mathsf{x})[\phi_1(\mathsf{x})\partial_{\mu} \phi_2^*(\mathsf{x})-\phi_2^*(\mathsf{x})\partial_{\mu}\phi_1(\mathsf{x})]
\end{equation}
where $\Sigma$ is, in fact, a Cauchy hypersurface (it can be shown that the value of the scalar product does not depend on the Cauchy hypersurface one chooses) and $n^{\mu}(\mathsf{x})$ is the normal versor to $\Sigma$ at $\mathsf{x}$;
\item assume that the normal modes of Eq. (\ref{eq_mot_sc_g}) can be determined.
\end{enumerate}

Now, I remind that also in Minkowski spacetime when a matter field has to be quantized in the presence of an external (for example, electric or magnetic) time-dependent field, then the definition of the ``energy'' of the particles and then of the particles themselves is obscure \cite{Greiner,Fradkin}. An analogous problem arises in quantum field theory in curved spacetimes when the metric tensor $g_{\mu\nu}(\mathsf{x})$ depends on the time coordinate $x^0$. It can be shown that, in order to avoid these further difficulties, one has to assume the spacetimes to admit a Killing vector which is everywhere timelike \cite{Weinberg,Birrell}. In particular, this feature implies that the spacetime is \emph{static} meaning that it is possible to choose the coordinates $\mathsf{x}$ in such a way that $g_{0i}(\mathsf{x})=0$ and $\partial_0g_{\mu\nu}(\mathsf{x})=0$. At this point the quantization procedure is completely analogous to that in Minkowski spacetime and I will not repeat the remaining steps. I only point out that the definition of the Hamiltonian density of a quantum field is a controversial operation in curved spacetime \cite{Fulling_p_1,Grib,Castagnino}. For definiteness, I shall adopt the same definition given in \cite{Fulling_p_2} that is, actually, the same one gives in Minkowski spacetime. In particular, in the present case one firstly introduces the momentum canonically conjugated to the field $\phi(\mathsf{x})$ as
\begin{equation}
\pi(\mathsf{x})\equiv\frac{\partial \mathscr{L}}{\partial(\partial_0\phi)}
\end{equation}
and then defines the Hamiltonian density as
\begin{equation}
\mathscr{H}\equiv\pi\partial_0\phi-\mathscr{L}.
\end{equation}
%
%
%
\subsection{Spinor fields in general relativity: the tetrad formalism}
\label{II_III_II}
We know that by using the equivalence principle and the principle of general covariance, a recipe can be given in order to transform a Lorentz covariant equation into a general covariant equation \cite{Weinberg}:
\begin{enumerate}
\item replace the metric tensor $\eta_{\alpha\beta}$ with $g_{\mu\nu}(\mathsf{x})$;
\item replace the derivatives with the covariant derivatives.
\end{enumerate}
Actually, in doing so one tacitly assumes to interpret all the \emph{Lorentz} tensor fields as \emph{general} tensor fields and this can be done only because a tensor under general coordinate transformations is also a tensor under Lorentz transformations. Now, this procedure can not be carried out when equations contain spinor quantities because \emph{there is not a representation of the group of the general coordinate transformations that behaves like a finite dimensional spinor representation under the subgroup of the Lorentz transformations} \cite{Weinberg}. For this reason the previous recipe to transform a Lorentz covariant equation into a general covariant equation becomes inapplicable if the equation itself contains spinor quantities and another procedure has to be followed. This procedure is called \emph{vierbein} or \emph{tetrad} formalism and it consists, firstly, in introducing normal coordinates $y^{\alpha}_{\mathsf{X}}$ at any spacetime point $\mathsf{X}$. By means of these coordinates the metric tensor $g_{\mu\nu}(\mathsf{x})$ can be written in terms of another set of coordinates $\mathsf{x}=(x^0,\dots,x^3)$ as
\begin{equation}
\label{g_tetrad}
g_{\mu\nu}(\mathsf{x})=e^{\alpha}_{\mu}(\mathsf{x})e^{\beta}_{\nu}(\mathsf{x})\eta_{\alpha\beta}
\end{equation}
where the quantities
\begin{equation}
\label{tetrad}
e^{\alpha}_{\mu}(\mathsf{X})\equiv\left.\frac{\partial y^{\alpha}_{\mathsf{X}}}{\partial x^{\mu}}\right\vert_{\mathsf{x}=\mathsf{X}}
\end{equation}
build up the so-called \emph{tetrad field}. The tetrad field has the peculiar property to transform as a fourvector both under a general coordinate transformation:
\begin{equation}
x^{\prime\mu}\rightarrow x^{\prime\mu}(\mathsf{x}) \qquad\Longrightarrow\qquad
e^{\alpha}_{\mu}(\mathsf{x})\rightarrow \frac{\partial x^{\nu}}{\partial x^{\prime\mu}}e^{\alpha}_{\nu}(\mathsf{x}) 
\end{equation}
and also under a local Lorentz transformation $\Lambda^{\alpha}_{\beta}(\mathsf{x})$:
\begin{equation}
y_{\mathsf{x}}^{\alpha}\rightarrow \Lambda^{\alpha}_{\beta}(\mathsf{x})y_{\mathsf{x}}^{\beta}\qquad\Longrightarrow\qquad  e^{\alpha}_{\mu}(\mathsf{x})\rightarrow \Lambda^{\alpha}_{\beta}(\mathsf{x})e^{\beta}_{\mu}(\mathsf{x}).
\end{equation}
For this reason, by contracting $e^{\alpha}_{\mu}(\mathsf{x})$ with a contravariant general fourvector $V^{\mu}(\mathsf{x})$ as
\begin{equation}
V^{\alpha}(\mathsf{x})\equiv e^{\alpha}_{\mu}(\mathsf{x})V^{\mu}(\mathsf{x})
\end{equation}
one obtains a set of four general scalars. Analogously, by using the metric tensor $\eta_{\alpha\beta}$ ($\eta^{\alpha\beta}$) to lower (rise) the indices $\alpha$, $\beta$ and so on, and the metric tensor $g_{\mu\nu}(\mathsf{x})$ ($g^{\mu\nu}(\mathsf{x})$) to lower (rise) the indices $\mu$, $\nu$ and so on, that is by putting
\begin{subequations}
\begin{align}
e_{\alpha\mu}(\mathsf{x}) &\equiv\eta_{\alpha\beta}e^{\beta}_{\mu}(\mathsf{x}),\\
e^{\alpha\mu}(\mathsf{x}) &\equiv g^{\mu\nu}(\mathsf{x})e^{\alpha}_{\nu}(\mathsf{x}),\\
e_{\alpha}^{\mu}(\mathsf{x}) &\equiv\eta_{\alpha\beta}g^{\mu\nu}(\mathsf{x})e^{\beta}_{\nu}(\mathsf{x}),
\end{align}
\end{subequations}
one can also transform, for example, general \emph{covariant} tensors $T_{\mu\nu}(\mathsf{x})$ into a set of general scalars $T_{\alpha\beta}(\mathsf{x})$ by simply defining
\begin{equation}
T_{\alpha\beta}(\mathsf{x})\equiv e_{\alpha}^{\mu}(\mathsf{x})e_{\beta}^{\nu}(\mathsf{x})T_{\mu\nu}(\mathsf{x}).
\end{equation}

Now, let consider only the particular problem of transforming a manifestly Lorentz scalar Lagrangean density into a manifestly general scalar Lagrangean density. At first sight, one can simply assume that all the fields appearing in the initial Lagrangean density are sets of general scalars fields like $V^{\alpha}(\mathsf{x})$ or $T_{\alpha\beta}(\mathsf{x})$, but this is not enough because a problem arises when derivative terms are present. In fact, let suppose that 
\begin{enumerate}
\item $\Psi(\mathsf{x})$ is a generic multicomponent field that
\begin{enumerate}
\item[1a.] is a set of general scalar fields [like the fields $V^{\alpha}(\mathsf{x})$ or $T_{\alpha\beta}(\mathsf{x})$];
\item[1b.] transforms as $\Psi(\mathsf{x}) \rightarrow D(\Lambda(\mathsf{x}))\Psi(\mathsf{x})$ under the local Lorentz transformation $\Lambda^{\alpha}_{\beta}(\mathsf{x})$ with $D(\Lambda(\mathsf{x}))$ a certain matrix representation of the Lorentz group;
\end{enumerate}
\item $\Psi(\mathsf{x})$ appears in the Lagrangean density to be transformed through a term like $\partial_{\alpha}\Psi(\mathsf{x})\equiv e_{\alpha}^{\mu}(\mathsf{x})\partial_{\mu}\Psi(\mathsf{x})$.
\end{enumerate}
Since $\partial_{\mu}\Psi(\mathsf{x})$ transform as a fourvector under general coordinate transformations then $\partial_{\alpha}\Psi(\mathsf{x})$ is a set of general scalars. But, the equivalence principle requires that special relativity should apply in locally inertial reference systems regardless the local inertial reference system one chooses. In other words, the final Lagrangean density must also be invariant under a local Lorentz transformation. Now, the quantity $\partial_{\alpha}\Psi(\mathsf{x})$ transforms under the local Lorentz transformation $\Lambda^{\alpha}_{\beta}(\mathsf{x})$ as
\begin{equation}
\label{part_trans}
\partial_{\alpha}\Psi(\mathsf{x})\rightarrow \Lambda^{\beta}_{\alpha}(\mathsf{x})D(\Lambda(\mathsf{x}))\partial_{\beta}\Psi(\mathsf{x})+\Lambda^{\beta}_{\alpha}(\mathsf{x})[\partial_{\beta}D(\Lambda(\mathsf{x}))]\Psi(\mathsf{x})
\end{equation}
and the presence of the second term would destroy the invariance of the Lagrangean density under the same transformation. For this reason, one introduces the ``covariant'' derivative
\begin{equation}
\mathscr{D}_{\alpha}\equiv e_{\alpha}^{\mu}(\mathsf{x})\left[\partial_{\mu}+\Gamma_{\mu}(\mathsf{x})\right]\equiv
e_{\alpha}^{\mu}(\mathsf{x})\mathscr{D}_{\mu}
\end{equation}
where the quantities $\Gamma_{\mu}(\mathsf{x})$ are matrices of the same dimension of the representation $D(\Lambda(\mathsf{x}))$ and are called \emph{connections}. By assumption, the connections $\Gamma_{\mu}(\mathsf{x})$ transform under the local Lorentz transformation $\Lambda^{\alpha}_{\beta}(\mathsf{x})$ as
\begin{equation}
\Gamma_{\mu}(\mathsf{x})\rightarrow D(\Lambda(\mathsf{x}))\Gamma_{\mu}(\mathsf{x})D^{-1}(\Lambda(\mathsf{x}))-[\partial_{\mu}D(\Lambda(\mathsf{x}))]D^{-1}(\Lambda(\mathsf{x}))
\end{equation}
in such a way that [see Eq. (\ref{part_trans})]
\begin{equation}
y_{\mathsf{x}}^{\prime\alpha}\rightarrow \Lambda^{\alpha}_{\beta}(\mathsf{x})y_{\mathsf{x}}^{\prime\beta} \qquad\Longrightarrow\qquad \mathscr{D}_{\alpha}\Psi(\mathsf{x})\rightarrow \Lambda^{\beta}_{\alpha}(\mathsf{x})D(\Lambda(\mathsf{x}))\mathscr{D}_{\beta}\Psi(\mathsf{x}).
\end{equation}
As a conclusion, \emph{by assuming the fields appearing in the initial Lorentz scalar Lagrangean density as sets of general scalars and by substituting the ordinary derivatives $\partial_{\alpha}$ with the covariant derivatives $\mathscr{D}_{\alpha}$, the resulting Lagrangean density is a manifestly general scalar quantity}. Actually, one requires that a Lagrangean density is a general \emph{scalar density} quantity that integrated on $d\mathsf{x}$ gives a general scalar action, then, after all, the resulting Lagrangean density must be multiplied by $\sqrt{-g(\mathsf{x})}$ with $g(\mathsf{x})\equiv\det(g_{\mu\nu}(\mathsf{x}))$ \cite{Landau2}. Obviously, by excluding this last point, the previous recipe can be followed to transform any Lorentz covariant equation into a general covariant one. In particular, it can be shown that when one deals only with tensor fields it coincides with the more usual recipe given at the beginning of the Paragraph.

Now, if the generic field $\Psi(\mathsf{x})$ is, actually, a \emph{fourcomponent spinor} field then it can be shown that the connections $\Gamma_{\mu}(\mathsf{x})$ are given by \cite{Weinberg}
\begin{equation}
\Gamma_{\mu}(\mathsf{x})=-\frac{i}{4}\sigma^{\alpha\beta}e_{\alpha}^{\nu}(\mathsf{x})[\partial_{\mu}+\Gamma^{\lambda}_{\mu\nu}(\mathsf{x})]e_{\beta\lambda}(\mathsf{x})
\end{equation}
where
\begin{equation}
\Gamma^{\lambda}_{\mu\nu}(\mathsf{x})\equiv\frac{g^{\lambda\rho}(\mathsf{x})}{2}\left[\partial_{\mu}g_{\nu\rho}(\mathsf{x})+\partial_{\nu}g_{\mu\rho}(\mathsf{x})-\partial_{\rho}g_{\mu\nu}(\mathsf{x})\right]
\end{equation}
are the Christoffel symbols \cite{Landau2} and where $\sigma^{\alpha\beta}=i[\gamma^{\alpha},\gamma^{\beta}]/2$ with $\gamma^0=\beta$, $\gamma^1=\beta\alpha_x$, $\gamma^2=\beta\alpha_y$ and $\gamma^3=\beta\alpha_z$ are the covariant Dirac matrices satisfying the anticommutation relations $\{\gamma^{\alpha},\gamma^{\beta}\}=2\eta^{\alpha\beta}$. In this way, by reminding that the Dirac Lagrangean density in Minkowski spacetime can be written as
\begin{equation}
\label{L_D}
\mathscr{L}=\frac{i}{2}\left[\bar{\Psi}\gamma^{\alpha}(\partial_{\alpha}\Psi)-
(\partial_{\alpha}\bar{\Psi})\gamma^{\alpha}\Psi\right]-m\bar{\Psi}\Psi
\end{equation}
and by following the rules just stated, the general scalar Dirac Lagrangean density is given by
\begin{equation}
\label{L_D_g}
\begin{split}
\mathscr{L} &=\sqrt{-g(\mathsf{x})}\left\{\frac{i}{2}\Big[\bar{\Psi}\gamma^{\alpha}e_{\alpha}^{\mu}(\mathsf{x})(\mathscr{D}_{\mu}\Psi)-
(\bar{\Psi}\overset{\shortleftarrow}{\mathscr{D}}_{\mu})e_{\alpha}^{\mu}(\mathsf{x})\gamma^{\alpha}\Psi\Big]-m\bar{\Psi}\Psi\right\}\\
&=\sqrt{-g(\mathsf{x})}\left\{\frac{i}{2}\Big[\bar{\Psi}\gamma^{\mu}(\mathsf{x})(\mathscr{D}_{\mu}\Psi)-(\bar{\Psi}\overset{\shortleftarrow}{\mathscr{D}}_{\mu})\gamma^{\mu}(\mathsf{x})\Psi\Big]-m\bar{\Psi}\Psi\right\}.
\end{split}
\end{equation}
It is worth doing some observations about Eq. (\ref{L_D_g}):
\begin{enumerate}
\item as in quantum field theory in Minkowski spacetime, the adjoint field $\bar{\Psi}(\mathsf{x})$ and the Hermitian conjugated field $\Psi^{\dag}(\mathsf{x})$ are connected here by the relation $\bar{\Psi}(\mathsf{x})=\Psi^{\dag}(\mathsf{x})\gamma^0$;
\item the ``left'' covariant derivative is defined as
\begin{equation}
\overset{\shortleftarrow}{\mathscr{D}}_{\mu}\equiv\overset{\shortleftarrow}{\partial}_{\mu}-\Gamma_{\mu}(\mathsf{x})
\end{equation}
where the symbol ``$\overset{\shortleftarrow}{\partial}_{\mu}$'' means that the partial derivative acts on the spinor on its left and the minus sign comes from the fact that the Lagrangean density must be real and that $\Gamma^{\dag}_{\mu}(\mathsf{x})=-\gamma^0\Gamma_{\mu}(\mathsf{x})\gamma^0$;
\item the matrices $\gamma^{\mu}(\mathsf{x})\equiv e_{\alpha}^{\mu}(\mathsf{x})\gamma^{\alpha}$ are called \emph{general covariant Dirac matrices} and they satisfy the anticommutation rules $\{\gamma^{\mu}(\mathsf{x}),\gamma^{\nu}(\mathsf{x})\}=2g^{\mu\nu}(\mathsf{x})$.
\end{enumerate}
By applying the usual variational method to the action $S=\int d\mathsf{x}\mathscr{L}$, one obtains the general covariant Dirac equation in the form
\begin{equation}
i\gamma^{\mu}(\mathsf{x})\mathscr{D}_{\mu}\Psi-m\Psi=0.
\end{equation}
Finally, the scalar product between two spinors $\Psi_1(\mathsf{x})$ and $\Psi_2(\mathsf{x})$ is defined in a general covariant way as
\begin{equation}
\label{Sc_prod_g}
(\Psi_1,\Psi_2)\equiv\int_{\Sigma}d\Sigma n^{\mu}(\mathsf{x})\bar{\Psi}_1(\mathsf{x})\gamma_{\mu}(\mathsf{x})\Psi_2(\mathsf{x})
\end{equation}
with $\Sigma$ a Cauchy hypersurface, $n^{\mu}(\mathsf{x})$ the normal versor to $\Sigma$ at $\mathsf{x}$ and $\gamma_{\mu}(\mathsf{x})\equiv g_{\mu\nu}(\mathsf{x})\gamma^{\nu}(\mathsf{x})$.

\chapter{Pair production in a strong, uniform and slowly-varying magnetic field}
In this Chapter I present my original results about the electron-positron pair production in the presence of a strong, uniform and slowly-varying magnetic field. The Chapter is divided in two parts. In the first one the theoretical assumptions and the general features of my approach are given (Sect. \ref{The_mod}) while in the second one (Sect. \ref{Pai_prod}) the presence probability of a single pair is calculated by considering two different magnetic field time variations.

\section{Theoretical model}
\label{The_mod}
As I have said in the Introduction and in Chap. 1, my work mainly concerns the production of pairs in the presence of strong ($\gg B_{cr}=4.4\times 10^{13}\;\text{gauss}$), uniform and slowly-varying magnetic fields. I will show in the next Paragraph that these assumptions about the structure of the magnetic field are justified in the astrophysical scenario I imagine to apply to my calculations and that I have described in Sect. \ref{GRB}. Since the magnetic field is assumed to be slowly-varying, I will perform in Par. \ref{Prel_calc} some preliminary calculations in the framework of the first-order adiabatic perturbation theory that I will use in Sect. \ref{Pai_prod}.
%
%
\subsection{General assumptions}
\label{Gen_ass}
As I have said in Par. \ref{Centr_eng}, I imagine the magnetic field I deal with to be produced by a (forming or collapsing) magnetar or by a forming black hole. In particular, in this Chapter I want restrict my considerations to the production of pairs around magnetars in order to simplify the treatment by neglecting all the effects due to the compact object gravitational field.\footnote{The gravitational force exerted on an electron (positron) by a typical $2$ solar masses magnetar is several orders of magnitudes weaker than that exerted by its magnetic field. Nevertheless, the gravitational effects will be taken into account in Chap. 5 where the pair production around massive black holes is considered.} Now, we have seen that indirect estimates suggest that the magnetic field of a magnetar can be safely assumed to be one or two orders of magnitude larger than $B_{cr}$. In general, the magnetic field produced by a magnetar is a dipole field and then it is not uniform in space. Also, since the production of the fireball happens during the formation or the collapse of the magnetar it is reasonable to assume the magnetic field produced by the magnetar itself to be time-varying. Now, the physical problem I want to solve clearly has two different scales of length and time: a \emph{macroscopic} one related to the typical linear dimension of a magnetar ($\sim 10^6\;\text{cm}$) and to its typical time formation or collapse ($\sim 1\;\text{s}$) and a \emph{microscopic} one related to the elementary particles to be produced which is given by the Compton length of the electron $\lambdabar\sim 10^{-11}\;\text{cm}$. Clearly, \emph{the order of magnitude of the macroscopic scale is much larger than that of the microscopic one and this allows me to consider during the pair production process the magnetic field of the magnetar as uniform in space and slowly-varying in time}. Below, I will consider two different kinds of magnetic field time evolutions and they are both particular cases of the following one:\footnote{Concerning the magnetic field and its components, I use, apart from the time dependence, the same notation I have used in the previous Chapter just to have the possibility to exploit many results obtained there.}
\begin{equation}
\label{B_t}
\mathbf{B}(t)=\begin{pmatrix}
0\\
B_y(t)\\
B_z(t)
\end{pmatrix}=
B(t)\begin{pmatrix}
0\\     
\sin\vartheta(t)\\
\cos\vartheta(t)
\end{pmatrix}
\end{equation}
where
\begin{subequations}
\begin{align}
\label{B_mod_t}
B(t) &=\sqrt{B_y^2(t)+B_z^2(t)},\\
\label{theta_t}
\tan\vartheta(t) &=\frac{B_y(t)}{B_z(t)}.
\end{align}
\end{subequations}

From a macroscopic point of view the time variation of $\mathbf{B}(t)$ implies the presence of a nonuniform electric field $\mathbf{E}(t,\mathbf{r})$. In the symmetric gauge in which
\begin{equation}
\label{A_t}
\mathbf{A}(t,\mathbf{r})=-\frac{1}{2}\left[\mathbf{r}\times\mathbf{B}(t)\right],
\end{equation}
the electric field is given by
\begin{equation}
\label{E}
\mathbf{E}(t,\mathbf{r})=-\partial_t\mathbf{A}(t,\mathbf{r})= 
\frac{1}{2}[\mathbf{r}\times\dot{\mathbf{B}}(t)].
\end{equation}
Apart from the particular case in which the magnetic field depends linearly on time, the electric field also depends on time. Now, since I am interested in the production of electrons and positrons \emph{from vacuum} I assume the calculations to be applied to regions where conduction currents are not present.\footnote{The electric field (\ref{E}) is divergenceless then the electric charge density vanishes everywhere.} In this case, the Amp\`ere-Maxwell equation $\boldsymbol{\partial}\times \mathbf{B}-\partial_t\mathbf{E}=\mathbf{0}$ would require the existence of a correction to $\mathbf{B}(t)$ proportional to $\ddot{\mathbf{B}}(t)$ and nonuniform in space. But, reminding that $\mathbf{B}(t)$ is assumed to be slowly-varying, I am allowed to carry out all the calculations up to first order in $\dot{\mathbf{B}}(t)$, so, consistently, to neglect any contribution proportional to $\ddot{\mathbf{B}}(t)$. On the other hand, as we will see, the presence of the electric field $\mathbf{E}(t,\mathbf{r})$ plays an important role in the pair production process. Actually, it can be said that just the presence of the electric field $\mathbf{E}(t,\mathbf{r})$ accounts for the pair production process because it supplies the energy to create the pair. Nevertheless, in this respect, a conceptual difference between the present model and the Schwinger pair production mechanism described in Sect. \ref{QEDPEF} must be pointed out: here the electric field (\ref{E}) is rotational, then it does not admit a scalar potential and the interpretation of the pair production process as a tunnel effect is not straightforward.
%
%
\subsection{Application of the first-order adiabatic perturbation theory to the pair production process: preliminary calculations}
\label{Prel_calc}
I have just said that the interpretation of the pair production process in terms of the Schwinger mechanism is not straightforward in the problem under consideration. Also, since the astrophysical models generally refer to the strong magnetic fields present around magnetars and not to their induced electric fields, I want to put into evidence in the final results the role of the magnetic field $\mathbf{B}(t)$. For this reason and being the magnetic field $\mathbf{B}(t)$ a slowly-varying quantity, I will interpret the pair production process as a transition induced by the external \emph{adiabatic} perturbation $\mathbf{B}(t)$ on the quantum system represented by a second-quantized Dirac field $\psi(t,\mathbf{r})$. The Hamiltonian of this system is given by
\begin{equation}
\label{H_t}
H(t)=\int d\mathbf{r}\psi^{\dag}(t,\mathbf{r})\mathcal{H}(t)\psi(t,\mathbf{r}) 
\end{equation}
with
\begin{equation}
\label{H_1p_t}
\mathcal{H}(t)=\boldsymbol{\alpha}\cdot\left[-i\boldsymbol{\partial}+e\mathbf{A}(t,\mathbf{r})\right]+\beta m
\end{equation}
and it depends on time through the slowly-varying magnetic field $\mathbf{B}(t)$ hidden in $\mathbf{A}(t,\mathbf{r})$. In order to apply the first-order adiabatic perturbation theory to the problem at hand, I have to determine the instantaneous eigenvalues and eigenstates of the Hamiltonian (\ref{H_t}) (see Sect. \ref{Ad_pert_th}). As a first step, I address the problem of quantizing the Dirac field $\psi(t,\mathbf{r})$ in the presence of the magnetic $\mathbf{B}$ introduced in the previous Chapter [see Eq. (\ref{B})] which is identical to $\mathbf{B}(t)$ just apart from the time dependence. Since I have already determined the one-particle orthonormal basis $\{u_j(\mathbf{r}),v_j(\mathbf{r})\}$ with $j\equiv\{n_d,k,\sigma,n_g\}$ [see Eqs. (\ref{u_v}) and (\ref{u_p_v_p})], I can expand the Dirac field $\psi(t,\mathbf{r})$ as
\begin{equation}
\label{psi_esp}
\psi (t,\mathbf{r})=\sum_j [c_j\exp(-iw_jt) u_j(\mathbf{r})+d^*_j\exp(i\tilde{w}_jt)v_j (\mathbf{r})]
\end{equation}
and impose the usual anticommutation rules
\begin{subequations}
\begin{align}
\{c_j,c^{\dag}_{j'}\} &=\{d_j,d^{\dag}_{j'}\}=\delta_{j,j'},\\
\{c_j,c_{j'}\} &=\{c_j,d^{\dag}_{j'}\}=\{c_j,d_{j'}\}=\{d_j,d_{j'}\}=0
\end{align}
\end{subequations}
among the coefficients $c_j$ and $d_j$ that are now operators. By exploiting the orthonormalization rules (\ref{ort_uv}) among the spinors $u_j(\mathbf{r})$ and $v_j(\mathbf{r})$ and the previous anticommutators it is easy to see that the second quantized Hamiltonian $H=\int d\mathbf{r}\;\psi^{\dag} (t,\mathbf{r})\mathcal{H}\psi (t,\mathbf{r})$ with $\mathcal{H}$ given by Eq. (\ref{H_1p}) [that is by Eq. (\ref{H_1p_t}) without the time dependence] becomes
\begin{equation}
\label{H}
H=\sum_j(w_jN_j+\tilde{w}_j\tilde{N}_j)
\end{equation}
where $N_j=c_j^{\dag}c_j$ and $\tilde{N}_j=d_j^{\dag}d_j$ and where the zero-point energy has been set to zero. Obviously, if $|0\rangle$ is the vacuum state defined by the relations $c_j|0\rangle=d_j|0\rangle=0$, the eigenstates of the Hamiltonian (\ref{H}) are the Fock states
\begin{equation}
\label{n_n_t}
|\{n_j\};\{\tilde{n}_{\tilde{j}}\}\rangle\equiv
\big(c^{\dag}_{j_1}\big)^{n_{j_1}}
\big(c^{\dag}_{j_2}\big)^{n_{j_2}}\cdots
\big(d^{\dag}_{\tilde{j}_1}\big)^{\tilde{n}_{\tilde{j}_1}}
\big(d^{\dag}_{\tilde{j}_2}\big)^{\tilde{n}_{\tilde{j}_2}}\cdots|0\rangle,
\end{equation}
while the corresponding eigenvalues are given by
\begin{equation}
\label{W}
W=\sum_l (w_{j_l}n_{j_l}+\tilde{w}_{\tilde{j}_l}\tilde{n}_{\tilde{j}_l})
\end{equation}
with $l$ a generic integer index. At this point, the instantaneous eigenvalues and eigenstates of the time-dependent Hamiltonian (\ref{H_t}) are simply given by the previous expressions (\ref{n_n_t}) and (\ref{W}) by adding the dependence on time. Also, the creation from vacuum of a pair with the electron and the positron in the states $j$ and $j'$ respectively at time $t$ is represented by the adiabatic transition
\begin{equation}
|0(t)\rangle\xrightarrow{\dot{\mathbf{B}}(t)}|jj'(t)\rangle\equiv c^{\dag}_j(t)d^{\dag}_{j'}(t)|0(t)\rangle.
\end{equation}
The probability associated with this transition is given, in the framework of the adiabatic perturbation theory up to first order in $\dot{\mathbf{B}}(t)$, by $P_{jj'}(t)=|\gamma_{jj'}(t)|^2$ where [see Eq. (\ref{a_m_1_f})]
\begin{equation}
\label{gamma_jjp}
\gamma_{jj'}(t)=\int_0^tdt' \frac{\dot{H}_{jj'}(t')}{w_j(t')+ 
\tilde{w}_{j'}(t')}\exp\left\{i\int_0^{t'}dt''\left[w_r(t'')+ \tilde{w}_{q'}(t'')\right]\right\}
\end{equation}
with
\begin{equation}
\label{H_der_jjp}
\dot{H}_{jj'}(t')\equiv\langle jj'(t')|\dot{H}(t')|0(t')\rangle.
\end{equation}
It is worth doing a couple of observations about this formula. Firstly, the initial time $t=0$ when the magnetic field is supposed to start changing represents, in the astrophysical scenario I imagine to apply my calculations, the beginning of the magnetar collapse into a black hole or the beginning of the magnetar (black hole) formation. Secondly, the usual physical interpretation of the square modulus of the coefficients $a_n(t)$ in Eq. (\ref{t}) forces to interpret $P_{jj'}(t)$ not as a \emph{creation} probability at time $t$ but as a \emph{presence} probability at time $t$ that is as the probability that, being the system in the vacuum state at the initial time $t=0$, a pair is \emph{present} at time $t>0$ with the electron in the state $j$ and the positron in the state $j'$.

Now, the time derivative of the Hamiltonian $H(t)$ is given by [see Eqs. (\ref{H_t}) and (\ref{H_1p_t})]
\begin{equation}
\label{dot_H}
\begin{split}
\dot{H}(t) &=\dot{\mathbf{B}}(t)\cdot\boldsymbol{\partial}_{\mathbf{B}}(H(t))=
\dot{\mathbf{B}}(t)\cdot\int d\mathbf{r}\psi^{\dag}(t,\mathbf{r}) \boldsymbol{\partial}_{\mathbf{B}}
(\mathcal{H}(t))\psi(t,\mathbf{r})\\
&=-\frac{e\dot{\mathbf{B}}(t)}{2}\cdot\int d\mathbf{r}\psi^{\dag}(t,\mathbf{r})(\mathbf{r}\times \boldsymbol{\alpha})
\psi (t,\mathbf{r})
\end{split}
\end{equation}
and it can be written in terms of the electric field (\ref{E}) as
\begin{equation}
\label{dot_H_E}
\dot{H}(t)=-e\int d\mathbf{r}\psi^{\dag}(t,\mathbf{r}) \boldsymbol{\alpha}\cdot\mathbf{E}(t,\mathbf{r})\psi(t,\mathbf{r}).
\end{equation}
This expression points out the fundamental role of the induced electric field $\mathbf{E}(t,\mathbf{r})$ in the production process and, by reasoning as in first quantization, it has a clear physical meaning. In fact, by reminding that the vector
$\boldsymbol{\alpha}$ can be interpreted as
the one-particle relativistic operator corresponding
to the velocity of the electron, Eq. (\ref{dot_H_E}) expresses the fact that the variation per unit time of the energy of the electron is equal to the mean value of the power supplied to the electron itself by the external electric field $\mathbf{E}(t,\mathbf{r})$. On the other hand, the time derivative in Eq. (\ref{dot_H}) can be written in terms of the rotated Dirac field $\psi'(t,\mathbf{r})=\mathcal{R}^{(1/2)}_x(\vartheta(t))\psi(t,\mathbf{r})$, which is the field operator when the magnetic field is directed along the $z$ axis and its strength is $B(t)$ [see Eqs. (\ref{u_v})]:
\begin{equation}
\label{T_op}
\begin{split}
\dot{H}(t) &=-\frac{e\dot{\mathbf{B}}(t)}{2}\cdot\int d\mathbf{r}\psi^{\dag}(t,\mathbf{r})(\mathbf{r}\times \boldsymbol{\alpha})
\psi (t,\mathbf{r})\\
&=-\frac{e\dot{\mathbf{B}}(t)}{2}\cdot\int d\mathbf{r}\psi'^{\dag}(t,\mathbf{r})\mathcal{R}^{(1/2)}_x(\vartheta(t))
(\mathbf{r}\times\boldsymbol{\alpha})\mathcal{R}_x^{(1/2)\dag}(\vartheta(t))\psi'(t,\mathbf{r})\\
&=-\frac{e\dot{\mathbf{B}}'(t)}{2}\int d\mathbf{r}\psi'^{\dag}(t,\mathbf{r})(\mathbf{r}\times \boldsymbol{\alpha})\psi'(t,\mathbf{r})
\end{split}
\end{equation}
where I used the fact that both $\boldsymbol{\alpha}$ and $\mathbf{r}$ are vector operators and where
\begin{equation}
\label{B_dot}
\dot{\mathbf{B}}'(t)=\begin{pmatrix}
0\\                           
\cos \vartheta (t)\dot{B}_y (t)-\sin \vartheta (t)\dot{B}_z (t)\\
\sin \vartheta (t)\dot{B}_y (t)+\cos \vartheta (t)\dot{B}_z (t)\\
\end{pmatrix}=\begin{pmatrix}                             
0\\    
B(t)\dot{\vartheta}(t)\\                             
\dot{B}(t)\\                             
\end{pmatrix}.
\end{equation}
The form of the vector $\dot{\mathbf{B}}'(t)$ shows that the effects of the change of the direction and of the 
change of the strength of the magnetic field have been completely disentangled. In this way, one is able to connect the time variation of the magnetic field with some selection rules concerning the states in which the pair can be created. To do this I define the operator
\begin{equation}
\label{T_p}
\mathbf{T}'(t)=-\frac{e}{2}\int d\mathbf{r} \psi'^{\dag}(t,\mathbf{r})(\mathbf{r}\times\boldsymbol{\alpha})
\psi'(t,\mathbf{r}).
\end{equation}
Since $\mathbf{T}'(t)$ is a vector operator and the states $|0(t)\rangle$ and $|jj'(t)\rangle$ are instantaneous eigenstates of the total angular momentum operator $J^{(1/2)}_{\parallel}(t)\equiv\int d\mathbf{r}\psi^{\dag}(t,\mathbf{r})\mathcal{J}^{(1/2)}_{\parallel}(t)\psi(t,\mathbf{r})$, then the Wigner-Eckart theorem allows to state the selection rules resumed in Tab. \ref{Tab_sel_rul} where $j_{\parallel}=n_d-n_g+\sigma/2$ and $j'_{\parallel}=n'_d-n'_g+\sigma'/2$ [see Eq. (\ref{B_dot}) and the last line of Eq. (\ref{T_op})].
\begin{table}[ht]
\begin{center}
\begin{tabular}{|c|c|c|}
\hline& Transition operator & Selection rule \\
\hline $\dot{\vartheta}(t)=0$ & $T'_z(t)$ & $j_{\parallel}+j'_{\parallel}=0$ \\
$\dot{B}(t)=0$ & $T'_y(t)$ & $j_{\parallel}+j'_{\parallel}=\pm 1$ \\
\hline
\end{tabular}
\end{center}
\caption{Selection rules}
\label{Tab_sel_rul}
\end{table}

The first selection rule in the previous table has an interesting classical counterpart. In fact, following the interpretation of the pair creation in the Dirac hole theory, this selection rule means that an electron in the presence of a slowly-varying magnetic field with constant direction conserves its total angular momentum going from a negative energy level to a positive one. Analogously, in classical mechanics the angular momentum of the electron is, in the same physical situation, an adiabatic invariant \cite{Jackson}. Apart from those in Tab. \ref{Tab_sel_rul}, another important selection rule can be obtained. In fact, it can easily be shown that
\begin{equation}
\label{gamma_jj_p}
\begin{split}
\dot{H}_{jj'}(t') &=
\dot{\mathbf{B}}'(t)\cdot\langle jj'(t)|\mathbf{T}'(t)|0(t)\rangle
\\
&=-\frac{e\dot{\mathbf{B}}'(t)}{2}\cdot\int d\mathbf{r} u_j^{\prime\dag}(t,\mathbf{r})(\mathbf{r}\times\boldsymbol{\alpha})v'_{j'}(t,\mathbf{r})\\
&=\dot{\mathbf{B}}'(t)\cdot\mathbf{T}'_{jj'}(t)
\end{split}
\end{equation}
where I have defined the one-particle transition matrix elements
\begin{equation}
\label{T_jj_p}
\mathbf{T}'_{jj'}(t)=-\frac{e}{2}\int d\mathbf{r} u_j^{\prime\dag}(t,\mathbf{r})(\mathbf{r}\times 
\boldsymbol{\alpha})v'_{j'}(t,\mathbf{r}).
\end{equation}
If the magnetic field has a constant direction then $\dot{\vartheta}(t)=0$ and only the matrix elements $T'_{jj'z}(t)$ play a role (see Tab. \ref{Tab_sel_rul}). Also, it can be seen that 
$(\mathbf{r}\times \boldsymbol{\alpha})_z$ anticommutes with the one-particle spin operator $\sigma_z/2$:
\begin{equation}
\{(\mathbf{r}\times \boldsymbol{\alpha})_z,\sigma_z\}=0.
\end{equation}
Then, if two states $u'_j(t,\mathbf{r})$ and $v'_{j'}(t,\mathbf{r})$ are eigenstates of $\sigma_z$ with eigenvalues $\sigma$ and $\sigma'$ respectively then the transition matrix element $T'_{jj'z}(t)$ is different from zero only if
\begin{equation}
\label{sel_rule}
\sigma+\sigma'=0.
\end{equation}
Now, even if the magnetic field changes with time, the instantaneous TGSs $u_{n,k}'(t,\mathbf{r})$ and $v'_{n',k'}(t,\mathbf{r})$ given in Eqs. (\ref{u_v_p_g}) are eigenstates of $\sigma_z$ with the same eigenvalue $-1$ and for every $t$ [see the discussion below Eq. (\ref{w_g})]. This implies that if $\mathbf{B}(t)$ changes only in strength the creation of a pair in which the electron and the positron are both in a TGS is forbidden because Eq. (\ref{sel_rule}) is not satisfied. In other words, up to first-order adiabatic perturbation theory \emph{only the rotation of the magnetic field may allow the creation of a pair with both the electron and the positron in a TGS}.
%
%
\section{Pair production in the presence of a strong, uniform and slowly-rotating magnetic field}
\label{Pai_prod}
In the first analysis on the production of pairs in the presence of strong, uniform and slowly-varying magnetic fields it was considered the case in which only the magnetic field strength changes with time \cite{Calucci}. In order to avoid the presence of nonuniform corrections to the magnetic field $\mathbf{B}(t)$ its strength was supposed to change linearly with time [see discussion below Eq. (\ref{E})] that is
\begin{equation}
\label{B_lin}
\mathbf{B}(t)=\mathbf{B}^{\text{lin}}_{\shortuparrow}(t)=
\begin{pmatrix}
0\\
0\\
B^{\text{lin}}_{\shortuparrow}(t)
\end{pmatrix}=
\begin{pmatrix}
0\\
0\\
B_0+bt
\end{pmatrix}
\end{equation}
with $B_0,b>0$ and $B_0\gg B_{cr}$. Since this case was analyzed before my own work, I only quote in my notation the final result that is the total probability per unit volume that a pair is present at a time $t$:
\begin{equation}
\label{dP_lin}
\frac{dP^{\text{lin}}_{\shortuparrow}(t)}{dV}\sim\frac{1}{16\pi\sqrt{2}}\left[\frac{\zeta(1.5)}{256}+\frac{1}{15\pi}\right]e^{3/2}(bR_{\perp M})^2 \left[\frac{B^{\text{lin}}_{\shortuparrow}(t)}{B_0^{3/2}}+\frac{1}{\sqrt{B^{\text{lin}}_{\shortuparrow}(t)}}\right]
\end{equation}
where $\zeta(x)$ is the Riemann function with $\zeta(1.5)=2.61\ldots$ \cite{Abramowitz} and where the symbol $\sim$ indicates that the result holds asymptotically at large times. In Eq. (\ref{dP_lin}) the square of the quantity $R_{\perp M}$ also appears. Its origin and its exact meaning will be explained below when the analogous presence probability will be calculated in the presence of a purely rotating magnetic field. At the moment, I only point out that
\begin{enumerate}
\item $R_{\perp M}$ can be interpreted as the typical radius of a circle in the plane perpendicular to $\mathbf{B}^{\text{lin}}_{\shortuparrow}(t)$ within which the magnetar magnetic field can be assumed as uniform;
\item the appearance of $R_{\perp M}$ in Eq. (\ref{dP_lin}) can be understood in terms of the nonuniform electric field induced by the time-varying magnetic field $\mathbf{B}^{\text{lin}}_{\shortuparrow}(t)$.
\end{enumerate}
Finally, another observation concerns the dependence of the presence probability (\ref{dP_lin}) on the square of the time derivative $b$ of the magnetic field $\mathbf{B}^{\text{lin}}_{\shortuparrow}(t)$ that is, in fact, typical when the first-order adiabatic treatment is used to calculate a transition probability (the transition amplitude clearly is proportional to the time derivative of the perturbation and it has to be squared to obtain the corresponding transition probability).

Now, already in \cite{Calucci} it was observed that if the magnetic field direction does not change with time the production of pairs in which both the electron and the positron are in a  TGS is forbidden. As I have said in Par. \ref{TGSs}, since in the strong magnetic field regime the energies of the TGSs are much smaller than those of the other excited Landau levels, it is reasonable to imagine that the production of pairs with both the electron and the positron in a TGS is strongly favoured (when it is not forbidden by any selection rule). For this reason, by remembering the last statement in the previous Paragraph [see the discussion below Eq. (\ref{sel_rule})], the most natural continuation of the analysis started in \cite{Calucci} has been the treatment of the pair production process in the presence of a changing-direction magnetic field. In particular, the purely rotating magnetic field configuration \cite{DiPiazza2}: 
\begin{equation}
\label{B_rot}
\mathbf{B}_{\nnearrow}(t)=
B_{\nnearrow}\begin{pmatrix}  
0 \\
\sin\Omega t\\         
\cos\Omega t                             
\end{pmatrix}
\end{equation}
with $B_{\nnearrow}>0$ and $B_{\nnearrow}\gg B_{cr}$ allows an easier mathematical treatment of the problem. In fact, from Eq. (\ref{B_dot}) we have that
\begin{equation}
\dot{\mathbf{B}}'_{\nnearrow}(t)=\begin{pmatrix}
0\\
\Omega B_{\nnearrow}\\
0
\end{pmatrix}
\end{equation}
and that only the matrix elements
\begin{equation}
\label{T_jjp_y}
\begin{split}
\mathbf{T}'_{jj'y}(t) &=-\frac{e}{2}\int d\mathbf{r} u_j^{\prime\dag}(t,\mathbf{r})(\mathbf{r}\times 
\boldsymbol{\alpha})_yv'_{j'}(t,\mathbf{r})\\
&=-\frac{e}{2}\int d\mathbf{r} u_j^{\prime\dag}(t,\mathbf{r})(z\alpha_x-x\alpha_z)_yv'_{j'}(t,\mathbf{r})
\end{split}
\end{equation}
have to be calculated. Now, I have checked that if $P_{gg'}(t)=P(|0(t)\rangle\rightarrow |gg'(t)\rangle)$ is the probability that a pair is present at time $t$ with the electron and the positron both in a TGS and $P_{jj'}(t)$ is the probability that a pair is present at time $t$ with the electron and/or the positron in another state, then
\begin{equation}
\frac{P_{jj'}(t)}{P_{gg'}(t)}\lesssim\left(\frac{B_{cr}}{B_{\nnearrow}}\right)^{3/2}\ll 1.
\end{equation}
For this reason, I am allowed in first approximation to neglect all the transitions to pair states in which at least the electron or the positron is not in a TGS.\footnote{In any case, in Appendix A I study the general features of the amplitudes of these remaining transitions.} There are only two transition amplitudes different from zero that contribute to the creation of a pair with the electron and the positron both in a TGS. In fact, since the TGSs are eigenstates of $\sigma_z$ the term in Eq. (\ref{T_jjp_y}) containing $\alpha_x$ vanishes. Also, reminding that the electron (positron) TGSs are those with $n_d=0$ and $\sigma=-1$ ($n_g=0$ and $\sigma=+1$), the only allowed final pair states are the states [see also the expression (\ref{x}) of the coordinate $x$ in terms of $a_d$, $a^{\dag}_d$, $a_g$, $a^{\dag}_g$]
\begin{equation}
\label{states}
|n,k;n\pm 1,-k\rangle\equiv|0,k,-1,n;n\pm 1,-k,+1,0\rangle
\end{equation}
where, for simplicity, I have omitted the time dependence. The transition amplitudes corresponding to these states are given by
\begin{subequations}
\label{gamma}
\begin{align}
\gamma_{n,k;n+1,-k}(t) &=\sqrt{\frac{eB_{\nnearrow}(n+1)}{32}} \frac{m\Omega}{\varepsilon_k^3}
\exp(i\varepsilon_k t)\sin \varepsilon_k t, \\
\gamma_{n,k;n-1,-k}(t) &=\sqrt{\frac{eB_{\nnearrow} n}{32}} \frac{m\Omega}{\varepsilon_k^3}
\exp(i\varepsilon_k t)\sin \varepsilon_k t
\end{align}
\end{subequations}
where [see Eq. (\ref{w_g})] $\varepsilon_k=\sqrt{m^2+k^2}$. Actually, for later convenience, I want to consider the probability that the electron and the positron are present in wave packets that are linear superpositions of TGSs with the same energy. In particular, I will consider the following pair state
\begin{equation}
\label{pair_coh_st}
|\alpha_e\alpha_p\rangle=|\alpha_e\rangle|\alpha_p\rangle
\end{equation}
where
\begin{subequations}
\label{coh_st}
\begin{align}
|\alpha_e\rangle &=\sum_{n=0}^{\infty}b_n(\alpha_e)|n,k\rangle\equiv\sum_{n=0}^{\infty}b_n(\alpha_e)|0,k,-1,n\rangle, \\
|\alpha_p\rangle &=\sum_{n'=0}^{\infty}b_{n'}(\alpha_p)|n',-k\rangle\equiv\sum_{n'=0}^{\infty}b_{n'}(\alpha_p)|n',-k,+1,0\rangle
\end{align}
\end{subequations}
and where
\begin{equation}
\label{b}
b_n(z)\equiv\frac{z^n}{\sqrt{n!}}\exp\left(-\frac{1}{2}|z|^2\right)
\end{equation}
with $z$ a generic complex number. As it is evident from these equations, I have chosen as final states a double plane wave state concerning the longitudinal motion and a double coherent state concerning the transverse motion \cite{Cohen}. In the coherent states with $|\alpha_e|\gg 1$ ($|\alpha_p|\gg 1$) the electron (positron) has a good spatial localization in the plane orthogonal to the magnetic field and this will give me the possibility to understand the role of the induced \emph{nonuniform} electric field 
\begin{equation}
\label{E_rot}
\mathbf{E}_{\nnearrow}(t,\mathbf{r})=\frac{1}{2}[\mathbf{r}\times\dot{\mathbf{B}}_{\nnearrow}(t)]
\end{equation}
in the process of pair creation. 

Now, it can easily be shown from Eqs. (\ref{gamma}) that the 
transition amplitude from vacuum to the state $|\alpha_e\alpha_p\rangle$ can be written in the form
\begin{equation}
\label{gamma_ee}
\gamma_k(\alpha_e,\alpha_p;t)=\gamma_{\parallel k}(t)\gamma_{\perp}(\alpha_e,\alpha_p)
\end{equation}
with the longitudinal and the transverse amplitudes given by
\begin{subequations}
\begin{align}
\gamma_{\parallel k}(t) &=\sqrt{\frac{eB_{\nnearrow}}{32}} \frac{m\Omega}{\varepsilon_k^3}\exp(i\varepsilon_kt)
\sin \varepsilon_{k} t,\\
\gamma_{\perp}(\alpha_e,\alpha_p) &=\sum_{n=0}^{\infty}
\left[b_n^*(\alpha_e)b_{n+1}^*(\alpha_p)\sqrt{n+1}+b_n^*(\alpha_e)b_{n-1}^*(\alpha_p)\sqrt{n}\right].
\end{align}
\end{subequations}
In this way the longitudinal and the transverse part of the presence amplitude have been disentangled and only the longitudinal part depends on $t$. Obviously, the corresponding presence probability $|\gamma_k(\alpha_e,\alpha_p;t)|^2$ can also be divided into a longitudinal part and a transverse part. The longitudinal part is given by
\begin{equation}
dP_{\parallel}(k,t)=\left|\gamma_{\parallel}(k,t)\right|^2\frac{L_z}{2\pi}dk= 
\frac{eB_{\nnearrow}}{32} (m\Omega)^2\frac{\sin^2 \varepsilon (k) t}{\varepsilon^6(k)}\frac{L_z}{2\pi}dk
\end{equation}
where I multiplied by the number of states $L_zdk/(2\pi)$ [see Eqs. (\ref{k_discr})] and where $\varepsilon_k\to \varepsilon (k)$ for continuous $k$. The corresponding integrated probability per unit time is given by
\begin{equation}
\label{dP_L_dt}
\frac{dP_{\parallel}(t)}{dt}=\frac{eB_{\nnearrow}L_z}{64\pi}(m\Omega)^2 \int_{-\infty}^{\infty} 
dk\frac{\sin 2\varepsilon (k)t}{\varepsilon^5 (k)}
\end{equation}
and I can give an asymptotic estimate of this integral by assuming to be interested only in times $t$ such that $mt=t/\lambdabar\gg 1$ (in the astrophysical system under consideration this assumption is very realistic). The asymptotic estimate can be performed by using the method proposed in \cite{Lighthill} and the result is
\begin{equation}
\label{dP_L_dt_a}
\frac{dP_{\parallel}(t)}{dt}\sim \frac{eB_{\nnearrow}L_z}{64\pi}\left(\frac{\Omega}{m}\right)^2 \sqrt{\frac{\pi}{mt}}\sin\left(2mt+\frac{\pi}{4}\right).
\end{equation}

Now, I will calculate the transverse part of Eq. (\ref{gamma_ee}) or, equivalently, its complex conjugate
\begin{equation}
\gamma_{\perp}^*(\alpha_e,\alpha_p)=\sum_{n=0}^{\infty} \left[b_n(\alpha_e)b_{n+1}(\alpha_p)\sqrt{n+1}+ 
b_n(\alpha_e)b_{n-1}(\alpha_p)\sqrt{n}\right].
\end{equation}
From Eqs. (\ref{b}) I obtain
\begin{equation}
\begin{split}
\gamma_{\perp}^*(\alpha_e,\alpha_p) &= \exp\left(-\frac{|\alpha_e|^2+|\alpha_p|^2}{2}\right)\sum_{n=0}^{\infty} 
\left[\alpha_e\frac{(\alpha_e\alpha_p)^n}{n!}+ \alpha_p\frac{(\alpha_e\alpha_p)^n}{n!}\right]\\
&=(\alpha_e+\alpha_p) \exp\left(-\frac{|\alpha_e|^2+|\alpha_p|^2}{2}+\alpha_e\alpha_p\right)
\end{split}
\end{equation}
then the transverse presence probability of a pair with the electron in a coherent state between 
$|\alpha_e\rangle$ and $|\alpha_e+d\alpha_e\rangle$ and the positron in a coherent state between $|\alpha_p\rangle$ and $|\alpha_p+d\alpha_p\rangle$ is
\begin{equation}
dP_{\perp}(\alpha_e,\alpha_p)=|\alpha_e+\alpha_p|^2 \exp[-(|\alpha_e|^2+|\alpha_p|^2)+2\mathrm{Re}(\alpha_e\alpha_p)]\frac{d\alpha_p}{\pi}\frac{d\alpha_p}{\pi}.
\end{equation}
It can be shown \cite{Cohen} that the phase of $\alpha_e$ ($\alpha_p$) is connected with the azimuth of the mean position of the electron (positron) in the plane orthogonal to $\mathbf{B}_{\nnearrow}(t)$ while the modulus $|\alpha_e|$ ($|\alpha_p|$) is connected with its mean distance from the origin in the same plane. Then, since I am not interested in the exact position of the pair, I put
\begin{subequations}
\begin{align}
\alpha_e&=|\alpha_e|\exp(i\phi_e)\equiv\eta_e\exp(i\phi_e),\\
\alpha_p&=|\alpha_p|\exp(i\phi_p)\equiv\eta_p\exp(i\phi_p)
\end{align}
\end{subequations}
and I integrate on the angles $\phi_e$ and $\phi_p$. The result is
\begin{equation}
\begin{split}
dP_{\perp}(\eta_e,\eta_p) &=4\eta_e\eta_p (\eta_e^2+\eta_p^2) 
\exp[-(\eta_e^2+\eta_p^2)]\mathrm{I}_0(2\eta_e\eta_p)d\eta_ed\eta_p
\end{split}
\end{equation}
where $\mathrm{I}_0(x)$ is the zero-order modified Bessel function \cite{Abramowitz}. In order to obtain a more transparent formula I define the variables $\eta$ and $\phi$ by means of the equations
\begin{subequations}
\begin{align}
\eta_e&=\eta \cos \phi, \\
\eta_p&=\eta \sin \phi
\end{align}
\end{subequations}
and I integrate on the phase $\phi$. Since $\eta_e$ and $\eta_p$ vary from $0$ to 
$\infty$, $\eta$ and $\phi$ vary from $0$ to $\infty$ and from $0$ to $\pi/2$ respectively, then, after performing the integral on $\phi$, I obtain the differential probability
\begin{equation}
\label{dP_T_dalpha}
dP_{\perp}(\eta)=\eta^3\left[1-\exp(-2\eta^2)\right]d\eta.
\end{equation}
Now, there is an interesting relation between $\eta$ and the mean value in the states $|\alpha_e\rangle$ and $|\alpha_p\rangle$ of the operator $R_{\perp}^2$ defined in Eq. (\ref{R_perp^2}) [see also Eq. (\ref{R_xy^2})]. In fact, by using the eigenvalue equations (\ref{eig_R_xy^2_e}), (\ref{eig_R_xy^2_p}) and the definitions (\ref{coh_st}), it is easy to show that
\begin{subequations}
\begin{align}
R_{\perp}^2(\alpha_e)&\equiv \langle \alpha_e|R_{\perp}^2|\alpha_e\rangle 
=\frac{2\eta_e^2+1}{eB_{\nnearrow}},\\
R_{\perp}^2(\alpha_p)&\equiv \langle \alpha_p|R_{\perp}^2|\alpha_p\rangle 
=\frac{2\eta_p^2+1}{eB_{\nnearrow}}.
\end{align}
\end{subequations}
But, since the one-particle energy of the electron (positron) do not depend on the quantum number $n_g$ ($n_d$) and then on the quantity $\eta_e$ ($\eta_p$), there are no dynamical constraints on the maximum value that $\eta_e$ ($\eta_p$) can assume. Also, since the pair production process is imagined to take place in a macroscopic astrophysical environment, I can assume that $R_{\perp}(\alpha_e)\gg \lambdabar$ and $R_{\perp}(\alpha_p)\gg\lambdabar$ in such a way
\begin{subequations}
\label{alpha_e_p_mod}
\begin{align}
\eta_e^2&\simeq \frac{eB_{\nnearrow}}{2}R_{\perp}^2(\alpha_e)\gg 1,\\
\eta_p^2&\simeq \frac{eB_{\nnearrow}}{2}R_{\perp}^2(\alpha_p)\gg 1
\end{align}
\end{subequations}
where I exploited the fact that in the strong magnetic field limit $\sqrt{eB_{\nnearrow}}\gg 1/\lambdabar$. Even if the previous inequalities hold, the internal consistency of the model requires that the quantities $R_{\perp}^2(\alpha_e)$ and $R_{\perp}^2(\alpha_p)$ can not be too large. In fact, on the one hand the magnetar magnetic field must be approximatively uniform inside a region with typical linear length $R_{\perp}(\alpha_e)$ or $R_{\perp}(\alpha_p)$ than $R_{\perp}(\alpha_e)\ll 10^6\;\text{cm}$ and $R_{\perp}(\alpha_p)\ll 10^6\;\text{cm}$. On the other hand, it can be shown, by applying Eq. (\ref{ad_cond}) to the present case, that the first-order adiabatic treatment can be safely used only if the conditions
\begin{subequations}
\label{lim_ad_pert}
\begin{align}
\Omega R_{\perp}(\alpha_e)\frac{B_{\nnearrow}}{B_{cr}}&\ll 1
,\\
\Omega R_{\perp}(\alpha_p)\frac{B_{\nnearrow}}{B_{cr}}&\ll 1
\end{align}
\end{subequations}
are satisfied. I observe that these conditions do not constraint very much the allowed values of $R_{\perp}(\alpha_e)$ and $R_{\perp}(\alpha_p)$. In fact, by assuming $B_{\nnearrow}=10^{15}\;\text{gauss}$ and $\Omega=1\;\text{s$^{-1}$}$ [see the discussion before Eq. (\ref{B_t})], then the conditions (\ref{lim_ad_pert}) are satisfied if $R_{\perp}(\alpha_e)\ll 10^9\;\text{cm}$ and, analogously, if $R_{\perp}(\alpha_e)\ll 10^9\;\text{cm}$ while we already know that the strong inequalities $R_{\perp}(\alpha_e)\ll 10^6\;\text{cm}$ and $R_{\perp}(\alpha_p)\ll 10^6\;\text{cm}$ must hold. Also, it can easily be checked that these previous strong inequalities do not contradict Eqs. (\ref{alpha_e_p_mod}).

Now, if I define the ``mean'' quantity
\begin{equation}
R_{\perp m} =\sqrt{\frac{R_{\perp}^2(\alpha_e)+ R_{\perp}^2(\alpha_p)}{2}}
\end{equation}
then [see Eqs. (\ref{alpha_e_p_mod})]
\begin{equation}
\eta=\sqrt{\eta_e^2+\eta_p^2}\simeq \sqrt{\frac{eB_{\nnearrow}}{2}\left[R_{\perp}^2(\alpha_e)+ 
R_{\perp}^2(\alpha_p)\right]}=\sqrt{eB_{\nnearrow}}R_{\perp m}
\end{equation}
and I can write the transverse probability per unit area as [see Eq. (\ref{dP_T_dalpha})]
\begin{equation}
\label{P_T}
\frac{dP_{\perp}(R_{\perp m})}{dA_{\perp}}\simeq \frac{(eB_{\nnearrow})^2}{2\pi}R_{\perp m}^2 
\end{equation}
where $dA_{\perp}=\pi dR_{\perp m}^2$ is the differential transverse area. Since Eq. (\ref{P_T}) does not depend on $t$, by putting together Eqs. (\ref{dP_L_dt_a}) and (\ref{P_T}) and by dividing by $L_z$, I obtain the presence probability per unit time and unit volume $dV=L_zdA_{\perp}$, as
\begin{equation}
\label{dP_rot}
\frac{dP_{\nnearrow} (t)}{dVdt}\sim \frac{m^4}{2\pi}\left(\frac{B_{\nnearrow}}{4B_{cr}}\right)^3 
(R_{\perp m}\Omega)^2 \frac{\sin(2mt+\pi/4)}{\sqrt{\pi mt}} 
\end{equation}
where the symbol $\sim$ reminds that this is an asymptotic formula valid for $mt\gg 1$. As I have said below Eq. (\ref{dP_lin}), the fact that the probability (\ref{dP_rot}) grows quadratically with $R_{\perp m}$ can be understood in terms of the induced electric field 
$\mathbf{E}_{\nnearrow}(t,\mathbf{r})$ [see Eq. (\ref{E_rot})]. In fact, for a purely rotating magnetic field
\begin{equation}
\label{E_y_E_z}
E_{\nnearrow y}^2(t,\mathbf{r})+E_{\nnearrow z}^2(t,\mathbf{r})=\frac{\Omega^2B_{\nnearrow}^2}{4}x^2.
\end{equation}
In order to connect this quantity with $R_{\perp m}$ I observe that
\begin{equation}
\label{x_R_phi}
x=x_{\perp}=R_{\perp} (t)\cos \phi (t) 
\end{equation}
where $R_{\perp} (t)$ and $\phi (t)$ are the polar coordinates in the plane orthogonal to $\mathbf{B}_{\nnearrow} (t)$. If I define the function $E_{\nnearrow yz}^2(R_{\perp})$ as the average of 
$E_{\nnearrow y}^2(t,\mathbf{r})+E_{\nnearrow z}^2(t,\mathbf{r})$ on the angle $\phi (t)$, I obtain from Eqs. (\ref{E_y_E_z}) and (\ref{x_R_phi})
\begin{equation}
E_{\nnearrow yz}^2(R_{\perp})\equiv \langle E_{\nnearrow y}^2+E_{\nnearrow z}^2\rangle_{\phi}= 
\frac{\Omega^2B_{\nnearrow}^2}{8}R_{\perp}^2.
\end{equation}
With this formula the probability (\ref{dP_rot}) can be written in the different form
\begin{equation}
\label{dP_rot_f}
\frac{dP_{\nnearrow} (t)}{dVdt}\sim \frac{e^2}{16\pi}\frac{B_{\nnearrow}E_{\nnearrow yz}^2(R_{\perp m})}{B_{cr}}\frac{\sin(2mt+\pi/4)}{\sqrt{\pi mt}}
\end{equation}
that points out which is the role of the induced electric field in the pair production process and why the final presence probability per unit volume and unit time depends on $R_{\perp m}^2$. The fact that the probability itself depends only on $E_{\nnearrow y}(t,\mathbf{r})$ and $E_{\nnearrow z}(t,\mathbf{r})$ can be explained by rewriting the transition matrix elements (\ref{gamma_jj_p}) as [see Eq. (\ref{dot_H_E})]
\begin{equation}
\label{m_e_E}
\dot{H}_{jj'}(t)=-e\int d\mathbf{r}\mathbf{E}'_{\nnearrow}(t,\mathbf{r})\cdot \left[u_j^{\prime\dag}(t,\mathbf{r})\boldsymbol{\alpha}v'_{j'}(t,\mathbf{r})\right]
\end{equation}
where
\begin{equation}
\label{E_p}
\mathbf{E}'_{\nnearrow}(t,\mathbf{r})=\frac{1}{2}[\mathbf{r}\times\dot{\mathbf{B}}'_{\nnearrow}(t)]
\end{equation}
is the induced electric field seen from the frame which rotates around the $x$ axis and whose $z$ axis is instantaneously parallel to $\mathbf{B}_{\nnearrow}(t)$. Now, the particular structure of the TGSs [see Eqs. (\ref{u_v_p_g})] makes so that only the $\alpha_z$ term contributes to the transition and this term contains only $E'_{\nnearrow z}(t,\mathbf{r})$ which is a linear superposition only of $E_{\nnearrow y}(t,\mathbf{r})$ and $E_{\nnearrow z}(t,\mathbf{r})$:
\begin{equation}
E'_{\nnearrow z}(t,\mathbf{r})=\cos \Omega t\; E_{\nnearrow z}(t,\mathbf{r})+\sin \Omega t\; E_{\nnearrow y}(t,\mathbf{r}).
\end{equation}

Obviously, it would be interesting to compare the final presence probability (\ref{dP_rot}) obtained here and that obtained in \cite{Calucci} [see Eq. (\ref{dP_lin})]. Nevertheless, this comparison is made hard because
\begin{enumerate}
\item the time evolutions of the magnetic fields (\ref{B_lin}) and (\ref{B_rot}) are very different from each other;
\item Eq. (\ref{dP_lin}) is a presence probability per unit volume while Eq. (\ref{dP_rot}) represents a presence probability per unit volume and unit time;
\item the physical meaning of the quantities $R_{\perp M}$ and $R_{\perp m}$ appearing in Eq. (\ref{dP_lin}) and in Eq. (\ref{dP_rot}) respectively is slightly different.
\end{enumerate}
For these reasons, in order to show explicitly that the pair production process is much more efficient in the presence of a changing-direction magnetic field, I considered an analogous problem to that just treated but in the presence of the magnetic field \cite{DiPiazza}:
\begin{equation}
\label{B_lin_rot}
\mathbf{B}^{\text{lin}}_{\nnearrow}(t)=\begin{pmatrix}
0\\
bt\\
B_0
\end{pmatrix}
\end{equation}
with $b$ and $B_0$ the same parameters as in Eq. (\ref{B_lin}). This magnetic field depends linearly on time as $\mathbf{B}^{\text{lin}}_{\shortuparrow}(t)$ but it also changes direction with time. Also, as in \cite{Calucci}, I have calculated only the presence probability of a pair with the electron and the positron in \emph{pure} Landau levels and not in coherent states as before. Obviously, unlike the case treated in \cite{Calucci}, the dominant contribution to the presence probability is given here by the presence probability that a pair is created with both the electron and the positron in a pure TGS. The calculations are similar to those I have just presented but it is instructive to quote some steps. As in the presence of the purely rotating magnetic field, if I take into account only the production of a pair with the electron and the positron in a TGS, the pair itself can be created here only in the states (\ref{states}). In this case the presence amplitudes corresponding to Eqs. (\ref{gamma}) are given, in the asymptotic limit $mt\gg 1$, by
\begin{subequations}
\label{gamma_lin}
\begin{align}
\gamma^{\text{lin}}_{n,k;n+1,-k}(t) &\sim i\sqrt{\frac{e(n+1)}{128B_0}} \frac{mb}{\varepsilon_k^3}, \\
\gamma^{\text{lin}}_{n,k;n+1,-k}(t) &\sim i\sqrt{\frac{en}{128B_0}} \frac{mb}{\varepsilon_k^3}.
\end{align}
\end{subequations}
By using the previous amplitudes, I obtain the total presence probability in the form
\begin{equation}
\label{P_rot_lin}
P^{\text{lin}}_{\nnearrow}(t)\sim \frac{em^2b^2}{128B_0}\left[\sum_{n=0}^{N^{\text{lin}}_{\nnearrow}(t)}(2n+1)\right]\frac{L_z}{2\pi}\int_{-\infty}^{\infty} \frac{dk}{(m^2+k^2)^3}
\end{equation}
where I have summed on the transverse quantum number $n$ and I have integrated on the longitudinal momentum $k$. Now, the sum on the quantum number $n$ can not be coherently extended to infinity. In fact, it must be stopped to a given value $N^{\text{lin}}_{\nnearrow}(t)$ corresponding, through the eigenvalue equations (\ref{eig_R_xy^2_e}) and (\ref{eig_R_xy^2_p}), to the fixed value $R_{\perp M}^2$ [see also \cite{Calucci}]
\begin{equation}
\label{R_perp_M^2}
R_{\perp M}^2=\frac{2N^{\text{lin}}_{\nnearrow}(t)+1}{eB^{\text{lin}}_{\nnearrow}(t)}
\end{equation}
with $B^{\text{lin}}_{\nnearrow}(t)=\sqrt{B_0^2+(bt)^2}$. This value of $R_{\perp M}^2$ must satisfy analogous upper limit conditions to those satisfied by $R_{\perp}^2(\alpha_e)$ and $R_{\perp}^2(\alpha_e)$, that is the stellar magnetic field must be approximatively uniform inside a region with typical linear length $R_{\perp M}$ and
\begin{equation}
\label{lim_ad_pert_R_M}
\Omega R_{\perp M}\frac{B_0}{B_{cr}}\ll 1.
\end{equation}
Also, since the dominant contribution to $P^{\text{lin}}_{\nnearrow}(t)$ comes from the terms with large $n$, the sum on $n$ can be calculated approximatively by means of the ``semiclassical'' substitutions
\begin{subequations}
\begin{align}
n&\longrightarrow \frac{eB^{\text{lin}}_{\nnearrow}(t)}{2}R_{\perp}^2,\\
\sum_{n=0}^{N^{\text{lin}}_{\nnearrow}(t)}&\longrightarrow \frac{eB^{\text{lin}}_{\nnearrow}(t)}{2}\int_0^{R_{\perp M}^2}dR_{\perp}^2.
\end{align}
\end{subequations}
Finally, by performing the remaining integral on $k$ in Eq. (\ref{P_rot_lin}) and by dividing by the total volume $V=L_z\pi R_{\perp M}^2$, I obtain the final presence probability per unit volume as
\begin{equation}
\label{dP_rot_lin}
\frac{dP^{\text{lin}}_{\nnearrow}(t)}{dV}\sim\frac{3}{\pi}\left[\frac{B^{\text{lin}}_{\nnearrow}(t)}{B_{cr}}\right]^2\left(\frac{bR_{\perp M}}{64B_0}\right)^2meB_0.
\end{equation}
Clearly, the procedure to obtain this quantity is less rigorous than that used to obtain Eq. (\ref{dP_rot}) because the pure TGSs are not well-localized states while the definition of a probability per unit volume is, in general, local. Nevertheless, it can be shown that by using here the more rigorous procedure used in the case of a purely rotating magnetic field, the final result would be again Eq. (\ref{dP_rot_lin}) but with $R_{\perp m}$ instead of $R_{\perp M}$.

At this point, to show that the production of pairs is much more efficient in the presence of a changing-direction magnetic field, I divide Eq. (\ref{dP_lin}) by Eq. (\ref{dP_rot_lin}) and, apart from numerical factors of the order of one, I obtain
\begin{equation}
\label{comp_rot_fix}
\frac{dP^{\text{lin}}_{\shortuparrow}(t)}{dP^{\text{lin}}_{\nnearrow}(t)}\sim\frac{B_{cr}B^{\text{lin}}_{\shortuparrow}(t)}{[B^{\text{lin}}_{\nnearrow}(t)]^2}\sqrt{\frac{B_{cr}}{B_0}}<\sqrt{2}\left(\frac{B_{cr}}{B_0}\right)^{3/2}\ll 1
\end{equation}
which shows, in fact, that \emph{in the strong magnetic field regime the pair presence probability is much larger in the presence of a changing-direction magnetic field}.
%
%
\section{Summary and conclusions}
In this Chapter I have continued the study started in \cite{Calucci} about the production of electron-positron pairs in the presence of a strong, uniform and slowly-varying magnetic field. 

We had already seen in the previous Chapter that, in the presence of a strong magnetic field, the so-called electron and positron TGSs are ``privileged'' states because their energies are independent of the magnetic field and then are much less than the other excited Landau levels. Now, by using the first-order adiabatic perturbation theory, I have shown in Par. \ref{Prel_calc} that the time-variation of the magnetic field can be connected with some selection rules concerning the state in which the pair can be created [see Tab. \ref{Tab_sel_rul} and Eq. (\ref{sel_rule})]. In particular, I have obtained that \emph{only if the direction of the magnetic field changes with time it is possible to create a pair with the electron and the positron both in a TGS}. This fact gives the changing-direction magnetic field configurations a particular relevance. In other words, \emph{the possibility of producing pairs with the electron and the positron both in a TGS makes a changing-direction magnetic field much more efficient than a fixed-direction magnetic field from the point of view of pair production}. This fact has been verified quantitatively by comparing the two pair presence probabilities per unit volume (\ref{dP_lin}) and (\ref{dP_rot_lin}). Both these probabilities have been calculated by means of the first-order adiabatic perturbation theory and in the presence of a magnetic field with a linear dependence on time. But, the probability (\ref{dP_lin}) refers to the fixed direction magnetic field (\ref{B_lin}) while the probability (\ref{dP_rot_lin}) refers to changing-direction magnetic field (\ref{B_lin_rot}) and, actually, the ratio (\ref{comp_rot_fix}) shows explicitly how the pair presence probability in the second case is much larger than the corresponding quantity calculated in the first case.

Finally, the pair presence probability has also been calculated in the presence of a purely rotating magnetic field. Actually, I have calculated the presence probability per unit volume and unit time of a pair with the electron and the positron both in a \emph{coherent} TGS [see Eqs. (\ref{pair_coh_st}) and (\ref{coh_st})]. These states are well spatially localized in the plane orthogonal to the magnetic field and this fact gave me the possibility to better understand the role of the \emph{nonuniform} electric field induced by the rotating magnetic field. In particular, I have shown that \emph{the pair presence probability per unit volume and unit time depends on the square of the electric field} [see Eq. (\ref{dP_rot_f})]. Actually, from a physical point of view it was clear a priori that the production of a pair would have vanished in the absence of an electric field. Nevertheless, in this respect, Eq. (\ref{dP_rot_f}) also shows \emph{an additional dependence on $B_{\nnearrow}/B_{cr}$ that can be interpreted as a \emph{direct} effect of the presence of the magnetic field}.

\chapter{Photon production in a strong, uniform and slowly-rotating magnetic field}
Even if the results of the previous Chapter about the production of electron-positron pairs are self consistent, I want to use them now to study some processes through that the electrons and the positrons created may produce photons. In fact, as I have said in Sect. \ref{GRB}, a possible application of my theoretical calculations is the study of GRBs and of their properties. In particular, in Sect. \ref{Ann} I will calculate the energy spectrum per unit time of the photons produced through the annihilation of already created electrons and positrons \cite{DiPiazza3}. Analogously, in Sect. \ref{Syn} I will evaluate the energy spectrum of the photons emitted as synchrotron radiation by already created electrons and positrons \cite{DiPiazza6}. Since one of the conclusions of the previous Chapter was that a rotating magnetic field primes very efficient mechanisms of electron-positron pair production, I will consider in this Chapter only the production of photons in the presence of this kind of magnetic field. Finally, I have also calculated in \cite{DiPiazza3} the energy spectrum of photons emitted directly from vacuum in the presence of the same magnetic field configuration. Nevertheless, this photon production mechanism is very inefficient and I will not report here all the calculations but only some intermediate steps and the final photon spectrum (Sect. \ref{Dir}). 

%
%
\section{Photon production through pairs annihilation}
\label{Ann}
Before starting the calculations I want to make a general observation concerning the approximations made about the spatiotemporal structure of the magnetic fields I deal with. As it will be clear, this observation is independent of the photon production process at hand. In fact, as I have just mentioned, I will consider here the production of photons in the presence of the already introduced strong, uniform and slowly-rotating magnetic field given in Eq. (\ref{B_rot}) and that, for the sake of clarity, I write again:
\begin{equation}
\label{B_rot_4}
\mathbf{B}_{\nnearrow}(t)=
B_{\nnearrow}\begin{pmatrix}  
0 \\
\sin\Omega t\\         
\cos\Omega t
\end{pmatrix}.
\end{equation}
Now, in the previous Chapter I was allowed to assume the magnetar magnetic field as uniform (in the production region) and slowly-varying (during the production process) because the nonvanishing mass $m$ of the electron (positron) provided a microscopic length and time scale given by the Compton length $\lambdabar=1/m$. Instead, here I will deal with photons that are massless then, in order to keep the previous approximations, \emph{I have to assume explicitly to restrict my attention to photons with energies $\omega$ such that $\Omega\ll\omega$}. Actually, this strong inequality does not constraint at all the following results from a physical point of view. In fact, we have seen that in the astrophysical context under consideration $\Omega\sim 1\;\text{s}^{-1}$ while the energies of the photons in a GRB pulse are typically larger than $10^{-2}\;\text{MeV}$ corresponding to an angular frequency of $1.6\times 10^{19}\;\text{s}^{-1}$.

Now, in Sect. \ref{Pai_prod} I have shown that pairs can be created in the presence of a strong, uniform and slowly-rotating magnetic field with the electron and the positron both in a \emph{coherent} TGS. By using the same technique sketched at the end of the previous Chapter, one can also obtain the probability per unit volume that a pair is present at time $t$ with the electron (positron) in any \emph{pure} TGS with longitudinal momentum between $k$ and $k+dk$ ($-k$ and $-k-dk$). By indicating this probability as $f(k,t)dk$ it can easily be shown that
\begin{equation}
\label{f}
f(k,t)=\frac{1}{2\pi^2}\left[\frac {eB_{\nnearrow}}{4\varepsilon^2(k)}\right]^3 m^2(\Omega R_{\perp M})^2\sin^2\varepsilon(k)t
\end{equation}
with $\varepsilon(k)=\sqrt{m^2+k^2}$. It is worth pointing out the dependence of $f(k,t)$ on $\varepsilon^{-6}(k)$. In fact, this means that in the physical situation under consideration the production of high-energy electrons and positrons is strongly suppressed and this fact will also affect the production of high energetic photons. 

Now, in Appendix B I show that the quantity $f(k,t)dk$ can be interpreted as the mean number of electrons or, symmetrically, positrons per unit volume present at time $t$ with a longitudinal momentum between $k$ and $k+dk$. This allows me to use the distribution (\ref{f}) to calculate the annihilation photon spectrum per unit time, \textit{i. e.} the number of photons per unit energy, unit volume and unit time that are produced as a consequence of the annihilations of the electrons and of the positrons previously created, by using the formula \cite{Akhiezer,Landau2}
\begin {equation}
\label{dN_domega}
\frac{dN^{(\text{ann})}(\omega,t)}{d\omega dV dt}=\int dk dk'
\frac{d\sigma (k,k',\omega)}{d\omega}\tilde{v}(k,k')f(k,t) f(k',t).
\end{equation}
In this formula, $d\sigma (k,k',\omega)/d\omega$ is the cross section per unit of photon energy of the pair annihilation into two photons process and $\tilde{v}(k,k')$ is the flux factor of the colliding particles defined as:
\begin{equation}
\tilde{v}(k,k')\equiv J(k,k')V
\end{equation}
where $J(k,k')$ is the flux density of the colliding particles and $V$ is the quantization volume. I will give below a more precise definition of the quantity $d\sigma (k,k',\omega)/d\omega$ and its explicit form. At the moment, I observe that this picture is already approximated. In fact, I am considering the distribution $f(k,t)$ as a given function while its time dependence is affected by the annihilation process itself. Nevertheless, I neglect this fact for the moment and I will deal with it at the end of this Section.

In order to calculate the quantity $d\sigma (k,k',\omega)/d\omega$, I start by writing the cross section $d\sigma$ of the annihilation of an electron with fourmomentum $(\varepsilon,\mathbf{k})$ and spin $s$ and a positron with fourmomentum $(\varepsilon',\mathbf{k}')$ and spin $s'$ into two photons with fourmomenta $(\omega,\mathbf{q})$ and $(\omega',\mathbf{q}')$ and polarizations $\lambda$ and $\lambda'$ respectively. Actually, the process of pair annihilation will take place in the magnetic field, which therefore affects all the dynamical processes. However, in computing the photon production rate I consider that the main dynamical effect of the magnetic field is the pair production. The rest of the process will be therefore calculated neglecting the effects of the magnetic field both on the annihilation process and on the photon final states. Following this approximation, I need the cross section $d\sigma$ of the pair annihilation into two photons process in the vacuum. This quantity may be found in many textbooks and I will quote (in my notation) Eq. (33.2) in \cite{Akhiezer}:
\begin {equation}
\label{dsigma}
d\sigma=\frac{4\pi^2\alpha^2_{em}}{\omega\omega'\tilde{v}(k,k')}|\bar{u}Qv|^2\frac{d\mathbf{q}d\mathbf{q}'}{(2\pi)^2}\delta(\mathbf{k}+\mathbf{k}'-\mathbf{q}-\mathbf{q}')\delta(\varepsilon+\varepsilon'-\omega-\omega')
\end{equation}
where $\bar{u}Qv$ is the annihilation matrix element and $\alpha_{em}=e^2/(4\pi)$ the fine-structure constant. In particular, even if the electrons and the positrons in the present case have a preferential direction of the spin [see Eq. (\ref{states})], I will also neglect this aspect and I will use the cross section summed over the photon polarizations and averaged over the electron and positron spins. Eq. (\ref{dsigma}) remains unchanged but the quantity $|\bar{u}Qv|^2$ must be replaced by \cite{Akhiezer}
\begin{equation}
\label{ME}
\overline{\sum|\bar{u}Qv|^2}=-\frac{1}{2\varepsilon\varepsilon'}\left[4\left(\frac{1}{\varkappa}+\frac{1}{\varkappa'}\right)^2-4\left(\frac{1}{\varkappa}+\frac{1}{\varkappa'}\right)-\left(\frac{\varkappa}{\varkappa'}+\frac{\varkappa'}{\varkappa}\right)\right]
\end{equation}
with
\begin{subequations}
\begin{align}
\varkappa &=\frac{2\omega}{m^2}(\varepsilon-k\cos\theta),\\
\varkappa' &=\frac{2\omega}{m^2}(\varepsilon'-k'\cos\theta)
\end{align}
\end{subequations}
where I have assumed that the colliding particles have momenta along the $z$ axis and that $\theta$ is the angle between the photon momentum $\mathbf{q}$ and this axis (this means that $k$ and $k'$ can be positive or negative).\footnote{Since the time $t$ in Eq. (\ref{dN_domega}) is fixed, I assume to work in a reference system which has the $z$ axis along the instantaneous direction of the magnetic field $\mathbf{B}_{\nnearrow}(t)$.} 

By using the distribution (\ref{f}) in Eq. (\ref{dN_domega}), I take into account, from a kinematical point of view, the anisotropy induced by the presence of the magnetic field. From the dynamical point of view, I should take into account this anisotropy by substituting the matrix element $\bar{u}Qv$ with the corresponding one calculated in the presence of the strong magnetic field $\mathbf{B}'_{\nnearrow}=(0,0,B_{\nnearrow})$. The calculation is very complicated and it involves the use of the \emph{Schwinger propagator} \cite{Schwinger}, \textit{i. e.} the electron propagator in the presence of a constant and uniform magnetic field. In Appendix C I will analyze the matrix element so calculated in order to give the correction induced by the use of the Schwinger propagator and to find the physical conditions in which the previous approximated treatment is correct.

Now, the threedimensional $\delta$ function in Eq. (\ref{dsigma}) can be exploited to perform the integral on $\mathbf{q}'$. Also, since I need the cross section as a function of the energy of one of the photons created (the other being fixed by the energy $\delta$ function), I integrate the cross section itself with respect to the angular variables. The result is
\begin{equation}
\label{dsigma_domega}
\frac{d\sigma (k,k',\omega)}{d\omega}=\frac{\omega\alpha^2_{em}}{\tilde{v}(k,k')}\int_{-1}^1 d(\cos\theta)\frac{\overline{\sum|\bar{u}Qv|^2}}{\omega'}\delta(\varepsilon+\varepsilon'-\omega-\omega')\pi
\end{equation}
where the integral on the azimuthal angle gives a factor $\pi$ to take into account that the two final photons are indistinguishable and where
\begin{equation}
\omega'\equiv|\mathbf{q}'|=|\mathbf{k}+\mathbf{k}'-\mathbf{q}|=
\sqrt{(k+k')^2+\omega^2-2\omega(k+k')\cos\theta}.
\end{equation}
By substituting this expression of $\omega'$ in the argument of the $\delta$ function in Eq. (\ref{dsigma_domega}) I can perform the integral on $\cos\theta$ and the final result is
\begin{equation}
\label{dsigma_domega_f}
\frac{d\sigma (k,k',\omega)}{d\omega}=
\frac{\pi\alpha^2_{em}}{\tilde{v}(k,k')}\frac{\overline{\sum|\bar{u}Qv|^2}\big\vert_{\cos\theta=\cos\theta_0}}{|k+k'|}\vartheta(1-|\cos\theta_0|)
\end{equation}
where $\vartheta(x)$ is the step function and where
\begin{equation}
\label{cos}
\cos\theta_0=\frac{(k+k')^2+\omega^2-(\varepsilon+\varepsilon'-\omega)^2}{2\omega(k+k')}.
\end{equation}

At this point I have all the ingredients to calculate the photon energy spectrum per unit time by means of Eq. (\ref{dN_domega}).\footnote{Note that the integral in Eq. (\ref{dN_domega}) seems to be divergent because of the factor $1/|k+k'|$ in Eq. (\ref{dsigma_domega_f}), but, actually, the constraint $|\cos\theta_0|\le 1$ prevents this divergence.} The region of integration in Eq. (\ref{dN_domega}) can be divided into the four sectors:
\begin{subequations}
\begin{align}
& k\ge 0, k'\ge 0,\\
& k\ge 0, k'< 0,\\
& k< 0, k'\ge 0,\\
& k< 0, k'< 0  
\end{align}
\end{subequations}
and it can easily be seen that the integrals in the sectors I and II are equal to those in the sectors IV and III respectively. Also, the time dependence of the spectrum (\ref{dN_domega}) is carried by the oscillating functions in the distribution $f(k,t)$ [see Eq. (\ref{f})]. Since I am interested in macroscopic times such that $mt\gg 1$, I can calculate directly the mean value of Eq. (\ref{dN_domega}) for large times by means of the substitutions $\langle\sin^2\varepsilon(k)t\rangle=\langle\sin^2\varepsilon(k')t\rangle\sim 1/2$. Finally, by using the adimensional variables $\eta=m/\varepsilon=m/\varepsilon(k)$ and $\eta'=m/\varepsilon'=m/\varepsilon(k')$ that vary in the unit square, Eq. (\ref{dN_domega}) can be written in the form
\begin{equation}
\label{dN_domega_a}
\begin{split} 
\left\langle\frac{dN^{(\text{ann})}(\omega,t)}{d\omega dV dt}\right\rangle\sim &\frac{\alpha^2_{em}m^3}{2\pi^3}\left(\frac{\Omega R_{\perp M}}{2\pi}\right)^4\left(\frac{B_{\nnearrow}}{B_{cr}}\right)^6\\
&\times\int_0^1 \int_0^1 d\eta d\eta'\frac{M(\eta,\eta',w,+1)+M(\eta,\eta',w,-1)}{\sqrt{(1-\eta^2)(1-\eta'^2)}}
\end{split}
\end{equation}
where $w=\omega/m$ and
\begin{equation}
\begin{split}
M(\eta,\eta',w,\zeta)=&
\frac{(\eta\eta')^6}{\left|\eta\sqrt{1-\eta'^2}+\zeta\eta'\sqrt{1-\eta^2}\right|}\vartheta(1-|\cos\theta_0(\eta,\eta',w,\zeta)|)\\
&\qquad\times\bigg\{\frac{\varkappa(\eta,\eta',w,\zeta)}{\varkappa'(\eta,\eta',w,\zeta)}+\frac{\varkappa'(\eta,\eta',w,\zeta)}{\varkappa(\eta,\eta',w,\zeta)}\\
&\qquad\qquad-4\left[\frac{1}{\varkappa(\eta,\eta',w,\zeta)}+\frac{1}{\varkappa'(\eta,\eta',w,\zeta)}\right]^2\\
&\qquad\qquad+4\left[\frac{1}{\varkappa(\eta,\eta',w,\zeta)}+\frac{1}{\varkappa'(\eta,\eta',w,\zeta)}\right]\bigg\}
\end{split}
\end{equation}
with
\begin{subequations}
\begin{align}
\varkappa (\eta,\eta',w,\zeta) &=\frac{2w}{\eta}\left[1-\sqrt{1-\eta^2}\cos\theta_0(\eta,\eta',w,\zeta)\right],\\
\varkappa'(\eta,\eta',w,\zeta) &=\frac{2w}{\eta'}\left[1-\zeta\sqrt{1-\eta'^2}\cos\theta_0(\eta,\eta',w,\zeta)\right],\\ 
\cos\theta_0(\eta,\eta',w,\zeta)
&
=\frac{w(\eta+\eta')+\zeta\sqrt{1-\eta^2}\sqrt{1-\eta'^2}-\eta\eta'-1}{w\left(\eta'\sqrt{1-\eta^2}+\zeta\eta\sqrt{1-\eta'^2}\right)}.
\end{align}
\end{subequations}

\begin{figure}[ht]
\begin{center}
\includegraphics[angle=90,width=\textwidth]{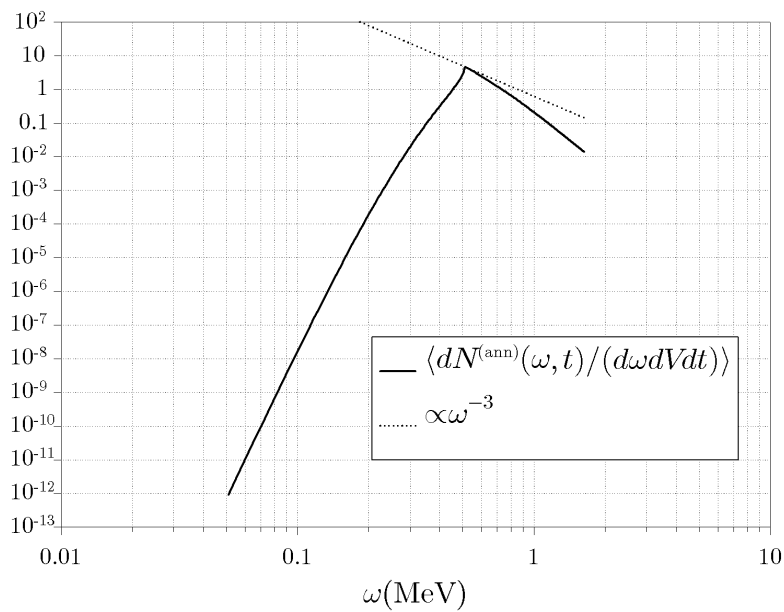}
\end{center}
\caption{Photon spectrum per unit time $\left\langle dN^{(\text{ann})}(\omega,t)/(d\omega dV dt)\right\rangle$ in arbitrary units. The 
dotted curve represents a function proportional to $\omega^{-3}$.}
\label{Spec_ann}
\end{figure}

The analytical integrations in Eq. (\ref{dN_domega_a}) are very difficult because the trivial condition $|\cos\theta_0(\eta,\eta',w,\zeta)|\le 1$ is a complicated condition on the integration domain over $\eta$ and $\eta'$, so I shall present the result of a numerical calculation (see Fig. \ref{Spec_ann}).
Although, the quantity $\left\langle dN^{(\text{ann})}(\omega,t)/(d\omega dV dt)\right\rangle$ is a spectrum per unit time I want to compare it at least qualitatively with the two typical GRBs energy spectra shown in Fig. \ref{GRB_en_sp}. Unfortunately, the shape of the theoretical spectrum (\ref{dN_domega_a}) results also qualitatively very different from those of GRBs (see Fig. \ref{GRB_en_sp}) and then I can conclude that \emph{the pair annihilation mechanism can not be the dominant one giving rise to a GRB}. Nevertheless, similarly to experimental GRBs energy spectra, it shows a break energy $\omega_b^{(\text{ann})}$ of the order of $0.1\;\text{MeV}$ (see the end of Par. \ref{GRB_char}). Actually, the value of $\omega_b^{(\text{ann})}$ is precisely around the value $0.51\;\text{MeV}$ of the electron mass and this result can be interpreted in the following way. In the center-of-momentum system the pair will yield two photons with equal energy and thus with $\omega\ge m$. In order to get a soft photon one needs a large boost of the pair system and this would unavoidably result in a very energetic electron (or positron), but the distribution $f(k,t)$ decreases rapidly for large $k$ and so this process has small probability [see Eq. (\ref{f})]. For the same reason the production of very energetic photons is also unlikely and the high-energy region of the spectrum in Fig. \ref{Spec_ann} decreases more rapidly than $\omega^{-3}$. I point out that the comparison with a function proportional to $\omega^{-3}$ is motivated because, as we have seen at the end of Par. \ref{GRB_char}, the high-energy part of the experimental GRBs spectra are well fitted by a function proportional to $\omega^{-\beta}$ with $\beta\simeq 2\text{--}3$. Finally, I want to observe that the sharpness of the peak at $0.51\;\text{MeV}$ is also due to the fact that only the TGSs have been put into the electron and positron distributions $f(k,t)$. In fact, the other possible states correspond to excited Landau levels and the energy of the electron (or of the positron) in such states has in the center-of-momentum system a minimum which is higher than $m$. Consequently, taking into account these states would make the maximum in Fig. \ref{Spec_ann} less sharp.

As I have mentioned before, the production of photons through pair annihilation has as a consequence, of course, the depletion of the electron and positron populations. In fact, even if the two processes of pair production and pair annihilation are not disjoint in time, I performed the calculation keeping them separated because, in the presence of a purely rotating magnetic field, the electron (positron) population does not grow continuously but it reaches a stationary density with superimposed rapid fluctuations [see Eq. (\ref{f})]. Now, the electron population is the same as the positron population and this equality is clearly kept by the annihilation process. However, the rate of change of these populations depends in general on $k$ and the time evolution may be very complicated.\footnote{For example, in this case one should also take into account the scattering processes among the electrons and the positrons that do not change the number of particles but that do change their momenta.} An overall indication of the time variation may be obtained by assuming that the momentum distributions of the electron and positron populations do not depend on time. In this approximation the electron (positron) distribution, that I also indicate as $f(k,t)$, can be written in the factorized form
\begin{equation}
f(k,t)=K(k)T(t).
\end{equation}
Now, it can always assumed that $\int dk K(k)=1$ in such a way the function $T(t)$ is given by
\begin{equation}
T(t)=\int dk f(k,t).
\end{equation}
Also, I observe that the photons production rate is twice with respect to the rate of the electrons annihilation, then, introducing the quantity $\sigma (k,k')=\int d\omega d\sigma(k,k',\omega)/d\omega$ that is the total cross section of the annihilation of an electron and a positron with energies $\varepsilon(k)$ and $\varepsilon(k')$ respectively into two photons, I get from Eq. (\ref{dN_domega})
\begin {equation}
\label{Eq_time_dep}
\dot{T}(t)=\int dk \partial_t f(k,t)=-\frac{1}{2}\int d\omega\frac{dN^{(\text{ann})}(\omega,t)}{d\omega dVdt}=-s^2T^2(t)
\end{equation}
with
\begin {equation}
s^2=\frac{1}{2}\int dk dk' \sigma (k,k')\tilde{v}(k,k')K(k)K(k').
\end{equation}

The solution of the differential equation (\ref{Eq_time_dep}) is
\begin{equation}
T(t)=\frac{T_0}{1+s^2T_0t}
\end{equation}
with $T_0=T(0)$, so I can conclude that the photon production rate $-2\dot{T}(t)$ decreases at large $t$ as $t^{-2}$. It is not difficult to verify that, in the same hypotheses, if I started with unbalanced populations (more electrons than positrons, or the opposite), then the photons production rate would show an exponential decay with time. 
%
%
\section{Photon production as synchrotron emission}
\label{Syn}
As I have said in Par. \ref{Fireball_model}, there are strong evidences that the photons forming a GRB are emitted as synchrotron radiation by the electrons and the positrons in the fireball. I have also said that the fact that the gamma-ray radiation of the GRBs is high linearly polarized leads to think that the magnetic field in which the radiation itself is emitted is produced by the central engine of the GRB that is by a magnetar or a black hole. For this reason, I want to calculate here the energy spectrum of the photons emitted as synchrotron radiation by electrons and positrons in the presence of the rotating magnetic field (\ref{B_rot_4}) \cite{DiPiazza6}. As in the previous Section, the choice of the rotating field configuration ensures an easier mathematical treatment but it is also justified because the production of electron-positron pairs (and then of photons) is much more efficient with respect to the production in the presence of a magnetic field 
varying only in strength. By contrast, in the case of photon production through pair annihilation I have used the electron and positron distributions previously calculated to obtain, by means of the pair annihilation cross section, the photon spectrum per unit time. In this case it is easier to start from the beginning and to calculate the matrix elements corresponding to the whole process ``pair production+photon emission''.
%
%
\subsection{Theoretical model}
By considering the process I want to study, a good theoretical starting point is the Lagrangean density $\mathscr{L}$ of QED in the presence of an external electromagnetic field. If $A_{\nnearrow\alpha}(t,\mathbf{r})=[0,-\mathbf{A}_{\nnearrow}(t)]$ with
\begin{equation}
\label{A_rot}
\mathbf{A}_{\nnearrow}(t,\mathbf{r})=-\frac{1}{2}[\mathbf{r}\times\mathbf{B}_{\nnearrow}(t)]
\end{equation}
is the fourpotential describing the external rotating magnetic field then
\begin{equation}
\label{L_syn}
\mathscr{L} =\bar{\psi}
\left\{\gamma^{\alpha}[i\partial_{\alpha}+eA_{\nnearrow\alpha}(t,\mathbf{r})+
e\Euler{A}_{\alpha}]-m\right\}\psi-\frac{1}{4}\Euler{F}_{\alpha\beta}\Euler{F}^{\alpha\beta}
\end{equation}
where the radiation field $\Euler{A}_{\alpha}(t,\mathbf{r})=
\bigl[\Euler{V}(t,\mathbf{r}),-\boldsymbol{\Euler{A}}(t,\mathbf{r})\bigr]$ is assumed in the Coulomb gauge 
\begin{subequations}
\begin{align}
\Euler{V}(t,\mathbf{r}) &=0,\\
\boldsymbol{\partial}\cdot\boldsymbol{\Euler{A}}(t,\mathbf{r}) &=0
\end{align}
\end{subequations}
and where
\begin{equation}
\Euler{F}_{\alpha\beta}(t,\mathbf{r})=
\partial_{\alpha}\Euler{A}_{\beta}(t,\mathbf{r})-\partial_{\beta}\Euler{A}_{\alpha}(t,\mathbf{r}).
\end{equation}
In Eq. (\ref{L_syn}) the two terms proportional to $F_{\nnearrow\alpha\beta}F_{\nnearrow}^{\alpha\beta}$ and to $F_{\nnearrow\alpha\beta}\Euler{F}^{\alpha\beta}$ with $F_{\nnearrow\alpha\beta}(t,\mathbf{r})=
\partial_{\alpha}A_{\nnearrow\beta}(t,\mathbf{r})-\partial_{\beta}A_{\nnearrow\alpha}(t,\mathbf{r})$ have been omitted because they do not give any significant contribution to the equations of motion of the fields $\psi(t,\mathbf{r})$ and $\Euler{A}(t,\mathbf{r})=[\Euler{A}^0(t,\mathbf{r}),\ldots,\Euler{A}^3(t,\mathbf{r})]$ and to the process I want to study.

Now, the Lagrangean density (\ref{L_syn}) is not in the most suitable form for the physical scenario I want to describe. In fact, as it stands it would be suitable for dealing with 
a \emph{weak} external magnetic field because the usual perturbation theory could 
be used, while I am dealing with a \emph{strong} magnetic field. Nevertheless, we also know that the perturbation induced by the macroscopic magnetic field (\ref{B_rot_4}) can be assumed 
to be \emph{adiabatic}. For this reason, I first perform the time-dependent rotation
\begin{equation}
\label{rot}
\mathbf{r}'(t)\equiv[x'(t),y'(t),z'(t)]=(x,y\cos\Omega t-z\sin\Omega t,y\sin\Omega t+z\cos\Omega t).
\end{equation}
As a consequence, the spinor field $\psi(t,\mathbf{r})$ and the fourvector field $\Euler{A}(t,\mathbf{r})$ transform as
\begin{subequations}
\label{psi_A_p}
\begin{align}
\label{psi_p}
\psi'(t,\mathbf{r}'(t)) &=\exp\left(-i\dfrac{\sigma_x}{2}\Omega t\right)\psi(t,\mathbf{r}),\\
\label{A_p_r}
\Euler{A}'(t,\mathbf{r}'(t)) &=
\exp\left(-iS_x\Omega t\right)\Euler{A}(t,\mathbf{r})
\end{align}
\end{subequations}
where the matrices $\sigma_x$ and $S_x$ are given by
\begin{align}
\sigma_x=
\left(\begin{matrix}
0 & 1 & 0 & 0\\
1 & 0 & 0 & 0\\
0 & 0 & 0 & 1\\
0 & 0 & 1 & 0
\end{matrix}\right), &&
S_x=
\left(\begin{matrix}
0 & 0 & 0 & 0\\
0 & 0 & 0 & 0\\
0 & 0 & 0 & i\\
0 & 0 & -i & 0
\end{matrix}\right).
\end{align}
I point out that, although $\sigma_x$ and $S_x$ are two $4\times 4$ matrices, 
they act on two different spaces: the first one acts on the spinor space and the second one 
acts on the fourvector space labeled by the Lorentz indices $\{0,\ldots,3\}$.

Now, by using the equality
\begin{equation}
\exp\left(-i\dfrac{\sigma_x}{2}\Omega t\right)=
\cos\left(\frac{\Omega t}{2}\right)-i\sigma_x\sin\left(\frac{\Omega t}{2}\right)
\end{equation}
and by exploiting the usual commutation rules among the Dirac $\gamma$ matrices, 
it can easily be shown that
\begin{equation}
\label{gamma_p}
\exp\left(-i\dfrac{\sigma_x}{2}\Omega t\right)\gamma
\exp\left(i\dfrac{\sigma_x}{2}\Omega t\right)=\exp\left(iS_x\Omega t\right)\gamma
\end{equation}
with $\gamma\equiv(\gamma^0,\ldots,\gamma^3)$. This equation is nothing but the mathematical expression of the fact that 
the current density $\bar{\psi}(t,\mathbf{r})
\gamma\psi(t,\mathbf{r})$ transforms under the rotation (\ref{rot}) as a fourvector 
[see Eq. (\ref{A_p_r})]:
\begin{equation}
\label{j_p}
\bar{\psi}\gamma\psi=\bar{\psi}'
\exp\left(-i\dfrac{\sigma_x}{2}\Omega t\right)\gamma
\exp\left(i\dfrac{\sigma_x}{2}\Omega t\right)\psi'=
\exp\left(iS_x\Omega t\right)\bar{\psi}'\gamma\psi'
\end{equation}
where, from now on, the ``primed'' fields are always intended to be calculated at the ``primed'' coordinates $\mathbf{r}'(t)$.

These previous equations can be exploited to rewrite the 
Lagrangean density (\ref{L_syn}) in terms of the primed variables and fields. 
In particular, it is evident from Eqs. (\ref{A_p_r}) and (\ref{j_p}) that
\begin{equation}
\label{int_rad_p}
\bar{\psi}\gamma^{\alpha}\psi \Euler{A}_{\alpha}=\bar{\psi}'
\gamma^{\alpha}\psi'\Euler{A}'_{\alpha}.
\end{equation}
The transformation of the terms involving the external magnetic 
field are more complicated. In fact, from Eqs. (\ref{B_rot_4}), (\ref{A_rot}) and (\ref{gamma_p}) I obtain
\begin{equation}
\label{int_ext_p}
\begin{split}
\bar{\psi}\gamma^{\alpha}\psi A_{\nnearrow\alpha}(t,\mathbf{r}) &=\frac{B_{\nnearrow}}{2}\left[\bar{\psi}'\gamma^1\psi'y'(t)-\bar{\psi}'\gamma^2\psi'x'(t)\right]=\bar{\psi}'\gamma^{\alpha}\psi'A'_{\nnearrow\alpha}(\mathbf{r}'(t))
\end{split}
\end{equation}
where I have introduced the fourpotential $A'_{\nnearrow\alpha}(\mathbf{r}'(t))=
\big[0,-\mathbf{A}'_{\nnearrow}(\mathbf{r}'(t))\big]$ with $\mathbf{A}'_{\nnearrow}(\mathbf{r}'(t))=-[\mathbf{r}'(t)\times\mathbf{B}'_{\nnearrow}]/2$ and $\mathbf{B}'_{\nnearrow}=(0,0,B_{\nnearrow})$. 

Now, I want to show in detail how the terms containing derivatives of the fields in the Lagrangean density (\ref{L_syn}) transform. I 
first transform separately the time derivatives $\partial_t\psi(t,\mathbf{r})$ and 
$\partial_t \Euler{A}(t,\mathbf{r})$. From Eqs. (\ref{psi_A_p}) and 
by reminding that the variables $\mathbf{r}'(t)$ depend on time [see Eq. (\ref{rot})], I have
\begin{subequations}
\begin{align}
\label{psi_t_p}
\partial_t\psi(t,\mathbf{r}) &=\exp\left(i\dfrac{\sigma_x}{2}\Omega t\right)\left[i\Omega\mathcal{J}^{(1/2)\prime}_x\psi'(t,\mathbf{r}'(t))+\partial_t\psi'(t,\mathbf{r}'(t))\right],\\
\label{A_t_p}
\partial_t \Euler{A}(t,\mathbf{r}) &=\exp\left(iS_x\Omega t\right)\left[i\Omega \mathcal{J}^{(1)\prime}_x\Euler{A}'(t,\mathbf{r}'(t))+\partial_t \Euler{A}'(t,\mathbf{r}'(t))\right].
\end{align}
\end{subequations}
where I have introduced the $x$ components 
\begin{subequations}
\begin{align}
\label{j_1/2}
\mathcal{J}^{(1/2)\prime}_x &=\mathcal{L}'_x+\frac{\sigma_x}{2},\\
\label{j_1}
\mathcal{J}^{(1)\prime}_x &=\mathcal{L}'_x+S_x
\end{align}
\end{subequations}
of the one-particle electron (positron) and photon total angular momentum operators respectively. On the one hand, by means of Eq. (\ref{psi_t_p}) it can be seen that
\begin{equation}
\label{der_p}
\begin{split}
\bar{\psi}\gamma^{\alpha}\partial_{\alpha}\psi &=\bar{\psi}'
\exp\left(-i\dfrac{\sigma_x}{2}\Omega t\right)\gamma^0\partial_0
\left[\exp\left(i\dfrac{\sigma_x}{2}\Omega t\right)\psi'\right]+\bar{\psi}'\gamma^1\partial'_1\psi'\\
&\quad+\bar{\psi}'\left(\gamma^2\cos\Omega t+\gamma^3\sin\Omega t\right)\left(\cos\Omega t\;\partial'_2+\sin\Omega t\;\partial'_3\right)\psi'\\
&\quad+\bar{\psi}'\left(-\gamma^2\sin\Omega t+\gamma^3\cos\Omega t\right)\left(-\sin\Omega t\;\partial'_2+\cos\Omega t\;\partial'_3\right)\psi'\\
&=\bar{\psi}'\gamma^{\alpha}\partial'_{\alpha}\psi'
+i\Omega\bar{\psi}'\gamma^0\mathcal{J}^{(1/2)\prime}_x\psi',
\end{split}
\end{equation}
where, for notational simplicity, I do not indicate the dependence on time of the derivatives with respect to the primed coordinates. On the other hand, Eq. (\ref{A_t_p}) can be used to see how the free radiation field Lagrangean density in Eq. (\ref{L_syn}) transforms. Firstly, I write it in the more useful noncovariant form 
\begin{equation}
\label{L_r_fr}
-\frac{1}{4}\Euler{F}_{\alpha\beta}\Euler{F}^{\alpha\beta}=
\frac{1}{2}\big[\left(\partial_t\boldsymbol{\Euler{A}}\right)^2-\left(\boldsymbol{\partial}\times \boldsymbol{\Euler{A}}\right)^2\big].
\end{equation}
Now, it is evident that
\begin{equation}
\label{rot_p}
\left(\boldsymbol{\partial}\times \boldsymbol{\Euler{A}}\right)^2=\left(\boldsymbol{\partial}'\times \boldsymbol{\Euler{A}}'\right)^2.
\end{equation}
Also, in the Coulomb gauge where the scalar potential vanishes identically one has that $(\partial_t\boldsymbol{\Euler{A}})^2=
-(\partial_t \Euler{A}^{\alpha})(\partial_t \Euler{A}_{\alpha})$, then, by using Eq. (\ref{A_t_p}), I obtain
\begin{equation}
\label{A_t_2_p}
\left(\partial_t\Euler{A}\right)^2=
\left(\partial_t\Euler{A}'\right)^2+
2i\Omega(\partial_t\boldsymbol{\Euler{A}}')\cdot\mathcal{J}^{(1)\prime}_x\boldsymbol{\Euler{A}}'+O(\Omega^2)
\end{equation}
where, being in my approximations the rotational frequency $\Omega$ a small quantity, I have neglected the terms proportional to $\Omega^2$. By substituting Eqs. (\ref{rot_p}) and (\ref{A_t_2_p}) in Eq. (\ref{L_r_fr}) I can write it as
\begin{equation}
\label{free_rad_p}
-\frac{1}{4}\Euler{F}_{\alpha\beta}\Euler{F}^{\alpha\beta}=
-\frac{1}{4}\Euler{F}'_{\alpha\beta}\Euler{F}^{\prime\alpha\beta}-i\Omega (\partial_t \Euler{A}'_{\alpha})\mathcal{J}^{(1)\prime}_x \Euler{A}^{\prime\alpha}+O(\Omega^2)
\end{equation}
with
\begin{equation}
\Euler{F}'_{\alpha\beta}(t,\mathbf{r}'(t))=
\partial'_{\alpha}\Euler{A}'_{\beta}(t,\mathbf{r}'(t))-
\partial'_{\beta}\Euler{A}'_{\alpha}(t,\mathbf{r}'(t)).
\end{equation}

By collecting Eqs. (\ref{int_rad_p}), (\ref{int_ext_p}), (\ref{der_p}) and (\ref{free_rad_p}) 
and by performing the remaining trivial transformation of the mass term in the 
Lagrangean density (\ref{L_syn}), it can be written in terms of the primed variables and fields 
as
\begin{equation}
\label{L_p}
\begin{split}
\mathscr{L}' &=\bar{\psi}'\left\{\gamma^{\alpha}\left[i\partial'_{\alpha}+eA'_{\nnearrow\alpha}(\mathbf{r}'(t))+e\Euler{A}'_{\alpha}\right]-m\right\}\psi'-\frac{1}{4}\Euler{F}'_{\alpha\beta}\Euler{F}^{\prime\alpha\beta}\\
&\quad+i\Omega\bar{\psi}'\gamma^0\mathcal{J}^{(1/2)\prime}_x\psi'-i\Omega (\partial_t \Euler{A}'_{\alpha})\mathcal{J}^{(1)\prime}_x \Euler{A}^{\prime\alpha}+O(\Omega^2)
\end{split}
\end{equation}
or, by removing the now useless prime and the time dependence on $\mathbf{r}'(t)$,\footnote{The elimination 
of the time-dependence on the variables $\mathbf{r}'(t)$ can be safely done because it does not change the equation of motion of the radiation field. In fact, on the one hand, being the transformation $\mathbf{r}'(t)=\mathbf{r}'(t,\mathbf{r})$ a rotation, then 
$d\mathbf{r}=d\mathbf{r}'(t)$. On the other hand, the integral of the Lagrangean density $\mathscr{L}'$ on $\mathbf{r}'(t)$ to obtain the total Lagrangean (and then the action) extends over the whole space and the coordinates $\mathbf{r}'(t)$ are dumb variables.} as
\begin{equation}
\label{L_eff}
\mathscr{L}'_{\text{eff}}=\mathscr{L}'_0+\mathscr{L}'_I
\end{equation}
with
\begin{equation}
\label{L_0}
\mathscr{L}'_0=\bar{\psi}'
\left\{\gamma^{\alpha}\left[i\partial_{\alpha}+
eA'_{\nnearrow\alpha}(\mathbf{r})\right]-m\right\}\psi'-
\frac{1}{4}\Euler{F}'_{\alpha\beta}\Euler{F}^{\prime\alpha\beta}
\end{equation}
and
\begin{equation}
\label{L_I}
\mathscr{L}'_I=e\bar{\psi}'\gamma^{\alpha}\psi' \Euler{A}'_{\alpha}+i\Omega\bar{\psi}'\gamma^0\mathcal{J}^{(1/2)}_x\psi'-i\Omega (\partial_t \Euler{A}'_{\alpha})\mathcal{J}^{(1)}_x \Euler{A}^{\prime\alpha}+O(\Omega^2).
\end{equation}
In this way the original time-dependent Lagrangean density (\ref{L_syn}) has 
been transformed into an effective Lagrangean density that does not depend 
explicitly on time and that embodies the effects of the rotation of the 
external magnetic field in the interaction terms proportional to 
its rotational frequency $\Omega$. I note that the Lagrangean density (\ref{L_eff}) 
is just the Lagrangean density of QED in the presence 
of the external static magnetic field $\mathbf{B}'_{\nnearrow}=(0,0,B_{\nnearrow})$ plus other 
extra interaction terms that are proportional to $\Omega$ (or to $\Omega^2$).\footnote{As in the rest of the thesis, I continue to use the ``prime'' to indicate all the fields and the related quantities when the external magnetic field lies in the $z$ direction.}

In order to build the Hamiltonian density I calculate now the momenta conjugated to the Dirac and to the radiation field. From Eqs. (\ref{L_eff})-(\ref{L_I}) I obtain
\begin{subequations}
\begin{align}
\pi'_{\psi}(t,\mathbf{r}) &\equiv\frac{\partial \mathscr{L}'_{\text{eff}}}{\partial (\partial_t\psi')}=i\psi^{\prime\dag}(t,\mathbf{r}),\\
\label{Pi_A}
\boldsymbol{\pi}'_{\boldsymbol{\Euler{A}}'}(t,\mathbf{r}) &\equiv\frac{\partial \mathscr{L}'_{\text{eff}}}{\partial (\partial_t\boldsymbol{\Euler{A}}')}=\partial_t\boldsymbol{\Euler{A}}'(t,\mathbf{r})+i\Omega \mathcal{J}^{(1)}_x \boldsymbol{\Euler{A}}'(t,\mathbf{r})
\end{align}
\end{subequations}
and the Hamiltonian density can be written in the form
\begin{equation}
\mathscr{H}'_{\text{eff}}\equiv\pi'_{\psi}\partial_t\psi'+\boldsymbol{\pi}'_{\boldsymbol{\Euler{A}}'}\cdot\partial_t\boldsymbol{\Euler{A}}'-\mathscr{L}'_{\text{eff}}=\mathscr{H}'_0+\mathscr{H}'_I
\end{equation}
with
\begin{equation}
\label{H_0}
\mathscr{H}'_0 =\psi^{\prime\dag}\left\{\boldsymbol{\alpha}\cdot\left[-i\boldsymbol{\partial}+
e\mathbf{A}'_{\nnearrow}(\mathbf{r})\right]+\beta m\right\}\psi'+\frac{1}{2}\left[\boldsymbol{\pi}^{\prime 2}_{\boldsymbol{\Euler{A}}'}+\left(\boldsymbol{\partial}\times \boldsymbol{\Euler{A}}'\right)^2\right]
\end{equation}
and
\begin{equation}
\label{H_I}
\mathscr{H}'_I =-e\psi^{\prime\dag}\boldsymbol{\alpha}\psi'\cdot\boldsymbol{\Euler{A}}'+
i\Omega\psi^{\prime\dag}\mathcal{J}^{(1/2)}_x\psi'-i\Omega \boldsymbol{\pi}'_{\boldsymbol{\Euler{A}}'}\cdot\mathcal{J}^{(1)}_x \boldsymbol{\Euler{A}}'+O(\Omega^2).
\end{equation}

\begin{figure}[ht]
\begin{center}
\includegraphics[height=5.5cm]{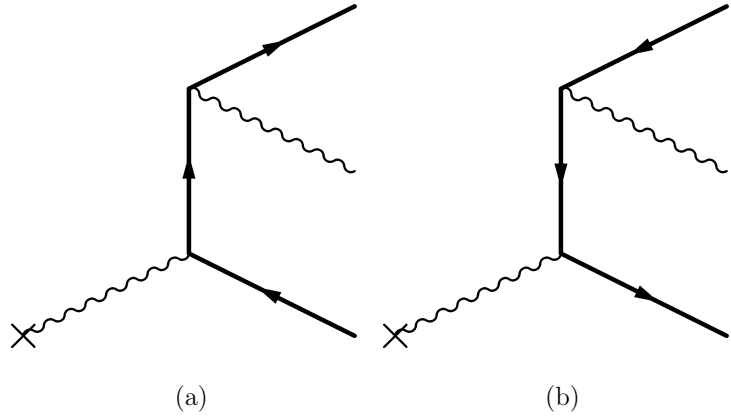}
\end{center}
\caption{Tree-level Feynman diagrams of the photon emission by an 
electron [part (a)] or by a positron [part (b)] created in the presence of the strong, uniform and slowly-rotating magnetic field $\mathbf{B}_{\nnearrow}(t)$ given in Eq. (\ref{B_rot_4}). The thick fermion lines indicate that the calculations of the corresponding $S$-matrix elements have been performed by using the electron and positron one-particle states in the presence of the magnetic field.}
\label{Feyn_dia}
\end{figure}

Now, since the perturbation induced by the magnetic field $\mathbf{B}_{\nnearrow}(t)$ is adiabatic, I can consider all the terms in the interaction Hamiltonian density (\ref{H_I}) as small perturbations of the free Hamiltonian density (\ref{H_0}). At this point, I can use the machinery of the ordinary perturbation theory to calculate the matrix elements corresponding to the process under study: the emission of photons by the electrons and the positrons created in the strong slowly-rotating magnetic field $\mathbf{B}_{\nnearrow}(t)$. By 
neglecting all the radiative corrections and taking into account only the tree-level 
contributions, the Feynman diagrams accounting for the mentioned process are those shown in 
Fig. \ref{Feyn_dia} where the lower vertices actually represent the interaction with the \emph{time derivative} of the external magnetic field and correspond to the term $i\Omega\psi^{\prime\dag}\mathcal{J}^{(1/2)}_x\psi'$ in $\mathscr{H}'_I$. 
Now, as in ordinary QED, in order to calculate the $S$-matrix elements corresponding to the Feynman diagrams in Fig. \ref{Feyn_dia} I quantize the Dirac field and the photon field in the 
interaction picture. Since, as I have said, the Lagrangean density (\ref{L_0}) is the 
free Lagrangean density of QED in the presence of a uniform and static magnetic field in the $z$ direction with strength $B_{\nnearrow}$, we already know that the Dirac field can be expanded as  [see Eqs. (\ref{psi_esp}) and (\ref{u_p_v_p})]
\begin{equation}
\psi'(t,\mathbf{r})=\sum_j\left[c_ju'_j(\mathbf{r})\exp(-iw_jt)+
d^{\dag}_jv'_j(\mathbf{r})\exp(i\tilde{w}_jt)\right].
\end{equation}

Now, we have seen in the previous Chapter that pairs are much more likely created in the presence of a strong and slowly-rotating magnetic field with both the electron and the positron in a TGS. Analogously, I assume here that all the electrons and the positrons entering the game are in TGSs [see Eqs. (\ref{u_v_p_g})]. In the same approximation I sum only on the electron and positron TGSs to build the propagator $G'(t,\mathbf{r},t',\mathbf{r}')$ that is \cite{Kuznetzov}
\begin{equation}
\label{G}
\begin{split}
iG'(t,\mathbf{r},t',\mathbf{r}')=\sum_{n,k}&\{\vartheta(t-t')
u'_{n,k}(\mathbf{r})\bar{u}'_{n,k}(\mathbf{r}')\exp[-i\varepsilon_k(t-t')]\\
&\;-\vartheta(t'-t)
v'_{n,k}(\mathbf{r})\bar{v}'_{n,k}(\mathbf{r}')\exp[i\varepsilon_k(t-t')]\}
\end{split}
\end{equation}
where the coordinates $\mathbf{r}'$ have obviously nothing to do with those introduced in Eq. (\ref{rot}) that, on the other hand, depended on time.

I pass now to the second quantization of the radiation field. The 
presence in the interaction Lagrangean density (\ref{L_I}) of terms containing the time derivative of the radiation field would make the quantization 
procedure very complicated. Nevertheless, I observe that
\begin{enumerate}
\item these additional terms are proportional to the rotational frequency $\Omega$;
\item the matrix elements that I will calculate are already proportional to $\Omega$ through the factor corresponding in Fig. \ref{Feyn_dia} to the interaction vertex with the time derivative of the external magnetic field.
\end{enumerate}
For these reasons, since I am not interested in higher order corrections in $\Omega$, all the other factors in the matrix elements 
can be evaluated neglecting the interaction with the external field. In this way, I can quantize the radiation field as it were free and then I can indicate the vector radiation field simply as $\boldsymbol{\Euler{A}}(t,\mathbf{r})$ and I have only to expand it into the usual plane-wave basis as
\begin{equation}
\boldsymbol{\Euler{A}}(t,\mathbf{r})=\sum_{\mathbf{q},\lambda}
\frac{\boldsymbol{\Euler{e}}_{\mathbf{q},\lambda}}{\sqrt{2V\omega}}
\left\{\Euler{a}_{\mathbf{q},\lambda}\exp[-i(\omega t-\mathbf{q}\cdot\mathbf{r})]+
\Euler{a}^{\dag}_{\mathbf{q},\lambda}\exp[i(\omega t-\mathbf{q}\cdot\mathbf{r})]\right\}
\end{equation}
where $V$ is the quantization volume, $\omega=|\mathbf{q}|$ is the photon energy 
and $\boldsymbol{\Euler{e}}_{\mathbf{q},\lambda}$ with $\lambda=1,2$ are the polarization versors \cite{Mandl}.

At this point I have all the quantities I need to calculate the matrix elements corresponding to the Feynman diagrams in Fig. \ref{Feyn_dia} and this is the subject of the next Paragraph. 
%
%
\subsection{Calculation of the photon energy spectrum}
\label{III}
By looking at the interaction Hamiltonian density $\mathscr{H}'_I$ it is clear that if the final state is the state $|e^-e^+\gamma\rangle\equiv
|k_0,n_0;k'_0,n'_0;\mathbf{q},\lambda\rangle$ (the initial state is obviously 
the vacuum $|0\rangle$) then the matrix element at time $t$ corresponding to the 
Feynman diagram in Fig. \ref{Feyn_dia}(a) can be written as\footnote{The Feynman diagram in Fig. \ref{Feyn_dia}(b) represents the emission of the photon by a positron but this process can be taken into account by simply multiplying by two the final spectrum 
of the photons emitted only by an electron.}
\begin{equation}
\label{S_fi_1}
\begin{split}
S'_{k_0,n_0,k'_0,n'_0,\mathbf{q},\lambda}(t)=&\int d\mathbf{r}'\int_{-\infty}^tdt'\int d\mathbf{r}''\int_0^tdt''\\
&\quad\times u^{\prime\dag}_{n_0,k_0}(\mathbf{r}')\exp(i\varepsilon_{k_0}t')
\frac{e\boldsymbol{\alpha}\cdot\boldsymbol{\Euler{e}}_{\mathbf{q},\lambda}}{\sqrt{2V\omega}}
\exp[i(\omega t'-\mathbf{q}\cdot\mathbf{r}')]\\
&\qquad\times iG'(t',\mathbf{r}',t'',\mathbf{r}'')i\Omega\gamma^0\mathcal{J}^{(1/2)\prime\prime}_xv'_{n'_0,k'_0}(\mathbf{r}'')\exp(i\varepsilon_{k'_0}t'')
\end{split}
\end{equation}
where I have pointed out that, while the electromagnetic interaction 
between the Dirac field and the radiation field is always present, the external field starts rotating at an arbitrary finite time $t''$ I called zero.\footnote{I remind (see Sects. \ref{Centr_eng} and \ref{Gen_ass}) that the emission of a GRB is supposed to happen during the collapse of a magnetar into a black hole or during the formation of the magnetar itself. In this framework the instant $t''=0$ in Eq. (\ref{S_fi_1}) indicates the beginning of the magnetar collapse or of its formation.}

Now, the term in Eq. (\ref{S_fi_1}) corresponding to the lower vertex in Fig. \ref{Feyn_dia}(a) will be calculated by means of the first-order adiabatic perturbation theory. In order to do this, I use the expression (\ref{G}) of the electron propagator to write the previous 
matrix element in the more useful form
\begin{equation}
\label{S_fi_2}
\begin{split}
&S'_{k_0,n_0,k'_0,n'_0,\mathbf{q},\lambda}(t)\\
&\quad =-\frac{e(\boldsymbol{\Euler{e}}_{\mathbf{q},\lambda})_z}{\sqrt{2V\omega}}
\sum_{n,k}\int_{-\infty}^tdt'\exp[i(\varepsilon_{k_0}+\omega-\varepsilon_k)t']\\
&\qquad\qquad\times\int d\mathbf{r}'u^{\prime\dag}_{n_0,k_0}(\mathbf{r}')
\alpha_z u'_{n,k}(\mathbf{r}')\exp(-i\mathbf{q}\cdot\mathbf{r}')\\
&\qquad\qquad\times\int_0^{t'}dt''\exp[i(\varepsilon_{k'_0}+\varepsilon_k)t'']\int d\mathbf{r}''u^{\dag}_{n,k}(t'',\mathbf{r}'')\partial_{t''}
v_{n'_0,k'_0}(t'',\mathbf{r}'')\\
&\qquad+\frac{e(\boldsymbol{\Euler{e}}_{\mathbf{q},\lambda})_z}{\sqrt{2V\omega}}
\sum_{n,k}\int_{-\infty}^tdt'\exp[i(\varepsilon_{k_0}+\omega+\varepsilon_k)t']\\
&\qquad\qquad\times\int d\mathbf{r}'u^{\prime\dag}_{n_0,k_0}(\mathbf{r}')
\alpha_z v'_{n,k}(\mathbf{r}')\exp(-i\mathbf{q}\cdot\mathbf{r}')\\
&\qquad\qquad\times\int_{t'}^tdt''\exp[i(\varepsilon_{k'_0}-\varepsilon_k)t'']\int d\mathbf{r}''v^{\dag}_{n,k}(t'',\mathbf{r}'')\partial_{t''}
v_{n'_0,k'_0}(t'',\mathbf{r}'')
\end{split}
\end{equation}
where I used the fact that the TGSs are eigenstates of $\sigma_z$ in such a way $\alpha_x$ and $\alpha_y$ can not couple two of them and where I introduced the ``rotating'' states [see Eqs. (\ref{u_v})]
\begin{subequations}
\begin{align}
u_{n,k}(t,\mathbf{r})&=\exp(i\Omega t\mathcal{J}_x^{(1/2)})u'_{n,k}(\mathbf{r}),\\
\label{v_rot}
v_{n,k}(t,\mathbf{r})&=\exp(i\Omega t\mathcal{J}_x^{(1/2)})v'_{n,k}(\mathbf{r})
\end{align}
\end{subequations}
that are the instantaneous eigenstates at time $t$ of the one-particle Hamiltonian
\begin{equation}
\mathcal{H}(t)=
\boldsymbol{\alpha}\cdot\left[\boldsymbol{\mathcal{P}}+
e\mathbf{A}_{\nnearrow}(t,\mathbf{r})\right]+\beta m
\end{equation}
with $\mathbf{A}_{\nnearrow}(t,\mathbf{r})$ given in Eq. (\ref{A_rot}). In this way, by applying Eq. (\ref{m_d_n}) of the adiabatic perturbation theory I obtain
\begin{subequations}
\begin{align}
\int d\mathbf{r}''u^{\dag}_{n,k}(t'',\mathbf{r}'')
\partial_{t''}v_{n'_0,k'_0}(t'',\mathbf{r}'')&=-
\int d\mathbf{r}''\frac{u^{\dag}_{n,k}(t'',\mathbf{r}'')\dot{\mathcal{H}}(t'')v_{n'_0,k'_0}(t'',\mathbf{r}'')}{\varepsilon_{k'_0}+\varepsilon_k}
,\\
\int d\mathbf{r}''v^{\dag}_{n,k}(t'',\mathbf{r}'')
\partial_{t''}v_{n'_0,k'_0}(t'',\mathbf{r}'')&=-\int d\mathbf{r}''\frac{v^{\dag}_{n,k}(t'',\mathbf{r}'')\dot{\mathcal{H}}(t'')
v_{n'_0,k'_0}(t'',\mathbf{r}'')}{\varepsilon_{k'_0}-\varepsilon_k}.
\end{align}
\end{subequations}
These kinds of matrix elements can be easily calculated and the result is
\begin{subequations}
\begin{align}
\label{u_u}
\int d\mathbf{r}''u^{\dag}_{n,k}(t'',\mathbf{r}'')
\partial_{t''}v_{n'_0,k'_0}(t'',\mathbf{r}'') &=
-\frac{e\Omega B_{\nnearrow}}{\varepsilon_k+\varepsilon_{k'_0}}
\int d\mathbf{r}''u^{\prime\dag}_{n,k}(\mathbf{r}'')\frac{x''\alpha_z}{2}
v'_{n'_0,k'_0}(\mathbf{r}''),\\
\label{v_v}
\int d\mathbf{r}''v^{\dag}_{n,k}(t'',\mathbf{r}'')
\partial_{t''}v_{n'_0,k'_0}(t'',\mathbf{r}'') &=
-\frac{e\Omega B_{\nnearrow}}{\varepsilon_{k'_0}-\varepsilon_k}
\int d\mathbf{r}''v^{\prime\dag}_{n,k}(\mathbf{r}'')
\frac{x''\alpha_z}{2}v'_{n'_0,k'_0}(\mathbf{r}'').
\end{align}
\end{subequations}
I observe that in both these matrix elements the integral on the $z$ variable gives 
a conservation of the longitudinal momentum and then of the energy. This does not cause 
any problem in the first matrix element, while the second one diverges when $k=k'_0$. For this 
reason, this particular matrix element will be calculated by writing the left-hand side of 
Eq. (\ref{v_v}) as [see Eq. (\ref{v_rot})]
\begin{equation}
\int d\mathbf{r}''u^{\dag}_{n,k'_0}(t'',\mathbf{r}'')
\partial_{t''}v_{n'_0,k'_0}(t'',\mathbf{r}'')=
-i\Omega\int d\mathbf{r}''u^{\prime\dag}_{n,k'_0}(\mathbf{r}'')
\mathcal{J}^{(1/2)\prime\prime}_xv'_{n'_0,k'_0}(\mathbf{r}'').
\end{equation}
By substituting the explicit expression of the one-particle electron total angular momentum 
(\ref{j_1/2}), I observe that, on the one hand, the term $z''\mathcal{P}''_y$ does not contribute because, by performing the integral on $z''$ from $-L_z/2$ to $L_z/2$, it vanishes and, on the other hand, the term $\sigma_x/2$ does not contribute either because the TGSs are eigenstates of $\sigma_z$. As a result, we have
\begin{equation}
\label{v_v_1}
\int d\mathbf{r}''v^{\dag}_{n,k'_0}(t'',\mathbf{r}'')
\partial_{t''}v_{n'_0,k'_0}(t'',\mathbf{r}'')=
-\Omega\int d\mathbf{r}''v^{\prime\dag}_{n,k'_0}(\mathbf{r}'')y''
\partial_{z''}v'_{n'_0,k'_0}(\mathbf{r}'').
\end{equation}
At this point, if I substitute the expressions (\ref{u_v_p_g}) of the TGSs I obtain that the matrix elements different from zero are the following ones:
\begin{subequations}
\begin{align}
&\int d\mathbf{r}''u^{\dag}_{n'_0-1,-k'_0}(t'',\mathbf{r}'')
\partial_{t''}v_{n'_0,k'_0}(t'',\mathbf{r}'') 
=\frac{m\Omega}{4\varepsilon^2_{k'_0}}\sqrt{\frac{eB_{\nnearrow}n'_0}{2}},\\
&\int d\mathbf{r}''u^{\dag}_{n'_0+1,-k'_0}(t'',\mathbf{r}'')
\partial_{t''}v_{n'_0,k'_0}(t'',\mathbf{r}'') 
=\frac{m\Omega}{4\varepsilon^2_{k'_0}}\sqrt{\frac{eB_{\nnearrow}(n'_0+1)}{2}},\\
&\int d\mathbf{r}''v^{\dag}_{n'_0-1,k'_0}(t'',\mathbf{r}'')
\partial_{t''}v_{n'_0,k'_0}(t'',\mathbf{r}'')=
k'_0\Omega\sqrt{\frac{n'_0}{2eB_{\nnearrow}}},\\
&\int d\mathbf{r}''v^{\dag}_{n'_0+1,k'_0}(t'',\mathbf{r}'')
\partial_{t''}v_{n'_0,k'_0}(t'',\mathbf{r}'')=
-k'_0\Omega\sqrt{\frac{n'_0+1}{2eB_{\nnearrow}}}.
\end{align}
\end{subequations}

By inserting the previous matrix elements in Eq. (\ref{S_fi_2}) and 
by omitting the now useless index ``0'' on $k_0$, $k'_0$, $n_0$ and $n'_0$, I obtain the two transition amplitudes
\begin{equation}
\label{S_fi_3_1}
\begin{split}
&S^{\prime(1)}_{k,n,k',n',\mathbf{q},\lambda}(t)\\ &\;=-\frac{me\Omega(\boldsymbol{\Euler{e}}_{\mathbf{q},\lambda})_z}{8\varepsilon^2_{k'}}\sqrt{\frac{eB_{\nnearrow}n'}{V\omega}}\int d\mathbf{r}'u^{\prime\dag}_{n,k}(\mathbf{r}')\alpha_zu'_{n'-1,-k'}(\mathbf{r}')
\exp(-i\mathbf{q}\cdot\mathbf{r}')\\
&\qquad\;\times\int_{-\infty}^tdt'\exp[i(\varepsilon_k-\varepsilon_{k'}+\omega)t']
\int_0^{t'}dt''\exp(2i\varepsilon_{k'}t'')\\
&\quad\;+\frac{e\Omega k'(\boldsymbol{\Euler{e}}_{\mathbf{q},\lambda})_z}{2}\sqrt{\frac{n'}{eB_{\nnearrow}V\omega}}\int d\mathbf{r}'u^{\prime\dag}_{n,k}(\mathbf{r}')\alpha_zv'_{n'-1,k'}(\mathbf{r}')
\exp(-i\mathbf{q}\cdot\mathbf{r}')\\
&\qquad\;\times\int_{-\infty}^tdt'\exp[i(\varepsilon_k+\varepsilon_{k'}+\omega)t']
\int_{t'}^tdt''
\end{split}
\end{equation}
and
\begin{equation}
\label{S_fi_3_2}
\begin{split}
&S^{\prime(2)}_{k,n,k',n',\mathbf{q},\lambda}(t)\\
&\;=-\frac{me\Omega(\boldsymbol{\Euler{e}}_{\mathbf{q},\lambda})_z}{8\varepsilon^2_{k'}}\sqrt{\frac{eB_{\nnearrow}(n'+1)}{V\omega}}\int d\mathbf{r}'u^{\prime\dag}_{n,k}(\mathbf{r}')\alpha_zu'_{n'+1,-k'}(\mathbf{r}')
\exp(-i\mathbf{q}\cdot\mathbf{r}')\\
&\qquad\;\times\int_{-\infty}^tdt'\exp[i(\varepsilon_k-\varepsilon_{k'}+\omega)t']
\int_0^{t'}dt''\exp(2i\varepsilon_{k'}t'')\\
&\quad\;-\frac{e\Omega k'(\boldsymbol{\Euler{e}}_{\mathbf{q},\lambda})_z}{2}\sqrt{\frac{n'+1}{eB_{\nnearrow}V\omega}}\int d\mathbf{r}'u^{\prime\dag}_{n,k}(\mathbf{r}')\alpha_zv'_{n'+1,k'}(\mathbf{r}')
\exp(-i\mathbf{q}\cdot\mathbf{r}')\\
&\qquad\;\times\int_{-\infty}^tdt'\exp[i(\varepsilon_k+\varepsilon_{k'}+\omega)t']
\int_{t'}^tdt''.
\end{split}
\end{equation}
Now, by using the expressions (\ref{u_v_p_g}) of the TGSs it is easy to show that
\begin{align}
\label{I_r_1}
\begin{split}
&\int d\mathbf{r}'u^{\prime\dag}_{n,k}(\mathbf{r}')\alpha_zu'_{n'\pm 1,-k'}(\mathbf{r}')
\exp(-i\mathbf{q}\cdot\mathbf{r}')\\
&\qquad=\sqrt{\frac{(\varepsilon_k+m)(\varepsilon_{k'}+m)}
{4\varepsilon_k\varepsilon_{k'}}}
\left(\frac{k}{\varepsilon_k+m}-\frac{k'}{\varepsilon_{k'}+m}\right)I'_{n,n'\pm 1,q_x,q_y}\delta_{k+q_z+k',0},
\end{split}\\
\label{I_r_2}
\begin{split}
&\int d\mathbf{r}'u^{\prime\dag}_{n,k}(\mathbf{r}')\alpha_zv'_{n'\pm 1,k'}(\mathbf{r}')
\exp(-i\mathbf{q}\cdot\mathbf{r}')\\
&\qquad=-\sqrt{\frac{(\varepsilon_k+m)(\varepsilon_{k'}+m)}
{4\varepsilon_k\varepsilon_{k'}}}
\left(1+\frac{k}{\varepsilon_k+m}\frac{k'}{\varepsilon_{k'}+m}\right) I'_{n,n'\pm 1,q_x,q_y}\delta_{k+q_z+k',0}
\end{split}
\end{align}
where
\begin{equation}
\label{I}
I'_{n,n',q_x,q_y}=\int dxdy\theta^{\prime *}_n(x,y)\theta'_{n'}(x,y)\exp[-i(q_x x+q_y y)]
\end{equation}
with the functions $\theta'_n(x,y)$ given in Eq. (\ref{theta_p_g}).

I proceed now by calculating the time integrals in Eqs. (\ref{S_fi_3_1}) and (\ref{S_fi_3_2}). The results are
\begin{equation}
\label{I_t_1}
\begin{split}
&\int_{-\infty}^tdt'\exp[i(\varepsilon_k+\omega-\varepsilon_{k'})t']
\int_0^{t'}dt''\exp(2i\varepsilon_{k'}t'')\\
&\qquad=\frac{1}{2i\varepsilon_{k'}}
\left\{\frac{\exp[i(\varepsilon_k+
\varepsilon_{k'}+\omega+is)t]}{i(\varepsilon_k+
\varepsilon_{k'}+\omega+is)}-\frac{\exp[i(\varepsilon_k-
\varepsilon_{k'}+\omega+is)t]}{i(\varepsilon_k-
\varepsilon_{k'}+\omega+is)}\right\}
\end{split}
\end{equation}
and 
\begin{equation}
\label{I_t_2}
\int_{-\infty}^tdt'\exp[i(\varepsilon_k+\omega+\varepsilon_{k'})t']
\int_{t'}^tdt''=-\frac{\exp[i(\varepsilon_k+
\varepsilon_{k'}+\omega-is)t]}{(\varepsilon_k+
\varepsilon_{k'}+\omega-is)^2}
\end{equation}
respectively and the $is$ terms with $s\to 0^+$ have been added in order to make the integrals 
convergent. Now, it is obvious that $\varepsilon_k+\varepsilon_{k'}+\omega>0$. Also, because of the overall conservation $k+q_z+k'=0$ of the longitudinal momentum coming from Eqs. (\ref{I_r_1}) and (\ref{I_r_2}) then, unless the trivial case $\mathbf{q}=\mathbf{0}$, it can be shown that $\varepsilon_k-\varepsilon_{k'}+\omega>0$. In this way, all the $is$ terms can be safely eliminated in the final results in Eqs. (\ref{I_t_1}) and (\ref{I_t_2}). 
Finally, by substituting the just calculated time integrals and Eqs. (\ref{I_r_1}) and (\ref{I_r_2}) in Eqs. (\ref{S_fi_3_1}) and (\ref{S_fi_3_2}), I obtain
\begin{equation}
\label{S_fi_4_1}
\begin{split}
S^{\prime(1)}_{k,n,k',n',\mathbf{q},\lambda}(t)=&\frac{e\Omega(\boldsymbol{\Euler{e}}_{\mathbf{q},\lambda})_z}{4}
\sqrt{\frac{(\varepsilon_k+m)(\varepsilon_{k'}+m)n'}
{\varepsilon_k\varepsilon_{k'}e\omega B_{\nnearrow}V}}
I'_{n,n'-1,q_x,q_y}\delta_{k+q_z+k',0}\\
&\times\left\{\frac{eB_{\nnearrow}m}{8\varepsilon^3_{k'}}
\left(\frac{k}{\varepsilon_k+m}-\frac{k'}{\varepsilon_{k'}+m}\right)
\frac{\exp[i(\varepsilon_k+
\varepsilon_{k'}+\omega)t]}{\varepsilon_k+\varepsilon_{k'}+\omega}\right.\\
&\qquad+\frac{k'\exp[i(\varepsilon_k+
\varepsilon_{k'}+\omega)t]}{(\varepsilon_k+\varepsilon_{k'}+\omega)^2}
\left(1+\frac{k}{\varepsilon_k+m}\frac{k'}{\varepsilon_{k'}+m}
\right)\\
&\qquad\left.-\frac{eB_{\nnearrow}m}{8\varepsilon^3_{k'}}
\left(\frac{k}{\varepsilon_k+m}-\frac{k'}{\varepsilon_{k'}+m}\right)
\frac{\exp[i(\varepsilon_k-
\varepsilon_{k'}+\omega)t]}{\varepsilon_k-\varepsilon_{k'}+\omega}\right\}
\end{split}
\end{equation}
and
\begin{equation}
\label{S_fi_4_2}
\begin{split}
S^{\prime(2)}_{k,n,k',n',\mathbf{q},\lambda}(t)=&\frac{e\Omega(\boldsymbol{\Euler{e}}_{\mathbf{q},\lambda})_z}{4}
\sqrt{\frac{(\varepsilon_k+m)(\varepsilon_{k'}+m)(n'+1)}
{\varepsilon_k\varepsilon_{k'}e\omega B_{\nnearrow}V}}
I'_{n,n'+1,q_x,q_y}\delta_{k+q_z+k',0}\\
&\times\left\{\frac{eB_{\nnearrow}m}{8\varepsilon^3_{k'}}
\left(\frac{k}{\varepsilon_k+m}-\frac{k'}{\varepsilon_{k'}+m}\right)
\frac{\exp[i(\varepsilon_k+
\varepsilon_{k'}+\omega)t]}{\varepsilon_k+\varepsilon_{k'}+\omega}\right.\\
&\quad\;\;\;-\frac{k'\exp[i(\varepsilon_k+
\varepsilon_{k'}+\omega)t]}{(\varepsilon_k+\varepsilon_{k'}+\omega)^2}
\left(1+\frac{k}{\varepsilon_k+m}\frac{k'}{\varepsilon_{k'}+m}
\right)\\
&\quad\;\;\;\left.-\frac{eB_{\nnearrow}m}{8\varepsilon^3_{k'}}
\left(\frac{k}{\varepsilon_k+m}-\frac{k'}{\varepsilon_{k'}+m}\right)
\frac{\exp[i(\varepsilon_k-
\varepsilon_{k'}+\omega)t]}{\varepsilon_k-\varepsilon_{k'}+\omega}\right\}.
\end{split}
\end{equation}
The probability that a photon is emitted at time $t$ with momentum between 
$\mathbf{q}$ and $\mathbf{q}+d\mathbf{q}$ by an electron or by a positron is given by
\begin{equation}
\label{dP_q}
dP^{(\text{syn})}(\mathbf{q},t)=2\frac{Vd\mathbf{q}}{(2\pi)^3}
\frac{L_z}{2\pi}\int dk \frac{L_z}{2\pi}\int dk'\sum_{i,\lambda=1}^2\sum_{n,n'=0}^{\infty}
\big|S^{\prime(i)}_{n,n',\lambda}(k,k',\mathbf{q},t)\big|^2
\end{equation}
where the limit of large $L_z$ and $V$ is understood and all the momenta are intended 
from now on to be continuous variables. As usual, I am interested in macroscopic 
times $t$ such that $mt\gg 1$ then I can neglect the oscillating 
terms coming from the square modulus of $S^{\prime(1)}_{n,n',\lambda}(k,k',\mathbf{q},t)$ and $S^{\prime(2)}_{n,n',\lambda}(k,k',\mathbf{q},t)$: 
\begin{equation}
\label{dP_q_1_2}
\begin{split}
&\left\langle dP^{(\text{syn})}(\mathbf{q},t)\right\rangle\\ &\quad\sim\frac{L_zd\mathbf{q}}{(2\pi)^4}\frac{eq_{xy}^2\Omega^2}{4\omega^3 B_{\nnearrow}}
\sum_{n,n'=0}^{\infty}\left[n'|I'_{n,n'-1}(q_x,q_y)|^2+
(n'+1)|I'_{n,n'+1}(q_x,q_y)|^2\right]\\
&\qquad\times\int dk\frac{[\varepsilon(k)+m][\varepsilon(k+q_z)+m]}
{\varepsilon(k)\varepsilon(k+q_z)}\\
&\qquad\qquad\times\left\{\left[\frac{eB_{\nnearrow}m}{8\varepsilon^3(k+q_z)}\right]^2\left[\frac{k}{\varepsilon(k)+m}+\frac{k+q_z}{\varepsilon(k+q_z)+m}\right]^2\right.\\
&\qquad\qquad\qquad\times
\frac{[\varepsilon(k)+\omega]^2+\varepsilon^2(k+q_z)}{\big[[\varepsilon(k)+\omega]^2-\varepsilon^2(k+q_z)\big]^2}\\
&\left.\qquad\qquad\qquad+\frac{(k+q_z)^2}{[\varepsilon(k)+\varepsilon(k+q_z)+\omega]^4}\left[1-\frac{k}{\varepsilon(k)+m}\frac{k+q_z}{\varepsilon(k+q_z)+m}
\right]^2\right\}
\end{split}
\end{equation}
where I exploited the longitudinal momentum conservation to perform the integration on $k'$ and where I substituted \cite{Mandl}
\begin{equation}
\label{sum_r}
\sum_{\lambda=1}^2\bigl|[\boldsymbol{\Euler{e}}_{\lambda}(\mathbf{q})]_z\bigr|^2=1-\frac{q_z^2}{\omega^2}=\frac{q_{xy}^2}{\omega^2}
\end{equation}
with $q_{xy}^2=q_x^2+q_y^2$. Concerning the sums on $n$ and $n'$, I 
will calculate them together with the integrals $I'_{n,n'\pm 1}(q_x,q_y)$. In fact,
\begin{equation}
\begin{split}
&\sum_{n,n'=0}^{\infty}\left[n'|I'_{n,n'-1}(q_x,q_y)|^2+
(n'+1)|I'_{n,n'+1}(q_x,q_y)|^2\right]\\
&\qquad\qquad=\sum_{n,n'=0}^{\infty}(2n'+1)|I'_{n,n'}(q_x,q_y)|^2.
\end{split}
\end{equation}
Now, from Eqs. (\ref{I}) and (\ref{theta_p_g}) I can write $I'_{n,n'}(q_x,q_y)$ as
\begin{equation}
\begin{split}
I'_{n,n'}(q_x,q_y)&=\frac{1}{\pi\sqrt{n!n'!}}\int d\xi d\eta(\xi+i\eta)^n(\xi-i\eta)^{n'}\\
&\qquad\qquad\qquad\times\exp\left\{-\left[\xi^2+\eta^2-i\sqrt{\frac{2}{eB_{\nnearrow}}}(q_x\xi+q_y\eta)\right]\right\}
\end{split}
\end{equation}
where the change of variable
\begin{subequations}
\label{ch_1}
\begin{align}
\label{ch_1_1}
\xi &=\sqrt{\frac{eB_{\nnearrow}}{2}}x,\\
\label{ch_1_2}
\eta &=\sqrt{\frac{eB_{\nnearrow}}{2}}y
\end{align}
\end{subequations}
has been performed. With this expression I can calculate explicitly the sums on $n$ and 
$n'$, in fact
\begin{equation}
\begin{split}
&\sum_{n,n'=0}^{\infty}(2n'+1)|I'_{n,n'}(q_x,q_y)|^2\\
&\qquad=\frac{1}{\pi^2}
\int d\xi d\eta d\xi'd\eta'
\exp[-(\xi^2+\eta^2+\xi^{\prime 2}+\eta^{\prime 2})]\\
&\qquad\quad\times\exp\left\{i\sqrt{\frac{2}{eB_{\nnearrow}}}[q_x(\xi'-\xi)+q_y(\eta'-\eta)]\right\}\\
&\qquad\quad\times\sum_{n=0}^{\infty}\frac{1}{n!}
[(\xi+i\eta)(\xi'-i\eta')]^n\sum_{n'=0}^{\infty}\frac{2n'+1}{n'!}
[(\xi-i\eta)(\xi'+i\eta')]^{n'}\\
&\qquad=\frac{1}{\pi^2}
\int d\xi d\eta d\xi'd\eta'
\exp[-(\xi'-\xi)^2-(\eta'-\eta)^2]\\
&\qquad\quad\times\exp\left\{i\sqrt{\frac{2}{eB_{\nnearrow}}}[q_x(\xi'-\xi)+q_y(\eta'-\eta)]\right\}[2(\xi-i\eta)(\xi'+i\eta')+1].
\end{split}
\end{equation}
If now I put
\begin{subequations}
\label{ch_2}
\begin{align}
\label{ch_2_1}
\xi_{\pm} &=\frac{\xi'\pm\xi}{\sqrt{2}},\\
\label{ch_2_2}
\eta_{\pm} &=\frac{\eta'\pm\eta}{\sqrt{2}}
\end{align}
\end{subequations}
I obtain
\begin{equation}
\label{sum_n}
\begin{split}
\sum_{n,n'=0}^{\infty}(2n'+1)|I'_{n,n'}(q_x,q_y)|^2=
&\frac{1}{\pi^2}
\int d\xi_+ d\xi_-d\eta_+ d\eta_-\exp[-2(\xi_-^2+\eta_-^2)]\\
&\qquad\times\exp\left[i\frac{2}{\sqrt{eB_{\nnearrow}}}(q_x\xi_-+q_y\eta_-)\right]\\
&\qquad\times[(\xi_++i\eta_-)^2-(\xi_-+i\eta_+)^2+1].
\end{split}
\end{equation}
As I have pointed out before, the 
presence of the external nonuniform electric field $\mathbf{E}_{\nnearrow}(t,\mathbf{r})=
-\partial_t\mathbf{A}_{\nnearrow}(t,\mathbf{r})$ [see Eq. (\ref{A_rot})] breaks the translational symmetry in the plane perpendicular to the magnetic field. In particular, as we have seen in the previous Chapter, this fact also made the presence probabilities per unit volume diverging in the regions far from the origin [see Eqs. (\ref{dP_lin}), (\ref{dP_rot}) and (\ref{dP_rot_lin})]. In the present case the divergence comes from the integrals on the variables $\xi_+$ and $\eta_+$ in Eq. (\ref{sum_n}). For this reason I will keep in (\ref{sum_n}) only the dominant terms in these integrals that is
\begin{equation}
\begin{split}
&\sum_{n,n'=0}^{\infty}(2n'+1)|I'_{n,n'}(q_x,q_y)|^2\sim\frac{1}{\pi^2}\int d\eta_+d\xi_+ (\xi_+^2+\eta_+^2)\\
&\qquad\qquad\qquad\quad\times\int d\xi_-d\eta_-\exp\left[\frac{2i}{\sqrt{eB_{\nnearrow}}}(q_x\xi_-+q_y\eta_-)-2(\xi_-^2+\eta_-^2)\right].
\end{split}
\end{equation}
Now, by passing to polar coordinates in the $\xi_+\text{--}\eta_+$ plane, I easily obtain 
[see Eqs. (\ref{ch_1}) and (\ref{ch_2})]
\begin{equation}
\int d\eta_+d\xi_+ (\xi_+^2+\eta_+^2)=
\frac{\pi}{2}\left(\sqrt{\frac{eB_{\nnearrow}}{2}}R_{\perp M}\right)^4
\end{equation}
where $R_{\perp M}$ is the transverse radius already introduced in Eq. (\ref{R_perp_M^2}). Instead, the integrals on the variables $\xi_-$ and $\eta_-$ are well-known exponential integrals and I only quote the final result:
\begin{equation}
\label{sum_n_f}
\begin{split}
\sum_{n,n'=0}^{\infty}(2n'+1)|I'_{n,n'}(q_x,q_y)|^2\sim
&\frac{1}{4}\left(\frac{eB_{\nnearrow}}{2}\right)^2R^4_{\perp M}
\exp\left(-\frac{q_{xy}^2}{2eB_{\nnearrow}}\right).
\end{split}
\end{equation}
By substituting Eqs. (\ref{sum_r}) and (\ref{sum_n_f}) in Eq. (\ref{dP_q_1_2}) I obtain the following expression of the probability $\left\langle dP^{(\text{syn})}(\mathbf{q},t)\right\rangle$:
\begin{equation}
\label{dP_q_2}
\begin{split}
&\left\langle dP^{(\text{syn})}(\mathbf{q},t)\right\rangle\\ &\quad\sim\frac{2eB_{\nnearrow}\Omega^2\alpha_{em}}{(8\pi)^3}
\frac{q^2_{xy}d\mathbf{q}}{\omega^3}L_zR^4_{\perp M}\exp\left(-\frac{q_{xy}^2}{2eB_{\nnearrow}}\right)\\
&\qquad\times\int dk\frac{[\varepsilon(k)+m][\varepsilon(k+q_z)+m]}
{\varepsilon(k)\varepsilon(k+q_z)}\\
&\qquad\quad\times\left\{\left[\frac{eB_{\nnearrow}m}{8\varepsilon^3(k+q_z)}\right]^2\left[\frac{k}{\varepsilon(k)+m}+\frac{k+q_z}{\varepsilon(k+q_z)+m}\right]^2\right.\\
&\qquad\qquad\quad\times
\frac{[\varepsilon(k)+\omega]^2+\varepsilon^2(k+q_z)}{\big[[\varepsilon(k)+\omega]^2-\varepsilon^2(k+q_z)\big]^2}\\
&\left.\qquad\qquad\quad+\frac{(k+q_z)^2}{2[\varepsilon(k)+\varepsilon(k+q_z)+\omega]^4}\left[1-\frac{k}{\varepsilon(k)+m}\frac{k+q_z}{\varepsilon(k+q_z)+m}
\right]^2\right\}.
\end{split}
\end{equation}
Finally, the photon spectrum per unit volume $V=L_z\pi R_{\perp M}^2$ is obtained 
by passing to photon momentum spherical coordinates $\{\omega,\theta,\phi\}$ and by integrating on the angular variables. Only the integral on the 
variable $\phi$ is trivial then by putting $u=\cos\theta$ I obtain
\begin{equation}
\label{dP_q_3}
\begin{split}
&\left\langle\frac{dN^{(\text{syn})}(\omega,t)}{d\omega dV}\right\rangle\sim\frac{\alpha_{em}wm^2(\Omega R_{\perp M})^2}{(4\pi)^3}\frac{B_{\nnearrow}}{B_{cr}}\\
&\qquad\times\int_{-1}^1 du(1-u^2)\exp\left[-\frac{B_{cr}}{2B_{\nnearrow}}w^2(1-u^2)\right]\\
&\qquad\quad\times\int_0^{\infty} dv 
\frac{[1+\epsilon(v)][1+\epsilon(v+uw)]}
{\epsilon(v)\epsilon(v+uw)}\\
&\qquad\qquad\times\Bigg\{\left(\frac{B_{\nnearrow}}{8B_{cr}}\right)^2
\left[\frac{v}{1+\epsilon(v)}+
\frac{v+uw}{1+\epsilon(v+uw)}\right]^2\\
&\qquad\qquad\quad\;\;\times \frac{1}{\epsilon^6(v+uw)}\frac{[\epsilon(v)+w]^2+\epsilon^2(v+uw)}{\big[[\epsilon(v)+w]^2-\epsilon^2(v+uw)\big]^2}\\
&\left.\qquad\qquad\qquad\;\;+\frac{(v+uw)^2}{2[\epsilon(v)+\epsilon(v+uw)+w]^4}\left[1-\frac{v}{1+\epsilon(v)}\frac{v+uw}{1+\epsilon(v+uw)}\right]^2\right\}
\end{split}
\end{equation}
where I introduced the adimensional quantities $w=\omega/m$, $v=k/m$ and  $\epsilon(v)=\sqrt{1+v^2}$. 

\begin{figure}[ht]
\begin{center}
\includegraphics[angle=90,width=\textwidth]{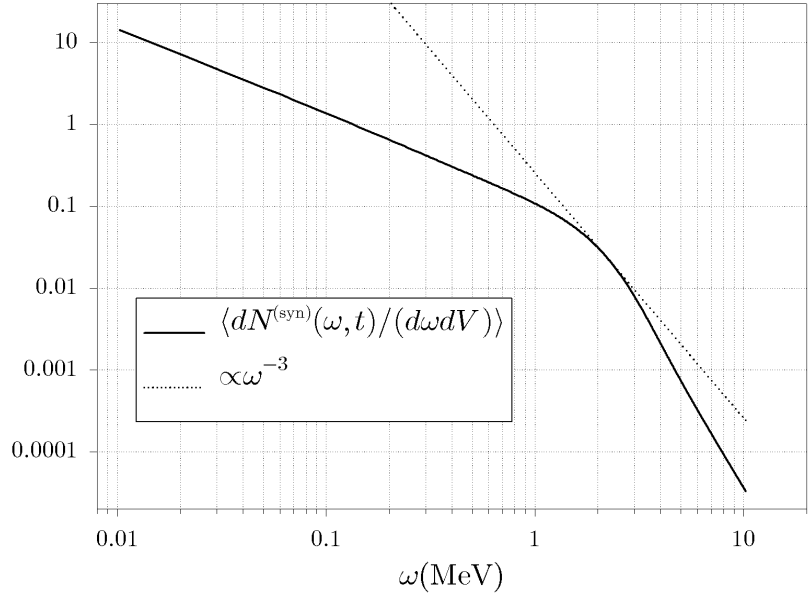}
\end{center}
\caption{Photon spectrum $\left\langle dN^{(\text{syn})}(\omega,t)/(d\omega dV)\right\rangle$ in arbitrary units. The magnetic field strength $B_{\nnearrow}$ is equal to $2.2\times 10^{14}\;\text{gauss}$. The dotted curve represents a function proportional to $\omega^{-3}$.}
\label{Spec_syn}
\end{figure}

Now, the integrals in Eq. (\ref{dP_q_3}) can not be performed analytically and, for this reason, I resort to a numerical integration. Fig. \ref{Spec_syn} shows the photon spectrum (\ref{dP_q_3}) in arbitrary units and with a magnetic field strength $B_{\nnearrow}=2.2\times 10^{14}\;\text{gauss}$. 
The qualitative form of this spectrum is very similar to those shown in Fig. \ref{GRB_en_sp}. In fact, as it is evident from the figure, the spectrum shows two different behaviours below and above the break energy $\omega_b^{(\text{syn})}\sim 1\text{--}3\;\text{MeV}$. I have checked numerically that, unlike the annihilation photon spectrum (see Fig. \ref{Spec_ann}) where $\omega_b^{(\text{ann})}\simeq m$ independently of the external magnetic field strength, the value of $\omega_b^{(\text{syn})}$ here depends on $B_{\nnearrow}$ and, in particular, the lower is $B_{\nnearrow}$ the lower is $\omega_b^{(\text{syn})}$. Nevertheless, it is important that by using a typical magnetar magnetic field strength such as $B_{\nnearrow}=2.2\times 10^{14}\;\text{gauss}$, then the value of $\omega_b^{(\text{syn})}$ is close to the break energies characterizing the experimental GRBs spectra. In fact, as I have said in Par. \ref{GRB_char}, the experimental break energies are typically below 1 MeV, but there are also cases of GRBs with $\omega_b>1\;\text{MeV}$. On the other hand, I have also said in Par. \ref{GRB_char} [see Eq. (\ref{GRB_spectr_eq}) and below] that the experimental spectra of GRBs are very well fitted by a function proportional to $\omega^{-1}$ in the low-energy region and by a function proportional to $\omega^{-\beta}$ with $\beta\sim 2\text{--}3$ in the high-energy region. In the present case, we see from the figure that the high-energy part of the spectrum decreases more rapidly than $\omega^{-3}$. This is be due to the fact that, as I have already pointed out in the previous Section, the production of high-energy pairs due to a slowly-rotating magnetic field is disfavored [see discussion below Eq. (\ref{f})]. Instead, concerning the low-energy region of the spectrum, I want to show analytically that the spectrum (\ref{dP_q_3}) goes just as $\omega^{-1}$ in the limit $w=\omega/m\ll 1$. In fact, 
all the terms in the integrals on $u$ and $v$ that are finite if calculated 
at $w=0$ give a linear dependence of the spectrum on the photon energy 
because of the presence of the overall factor proportional to $w$ in Eq. 
(\ref{dP_q_3}). By keeping the only diverging term in the low-energy limit, Eq. (\ref{dP_q_3}) becomes
\begin{equation}
\label{dP_q_3_l}
\begin{split}
\left\langle\frac{dN^{(\text{syn})}(\omega,t)}{d\omega dV}\right\rangle &\overset{w\ll 1}{\sim} 2\alpha_{em}wm^2(\Omega R_{\perp M})^2
\left(\frac{B_{\nnearrow}}{16\pi B_{cr}}\right)^3\\
&\quad\;\times\int_{-1}^1 du(1-u^2)\int_0^{\infty} dv 
\frac{v^2}{\epsilon^8(v)}\frac{1}{[\epsilon(v)-\epsilon(v+uw)+w]^2}.
\end{split}
\end{equation}
Now, I can write the diverging factor $[\epsilon(v)-\epsilon(v+uw)+w]^{-2}$ in the form
\begin{equation}
\frac{1}{[\epsilon(v)-\epsilon(v+uw)+w]^2}\simeq\frac{1}{w^2}\frac{\epsilon^2(v)}{\left[\epsilon(v)-v u\right]^2}
\end{equation}
and, by substituting this expression in Eq. (\ref{dP_q_3_l}), I finally have
\begin{equation}
\label{dP_q_3_l_2}
\begin{split}
&\left\langle\frac{dN^{(\text{syn})}(\omega,t)}{d\omega dV}\right\rangle\\
&\quad\overset{w\ll 1}{\sim}\frac{\alpha_{em}(m\Omega R_{\perp M})^2}{4w}
\left(\frac{B_{\nnearrow}}{8\pi B_{cr}}\right)^3\int_{-1}^1 du(1-u^2)\int_0^{\infty} dv \frac{v^2}{\epsilon^6(v)}\frac{1}{[\epsilon(v)-v u]^2}
\end{split}
\end{equation}
which is the desired result. In fact, since $\epsilon(v)-
v u>0$ in the integration domain, the two integrals are finite and then 
\begin{equation}
\left\langle\frac{dN^{(\text{syn})}(\omega,t)}{d\omega dV}\right\rangle \overset{\omega/m\ll 1}{\propto}\omega^{-1}.
\end{equation}
%
%
%
\section{Direct photon production from vacuum}
\label{Dir}
In the two previous Sections I have discussed two processes in which photons are produced by real electrons and positrons previously created in the presence of the magnetic field (\ref{B_rot_4}). Instead, I want to study here a process in which the photons are still produced in the presence of the slowly-rotating magnetic field (\ref{B_rot_4}) but directly from vacuum that is without an intermediate electromagnetic interaction with real electrons and positrons. As I have done previously, I will treat the dynamics of the system by means of the adiabatic perturbation theory. Since I want to analyze the direct production of photons, I have to build an Hamiltonian which describes the interaction between a quantized and a classical electromagnetic field. The use of the effective Lagrangian technique is particularly useful to this scope \cite{Dittrich1,Dittrich2} but, as a consequence, the results will be reliable only for photon energies much less than the electron mass $m$.\footnote{On the other hand, as I have said at the beginning of Sect. \ref{Ann}, in order to consider the external magnetic field as uniform and slowly-varying then the photon energies are always assumed to be much larger than $\Omega$.}

I start with a completely general expression of the effective Lagrangian density $\mathscr{L}$ of a free electromagnetic field $[\mathbf{E}_T(t,\mathbf{r}),\mathbf{B}_T(t,\mathbf{r})]$. It is known that, in order to be a \emph{true} Lorentz scalar, the effective Lagrangian density must be a function only of the relativistic invariants $F_T$ and $G_T^2$ (because $G_T$ is, actually, a \emph{pseudoinvariant}) with
\begin{subequations}
\begin{align}
F_T &=\frac{1}{2}(B_T^2-E_T^2), \\
G_T &=\mathbf{E}_T\cdot\mathbf{B}_T,
\end{align}
\end{subequations}
then, in general,
\begin{equation}
\label{L}
\mathscr{L}=\mathscr{L}(F_T,G_T^2).
\end{equation}
Now, suppose that the total electromagnetic field is the sum of a radiation field 
$[\boldsymbol{\Euler{E}}(t,\mathbf{r}),\boldsymbol{\Euler{B}}(t,\mathbf{r})]$ which will be quantized and of a classical field 
$[\mathbf{E}(t,\mathbf{r}),\mathbf{B}(t,\mathbf{r})]$ which, at the moment, I assume to be the constant and uniform magnetic field $\mathbf{B}$ given in Eq. (\ref{B}):
\begin{subequations}
\begin{align}
\mathbf{E}_T(t,\mathbf{r}) &=\boldsymbol{\Euler{E}}(t,\mathbf{r}), \\
\mathbf{B}_T(t,\mathbf{r}) &=\boldsymbol{\Euler{B}}(t,\mathbf{r})+\mathbf{B}.
\end{align}
\end{subequations}
Since I am not interested in the interactions of the radiation field with itself, I expand the effective Lagrangian density (\ref{L}) up to quadratic terms in $\boldsymbol{\Euler{E}}(t,\mathbf{r})$ and $\boldsymbol{\Euler{B}}(t,\mathbf{r})$ and I obtain the expression
\begin{equation}
\label{L_appr}
\mathscr{L}^{(2)}= 
\frac{c_1}{2}\left(\Euler{E}^2-\Euler{B}^2\right)+\frac{1}{2}\left[c_2(\mathbf{n}
\cdot\boldsymbol{\Euler{E}})^2+c_3(\mathbf{n}\cdot\boldsymbol{\Euler{B}})^2\right]
\end{equation}
where $\mathbf{n}=\mathbf{B}/B$ and where
\begin{subequations}
\label{c}
\begin{align}
\label{c_1}
c_1 &=-\left.\frac{\partial \mathscr{L}}{\partial F_T}\right\vert_{\boldsymbol{\Euler{E}}=\boldsymbol{\Euler{B}}=\mathbf{0}},\\
\label{c_2}
c_2 &=2B^2\left.\frac{\partial \mathscr{L}}{\partial 
G_T^2}\right\vert_{\boldsymbol{\Euler{E}}=\boldsymbol{\Euler{B}}=\mathbf{0}},\\
\label{c_3}
c_3 &=B^2\left.\frac{\partial^2 \mathscr{L}}{\partial 
F_T^{\;\;2}}\right\vert_{\boldsymbol{\Euler{E}}=\boldsymbol{\Euler{B}}=\mathbf{0}}.
\end{align}
\end{subequations}
In the Lagrangian density (\ref{L_appr}) the interaction between the radiation field and the classical 
field is described by the last two terms. The strength of the interaction depends on the 
coefficients $c_2$ and $c_3$ and I will give them later in the case in which the effective 
Lagrangian density is the Euler-Heisenberg Lagrangian density \cite{Heisenberg,Weisskopf,Schwinger}.

In order to build up the Hamiltonian density corresponding to the Lagrangian density (\ref{L_appr}) I have to introduce the fourpotential field $\Euler{A}^{\alpha}(t,\mathbf{r})=
[\Euler{V}(t,\mathbf{r}),\boldsymbol{\Euler{A}}(t,\mathbf{r})]$ relative to the radiation 
field. If I choose a gauge in which $\Euler{V}(t,\mathbf{r})=0$,\footnote{It can be shown that another condition is needed to fix unambiguously the radiation field gauge [see \cite{DiPiazza3} for a more detailed treatment].} the radiation field is given by
\begin{subequations}
\begin{align}
\boldsymbol{\Euler{E}}(t,\mathbf{r}) &=-\partial_t 
\boldsymbol{\Euler{A}}(t,\mathbf{r}), \\
\boldsymbol{\Euler{B}}(t,\mathbf{r}) &=\boldsymbol{\partial}\times 
\boldsymbol{\Euler{A}}(t,\mathbf{r}).
\end{align}
\end{subequations}
By expressing the Lagrangian density (\ref{L_appr}) in terms of the vector potential 
$\boldsymbol{\Euler{A}}(t,\mathbf{r})$ and of its derivatives, I can calculate the momenta 
$\boldsymbol{\pi}_{\boldsymbol{\Euler{A}}}(t,\mathbf{r})$ conjugated to $\boldsymbol{\Euler{A}}(t,\mathbf{r})$ as
\begin{equation}
\label{Pi}
\boldsymbol{\pi}_{\boldsymbol{\Euler{A}}}(t,\mathbf{r}) \equiv \frac{\partial \mathscr{L}^{(2)}}{\partial 
(\partial_t\boldsymbol{\Euler{A}})}=-c_1\boldsymbol{\Euler{E}}(t,\mathbf{r})-c_2[\mathbf{n}\cdot\boldsymbol{\Euler{E}}(t,\mathbf{r})]\mathbf{n}
\end{equation}
and, finally, the Hamiltonian density
\begin{equation}
\label{h}
\begin{split}
\mathscr{H}^{(2)} &\equiv\boldsymbol{\pi}_{\boldsymbol{\Euler{A}}}\cdot 
\partial_t\boldsymbol{\Euler{A}}-\mathscr{L}^{(2)}\\
&=\frac{\pi_{\boldsymbol{\Euler{A}}}^2}{2c_1}+\frac{c_1}{2}\Euler{B}^2-\frac{c_2}{2c_1(c_1+c_2)}(\mathbf{n}\cdot\boldsymbol{\pi}_{\boldsymbol{\Euler{A}}})^2-
\frac{c_3}{2}(\mathbf{n}\cdot\boldsymbol{\Euler{B}})^2.
\end{split}
\end{equation}

At this point in order to calculate the energy spectrum of the photons produced directly from vacuum in the presence of the slowly-rotating magnetic field (\ref{B_rot_4}), I should substitute the static magnetic field $\mathbf{B}$ with $\mathbf{B}_{\nnearrow}(t)$ and to apply to the total time-dependent Hamiltonian $H^{(2)}(t)=\int d\mathbf{r}\mathscr{H}^{(2)}(t)$ the adiabatic perturbation theory. Since, as I have mentioned, the production through this mechanism is low I only quote the final result in the particular case of the Euler-Heisenberg Lagrangean density \cite{Heisenberg,Weisskopf,Schwinger}:
\begin{equation}
\begin{split}
&\mathscr{L}^{(\text{E--H})}=-\frac{1}{4\pi}F_T\\
&\quad+\frac{1}{8\pi^2}\int_0^{\infty}\frac{ds}{s^3}\exp(-im^2s)\left[(es)^2|G_T|\cot\left(es\sqrt{\sqrt{F_T^2+G_T^2}+F_T}\right)\right.\\
&\left.\qquad\qquad\qquad\qquad\qquad\times\coth\left(es\sqrt{\sqrt{F_T^2+G_T^2}-F_T}\right)-1+\frac{2}{3}(es)^2F_T\right]
\end{split}
\end{equation}
where $F_T$ and $G_T$ are the relativistic invariants of a generic constant and uniform total electromagnetic field $(\mathbf{E}_T,\mathbf{B}_T)$. In our case, it can be shown that the following asymptotic estimates of the coefficients $c_1$, 
$c_2$ and $c_3$ hold for external magnetic field strengths $B_{\nnearrow}\gg B_{cr}$ [see Eqs. (\ref{c})]\footnote{Note that the coefficients $c_1$, 
$c_2$ and $c_3$ were defined starting from the Lagrangian density (\ref{L_appr}) where the magnetic field does not yet depend on time.}:
\begin{subequations}
\begin{align}
\label{c_1EH}
c_1^{(\text{E--H})} &\sim\frac{1}{4\pi}-\frac{\alpha_{em}}{3\pi}\log\left(\frac{B_{\nnearrow}}{B_{cr}}\right)\simeq\frac{1}{4\pi},\\
\label{c_2EH}
c_2^{(\text{E--H})} &\sim\frac{\alpha_{em}}{3\pi}\log\left(\frac{B_{\nnearrow}}{B_{cr}}\right),\\
\label{c_3EH}
c_3^{(\text{E--H})} &\sim\frac{\alpha_{em}}{3\pi}
\end{align}
\end{subequations}
where I put approximatively $c_1^{(\text{E--H})}\sim 1/(4\pi)$ since in every realistic physical situation the ``$\log$'' term is always negligible. By reminding that $\alpha_{em}^{-1}=137$, I can assume that the magnetic field $\mathbf{B}_{\nnearrow}(t)$ is such that $1\ll B_{\nnearrow}/B_{cr}\ll 1/\alpha_{em}$ (if $B_{\nnearrow}=10^{15}\;\text{gauss}$ then $B_{\nnearrow}/B_{cr}=22.7$). In this approximation it can be shown that the final energy spectrum of the photons produced directly from vacuum is given by
\begin{align}
\label{spec_dir_s}
\left\langle\frac{dN^{(\text{dir})}(\omega,t)}{d\omega dV}\right\rangle &\sim\frac{28}{135}\frac{\Omega^2}{(2\pi)^3}\left(\frac{\alpha_{em} B_{\nnearrow}}{B_{cr}}\right)^2 && \text{if $1\ll \frac{B_{\nnearrow}}{B_{cr}}\ll \alpha^{-1}_{em}$}.
\end{align}
As it is evident, the photon energy spectrum does not depend on the photon energy in contrast with the experimental GRBs spectra that, in the low-energy region $\omega\ll m$ are proportional to $\omega^{-1}$. Also, it can easily be checked that the direct production of photons from vacuum is low and completely negligible with respect to the production through synchrotron radiation (the comparison with the production through pair annihilation is harder because I have calculated not a photon spectrum but a photon spectrum per unit time). In fact, just to give an idea, by assuming $B_{\nnearrow}=10^{15}\;\text{gauss}$ and $R_{\perp M}=10^5\;\text{cm}$ [see the discussion below Eq. (\ref{alpha_e_p_mod})] then
\begin{equation}
\label{comp_dir_syn}
\left.\frac{\left\langle dN^{(\text{dir})}(\omega,t)/(d\omega dV)\right\rangle}{\left\langle dN^{(\text{syn})}(\omega,t)/(d\omega dV)\right\rangle}\right\vert_{\omega=1\;\text{MeV}}\sim \alpha_{em}\frac{B_{\nnearrow}}{B_{cr}}\left(\frac{\lambdabar}{R_{\perp M}}\right)^2=2.5\times 10^{-32}.
\end{equation}
%
%
\section{Summary and conclusions}
In this Chapter I have presented an analysis of the production of photons in the presence of a strong, uniform and slowly-rotating magnetic field. As I have said, this investigation has been motivated by suggestions coming from the study of GRBs. Nevertheless, the theoretical attitude has been to consider some very simplified versions of the processes so that the phenomenological parameters and the dynamical details could be reduced to the minimum. In fact, the external parameters entering the game are only two: the magnetic field strength $B_{\nnearrow}$ and its rotational frequency $\Omega$. The analysis has yielded some definite results, making evident, in quantitative form, the presence of a production of photons
\begin{enumerate}
\item through the annihilation of pairs previously created in the presence of the rotating magnetic field;
\item as synchrotron radiation by electrons and positrons previously created in the presence of the rotating magnetic field;
\item directly from vacuum.
\end{enumerate}
In the last two cases I have calculated the energy spectrum of the photons produced, while in the first one it resulted more natural to calculate the energy spectrum per unit time. In particular, only in the third case an analytical spectrum has been obtained. To do this, the effective Lagrangean technique has been used in such a way the resulting spectrum is reliable only for photon energies much less than the electron mass $m$. In this case I found that the spectrum becomes asymptotically independent of the photon energy $\omega$ [see Eq. (\ref{spec_dir_s})] while the experimental GRBs spectra go in the same limit as $\omega^{-1}$. Nevertheless, I have shown explicitly that \emph{the direct photon production mechanism is not quantitatively very efficient and it gives an energy spectrum completely negligible with respect to the corresponding synchrotron spectrum} [see Eq. (\ref{comp_dir_syn})]. The comparison with the photon production process through pair annihilation is less evident. In fact, in this case I have calculated the photon energy spectrum \emph{per unit time} resulting from the annihilation of two identical ``distributions'' of electrons and positrons created from vacuum in the presence of the slowly-rotating magnetic field (\ref{B_rot_4}) [see Eqs. (\ref{dN_domega}) and (\ref{f})]. The ``annihilation'' spectrum so obtained (see Fig. \ref{Spec_ann}) shows a sharp peak (break energy) around the electron mass and a rapid increasing (decreasing) below (above) $m$ and \emph{it is very different from the experimental GRB energy spectra} (see Fig. \ref{GRB_en_sp}). 

Instead, a much better qualitative agreement has been obtained between the synchrotron spectrum and the experimental ones (see Fig. \ref{Spec_syn}). In this case, I started from the Lagrangean density of QED in the presence of the external magnetic field $\mathbf{B}_{\nnearrow}(t)$ and I transformed it into a more useful ``effective'' Lagrangean density in order to exploit the fact that $\mathbf{B}_{\nnearrow}(t)$ had been assumed, in fact, \emph{slowly-rotating}. By calculating the $S$-matrix elements corresponding to the process of pair creation and subsequent photon emission by the electron or by the positron (see Fig. \ref{Feyn_dia}) I finally obtained the photon spectrum (\ref{dP_q_3}) that is shown in Fig. \ref{Spec_syn} when $B_{\nnearrow}=2.2\times 10^{14}\;\text{gauss}$. Similarly to the GRBs experimental spectra, the theoretical spectrum shows a double decreasing behaviour with two different slopes around the break energy. In this case the value of the break energy depends on the magnetic field strength and \emph{if $B_{\nnearrow}\sim 10^{14}\;\text{gauss}$ then the break energy lies between $1\;\text{MeV}$ and $3\;\text{MeV}$ as some experimental GRBs spectra}. Most important, \emph{the low-energy region of the synchrotron spectrum shows a linear dependence on the inverse of the photon energy exactly as the energy spectra of GRBs}. Finally, it is worth noting that, instead, in the high-energy region above the break energy both the annihilation and the synchrotron spectra decrease too rapidly with respect to the experimental GRB spectra. As I have said, this common feature is due to the fact that in my model the production of high-energy pairs (and then of high-energy photons) is not very efficient [see Eq. (\ref{f})].

\chapter{Pair production in a strong, uniform and slowly-varying magnetic field: the effect of a background gravitational field}
The aim of this Chapter is to continue the study of the production of electron-positron pairs in the presence of a strong, uniform and slowly-varying magnetic field started in Chap. 3. As I have pointed out in Sect. \ref{GRB}, the physical situation I have in mind is the creation of pairs
around astrophysical compact objects like magnetars or black holes. But, till now I have performed all the calculations in a flat spacetime, that is neglecting the effects of the gravitational field created by the
compact object. Even if there are situations in which
this can be safely done \cite{Gibbons1,Heyl}, it is interesting to study
what happens if the effects of the gravitational field are taken into
account. This is done in the present Chapter. In particular, after stating the general assumptions of the theoretical model (Sect. \ref{Gen_ass_g}), in Sect. \ref{weak} the structure of the gravitational field is assumed to be such that its
effects can be calculated perturbatively \cite{DiPiazza4} while, in Sect. \ref{strong} the effects of a strong gravitational field are taken into account nonperturbatively \cite{DiPiazza5}.
%
%
\section{General assumptions}
\label{Gen_ass_g}
In the previous Chapters I have performed all the calculations in Minkowski spacetime. By reminding the astrophysical scenario I described in Sect. \ref{GRB}, this means that the effects of the gravitational field of the central engine producing a GRB have been neglected. As I have said at the beginning of Par. \ref{Gen_ass}, if one considers the production of pairs around magnetars, this assumptions is very realistic. Nevertheless, we have seen in Par. \ref{Centr_eng} that many models identify the central engine of GRBs with a massive rotating black hole surrounded by an accretion disk. In this case, obviously, the strength of the gravitational field can be such large that its effects could be relevant. As I will explain in the following Sections, these effects are taken into account by working in the framework of quantum field theory in curved spacetime (see Sect. \ref{QFTCS}). Actually, in my model the main responsible of the pair creation is still the magnetic field and the fact that it varies with time. For this reason, the final presence probabilities will be calculated again by using the first-order adiabatic perturbation theory. Nevertheless, the presence of the black hole gravitational field will be taken into account in the determination of the one-particle electron and positron modes and energies and this fact will make the presence probabilities calculated here different from those obtained in Minkowski spacetime \cite{DiPiazza4,DiPiazza5}.

According to what I have just said, I assume that the spatial structure and
the time evolution of both the gravitational field and the magnetic field produced by the black hole are given. Nevertheless, in order to determine a
realistic form of these fields I should fix the physical properties of the source (its mass, its
eventual electric charge, its angular momentum and so on) and solve the system
built up by the Einstein equations and the general covariant Maxwell
equations. Clearly, this is a hopeless problem and a number of
approximations have to be done. In particular, I first assume that the
Einstein equations and the general covariant Maxwell equations are
disentangled. This corresponds to neglect the gravitational field produced
by the magnetic field and to assume that the spacetime metric is determined
only by the black hole. In order to further simplify the problem I assume that the spacetime metric is actually that produced by a spherical, uncharged, nonrotating black hole and that the corrections to this metric due to its eventual charge, to its rotation and to its magnetic field can be neglected. Nevertheless, the black hole itself should be capable to produce a \emph{time-varying} magnetic field. This request can be satisfied without changing what I have said about the spacetime metric, if the spherical body is collapsing but keeping its spherical symmetry and without rotating. In this case, in fact, the metric tensor outside the body does not change because of the Birkhoff theorem \cite{Weinberg} while the magnetic field is found to grow with time to compensate for the decreasing of the gravitational energy of the collapsing body.

In these approximations, as I have said, my starting point is the metric tensor corresponding to the field created by a spherical body with mass $M$ outside the body itself. If I call $t$, and $\mathbf{R}=(X,Y,Z)$ the four coordinates, this metric tensor is a function only of the radius $R=\sqrt{X^2+Y^2+Z^2}$ and it can be written as \cite{Landau2}
\begin{equation}
\label{g_mu_nu}
g_{\mu\nu}(R)=\text{diag}\left[\left(\frac{1-\frac{r_G}{4R}}{1+\frac{r_G}{4R}}\right)^2,
-\left(1+\frac{r_G}{4R}\right)^4,-\left(1+\frac{r_G}{4R}\right)^4,-\left(1+\frac{r_G}{4R}\right)^4\right]
\end{equation}
with $r_G=2GM$ the gravitational radius of the body ($G$ is the gravitational constant). I have chosen the so-called isotropic metric instead of the usual
(and equivalent) Schwarzshild metric, because from Eq. (\ref{g_mu_nu}) one sees
that the spatial distance is proportional to its Euclidean expression and this
will simplify future calculations. I point out that in this metric the
event horizon of the black hole is the spherical surface $R=r_G/4$.

Concerning the magnetic field, we have seen before that the rotating magnetic field configuration is very efficient from the point of view of pair production then it would be natural to consider here again the same configuration. Nevertheless, I have evaluated the corrections to the presence probabilities due to a weak gravitational field and, actually, there are not new interesting qualitative effects to be discussed [in any case they can be found in \cite{DiPiazza4}]. Instead, my first task in this Chapter is to show that also the presence of a \emph{weak} gravitational field superimposed to a time-varying magnetic field with fixed direction makes different from zero the probability that starting from vacuum a pair is present with the electron and the positron in a TGS. For this reason I will consider in this Chapter the following magnetic field time evolution
\begin{equation}
\label{B_exp}
\mathbf{B}^{\text{exp}}_{\shortuparrow}(t)=
\begin{pmatrix}
0\\
0\\
B^{\text{exp}}_{\shortuparrow}(t)
\end{pmatrix}=
\begin{pmatrix}
0\\
0\\
B_f+(B_i-B_f)\exp(-t/\tau)
\end{pmatrix}
\end{equation}
with $B_i<B_f$ and with $\tau$ a typical macroscopic time characterizing the time evolution of the black hole (for example, its collapse duration).

Since the spacetime is curved and the metric tensor is not simply $\eta_{\alpha\beta}$,
one must pay attention in defining the vector potential $A^{\text{exp}}_{\shortuparrow \mu}(t,\mathbf{R})$ that gives rise to $\mathbf{B}^{\text{exp}}_{\shortuparrow}(t)$. To this end, I assume that the
threedimensional components of the magnetic field $\mathbf{B}^{\text{exp}}_{\shortuparrow}(t)$ define the spatial-spatial components of the full covariant electromagnetic tensor
$F^{\text{exp}}_{\shortuparrow\mu\nu}(t,\mathbf{R})$ that then are uniform in space:
\begin{subequations}
\label{F_i_j}
\begin{align}
\label{F_i_j_x}
F^{\text{exp}}_{\shortuparrow 32}(t)=-F^{\text{exp}}_{\shortuparrow 23}(t) &\equiv 0,\\
F^{\text{exp}}_{\shortuparrow 13}(t)=-F^{\text{exp}}_{\shortuparrow 31}(t) &\equiv 0,\\
\label{F_i_j_z}
F^{\text{exp}}_{\shortuparrow 21}(t)=-F^{\text{exp}}_{\shortuparrow 12}(t) &\equiv B^{\text{exp}}_{\shortuparrow}(t).
\end{align}
\end{subequations}
All the other full covariant components of $F^{\text{exp}}_{\shortuparrow\mu\nu}(t,\mathbf{R})$ obviously vanish while the mixed or the full contravariant components are built by means of the
metric tensor (\ref{g_mu_nu}). Now, also in curved spacetime
\begin{equation}
\label{F_mu_nu}
F^{\text{exp}}_{\shortuparrow\mu\nu}(t)=\partial_{\mu}A^{\text{exp}}_{\shortuparrow \nu}(t,\mathbf{R})-\partial_{\nu}A^{\text{exp}}_{\shortuparrow \mu}(t,\mathbf{R}),
\end{equation}
then by means of Eqs. (\ref{F_i_j}), I can choose the
covariant vector $A^{\text{exp}}_{\shortuparrow \mu}(t,\mathbf{R})$ as
\begin{subequations}
\begin{align}
A^{\text{exp}}_{\shortuparrow 0}(t,\mathbf{R}) &=0,\\
A^{\text{exp}}_{\shortuparrow 1}(t,\mathbf{R}) &=\frac{1}{2}YB^{\text{exp}}_{\shortuparrow}(t),\\
A^{\text{exp}}_{\shortuparrow 2}(t,\mathbf{R}) &=-\frac{1}{2}XB^{\text{exp}}_{\shortuparrow}(t),\\
A^{\text{exp}}_{\shortuparrow 3}(t,\mathbf{R}) &=0.
\end{align}
\end{subequations}
These equations define a gauge corresponding to the symmetric gauge in Minkowski spacetime [see Eq. (\ref{A_p})]. Finally, it is convenient to define also the threedimensional vector potential $\mathbf{A}^{\text{exp}}_{\shortuparrow}(t,\mathbf{R})$ as
\begin{equation}
\label{A_exp}
\mathbf{A}^{\text{exp}}_{\shortuparrow}(t,\mathbf{R})=-\frac{1}{2}\big[\mathbf{R}\times\mathbf{B}^{\text{exp}}_{\shortuparrow}(t)\big]
\end{equation}
where the minus sign has been inserted to have $\boldsymbol{\partial}\times
\mathbf{A}^{\text{exp}}_{\shortuparrow}(t,\mathbf{R})=\mathbf{B}^{\text{exp}}_{\shortuparrow}(t)$.

Now, as I have done previously, I will calculate the probability that a pair is present at time $t$ by applying the first-order adiabatic perturbation theory. To do this, I have to build up the second-quantized Hamiltonian of a Dirac field $\Psi' (t,\mathbf{R})$ in the presence of the slowly-varying magnetic field (\ref{B_exp}) and in the curved spacetime with the static metric tensor (\ref{g_mu_nu}) and to determine its instantaneous eigenstates and
eigenenergies.\footnote{In this Chapter, I use the primed notation to indicate the Dirac field and the related quantities because I deal with a magnetic field varying only in strength and always lying in the $Z$ direction (see the notation used in Sect. \ref{Motion}).} The Lagrangian density of this system is given by [see Eq. (\ref{L_D_g})]:
\begin{equation}
\label{L_G}
\begin{split}
\mathscr{L}'(t)=
\sqrt{-g(R)}\bigg\{
&\frac{1}{2}\Big[\bar{\Psi}'\gamma^{\mu}(R)[i\mathscr{D}_{\mu}+eA^{\text{exp}}_{\shortuparrow\mu}(t,\mathbf{R})]\Psi'\\
&\left.-\bar{\Psi}'[i\overset{\shortleftarrow}{\mathscr{D}}_{\mu}-
eA^{\text{exp}}_{\shortuparrow\mu}(t,\mathbf{R})]\gamma^{\mu}(R)\Psi'\right]-m\bar{\Psi}'\Psi'\bigg\}
\end{split}
\end{equation}
where all the quantities representing the gravitational field are defined in Par. \ref{II_III_II}. Now, in general, a pair is created in a spatial volume with typical length given by the Compton length $\lambdabar=3.9\times 10^{-11}\;\text{cm}$ while for a typical 10 solar masses black hole $r_G=3.0\times 10^6\;\text{cm}$. In this way, since
$\lambdabar\ll r_G$, I am allowed to make some simplifications on the metric tensor (\ref{g_mu_nu}). In the following two Sections I will approximate the metric tensor (\ref{g_mu_nu}) and all the connected quantities appearing in Eq. (\ref{L_G}) in the two cases in which the pair is present far from (Sect. \ref{weak}) or near (Sect. \ref{strong}) the event horizon of the black hole. While in the first case the gravitational effects will be accounted for perturbatively, in the second one a nonperturbative approach will be followed.
%
%
\section{Weak-gravitational field case}
\label{weak}
In this Section I assume that the pair is produced in a small neighborhood of the space point labeled by the coordinates $(X_c,0,0)$ with $X_c>r_G/4+\Delta$ and $\Delta>0$. If
$(X_c+x,y,z)$ is a generic point in this neighborhood then
$|x|\lesssim\lambdabar$, $|y|\lesssim\lambdabar$ and $|z|\lesssim\lambdabar$
and I can calculate the metric tensor $g_{\mu\nu}(X_c+x,y,z)$ by keeping only the terms up to first order in $x/X_c$, $y/X_c$ and $z/X_c$. It can easily be seen that the resulting metric tensor depends only on $x$ and that it can be written as
\begin{equation}
\label{g_mu_nu_l}
g^{(1)}_{\mu\nu}(x)=g^{(0)}_{\mu\nu}+h_{\mu\nu}(x)
\end{equation}
where
\begin{subequations}
\begin{align}
\label{g_mu_nu_l_2}
g^{(0)}_{\mu\nu} &=\text{diag}(g_t,-g_s,-g_s,-g_s),\\
\label{g_mu_nu_l_2_h}
h_{\mu\nu}(x) &=\text{diag}(2f_t x,2f_s x,2f_s x,2f_s x)
\end{align}
\end{subequations}
with [see Eq. (\ref{g_mu_nu})]
\begin{subequations}
\label{Phi_c_g}
\begin{align}
\label{Phi_c}
g_t &=\left(\frac{1-r_G/4X_c}{1+r_G/4X_c}\right)^2,
& g_s &=\left(1+\frac{r_G}{4X_c}\right)^4,\\
\label{g}
f_t &=\frac{1-r_G/4X_c}{(1+r_G/4X_c)^3}\frac{r_G}{2X_c^2}, & f_s &=\left(1+\frac{r_G}{4X_c}\right)^3\frac{r_G}{2X_c^2}.
\end{align}
\end{subequations}

It is evident that, in order that $g^{(1)}_{\mu\nu}(x)$ is a good approximation
of $g_{\mu\nu}(X_c+x,y,z)$, $X_c$ can not be chosen to be too close to the critical
value $r_G/4$. Just to give an idea, it easy to see that, if $N\gg 1$ is a
large pure number, then
\begin{equation}
\label{h_on_g}
\left\vert \frac{g_{\mu\mu}(X_c+x,y,z)-g^{(1)}_{\mu\mu}(x)}{g_{\mu\mu}(X_c+x,y,z)}\right
\vert<\frac{1}{N}
\end{equation}
with $\mu=0,\ldots,3$, only if
\begin{equation}
\label{ineq}
X_c>\frac{r_G}{4}+\sqrt{N}\lambdabar.
\end{equation}
This condition automatically implies that
\begin{equation}
\label{h_on_g_2}
\left\vert \frac{h_{\mu\mu}(x)}{g^{(0)}_{\mu\mu}}\right\vert<\frac{1}{2\sqrt{N}}
\end{equation}
with $\mu=0,\ldots,3$ and then that $h_{\mu\nu}(x)$ can be considered as a
small correction of $g^{(0)}_{\mu\nu}$. In what follows, I assume that the
previous inequalities hold with sufficiently large $N$. I want to observe here that even for very large values of $N$, Eq. (\ref{ineq}) does not constraint very much the values of $X_c$: in this respect the expression ``far from the black hole event horizon'' is to be interpreted as ``\emph{microscopically} far from the black hole event horizon''.

Since $g^{(1)}_{\mu\nu}(x)$ has been split as in Eq. (\ref{g_mu_nu_l}) with
the matrix $h_{\mu\nu}(x)$ much smaller than the matrix $g^{(0)}_{\mu\nu}$,
I am allowed to keep in the Lagrangean density (\ref{L_G}) only the first-order terms in
$h_{\mu\nu}(x)$. To do this, I observe that
\begin{equation}
\label{g_l_expl}
g^{(1)}(x)\simeq g^{(0)}[1+h(x)]
\end{equation}
where
\begin{subequations}
\begin{align}
\label{g_0}
g^{(0)} &\equiv\det(g^{(0)}_{\mu\nu})=-g_tg_s^3,\\
\label{h_l}
h(x) &\equiv h_{\mu}^{\;\mu}(x)=2\left(\frac{f_t}{g_t}-\frac{3f_s}{g_s}\right)x.
\end{align}
\end{subequations}
Also, being the metric tensor $g^{(1)}_{\mu\nu}(x)$ diagonal, I can
choose a diagonal tetrad [see Eq. (\ref{g_tetrad})] with
\begin{subequations}
\begin{align}
\label{V_0^(1)1}
e_0^{(1)0}(x) &=\frac{1}{\sqrt{g^{(1)}_{00}(x)}}\simeq\frac{1}{\sqrt{g_t}}
\left(1-\frac{f_tx}{g_t}\right),\\
e_i^{(1)i}(x) &=\frac{1}{\sqrt{-g^{(1)}_{ii}(x)}}\simeq\frac{1}{\sqrt{g_s}}
\left(1+\frac{f_sx}{g_s}\right) && \text{(no sum)}.
\end{align}
\end{subequations}
By means of this tetrad it can be shown that the connections hidden in the covariant derivatives $\mathscr{D}_{\mu}$ and $\overset{\shortleftarrow}{\mathscr{D}}_{\mu}$ are already first-order quantities given by
\begin{equation}
\Gamma_{\mu}^{(1)}(x)=\frac{i}{4}\sigma^{1\beta}e_{1}^{(0)1}e_{\beta}^{(0)\rho}
\frac{dh_{\mu\rho}(x)}{dx}
\end{equation}
where $e_{\alpha}^{(0)\mu}$ is the diagonal zero-order tetrad with
\begin{subequations}
\begin{align}
e_0^{(0)0} &=\frac{1}{\sqrt{g^{(0)}_{00}}}=\frac{1}{\sqrt{g_t}},\\
e_i^{(0)i} &=\frac{1}{\sqrt{-g^{(0)}_{ii}}}=\frac{1}{\sqrt{g_s}}  && \text{(no sum)}.
\end{align}
\end{subequations}

Now, the procedure to calculate the Lagrangian density (\ref{L_G}) up to first
order in $h_{\mu\nu}(x)$ is identical to that used in the weak gravitational field
approximation \cite{Gupta,Ogievetskii} and I give only its final expression:
\begin{equation}
\label{L_lin}
\begin{split}
&\mathscr{L}^{\prime(1)}(t)\\
&\;=\frac{\sqrt{g_s^3}}{2}(1-f_Ex)
[\bar{\Psi}'\gamma^0(i\partial_0\Psi')-(i\partial_0\bar{\Psi}')\gamma^0\Psi']\\
&\quad\;+\frac{g_s\sqrt{g_t}}{2}(1-f_Px)\Big[\bar{\Psi}'\gamma^i[i\partial_i+
eA^{\text{exp}}_{\shortuparrow i}(t,\mathbf{r})]\Psi'-\bar{\Psi}'[i\overset{\shortleftarrow}{\partial}_i-eA^{\text{exp}}_{\shortuparrow i}(t,\mathbf{r})]\gamma^i\Psi'\Big]\\
&\quad\;-\sqrt{g_tg_s^3}(1-f_Mx)m\bar{\Psi}'\Psi'
\end{split}
\end{equation}
where $\mathbf{r}=(x,y,z)$ and where I defined the three couplings\footnote{I introduced three couplings
for later convenience because, actually, only two of them are independent.}
\begin{subequations}
\label{g_E_P_M}
\begin{align}
\label{g_E}
f_E &\equiv\frac{3f_s}{g_s}=\frac{3r_G}{2X_c^2}\bigg/\left(1+\frac{r_G}{4X_c}\right),\\
\label{g_P}
f_P &\equiv\frac{2f_s}{g_s}-\frac{f_t}{g_t}=\frac{r_G}{2X_c^2}
\left[2-\left(1-\frac{r_G}{4X_c}\right)^{-1}\right]\bigg/\left(1+\frac{r_G}{4X_c}\right),\\
\label{g_M}
f_M &\equiv\frac{3f_s}{g_s}-\frac{f_t}{g_t}=\frac{r_G}{2X_c^2}
\left[3-\left(1-\frac{r_G}{4X_c}\right)^{-1}\right]\bigg/\left(1+\frac{r_G}{4X_c}\right).
\end{align}
\end{subequations}
Note that the modifications induced by the metric tensor (\ref{g_mu_nu_l})
in the Lagrangian density (\ref{L_lin}) are linear in $f_t$ and $f_s$ (obviously)
but nonlinear in $g_t$ and $g_s$.

According to what I have said in Sect. \ref{QFTCS}, I define the Hamiltonian density of the Dirac field $\Psi'(t,\mathbf{r})$ as
\begin{equation}
\label{H_def}
\mathscr{H}^{\prime(1)}(t)\equiv \bar{\Pi}^{\prime(1)}(\partial_0\Psi')+(\partial_0
\bar{\Psi}')\Pi^{\prime(1)}-\mathscr{L}^{\prime(1)}(t)
\end{equation}
where
\begin{subequations}
\begin{align}
\bar{\Pi}^{\prime(1)}(t,\mathbf{r}) &\equiv
\frac{\partial \mathscr{L}^{\prime(1)}}{\partial (\partial_0\Psi')}=\sqrt{g_s^3}\frac{i(1-f_Ex)}{2}
\bar{\Psi}'(t,\mathbf{r})\gamma^0,\\
\Pi^{\prime(1)}(t,\mathbf{r}) &\equiv\frac{\partial \mathscr{L}^{\prime(1)}}{\partial (\partial_0\bar{\Psi}')}=-\sqrt{g_s^3}\frac{i(1-f_Ex)}{2}\gamma^0\Psi'(t,\mathbf{r})
\end{align}
\end{subequations}
are the first-order conjugated momenta to the fields $\Psi'(t,\mathbf{r})$ and
$\bar{\Psi}'(t,\mathbf{r})$ respectively. By using the previous equations it can easily be shown that, apart from derivatives terms, the Hamiltonian density (\ref{H_def}) can be written as
\begin{equation}
\label{H_a_1}
\mathscr{H}^{\prime(1)}(t)=\sqrt{g_s^3}
(1-f_Ex)\Psi^{\prime\dag}\mathcal{H}^{\prime(1)}(t)\Psi'
\end{equation}
where I introduced the one-particle first-order Hamiltonian
\begin{equation}
\label{H_1p_w}
\begin{split}
\mathcal{H}^{\prime(1)}(t) &= \sqrt{g_t}\bigg\{\frac{1}{2\sqrt{g_s}}\Big[(1-f_Px)\boldsymbol{\alpha}
\cdot[-i\boldsymbol{\partial}+e\mathbf{A}^{\text{exp}}_{\shortuparrow}(t,\mathbf{r})]\\
&\qquad\qquad +\boldsymbol{\alpha}\cdot[-i\boldsymbol{\partial}+
e\mathbf{A}^{\text{exp}}_{\shortuparrow}(t,\mathbf{r})](1-f_Px)\Big]+(1-f_Mx)\beta m\\
&\qquad\qquad +f_Ex\left[\frac{1}{\sqrt{g_s}}\boldsymbol{\alpha}
\cdot[-i\boldsymbol{\partial}+e\mathbf{A}^{\text{exp}}_{\shortuparrow}(t,\mathbf{r})]+\beta m\right]\bigg\}.
\end{split}
\end{equation}
Despite its appearance, the previous one particle Hamiltonian (\ref{H_1p_w}) is an Hermitian operator. In fact, the Hermiticity depends on the definition of the scalar product and I remind that the scalar product of two spinors $\Psi_1(t,\mathbf{r})$ and $\Psi_2(t,\mathbf{r})$ is defined in curved spacetime as in Eq. (\ref{Sc_prod_g}). By choosing the Cauchy surface $\Sigma$ as the $t=\text{const.}$ hypersurface, then $d\Sigma=d\mathbf{r}$, $n^{\mu}(t,\mathbf{r})=(1,0,0,0)$ \cite{Landau2} and the scalar product (\ref{Sc_prod_g}) becomes up to first-order terms
\begin{equation}
\label{s_p_f}
(\Psi_1,\Psi_2)^{(1)}=\int d\mathbf{r}\sqrt{g_s^3}(1-f_Ex)\Psi_1^{\dag}\Psi_2
\end{equation}
and the one-particle Hamiltonian (\ref{H_1p_w}) results, in fact, to be Hermitian. The fact that this one-particle Hamiltonian is a ``good'' Hamiltonian is also corroborated by the fact that the equation of motion of the field $\Psi'(t,\mathbf{r})$ given, in general, by
\begin{equation}
\partial_0\frac{\partial \mathscr{L}^{\prime(1)}}{\partial
(\partial_0\bar{\Psi}')}+\boldsymbol{\partial}\cdot\frac{\partial
\mathscr{L}^{\prime(1)}}{\partial (\boldsymbol{\partial}\bar{\Psi}')}=0,
\end{equation}
can be written in the present case as
\begin{equation}
\label{Eq_m}
i\partial_0\Psi'=\mathcal{H}^{\prime(1)}(t)\Psi'.
\end{equation}

Coherently, the total Hamiltonian of the Dirac field is given by
\begin{equation}
\label{H_tot_f}
H^{\prime(1)}(t)\equiv\int d\mathbf{r}\mathscr{H}^{\prime(1)}(t)=\int d\mathbf{r}\sqrt{g_s^3}(1-f_Ex)\Psi^{\prime\dag}\mathcal{H}^{\prime(1)}(t)\Psi'.
\end{equation}
Now, this Hamiltonian depends explicitly on time only through the dependence of
$\mathcal{H}^{\prime(1)}(t)$ on the magnetic field $\mathbf{B}^{\text{exp}}_{\shortuparrow}(t)$ hidden in the vector potential $\mathbf{A}^{\text{exp}}_{\shortuparrow}(t,\mathbf{r})$ [see Eq. (\ref{A_exp})], then it is a slowly-varying quantity. In this way, also in this case, in order to calculate the pair presence probability I can use the first-order adiabatic perturbation theory. Since the conceptual steps are the same as those followed in Chap. 3, I will not repeat all the details. In particular, I have to determine the one-particle electron and positron modes and energies of the time-independent Hamiltonian obtained from Eq. (\ref{H_1p_w}) by substituting the vector potential $\mathbf{A}^{\text{exp}}_{\shortuparrow}(t,\mathbf{r})$ with $\mathbf{A}'(\mathbf{r})$ given in Eq. (\ref{A_p}) and corresponding to the static magnetic field $\mathbf{B}'(0,0,B)$ [see Eq. (\ref{B_p})]. This one-particle Hamiltonian can be written as the sum
\begin{equation}
\label{H_1p_2}
\mathcal{H}^{\prime(1)}=\mathcal{H}^{\prime(0)}+\mathcal{I}'
\end{equation}
of the zero-order Hamiltonian
\begin{equation}
\label{H_1p_0}
\mathcal{H}^{\prime(0)}=\sqrt{\frac{g_t}{g_s}}
\left\{\boldsymbol{\alpha}\cdot[-i\boldsymbol{\partial}+e\mathbf{A}'(\mathbf{r})]+
\sqrt{g_s}\beta m\right\}
\end{equation}
and of the first-order interaction
\begin{equation}
\label{U_1p}
\begin{split}
\mathcal{I}'&=\sqrt{g_t}(f_P-f_M)\beta mx-\frac{f_P}{2}\{x,\mathcal{H}^{\prime(0)}\}+f_Ex\mathcal{H}^{\prime(0)}.
\end{split}
\end{equation}
In this way, in the present physical situation the one-particle electron and positron modes and energies can be determined perturbatively in the couplings $f_E$, $f_P$ and $f_M$ (or, equivalently, $f_t$ and $f_s$) by using the time-independent perturbation theory \cite{Landau3}. In particular, in the following Paragraph I will determine the zero-order electron and positron modes and the corresponding first-order energies. Instead, in Par. \ref{mod_1_ord} I will calculate the first-order corrections only to the zero-order TGSs previously obtained.
%
%
\subsection{Computation of the one-particle modes up to zero order and of the one-particle energies up to first order}
If $U_{\jmath}(\mathbf{r})$ and $V_{\jmath}(\mathbf{r})$ with $\jmath$ embodying all the needed quantum numbers are the electron and positron modes then they satisfy the equations
\begin{subequations}
\label{Eq_eig_u_v}
\begin{align}
\label{Eq_eig_u}
\mathcal{H}^{\prime(1)}U'_{\jmath} &=w_{\jmath} U'_{\jmath},\\
\label{Eq_eig_v}
\mathcal{H}^{\prime(1)}V'_{\jmath} &=-\tilde{w}_{\jmath} V'_{\jmath}
\end{align}
\end{subequations}
where $w_{\jmath}$ and $\tilde{w}_{\jmath}$ are the electron and positron one-particle energies.
The states $U'_{\jmath}(\mathbf{r})$ and $V'_{\jmath}(\mathbf{r})$ are assumed to be an
orthonormal basis with respect to the scalar product (\ref{s_p_f}), \textit{i.e.}
\begin{subequations}
\label{norm_1_2}
\begin{align}
\label{norm_1}
(U'_{\jmath},U'_{\jmath'})^{(1)} &=(V'_{\jmath},V'_{\jmath'})^{(1)}=\delta_{\jmath,\jmath'},\\
\label{norm_2}
(U'_{\jmath},V'_{\jmath'})^{(1)} &=0.
\end{align}
\end{subequations}

I first want to determine the zero-order solution of Eqs. (\ref{Eq_eig_u_v}). Up to this order those equations can be written as
\begin{subequations}
\begin{align}
\label{Eq_eig_u_0}
\sqrt{\frac{g_t}{g_s}}\left\{\boldsymbol{\alpha}\cdot[-i\boldsymbol{\partial}+
e\mathbf{A}'(\mathbf{r})]+\beta \sqrt{g_s}m\right\}U_{\jmath}^{\prime(0)} &=
w_{\jmath}^{(0)} U_{\jmath}^{\prime(0)},\\
\label{Eq_eig_v_0}
\sqrt{\frac{g_t}{g_s}}\left\{\boldsymbol{\alpha}\cdot[-i\boldsymbol{\partial}+
e\mathbf{A}'(\mathbf{r})]+\beta \sqrt{g_s}m\right\}V_{\jmath}^{\prime(0)} &=
-\tilde{w}_{\jmath}^{(0)} V_{\jmath}^{\prime(0)}.
\end{align}
\end{subequations}
These equations are the eigenvalue equations in Minkowski spacetime of a particle with mass $\sqrt{g_s}m$ and charge $-e<0$ in the presence of the magnetic field $\mathbf{B}'$, then their solutions can be written immediately. In particular, $\jmath=j\equiv\{n_d,k,\sigma,n_g\}$ and $w_{\jmath}^{(0)}$ and $\tilde{w}_{\jmath}^{(0)}$ are given by the modified Landau levels
\begin{subequations}
\label{w_0_t_0}
\begin{align}
\label{w_0}
w_j^{(0)} &=\sqrt{g_t m^2+\frac{g_t}{g_s}\left[k^2+eB(2n_d+1+\sigma)\right]},\\
\label{w_t_0}
\tilde{w}_j^{(0)} &=\sqrt{g_t m^2+\frac{g_t}{g_s}\left[k^2+eB(2n_g+1-\sigma)\right]}.
\end{align}
\end{subequations}
Instead, the corresponding modes will be indicated as $u^{\prime(0)}_j(\mathbf{r})$ and $v^{\prime(0)}_j(\mathbf{r})$ respectively and they are given by Eqs. (\ref{u_p_v_p}) with $\sqrt{g_s}m$ instead of $m$, $w_j^{(0)}$ instead of $w_j$ and with $\tilde{w}_j^{(0)}$ instead of $\tilde{w}_j$. Now, analogously to what I have said in Par. \ref{Motion_QM}, the one-particle electron and positron energies (\ref{w_0_t_0}) have two kinds of degenerations. The first one is due to the fact that they do not depend on the quantum number $n_g$ ($n_d$). The second one is due to the fact that the electron (positron) modes with quantum numbers
$j_+\equiv\{n_d,k,+1,n_g\}$ and $j_-\equiv\{n_d+1,k,-1,n_g\}$ ($\tilde{j}_+\equiv\{n_d,k,+1,n_g+1\}$ and
$\tilde{j}_-\equiv\{n_d,k,-1,n_g\}$) have
the same energy whatever $n_g$ ($n_d$). This means, following the time-independent perturbation theory, that the modes $u^{\prime(0)}_j(\mathbf{r})$ and $v^{\prime(0)}_j(\mathbf{r})$
will not represent, in general, the correct zero-order modes
$U_{\jmath}^{\prime(0)}(\mathbf{r})$ and $V_{\jmath}^{\prime(0)}(\mathbf{r})$ and, for this
reason, they have been indicated by means of the symbols $u^{\prime(0)}_j(\mathbf{r})$ and
$v^{\prime(0)}_j(\mathbf{r})$.

Now, in the following, I will compute explicitly only the zero-order electron modes and the
first-order electron energies, while the analogous results for the positron
modes and energies will be only quoted. Following the time-independent
perturbation theory for degenerate states, I write the zero-order solutions of
Eq. (\ref{Eq_eig_u}) with a given energy as linear combinations of all the
degenerate modes $u^{\prime(0)}_j(\mathbf{r})$ and $v^{\prime(0)}_j(\mathbf{r})$ with that
energy. Now, it can easily be shown that the perturbation $\mathcal{I}'$ does not remove the energy degeneracy of the modes characterized by the quantum numbers $j_-$ and $j_+$ [see \cite{DiPiazza4} for further details]. In other words, following the time-independent perturbation theory for degenerate states \cite{Landau3}, I have to diagonalize the perturbation $\mathcal{I}'$ inside every subspace labeled by the quantum numbers $j$ whatever $n_g$. In this way, the ``true'' zero-order modes are characterized by the quantum numbers $\{n_d,k,\sigma\}$ and by a new index that, for later convenience, will be indicated as $x_0$. If I call $U^{\prime(0)}_{n_d,k,\sigma,x_0}(\mathbf{r})$ the resulting zero-order modes then
\begin{equation}
\label{U_0}
U^{\prime(0)}_{n_d,k,\sigma,x_0}(\mathbf{r})=\sum_{n_g=0}^{\infty} \mathsf{P}^{(0)}_{n_d,k,\sigma,x_0;n_d,k,\sigma,n_g}u^{\prime(0)}_{n_d,k,\sigma,n_g}(\mathbf{r})
\end{equation}
where the coefficients $\mathsf{P}^{(0)}_{n_d,k,\sigma,x_0;n_d,k,\sigma,n_g}$ solve the secular equation \cite{Landau3}
\begin{equation}
\label{Sec_eq}
\sum_{n'_g=0}^{\infty}\left(\mathcal{I}'_{n_d,k,\sigma,n_g;n_d,k,\sigma,n'_g}- \epsilon_{n_d,k,\sigma,x_0}\delta_{n_g,n'_g}\right)
\mathsf{P}^{(0)}_{n_d,k,\sigma,x_0;n_d,k,\sigma,n'_g}=0
\end{equation}
with, in general,
\begin{equation}
\label{I_jjp}
\mathcal{I}'_{jj'}\equiv\int d\mathbf{r}\sqrt{g_s^3} u^{\prime\dag}_j(\mathbf{r})\mathcal{I}'u'_{j'}(\mathbf{r})
\end{equation}
and with $\epsilon_{n_d,k,\sigma,x_0}$ the first-order corrections (to be determined) to the energies of the modes $U^{\prime(0)}_{n_d,k,\sigma,x_0}(\mathbf{r})$.

Now, one can show that the matrix elements $\mathcal{I}'_{n_d,k,\sigma,n_g;n_d,k,\sigma,n'_g}$ can be written as
\begin{equation}
\label{m_e}
\begin{split}
&\mathcal{I}'_{n_d,k,\sigma,n_g;n_d,k,\sigma,n'_g}\\
&\qquad=\bigg[(f_P-f_M) \frac{g_tm^2}{w^{(0)}_j}+(f_E-f_P)
w^{(0)}_j\bigg]\int dx dy \theta^{\prime *}_{n_d,n_g}(x,y)x_0\theta'_{n_d,n'_g}(x,y)
\end{split}
\end{equation}
where the operator $x_0$ and the functions $\theta'_{n_d,n_g}(x,y)$ have been
defined in Eqs. (\ref{x_0}) and (\ref{theta_p}) respectively. From Eqs.
(\ref{u_p_v_p}) and (\ref{phi_p_chi_p}) we see that only the transverse
functions $\theta'_{n_d,n_g}(x,y)$ in $u^{\prime(0)}_j(\mathbf{r})$ depend on $n_g$, then
I have to determine the coefficients $\mathsf{P}^{(0)}_{n_d,k,\sigma,x_0;n_d,k,\sigma,n_g}$ in such a way that the linear combination $\sum_{n_g=0}^{\infty} \mathsf{P}^{(0)}_{n_d,k,\sigma,x_0;n_d,k,\sigma,n_g}
\theta'_{n_d,n_g}(x,y)$ diagonalizes the operator $x_0$.\footnote{Now, it is
clear why I have called the additional index $x_0$.} This linear combination is
given in \cite{Cohen} [it is Eq. (104) in Complement $\text{E}_{\text{VI}}$].
The coefficients $\mathsf{P}^{(0)}_{n_d,k,\sigma,x_0;n_d,k,\sigma,n_g}$ result independent of the quantum numbers $\{n_d,k,\sigma\}$ and they are given by
\begin{equation}
\mathsf{P}^{(0)}_{n_d,k,\sigma,x_0;n_d,k,\sigma,n_g}=\mathsf{P}^{(0)}_{n_g}(x_0)=\sqrt[4]{\frac{eB}{\pi}}\frac{1}{\sqrt{2^{n_g} n_g!}}\mathrm{H}_{n_g}(\sqrt{eB}x_0)\exp\left(-\frac{eBx_0^2}{2}\right)
\end{equation}
where I pointed out that $x_0$ is a continuous quantum number and where
\begin{equation}
\mathrm{H}_{n_g}(z)=\frac{2^{n_g}}{\sqrt{\pi}}\int_{-\infty}^{\infty}
ds(z+is)^{n_g}\exp\left(-s^2\right)
\end{equation}
is the $n_g$th-order Hermite polynomial \cite{Ryzhik}. As a conclusion, the spinors
$U^{\prime(0)}_{n_d,k,\sigma}(x_0;\mathbf{r})$ have the same form of the spinors $u^{\prime(0)}_j(\mathbf{r})$, but with the function $\theta'_{n_d,n_g}(x,y)$ substituted by the function
\begin{equation}
\Theta'_{n_d}(x_0;x,y)\equiv\sum_{n_g=0}^{\infty}\mathsf{P}^{(0)}_{n_g}(x_0)\theta'_{n_d,n_g}(x,y).
\end{equation}
By using the second expression of $\theta'_{n_d,n_g}(x,y)$ in Eq. (\ref{theta_p}) I obtain
\begin{equation}
\label{phi_t_xg}
\begin{split}
\Theta'_{n_d}(x_0;x,y) &=\frac{(a_d^{\dag})^{n_d}}{\sqrt{n_d!}}\sqrt[4]{\frac{eB}{\pi}}
\sqrt{\frac{eB}{2}}\exp\left[-\frac{eB}{2}\left(x_0^2+\frac{x^2+y^2}{2}\right)\right]
\frac{1}{\pi}\\
&\quad\times\int_{-\infty}^{\infty}ds\exp\left(-s^2\right)\sum_{n_g=0}^{\infty}\frac{1}{n_g!}
\left[(\sqrt{eB}x_0+is)\sqrt{eB}(x-iy)\right]^{n_g}\\
&=\frac{(a_d^{\dag})^{n_d}}{\sqrt{n_d!}}\sqrt[4]{\frac{eB}{\pi}}\sqrt{\frac{eB}{2}}
\exp\left[-\frac{eB}{2}\left(x_0^2+\frac{x^2+y^2}{2}\right)\right]\frac{1}{\pi}\\
&\quad\times  \exp\left[eBx_0(x-iy)\right]\int_{-\infty}^{\infty}ds
\exp\left[-s^2+i\sqrt{eB}(x-iy)s\right]\\
&=\frac{(a_d^{\dag})^{n_d}}{\sqrt{n_d!}}\sqrt[4]{\frac{eB}{\pi}}
\sqrt{\frac{eB}{2\pi}}\exp\left\{-\frac{eB}{2}\left[(x-x_0)^2-iy(x-2x_0)\right]\right\}.
\end{split}
\end{equation}

From now on, the quantum number $n_g$ completely disappears in the calculations
because it is essentially substituted by the quantum number $x_0$. This circumstance
allows me to simplify the notation. In fact, I can eliminate the index ``$d$''
from the remaining quantum number $n_d$ and from the related operators $a_d$ and $a_d^{\dag}$ [see Eqs. (\ref{a_d}) and (\ref{a_d_c})]. Also, later calculations will be simplified if I discretize the eigenvalues $x_0$. This can be done by requiring the functions $\Theta'_{n_d}(x_0;x,y)$ to satisfy the periodicity condition at $x=0$
\begin{equation}
\exp\left(i\frac{eBx_0L_y}{2}\right)=\exp\left(-i\frac{eBx_0L_y}{2}\right)
\end{equation}
where $L_y$ is the length of the quantization volume in the $y$ direction. In this
way, the allowed eigenvalues are given by the discrete values
\begin{align}
x_0=\frac{2\ell\pi}{eBL_y} && \ell=0,\pm 1,\dots.
\end{align}
I point out that if I imposed the periodicity condition at $x'\neq 0$, the allowed eigenvalues would change. Nevertheless, since at the end of the calculations I will perform the continuum limit $L_y\to\infty$, I am not interested in the exact values of the allowed eigenvalues but only in the eigenstate density $\varrho (x_0)$ which is
\begin{equation}
\label{varrho}
\varrho (x_0)\equiv \frac{d\ell}{d x_0}=\frac{eBL_y}{2\pi}
\end{equation}
independently of $x_0$. In this way, the function (\ref{phi_t_xg}) becomes
\begin{equation}
\label{phi_t_xg_d}
\Theta'_{n,x_0}(x,y)=\frac{(a^{\dag})^n}{\sqrt{n!}}\sqrt[4]{\frac{eB}{\pi L_y^2}}
\exp\left\{-\frac{eB}{2}\left[(x-x_0)^2-iy(x-2x_0)\right]\right\}.
\end{equation}
The numerical factors have been chosen in such a way that if I define [analogously
to the twodimensional spinors $\varphi'_j(\mathbf{r})$ given in Eq. (\ref{phi_p})]
the twodimensional spinors
\begin{equation}
\label{Phi}
\Phi'_J(\mathbf{r})\equiv\frac{\exp(ikz)}{\sqrt{L_z}}f'_{\sigma}\Theta'_{n,x_0}(x,y)
\end{equation}
with
\begin{equation}
J\equiv\{n,k,\sigma,x_0\},
\end{equation}
they result normalized as [see also Eq. (\ref{ort_phi_chi_1})]
\begin{equation}
\label{ort_Phi}
\int d\mathbf{r}\Phi^{\prime\dag}_J(\mathbf{r})\Phi'_{J'}(\mathbf{r})=\delta_{J,J'}
\end{equation}
with $\delta_{J,J'}\equiv\delta_{n,n'}\delta_{k,k'}\delta_{\sigma,\sigma'}
\delta_{x_0,x_0'}$. Finally, the zero-order spinors $U^{\prime (0)}_J(\mathbf{r})$
are given by [see Eq. (\ref{u_p})]
\begin{equation}
\label{U^P0_f}
U^{\prime (0)}_J(\mathbf{r})=\frac{1}{\sqrt[4]{g_s^3}}\sqrt{\frac{w^{(0)}_J+
\sqrt{g_t}m}{2w^{(0)}_J}}
\begin{pmatrix}
      \Phi'_J(\mathbf{r})\\ \sqrt{\dfrac{g_t}{g_s}}
\dfrac{\mathcal{V}'}{w_J^{(0)}+
\sqrt{g_t}m}\Phi'_J(\mathbf{r})
      \end{pmatrix}
\end{equation}
and they are normalized as [see Eq. (\ref{s_p_f})]
\begin{equation}
\int d\mathbf{r}\sqrt{g_s^3}U^{\prime(0)\dag}_J(\mathbf{r})U^{\prime(0)}_{J'}(\mathbf{r})=\delta_{J,J'}.
\end{equation}
In Eq. (\ref{U^P0_f}) I used the fact that the energies (\ref{w_0}) do not depend neither on $n_g$ nor on $x_0$ then $w_j^{(0)}\equiv w_{n_d,k,\sigma,n_g}^{(0)}=w_{n,k,\sigma,x_0}^{(0)}\equiv w_J^{(0)}$.

Instead, from Eq. (\ref{m_e}) the energy corresponding to the mode $U^{\prime (0)}_J(\mathbf{r})$ is, up to first order,
\begin{equation}
\label{w_1}
\begin{split}
w_J^{(1)} &=w_J^{(0)}+\left[(f_P-f_M) \frac{g_tm^2}{w^{(0)}_J}+
(f_E-f_P)w^{(0)}_J\right]x_0\\
&=\sqrt{g_t m^2+\frac{g_t}{g_s}\left[k^2+eB(2n+1+\sigma)\right]}\\
&\qquad\qquad\times\left\{1+\left[\frac{f_t}{g_t}+\frac{f_s}{g_s}\frac{k^2+eB(2n+1+\sigma)}{g_sm^2+k^2+eB(2n+1+\sigma)}\right]x_0\right\}
\end{split}
\end{equation}
where I used the definitions (\ref{g_E_P_M}) of the coefficients
$f_E$, $f_P$ and $f_M$. In an analogous way one can write the positron energies up to first order as
\begin{equation}
\label{w_t_1}
\begin{split}
\tilde{w}_J^{(1)}&=\sqrt{g_t m^2+\frac{g_t}{g_s}
\left[k^2+eB(2n+1-\sigma)\right]}\\
&\qquad\qquad\times\left\{1+\left[\frac{f_t}{g_t}+
\frac{f_s}{g_s}\frac{k^2+eB(2n+1-\sigma)}{g_sm^2+k^2+eB(2n+1-\sigma)}\right]x_0\right\}.
\end{split}
\end{equation}
I observe that, as in the case of the electron modes and
energies, the quantum number $n_d$ is here substituted by $x_0$, and
it is pointless to keep the index ``$g$'' in the remaining quantum number
$n_g$ because there is no more possibility of ambiguity.

Finally, the zero-order positron modes are given by [see Eq. (\ref{v_p})]
\begin{equation}
\label{V^P0_f}
V^{\prime (0)}_J(\mathbf{r})= \frac{\sigma}{\sqrt[4]{g_s^3}}
\sqrt{\frac{\tilde{w}^{(0)}_J+\sqrt{g_t}m}{2\tilde{w}^{(0)}_J}}
\begin{pmatrix}
-\sqrt{\dfrac{g_t}{g_s}}\dfrac{\mathcal{V}'}{\tilde{w}_J^{(0)}+\sqrt{g_t}m}\mathrm{X}'_{J}(\mathbf{r})\\
      \mathrm{X}'_J(\mathbf{r})
      \end{pmatrix}
\end{equation}
with [see Eq. (\ref{chi_p})]
\begin{equation}
\label{Chi}
\mathrm{X}'_J(\mathbf{r})\equiv\frac{\exp(-ikz)}{\sqrt{L_z}}f'_{-\sigma}\Theta'_{n,x_0}(x,y)
\end{equation}
and they satisfy the orthonormalization relations
\begin{subequations}
\begin{align}
\label{norm_V^0}
&\int d\mathbf{r}\sqrt{g_s^3}V^{\prime(0)\dag}_J(\mathbf{r})V^{\prime(0)}_{J'}(\mathbf{r})=\delta_{J,J'}\\
&\int d\mathbf{r}\sqrt{g_s^3}V^{\prime(0)\dag}_J(\mathbf{r})U^{\prime(0)}_{J'}(\mathbf{r})=0.
\end{align}
\end{subequations}
%
%
%
\subsection{Computation of the one-particle TGSs up to first order}
\label{mod_1_ord}
I should pass now to the determination of the one-particle states
$U'_{\jmath}(\mathbf{r})$ and $V'_{\jmath}(\mathbf{r})$ up to first order in the
perturbation $\mathcal{I}'$. But, from Eqs.
(\ref{w_1}) and (\ref{w_t_1}) we see that the first-order energies of the
TGSs, that are given by
\begin{equation}
\label{w_g_1}
\begin{split}
\varepsilon^{(1)}_{k,x_0}&\equiv w^{(1)}_{0,k,-1,x_0}=\tilde{w}^{(1)}_{0,k,+1,x_0}\\
&=\sqrt{g_t m^2+\frac{g_t}{g_s}k^2}\left[1+\left(\frac{f_t}{g_t}+
\frac{f_s}{g_s}\frac{k^2}{g_sm^2+k^2}\right)x_0\right],
\end{split}
\end{equation}
are also independent of the magnetic field strength as in Minkowski spacetime. As we have seen before, this fact gives the TGSs a particular relevance in the
presence of strong magnetic fields because their energies are much smaller than the excited Landau levels. For this reason, in the following I will only calculate the presence probability of a pair with the electron and the positron both in a TGS and then I need to compute only the one-particle TGSs corrected up to first order. As in the previous
Paragraph, I will present the calculations of the first-order corrections to the
electron TGSs and I will quote the analogous results for the positron TGSs.

The first-order corrections to a given zero-order electron TGS come from the coupling of this state with the following classes of states:
\begin{enumerate}
\item the zero-order electron TGSs with the same energy
(all but the state to be corrected);
\item the zero-order electron TGSs with different energy;
\item the zero-order electron states that are not TGSs;
\item the zero-order positron states.
\end{enumerate}
Now, I suppose I want to calculate the first-order corrections to the
TGS labeled by $J_0\equiv\{0,k,-1,x_0\}$. The states in the
first class are labeled by the quantum numbers $\{0,k,-1,x'_0\}$ with $x'_0\neq x_0$ because they have the same energy of the state labeled by $J_0$. But, all the contributions vanish
because, in general, the perturbation (\ref{U_1p}) can not couple two modes of
$\mathcal{H}^{\prime(0)}$ with the same $n_d$ and two different
$x_0$ and $x'_0$. Similarly, the states in the second class are
characterized by the quantum numbers $\{0,k',-1,x'_0\}$ with $k'\neq k$ and, since
$[\mathcal{I}',\mathcal{P}_z]=0$ they do not give any contribution. Instead, the contributions from the states of the remaining two classes are, in general, different from zero in such a way the state $U^{\prime(1)}_{J_0}(\mathbf{r})$ can be written as
\begin{equation}
U^{\prime(1)}_{J_0}(\mathbf{r})=U^{\prime(0)}_{J_0}(\mathbf{r})+\sideset{}{'}\sum_{J'}
\mathsf{P}^{(1)}_{J_0,J'}U^{\prime(0)}_{J'}(\mathbf{r})+
\sum_{J'}\mathsf{Q}^{(1)}_{J_0,J'}V^{\prime(0)}_{J'}(\mathbf{r})
\end{equation}
where the primed sum does not include the TGSs and where \cite{Landau3}
\begin{subequations}
\begin{align}
\mathsf{P}^{(1)}_{J_0,J'} &=\frac{1}{\varepsilon^{(0)}_k-w^{(0)}_{J'}}
\int d\mathbf{r}\sqrt{\phi^3_s}U^{\prime(0)\dag}_{J'}(\mathbf{r})
\mathcal{I}'U^{\prime(0)}_{J_0}(\mathbf{r}),\\
\mathsf{Q}^{(1)}_{J_0,J'} &=\frac{1}{\varepsilon^{(0)}_k+\tilde{w}^{(0)}_{J'}}
\int d\mathbf{r}\sqrt{\phi^3_s}V^{\prime(0)\dag}_{J'}(\mathbf{r})
\mathcal{I}'U^{\prime(0)}_{J_0}(\mathbf{r}).
\end{align}
\end{subequations}
In these equations I introduced the zero-order energies of the TGSs $\varepsilon^{(0)}_k$ defined as [see Eqs. (\ref{w_0_t_0})]
\begin{equation}
\label{E_0}
\varepsilon^{(0)}_k\equiv w^{(0)}_{0,k,-1,x_0}=\tilde{w}^{(0)}_{0,k,+1,x_0}=
\sqrt{g_t m^2+\frac{g_t}{g_s}k^2}.
\end{equation}
Now, I start by calculating the coefficients $\mathsf{P}^{(1)}_{J_0,J'}$. From the expression (\ref{U_1p}) of the interaction Hamiltonian and from Eq. (\ref{U^P0_f}) we have
\begin{equation}
\begin{split}
\mathsf{P}^{(1)}_{J_0,J'} &=\frac{1}{\varepsilon^{(0)}_k-w^{(0)}_{J'}}
\sqrt{\frac{w^{(0)}_{J'}+\sqrt{g_t}m}{2w^{(0)}_{J'}}}
\sqrt{\frac{\varepsilon^{(0)}_k+\sqrt{g_t}m}{2\varepsilon^{(0)}_k}}\\
&\quad\times\bigg\{\left[\sqrt{g_t}(f_P-f_M)m-\frac{f_P}{2}\big(w^{(0)}_{J'}+
\varepsilon^{(0)}_k\big)+f_E\varepsilon^{(0)}_k\right]\\
&\qquad\qquad\times\int d\mathbf{r}
\Phi^{\prime \dag}_{J'}(\mathbf{r})x\Phi'_{J_0}(\mathbf{r})\\
&\qquad\quad +\left[-\sqrt{g_t}(f_P-f_M)m-\frac{f_P}{2}\big(w^{(0)}_{J'}+
\varepsilon^{(0)}_k\big)+f_E\varepsilon^{(0)}_k\right]\frac{g_t}{g_s}\\
&\qquad\qquad \times\frac{1}{\varepsilon^{(0)}_k+\sqrt{g_t}m}\frac{1}{w^{(0)}_{J'}+
\sqrt{g_t}m}\int d\mathbf{r} \Phi^{\prime \dag}_{J'}(\mathbf{r})
\mathcal{V}'x\mathcal{V}'\Phi'_{J_0}(\mathbf{r})\bigg\}.
\end{split}
\end{equation}
By using the orthonormal properties (\ref{ort_Phi}) of the functions
$\Phi'_J(\mathbf{r})$, I can write the quantities $\mathsf{P}^{(1)}_{J_0,J'}$ as
\begin{equation}
\label{P_1_f}
\mathsf{P}^{(1)}_{J_0,J'}=\left(\mathsf{B}^{(1)}_{k,x_0}\delta_{n',+1}
\delta_{\sigma', -1}-i\mathsf{C}^{(1)}_{k,x_0}\delta_{n',0}\delta_{\sigma', +1}\right)
\delta_{k',k}\delta_{x'_0,x_0}
\end{equation}
where I defined the coefficients
\begin{subequations}
\begin{align}
\label{B_c}
\begin{split}
\mathsf{B}^{(1)}_{k,x_0} &\equiv\frac{1}{\mathcal{E}^{(0)}_k-
\varepsilon^{(0)}_k}\sqrt{\frac{\mathcal{E}^{(0)}_k+\sqrt{g_t}m}
{2\mathcal{E}^{(0)}_k}}\sqrt{\frac{\varepsilon^{(0)}_k+\sqrt{g_t}m}
{2\varepsilon^{(0)}_k}}\\
&\quad\times\left\{\sqrt{g_t}(f_M-f_P)\frac{m}{\sqrt{2eB}}
\left[1-\frac{g_t}{g_s}\frac{k^2}{\big(\mathcal{E}^{(0)}_k+
\sqrt{g_t}m\big)\big(\varepsilon^{(0)}_k+\sqrt{g_t}m\big)}\right]\right.\\
&\qquad\quad+\frac{1}{\sqrt{2eB}}\left[\frac{f_P}{2}\big(\mathcal{E}^{(0)}_k+
\varepsilon^{(0)}_k\big)-f_E\varepsilon^{(0)}_k\right]\\
&\qquad\qquad\quad\times\left.\left[1+
\frac{g_t}{g_s}\frac{k^2}{\big(\mathcal{E}^{(0)}_k+
\sqrt{g_t}m\big)\big(\varepsilon^{(0)}_k+\sqrt{g_t}m\big)}\right]\right\},
\end{split}\\
\label{C_c}
\begin{split}
\mathsf{C}^{(1)}_{k,x_0} &\equiv\frac{1}{\mathcal{E}^{(0)}_k-
\varepsilon^{(0)}_k}\sqrt{\frac{\mathcal{E}^{(0)}_k+\sqrt{g_t}m}
{2\mathcal{E}^{(0)}_k}}\sqrt{\frac{\varepsilon^{(0)}_k+\sqrt{g_t}m}
{2\varepsilon^{(0)}_k}}\\
&\quad\times\left[\sqrt{g_t}(f_M-f_P)m-\frac{f_P}{2}\big(\mathcal{E}^{(0)}_k+
\varepsilon^{(0)}_k\big)+f_E\varepsilon^{(0)}_k\right]\\
&\quad\times\frac{g_t}{g_s}
\frac{k}{\big(\mathcal{E}^{(0)}_k+\sqrt{g_t}m\big)\big(\varepsilon^{(0)}_k+
\sqrt{g_t}m\big)}
\end{split}
\end{align}
\end{subequations}
with [see Eqs. (\ref{w_0_t_0})]
\begin{equation}
\label{E_1}
\begin{split}
\mathcal{E}^{(0)}_k&\equiv w^{(0)}_{0,k,+1,x_0}=w^{(0)}_{1,k,-1,x_0}=
\tilde{w}^{(0)}_{0,k,-1,x_0}=\tilde{w}^{(0)}_{1,k,+1,x_0}\\
&=\sqrt{g_t m^2+
\frac{g_t}{g_s}\left(k^2+2eB\right)}
\end{split}
\end{equation}
the zero-order energy of the first-excited Landau level. In the same
way, I can write the coefficients $\mathsf{Q}^{(1)}_{J_0,J'}$ as
\begin{equation}
\label{Q_1_f}
\begin{split}
&\mathsf{Q}^{(1)}_{J_0,J'}\\
&\quad=\left(-\mathsf{D}^{(1)}_{k,x_0}\delta_{n',0}
\delta_{\sigma', +1}+i\mathsf{E}^{(1)}_{k,x_0}\delta_{n',0}
\delta_{\sigma', -1}-\mathsf{F}^{(1)}_{k,x_0}\delta_{n',+1}
\delta_{\sigma',+1}\right)\delta_{k',-k}\delta_{x'_0,x_0}
\end{split}
\end{equation}
with
\begin{subequations}
\begin{align}
\label{D_c}
\mathsf{D}^{(1)}_{k,x_0} &\equiv\frac{1}{2\left(\varepsilon^{(0)}_k\right)^2}
\sqrt{g_t}(f_M-f_P)m\sqrt{\frac{g_t}{g_s}}kx_0,\\
\label{E_c}
\begin{split}
\mathsf{E}^{(1)}_{k,x_0} &\equiv
\frac{1}{\mathcal{E}^{(0)}_k+\varepsilon^{(0)}_k}
\sqrt{\frac{\mathcal{E}^{(0)}_k+\sqrt{g_t}m}{2\mathcal{E}^{(0)}_k}}
\sqrt{\frac{\varepsilon^{(0)}_k+\sqrt{g_t}m}{2\varepsilon^{(0)}_k}}\\
&\quad\times\left[\sqrt{g_t}(f_M-f_P)m-\frac{f_P}{2}
\big(\mathcal{E}^{(0)}_k-\varepsilon^{(0)}_k\big)-f_E\varepsilon^{(0)}_k\right]
\sqrt{\frac{g_t}{g_s}}\frac{1}{\mathcal{E}^{(0)}_k+\sqrt{g_t}m},
\end{split}\\
\label{F_c}
\begin{split}
\mathsf{F}^{(1)}_{k,x_0} &\equiv
\frac{1}{\mathcal{E}^{(0)}_k+\varepsilon^{(0)}_k}
\sqrt{\frac{\mathcal{E}^{(0)}_k+\sqrt{g_t}m}{2\mathcal{E}^{(0)}_k}}
\sqrt{\frac{\varepsilon^{(0)}_k+\sqrt{g_t}m}{2\varepsilon^{(0)}_k}}
\sqrt{\frac{g_t}{g_s}}\frac{k}{\sqrt{2eB}}\\
&\quad\times\left\{\sqrt{g_t}(f_M-f_P)m\left(\frac{1}{\varepsilon^{(0)}_k+
\sqrt{g_t}m}+\frac{1}{\mathcal{E}^{(0)}_k+\sqrt{g_t}m}\right)\right.\\
&\qquad\;\left.+\left[\frac{f_P}{2}\big(\mathcal{E}^{(0)}_k-
\varepsilon^{(0)}_k\big)+f_E\varepsilon^{(0)}_k\right]
\left(\frac{1}{\varepsilon^{(0)}_k+\sqrt{g_t}m}-
\frac{1}{\mathcal{E}^{(0)}_k+\sqrt{g_t}m}\right)\right\}.
\end{split}
\end{align}
\end{subequations}
If I also define the coefficients $\mathsf{A}^{(1)}_{k,x_0}$ as
\begin{equation}
\label{A_c}
\mathsf{A}^{(1)}_{k,x_0}\equiv\frac{f_E}{2}x_0
\end{equation}
then, the first-order TGS $U^{\prime(1)}_{J_0}
(\mathbf{r})=U^{\prime(1)}_{0,k,-1,x_0}
(\mathbf{r})$ can be written simply as
\begin{equation}
\label{U^1_f}
\begin{split}
&U^{\prime(1)}_{0,k,-1,x_0}(\mathbf{r})\\
&\quad=\left(1+\mathsf{A}^{(1)}_{k,x_0}\right)U^{\prime(0)}_{0,k,-1,x_0}(\mathbf{r})+\mathsf{B}^{(1)}_{k,x_0}U^{\prime(0)}_{1,k,-1,x_0}(\mathbf{r})-i\mathsf{C}^{(1)}_{k,x_0}U^{\prime(0)}_{0,k,+1,x_0}(\mathbf{r})\\
&\qquad-\mathsf{D}^{(1)}_{k,x_0}V^{\prime(0)}_{0,-k,+1,x_0}(\mathbf{r})+i\mathsf{E}^{(1)}_{k,x_0}V^{\prime(0)}_{0,-k,-1,x_0}(\mathbf{r})-\mathsf{F}^{(1)}_{k,x_0}V^{\prime(0)}_{1,-k,+1,x_0}(\mathbf{r})
\end{split}
\end{equation}
(for the sake of clarity the explicit expression of the zero-order states appearing in the right hand side of the previous equation are given in Appendix D). The term $\mathsf{A}^{(1)}_{k,x_0}U^{\prime(0)}_{0,k,-1,x_0}(\mathbf{r})=
f_Ex_0U^{\prime(0)}_{0,k,-1,x_0}(\mathbf{r})/2$ has been added to compensate
for the factor $(1-f_Ex_0)$ in the scalar product (\ref{s_p_f}) and
then to have the states correctly normalized up to first order, as
\begin{equation}
(U^{\prime(1)}_{J_0},U^{\prime(1)}_{J'_0})^{(1)}=\delta_{J_0,J'_0}
\end{equation}
where $J_0\equiv\{0,k,-1,x_0\}$ and $J'_0\equiv\{0,k',-1,x'_0\}$.

Finally, with analogous calculations it can be shown that the first-order
positron TGS $V^{\prime(1)}_{0,k,+1,x_0}(\mathbf{r})$ can be written as
\begin{equation}
\label{V^1_f}
\begin{split}
&V^{\prime(1)}_{0,k,+1,x_0}(\mathbf{r})\\
&\quad=\left(1+\mathsf{A}^{(1)}_{k,x_0}\right)
V^{\prime(0)}_{0,k,+1,x_0}(\mathbf{r})+\mathsf{B}^{(1)}_{k,x_0}V^{\prime(0)}_{1,k,+1,x_0}
(\mathbf{r})-i\mathsf{C}^{(1)}_{k,x_0}V^{\prime(0)}_{0,k,-1,x_0}(\mathbf{r})\\
&\qquad-\mathsf{D}^{(1)}_{k,x_0}U^{\prime(0)}_{0,-k,-1,x_0}(\mathbf{r})+i\mathsf{E}^{(1)}_{k,x_0}U^{\prime(0)}_{0,-k,+1,x_0}(\mathbf{r})-\mathsf{F}^{(1)}_{k,x_0}U^{\prime(0)}_{1,-k,-1,x_0}(\mathbf{r}).
\end{split}
\end{equation}
and that the satisfy the orthonormalization relations
\begin{subequations}
\begin{align}
(V^{\prime(1)}_{J_0},V^{\prime(1)}_{J'_0})^{(1)}&=\delta_{J_0,J'_0}\\
(V^{\prime(1)}_{J_0},U^{\prime(1)}_{J'_0})^{(1)}&=0.
\end{align}
\end{subequations}
%
%
%
\subsection{Calculation of the presence probability}
\label{IV}
As I have said in the previous Section, I can calculate the pair presence probability in the presence of the slowly-varying magnetic field (\ref{B_exp}) and of the static gravitational field described by the metric tensor (\ref{g_mu_nu_l}) by means of the adiabatic perturbation theory up to first order in the time derivative of the magnetic field. In fact, from now on, the gravitational field will not play any further role: I took into account its presence by correcting the one-particle electron and positron modes and energies. Nevertheless, in order to avoid the possible confusion between the ``first-order'' relative to the adiabatic
perturbation theory and the ``first-order'' relative to the gravitational
couplings $f_E$, $f_P$ and $f_M$, I stress that in what follows I will always refer to
the second one. In particular, \emph{the superscript $(1)$ indicates quantities that
are first-order in the gravitational couplings}.

Now, I pointed out in Par. \ref{Prel_calc} that in Minkowski spacetime if the magnetic field changes with time only in strength, the probability that a pair is present with both the electron and positron in a TGS is zero. In the following we will see that the first-order corrections to the TGSs I calculated before will make this probability different from zero in the present physical situation. In fact, the gravitational field lies in the $x$ direction [see Eq. (\ref{g_mu_nu_l})] in such a way the first-order TGSs are not eigenstates of $\sigma_z$ as in Minkowski spacetime and then the selection rule (\ref{sel_rule}) does not hold. 

If the electron and the positron are assumed to be
present at time $t$ in the TGS state $U_{J_0}^{\prime(1)}(\mathbf{r},t)$
with $J_0=\{0,k,-1,x_0\}$ and $V_{\tilde{J}'_0}^{\prime(1)}(\mathbf{r},t)$ with $\tilde{J}'_0=\{0,k',+1,x'_0\}$ respectively, then the presence matrix element is given by
\begin{equation}
\label{M_E_cr}
\dot{H}^{(1)}_{J_0\tilde{J}'_0}(t)=\sqrt{\frac{g_t}{g_s}}\frac{e\dot{B}^{\text{exp}}_{\shortuparrow}(t)}{2}\int d\mathbf{r}
\sqrt{g_s^3}(1-f_Px)U_{J_0}^{\prime(1)\dag}(\mathbf{r},t)\left(\mathbf{r}\times
\boldsymbol{\alpha}\right)_zV^{\prime(1)}_{\tilde{J}'_0}(\mathbf{r},t).
\end{equation}
The presence amplitude at time $t$ can be calculated from this matrix element as
\begin{equation}
\label{cr_ampl}
\gamma^{(1)}_{J_0\tilde{J}'_0}(t)=\frac{1}{\varepsilon^{(1)}_{k,x_0}+
\varepsilon^{(1)}_{k',x'_0}}\int_0^tdt'\dot{H}^{(1)}_{J_0\tilde{J}'_0}(t')
\exp\left[i\left(\varepsilon^{(1)}_{k,x_0}+\varepsilon^{(1)}_{k',x'_0}\right)t'\right]
\end{equation}
where I used the fact that the first-order energies $\varepsilon^{(1)}_{k,x_0}$
of the TGSs do not depend on $B^{\text{exp}}_{\shortuparrow}(t)$ and then on
time [see Eq. (\ref{w_g_1})].

Now, the selection rule (\ref{sel_rule}) concerning the spin of two TGSs allows me to conclude that for the zero-order TGSs the following equalities hold:
\begin{subequations}
\begin{align}
\int d\mathbf{r}U_{J_0}^{\prime(0)\dag}(\mathbf{r},t)\left(\mathbf{r}\times
\boldsymbol{\alpha}\right)_zU^{\prime(0)}_{J'_0}(t,\mathbf{r}) &=0,\\
\label{U_z_V}
\int d\mathbf{r}U_{J_0}^{\prime(0)\dag}(t,\mathbf{r})\left(\mathbf{r}\times
\boldsymbol{\alpha}\right)_zV^{\prime(0)}_{\tilde{J}'_0}(t,\mathbf{r}) &=0,\\
\label{V_z_V}
\int d\mathbf{r}V_{\tilde{J}_0}^{\prime(0)\dag}(t,\mathbf{r})\left(\mathbf{r}
\times\boldsymbol{\alpha}\right)_zV^{\prime(0)}_{\tilde{J}'_0}(t,\mathbf{r}) &=0.
\end{align}
\end{subequations}
By exploiting these equations and by keeping only the terms up to first order,
I can write Eq. (\ref{M_E_cr}) as
\begin{equation}
\label{M_E_cr_fd_p}
\begin{split}
&\dot{H}^{(1)}_{J_0\tilde{J}'_0}(t)\\
&\simeq
\sqrt{\frac{g_t}{g_s}}\frac{e\dot{B}^{\text{exp}}_{\shortuparrow}(t)}{2}\int d\mathbf{r}\sqrt{g_s^3}
U_{0,k,-1,x_0}^{\prime(0)\dag}(t,\mathbf{r})\left(\mathbf{r}\times
\boldsymbol{\alpha}\right)_z\\
&\qquad\times\left[\mathsf{B}^{(1)}_{k',x'_0}(t)
V^{\prime(0)}_{1,k',+1,x'_0}(t,\mathbf{r})-i\mathsf{C}^{(1)}_{k',x'_0}(t)
V^{\prime(0)}_{0,k',-1,x'_0}(t,\mathbf{r})\right.\\
&\qquad\quad\;\left.+i\mathsf{E}^{(1)}_{k',x'_0}(t)
U^{\prime(0)\dag}_{0,-k',+1,x'_0}(t,\mathbf{r})-
\mathsf{F}^{(1)}_{k',x'_0}(t)U^{\prime(0)\dag}_{1,-k',-1,x'_0}(t,\mathbf{r})\right]\\
&\quad+\sqrt{\frac{g_t}{g_s}}
\frac{e\dot{B}^{\text{exp}}_{\shortuparrow}(t)}{2}\int d\mathbf{r}\sqrt{g_s^3}
\left[\mathsf{B}^{(1)}_{k,x_0}(t)U^{\prime(0)\dag}_{1,k,-1,x_0}
(t,\mathbf{r})\right.\\
&\qquad\qquad\qquad\qquad\quad+i\mathsf{C}^{(1)}_{k,x_0}(t)U^{\prime(0)\dag}_{0,k,+1,x_0}
(t,\mathbf{r})-i\mathsf{E}^{(1)}_{k,x_0}(t)
V^{\prime(0)\dag}_{0,-k,-1,x_0}(t,\mathbf{r})\\
&\qquad\qquad\qquad\qquad\quad\left.-\mathsf{F}^{(1)}_{k,x_0}(t)
V^{\prime(0)\dag}_{1,-k,+1,x_0}(t,\mathbf{r})\right]\left(\mathbf{r}\times\boldsymbol{\alpha}\right)_zV^{\prime(0)}_{0,k',+1,x'_0}(t,\mathbf{r}).
\end{split}
\end{equation}
This expression can be further simplified if I use the definitions (\ref{x})
and (\ref{y}) of the operators $x$ and $y$ and the fact that the operator $x_0$ is diagonal with respect to the basis $\big\{U^{\prime(0)}_J(t,\mathbf{r}),V^{\prime(0)}_J(t,\mathbf{r})\big\}$. In fact, in this way the operator $\left(\mathbf{r}\times\boldsymbol{\alpha}\right)_z$ becomes
\begin{equation}
\left(\mathbf{r}\times\boldsymbol{\alpha}\right)_z=i\alpha_-\left(x_0+iy_0+\sqrt{\frac{2}{eB}}a^{\dag}\right)-i\alpha_+\left(x_0-iy_0+\sqrt{\frac{2}{eB}}a\right)
\end{equation}
with $\alpha_{\pm}=(\alpha_x\pm i\alpha_y)/2$. Now, from the expressions (\ref{U_gr_0}) and (\ref{V_gr_0}) of the zero-order TGSs I conclude that only the operator $i\alpha_-(x_0+iy_0)$ gives a
nonvanishing contribution in the first integral in Eq. (\ref{M_E_cr_fd_p}) and
that, analogously, only the operator $-i\alpha_+(x_0-iy_0)$ gives a nonvanishing
contribution in the second one. In conclusion, the matrix element (\ref{M_E_cr_fd_p}) can
be written as [see Eq. (\ref{y_0_der})]
\begin{equation}
\label{M_E_cr_fd_pp}
\begin{split}
\dot{H}^{(1)}_{J_0\tilde{J}'_0}(t) &=i\sqrt{\frac{g_t}{g_s}}\frac{e\dot{B}^{\text{exp}}_{\shortuparrow}(t)}{2}\left\{\left[x_0+
\frac{1}{eB^{\text{exp}}_{\shortuparrow}(t)}\partial_{x_0}\right]\big\langle
U^{\prime(1)}_{J_0}(t)\big\vert\alpha_-\big\vert V^{\prime(1)}_{\tilde{J}'_0}(t)
\big\rangle\right.\\
&\left.\qquad\qquad\qquad\qquad\;\;-\left[x_0-\frac{1}{eB^{\text{exp}}_{\shortuparrow}(t)}
\partial_{x_0}\right]\big\langle U^{\prime(1)}_{J_0}(t)
\big\vert\alpha_+\big\vert V^{\prime(1)}_{\tilde{J}'_0}(t)\big\rangle\right\}
\end{split}
\end{equation}
where
\begin{subequations}
\label{M_-_+}
\begin{align}
\label{M_-}
\begin{split}
&\big\langle U^{\prime(1)}_{J_0}(t)\big\vert\alpha_-\big\vert
V^{\prime(1)}_{\tilde{J}'_0}(t)\big\rangle\\
&\qquad=\int d\mathbf{r}
\sqrt{g_s^3}U_{0,k,-1,x_0}^{(0)\dag}(t,\mathbf{r})\alpha_-\\
&\qquad\qquad\times\left[\mathsf{B}^{(1)}_{k',x'_0}(t)V^{\prime(0)}_{1,k',+1,x'_0}
(t,\mathbf{r})-i\mathsf{C}^{(1)}_{k',x'_0}(t)V^{\prime(0)}_{0,k',-1,x'_0}
(t,\mathbf{r})\right.\\
&\qquad\qquad\quad\;\left.+i\mathsf{E}^{(1)}_{k',x'_0}(t)U^{\prime(0)\dag}_{0,-k',+1,x'_0}
(t,\mathbf{r})-\mathsf{F}^{(1)}_{k',x'_0}(t)U^{\prime(0)\dag}_{1,-k',-1,x'_0}
(t,\mathbf{r})\right],
\end{split}\\
\label{M_+}
\begin{split}
&\big\langle U^{\prime(1)}_{J_0}(t)\big\vert\alpha_+\big\vert
V^{\prime(1)}_{\tilde{J}'_0}(t)\big\rangle\\
&\qquad=\int d\mathbf{r}\sqrt{g_s^3}
\left[\mathsf{B}^{(1)}_{k,x_0}(t)U^{\prime(0)\dag}_{1,k,-1,x_0}(t,\mathbf{r})+
i\mathsf{C}^{(1)}_{k,x_0}(t)U^{\prime(0)\dag}_{0,k,+1,x_0}(t,\mathbf{r})\right.\\
&\qquad\qquad\qquad\qquad\left. -i\mathsf{E}^{(1)}_{k,x_0}(t)V^{\prime(0)\dag}_{0,-k,-1,x_0}
(t,\mathbf{r})-\mathsf{F}^{(1)}_{k,x_0}(t)V^{\prime(0)\dag}_{1,-k,+1,x_0}
(t,\mathbf{r})\right]\\
&\qquad\qquad\qquad\times\alpha_+V^{\prime(0)}_{0,k',+1,x'_0}(t,\mathbf{r}).
\end{split}
\end{align}
\end{subequations}
The calculation of the matrix elements (\ref{M_-_+}) is quite tedious but straightforward. In particular, it can be shown that
\begin{equation}
\begin{split}
\big\langle U^{\prime(1)}_{J_0}(t)\big\vert\alpha_-\big\vert
V^{(1)}_{\tilde{J}'_0}(t)\big\rangle &=-
\big\langle U^{\prime(1)}_{J_0}(t)\big\vert\alpha_+\big\vert
V^{\prime(1)}_{\tilde{J}'_0}(t)\big\rangle\\
&=-i\frac{1}{eB^{\text{exp}}_{\shortuparrow}(t)}\frac{1}{4\varepsilon^{(0)}_k}\sqrt{g_t}(f_M-f_P)mk
\delta_{k,-k'}\delta_{x_0,x'_0}.
\end{split}
\end{equation}
For this reason, the terms in Eq. (\ref{M_E_cr_fd_pp}) with the derivative with respect to $x_0$
cancel each other and the final expression of $\dot{H}^{(1)}_{J_0J'_0}(t)$ is
\begin{equation}
\label{M_E_cr_fd_f}
\dot{H}^{(1)}_{J_0J'_0}(t)=\sqrt{\frac{g_t}{g_s}}
\frac{\dot{B}^{\text{exp}}_{\shortuparrow}(t)}{B^{\text{exp}}_{\shortuparrow}(t)}\frac{1}{2\varepsilon^{(0)}_k}
\sqrt{g_t}(f_M-f_P)mkx_0\delta_{k,-k'}\delta_{x_0,x'_0}.
\end{equation}

The creation amplitudes at time $t$ can be calculated by means of Eq. (\ref{cr_ampl})
and the only one different from zero is equal to
\begin{equation}
\begin{split}
&\gamma^{(1)}_{0,k,-1,x_0;0,-k,+1,x_0}(t)\\
&\quad=\sqrt{\frac{g_t}{g_s}}
\frac{k}{4\left(\varepsilon^{(0)}_k\right)^2}\sqrt{g_t}(f_M-f_P)mx_0\int_0^{t/\tau}ds'\frac{\exp\left[-\left(1-2i\varepsilon^{(0)}_k\tau\right)s'\right]}
{\exp(-s')-B_f/(B_f-B_i)}
\end{split}
\end{equation}
where only the first-order terms have been kept and where the time derivative of $B^{\text{exp}}_{\shortuparrow}(t)$ has been substituted. Now, as I have said below Eq. (\ref{B_exp}), $\tau$ is a macroscopic
time connected with the typical evolution times of a black hole,
then I can safely assume that $\varepsilon^{(0)}_k\tau\gg 1$. This allows me to
give an asymptotic estimate of the remaining integral for large times $t$.
The result is
\begin{equation}
\gamma^{(1)}_{0,k,-1,x_0;0,-k,+1,x_0}(t)\sim
\sqrt{\frac{g_t}{g_s}}\frac{k}{4\left(\varepsilon^{(0)}_k\right)^2}
\sqrt{g_t}(f_M-f_P)\frac{mx_0}{2i\varepsilon^{(0)}_k\tau}\frac{B_f-B_i}{B_i}.
\end{equation}
Finally, by squaring the modulus of this expression and by multiplying the result by
the number of states for large times [see Eq. (\ref{varrho})]
\begin{equation}
\frac{eB^{\text{exp}}_{\shortuparrow}(t)L_y}{2\pi}dx_0\frac{L_z}{2\pi}dk=\frac{eB_f}{4\pi^2\sqrt{g^3_s}}dV^{(0)}dk
\end{equation}
with $dV^{(0)}=\sqrt{g_s^3}L_yL_zdx_0$ the ``physical'' quantization volume up to zero order [see Eq. (\ref{g_mu_nu_l_2})], I obtain the differential probability that a pair is present with the electron (positron) between $x_0$ and $x_0+dx_0$ and with longitudinal momentum between $k$ and $k+dk$ ($-k$ and $-k-dk$) as
\begin{equation}
\label{dP_fd}
dP^{(1)}(x_0,k;t)\sim\left(\frac{f_s}{16\pi g_s}\right)^2
\frac{\sqrt{g_s}}{g_t}\frac{eB_fk^2}{(g_s m^2+k^2)^3}\left(\frac{B_f-B_i}{\tau B_i}\right)^2\left(mx_0\right)^2dV^{(0)}dk.
\end{equation}
In this equation the continuum limits $L_y\to\infty$ and $L_z\to\infty$ are understood and, since the probability is already proportional to $f_s$ and our calculations are exact up to first order in $f_s$ and $f_t$, it is enough to use the zero-order ``physical'' volume $dV^{(0)}$. Finally, the corresponding probability per unit volume and unit longitudinal momentum is given by
\begin{equation}
\label{dP_fd_f}
\frac{dP^{(1)}(x_0,k;t)}{dV^{(0)}dk}\sim\left(\frac{f_s}{16\pi g_s}\right)^2
\frac{\sqrt{g_s}}{g_t}\frac{eB_fk^2}{(g_s m^2+k^2)^3}\left(\frac{B_f-B_i}{\tau B_i}\right)^2\left(mx_0\right)^2.
\end{equation}
It is worth pointing out that the presence of the square of the length $x_0$ in Eq. (\ref{dP_fd_f}) is the counterpart of the presence of $R_{\perp M}$ or $R_{\perp m}$ in the presence probabilities calculated in Minkowski spacetime [see Eqs. (\ref{dP_lin}), (\ref{dP_rot}) and (\ref{dP_rot_lin})]. Also, note that, as it must be, due to the presence of $g_t$ in the denominator, the probability (\ref{dP_fd_f}) grows as the expansion center $X_c$ is moved towards the event horizon $r_G/4$ [see Eqs. (\ref{g})].

Finally, as I reminded in Sect. \ref{Pai_prod}, in \cite{Calucci} it was calculated the total presence probability of a pair in the presence of a slowly-varying magnetic field with fixed direction but in Minkowski spacetime [see Eq. (\ref{dP_lin})].\footnote{I remind that in that
case either the electron or the positron must be created in a state which is not
a TGS.} In order to have a quantity to be compared to the
total probability per unit volume (\ref{dP_lin}), I have to integrate Eq.
(\ref{dP_fd}) with respect to $k$. After this integration and indicating the resulting probability per unit volume as $dP^{(1)}(x_0;t)/dV^{(0)}$, it can
easily be seen that\footnote{We have to be satisfied of an
order-of-magnitude comparison because the time evolution of the magnetic
field used to obtain Eq. (\ref{dP_lin}) is different from that used here.}
\begin{equation}
\frac{dP^{(1)}(x_0;t)/dV^{(0)}}{dP^{\text{lin}}_{\shortuparrow}(t)/dV}\lesssim\frac{(f_s\lambdabar)^2}{g_tg_s^3}\sqrt{\frac{B_{cr}}{B_i}}
\end{equation}
where I assumed, for simplicity, that $B_i\sim B_f$, that $R_{\perp M}=x_0$ and that the magnetic field (\ref{B_lin}) is such that $B_0=B_i$ and $b=(B_f-B_i)/\tau$. Now, as I have mentioned below Eq. (\ref{h_on_g_2}), even very large values of $N$ in Eqs. (\ref{h_on_g})-(\ref{h_on_g_2}), Eq. (\ref{ineq}) suggests that $X_c$ can be chosen almost equal to $r_G/4$. In particular, I can evaluate the previous inequality approximatively by putting $g_t=4N(\lambdabar/r_G)^2$ $g_s=16$ and $f_s=64/r_G$ [see Eqs. (\ref{Phi_c_g})]:
\begin{equation}
\label{eta_g}
\frac{dP^{(1)}(x_0;t)/dV^{(0)}}{dP^{\text{lin}}_{\shortuparrow}(t)/dV}\lesssim\frac{1}{N}\sqrt{\frac{B_{cr}}{B_i}}.
\end{equation}
This quantity is in any case much less than one and then \emph{the gravitational effect is small in the present weak gravitational field approximation}. Nevertheless, the effect is there and \emph{it is reasonable to imagine that it can be amplified in the presence of a real gravitational field which is not restricted by the present assumptions}. In particular, we will see that this is true in the strong-field case treated in the following Section.
%
%
\section{Strong-gravitational field case}
\label{strong}
I have said that, as a general aim of my investigation I consider the production of particles by a nonstationary magnetic field, so I am interested in situations where the gravitational effects are not the dominant dynamical feature. In the previous Section, we have seen that if the electron-positron pair is present not too close to the event horizon of a Schwarzshild black hole, the gravitational effects may be treated perturbatively. In order to complete the investigation I suppose here that the pair production happens near the event horizon where, because of the singularity of the spacetime metric, a perturbative approach is inapplicable. The investigation is still possible because the isotropic metric (\ref{g_mu_nu}) can be approximated in a form [the \emph{Rindler metric} \cite{Rindler}], where the general covariant Dirac equation is also solvable in the presence of a uniform magnetic field, provided the magnetic and the gravitational fields are parallel \cite{Torres,Bautista}. In this sense, the physical situation is different from that treated in the previous Section where the gravitational and the magnetic field were perpendicular [see Eqs. (\ref{B_exp}) and (\ref{g_mu_nu_l})].

Following what I have just said, I want to consider here the case in which the pair is present microscopically speaking near the black hole event horizon lying at 
$R=r_G/4$ [see Eq. (\ref{g_mu_nu})]. Also, I choose the reference system in 
such a way the pair is created in a volume centered on the $z$ axis. In this way, the same considerations done at the beginning of the previous Section allow me to expand the metric tensor (\ref{g_mu_nu}) around the point $(0,0,r_G/4)$. In particular, 
\begin{subequations}
\begin{align}
g_{00}(x,y,r_G/4+z) &=\left(\frac{2z}{r_G}\right)^2+O\left[\left(\frac{z}{r_G}\right)^3\right], && \\
g_{ii}(x,y,r_G/4+z) &=-16+O\left(\frac{z}{r_G}\right)
\end{align}
\end{subequations}
with $z$ assumed to be positive. It is clear that I am only interested in the pairs created in the $(z>0)$-region because those created in the $(z<0)$-region will fall into the black hole.

Now, if I keep only the lowest-order nonzero term in $g_{\mu\mu}(x,y,r_G/4+z)$ then the initial 
metric tensor (\ref{g_mu_nu}) can be written approximatively in the form
\begin{equation}
\label{g_R}
g_{\mu\nu}(x,y,r_G/4+z)\simeq g_{\mu\nu}^{(R)}(z)=\text{diag}
\left[\left(\frac{2z}{r_G}\right)^2,-16,-16,-16\right].
\end{equation}
This metric tensor has the same form of a Rindler metric tensor describing 
an observer uniformly accelerated in the $z$ direction \cite{Rindler}.\footnote{I could have scaled the spatial coordinates in 
order to have exactly the Rindler metric tensor, but I prefer to 
work with $x$, $y$ and $z$ that are the Cartesian coordinates at infinity.} Actually, the physical meaning of the previous coordinates is 
very different from that of the coordinates in Rindler spacetime. For example, while here the coordinate $t$ is precisely the time coordinate in the region far from the black hole, the time coordinate in the Rindler spacetime is a combination of the Minkowski time coordinate and of the Minkowski spatial coordinate along the acceleration. Nevertheless, the fact that the two metric tensors have the same form allows me to conclude that the metric tensor (\ref{g_R}) describes a constant and uniform gravitational field in the $z$ direction. Observe that 
no assumption is needed about the strength of the gravitational field itself.

Now, I should pass to the mathematical description of the magnetic field that, actually, is identical to that I have done in Sect. \ref{Gen_ass_g} and it will not be repeated here.

As previously, in order to calculate the pair presence probability, I have to build the second quantized Hamiltonian of a Dirac field $\Psi' (t,\mathbf{r})$ in the presence of the just 
introduced gravitational field and of the magnetic field (\ref{B_exp}). By choosing the diagonal tetrad field $e_{\alpha}^{(R)\mu}(z)$ with
\begin{subequations}
\begin{align}
e_0^{(R)0}(z) &=\frac{r_G}{2z},\\
e_i^{(R)i}(z) &=\frac{1}{4} && (\text{no sum}),
\end{align}
\end{subequations}
the spatial connections $\Gamma^{(R)}_i(z)$ vanish while $\Gamma^{(R)}_0(z)$ is independent of $z$ and it is given by
\begin{equation}
\Gamma^{(R)}_{0}=\frac{1}{4r_G}\gamma^0\gamma^3.
\end{equation}
In this way, the Lagrangean density (\ref{L_G}) becomes
\begin{equation}
\label{L_R}
\begin{split}
\mathscr{L}^{\prime(R)}(t)&=32\big[\bar{\Psi}'(i\partial_0\Psi')-(i\partial_0\bar{\Psi}')\Psi'\big]\\
&\quad+\frac{16z}{r_G}\big\{\bar{\Psi}'\gamma^i[i\partial_i+
eA^{\text{exp}}_{\shortuparrow i}(t,\mathbf{r})]\Psi'-\bar{\Psi}'[i\overset{\shortleftarrow}{\partial}_i-eA^{\text{exp}}_{\shortuparrow i}(t,\mathbf{r})]\gamma^i\Psi'\big\}\\
&\quad-\frac{128z}{r_G}m\bar{\Psi}'\Psi'.
\end{split}
\end{equation}

By using the same definition used in the previous Section, the Hamiltonian density of the Dirac field $\Psi' (t,\mathbf{r})$ can be written, apart from total derivative terms, as
\begin{equation}
\label{H_R}
\begin{split}
\mathscr{H}^{\prime(R)}(t) &=
64\Psi^{\prime\dag}\mathcal{H}^{\prime(R)}(t)\Psi'
\end{split}
\end{equation}
with
\begin{equation}
\label{H_1p_R}
\begin{split}
\mathcal{H}^{\prime(R)}(t)&=\frac{2z}{r_G}
\left\{\frac{\alpha_x}{4}[-i\partial_x+eA^{\text{exp}}_{\shortuparrow x}(t,y)]+
\frac{\alpha_y}{4}[-i\partial_y+eA^{\text{exp}}_{\shortuparrow y}(t,x)]+\beta m\right\}\\
&\quad+\frac{\alpha_z}{4i}\frac{\{z,\partial_z\}}{r_G}
\end{split}
\end{equation}
the one-particle Hamiltonian of an electron in the presence of the magnetic field 
(\ref{B_exp}) in the spacetime with the metric tensor (\ref{g_R}).

Now, the scalar product (\ref{Sc_prod_g}) between two generic spinors $\Psi_1(t,\mathbf{r})$ and $\Psi_2(t,\mathbf{r})$ becomes here
\begin{equation}
\label{s_p^R_f}
(\psi_1,\psi_2)^{(R)}=\int d\mathbf{r} \frac{128z}{r_G}
\Psi^{\dag}_1(t,\mathbf{r})\gamma^0\gamma^0\frac{r_G}{2z}\Psi_2(t,\mathbf{r})=
64\int d\mathbf{r} \Psi^{\dag}_1(t,\mathbf{r})\Psi_2(t,\mathbf{r})
\end{equation}
and the one-particle Hamiltonian $\mathcal{H}^{(R)}(t)$ is Hermitian [this 
definition of the scalar product clarifies the presence of the numerical coefficient 
$64$ in Eq. (\ref{H_R})]. Finally, the total Hamiltonian of the system under study is
\begin{equation}
\label{H_tot_R}
H^{\prime(R)}(t)=64\int d\mathbf{r}\Psi^{\prime\dag}(t,\mathbf{r})
\mathcal{H}^{\prime(R)}(t)\Psi'(t,\mathbf{r})
\end{equation}
and it depends explicitly on time through the time-dependence of the magnetic field [see Eqs. (\ref{H_1p_R}) and (\ref{A_exp})]. Since also in this case I will apply the first-order adiabatic perturbation theory to calculate the pair presence probability, then in the next Paragraph I will determine the electron and positron modes of the time-independent counterpart of the one-particle Hamiltonian (\ref{H_1p_R}).
%
%
\subsection{Computation of the one-particle modes and energies}
In this Section, in order to compute the electron and positron one-particle modes and energies, I assume the magnetic field to be static, lying in the $z$ direction and, in particular, to be given by Eq. (\ref{B_p}). All the quantities that depend on time through the magnetic field $\mathbf{B}_{\shortuparrow}^{\text{exp}}(t)$, except the vector potential $\mathbf{A}_{\shortuparrow}^{\text{exp}}(t,\mathbf{r})$ and the magnetic field $\mathbf{B}_{\shortuparrow}^{\text{exp}}(t)$ itself that will be substituted by $\mathbf{A}'(\mathbf{r})$ and $\mathbf{B}'$ respectively [see Eqs. (\ref{A_p}) and (\ref{B_p})], will be indicated here with the same symbol used in the previous Paragraph but, of course, omitting the time-dependence.

In the following, I limit myself to the determination of the electron one-particle modes $U'_{\jmath}(\mathbf{r})$ and energies $w_{\jmath}$ where $\jmath$ embodies all the needed quantum numbers. Since the magnetic field $\mathbf{B}'$ is parallel to the $z$ axis, the eigenvalue equation 
\begin{equation}
\label{Eig_E_p}
\mathcal{H}^{\prime(R)}U'_{\jmath}=w_{\jmath}U'_{\jmath}
\end{equation}
that is [see Eq. (\ref{H_1p_R})]
\begin{equation}
\label{Eig_E}
\begin{split}
&\left[\frac{2z}{r_G}
\left\{\frac{\alpha_x}{4}[-i\partial_x+eA'_x(y)]+
\frac{\alpha_y}{4}[-i\partial_y+eA'_y(x)]+\beta m\right\}
+\frac{\alpha_z}{4i}\frac{\{z,\partial_z\}}{r_G}\right]U'_{\jmath}\\
&\qquad\qquad\qquad\qquad\qquad\qquad\qquad\qquad\qquad\qquad\qquad\qquad\qquad\qquad\quad\;\,=w_{\jmath}U'_{\jmath}
\end{split}
\end{equation}
can be solved exactly \cite{Torres}. In order to determine unambiguously the spinor basis, I 
require that the functions $U'_{\jmath}(\mathbf{r})$ are also eigenstates of the conserved 
spin operator \cite{Bagrov,Torres}
\begin{equation}
\label{S}
\mathcal{S}_z=\frac{\sigma_z}{2}-\frac{i\beta}{8m}
\left\{\alpha_x\left[-i\partial_y+eA'_y(x)\right]-
\alpha_y\left[-i\partial_x+eA'_x(y)\right]\right\}.
\end{equation}
It is useful to write the corresponding eigenvalue equation in the form
\begin{equation}
\label{Eig_S}
\mathcal{S}_zU'_{\jmath}=\sigma\frac{k_{\jmath}}{2m}U'_{\jmath}
\end{equation}
where $\sigma =\pm 1$ and $k_{\jmath}>0$ is a real parameter that, in general, depends on the 
various quantum numbers.

Now, Eqs. (\ref{Eig_E}) and (\ref{Eig_S}) 
have been solved together in \cite{Torres} and in \cite{Bautista} (where the 
effect of the anomalous magnetic moment of the electron is also 
taken into account). Actually, in those papers another electromagnetic gauge for the vector potential $\mathbf{A}'(\mathbf{r})$ is used but the calculations can be adapted straightforwardly to the present case. I only sketch the procedure to determine the electron modes $U'_{\jmath}(\mathbf{r})$ by quoting the relevant steps. The first goal is to decouple Eq. (\ref{Eig_E}) into a longitudinal part depending only 
on $z$ and a transverse part depending only on $x$ and $y$. This is achieved by 
multiplying Eq. (\ref{Eig_E}) by $\sigma_z\beta$ and by exploiting Eqs. (\ref{S}) and 
(\ref{Eig_S}). The resulting ``longitudinal'' equation is
\begin{equation}
\left(4\sigma k_{\jmath}-i\sigma_z\gamma^3\partial_z-i\frac{\sigma_z\gamma^3}{2z}-
\frac{\sigma_z\gamma^02w_{\jmath}r_G}{z}\right)U'_{\jmath}=0.
\end{equation}
Since $\sigma_z$ commutes with $\gamma^0$ and $\gamma^3$ and since 
$\{\gamma^{\alpha},\gamma^{\beta}\}=2\eta^{\alpha\beta}$, if I square the previous equation I obtain
\begin{equation}
\label{Eig_E_L}
\left[\partial^2_z+\frac{1}{z}\partial_z-16k_{\jmath}^2-\frac{1}{z^2}
\left(\frac{1}{4}-2iw_{\jmath}r_G\alpha_z-4w_{\jmath}^2r_G^2\right)\right]U'_{\jmath}=0.
\end{equation}
In order to satisfy this equation I write the spinor $U'_{\jmath}(\mathbf{r})$ as
\begin{equation}
\label{u_L}
U'_{\jmath}(\mathbf{r})=N_{\jmath}\left[P_-M'_{-,\jmath}(z)+P_+M'_{+,\jmath}(z)\right]\Xi'_{\jmath}(x,y)
\end{equation}
where $N_{\jmath}$ is a normalization factor, 
\begin{equation}
\label{P_pm}
P_{\pm}=\frac{1\pm\alpha_z}{2}
\end{equation}
are two $4\times 4$ projectors, $\Xi'_{\jmath}(x,y)$ is a spinor depending 
only on the transverse coordinates and $M'_{\pm,\jmath}(z)$ are two 
functions to be determined. By substituting Eq. (\ref{u_L}) 
in Eq. (\ref{Eig_E_L}) I obtain the following equations for the 
functions $M'_{\pm,\jmath}(z)$:
\begin{equation}
\label{Bess}
\left\{\frac{d^2}{dz^2}+\frac{1}{z}\frac{d}{dz}-
\left[16k_j^2+\frac{1}{z^2}\left(\frac{1}{2}\pm 2iw_{\jmath}r_G\right)^2
\right]\right\}M'_{\mp,\jmath}=0
\end{equation}
whose general solution is given by
\begin{equation}
\label{I_K}
M'_{\mp,\jmath}(z)=a_II_{1/2\pm 2iw_{\jmath}r_G}(4k_{\jmath}z)+a_KK_{1/2\pm 2iw_{\jmath}r_G}(4k_{\jmath}z)
\end{equation}
with $I_{\lambda}(\xi)$ and $K_{\lambda}(\xi)$ the modified Bessel 
functions \cite{Abramowitz}. Now, the functions $I_{\lambda}(\xi)$ 
diverge exponentially as $\xi\to \infty$ while 
the functions $K_{\lambda}(\xi)$ go to zero exponentially in the same limit. 
Now, my model is reliable only for $z\ll r_G$ then I have to choose the solution that 
coherently is very small in the region of large $z$. For this reason I put $a_I=0$ and $a_K=1$ in Eq. (\ref{I_K}) and I obtain\footnote{I point out that, since the behaviour of the functions $I_{\lambda}(\xi)$ is regular near $\xi=0$, they will be used again later [see Eq. (\ref{u_d})].}
\begin{equation}
\label{K}
M'_{\mp,\jmath}(z)=K_{1/2\pm 2iw_{\jmath}r_G}(4k_{\jmath}z).
\end{equation}

In order to determine the transverse spinors $\Xi'_{\jmath}(x,y)$ I observe that $\left[P_{\pm},\mathcal{S}_z\right]=0$, then by substituting the spinor (\ref{u_L}) in Eq. (\ref{Eig_S}) I have
\begin{equation}
\label{Eig_E_T_l}
\left[4m\sigma_z-i\beta
\left\{\alpha_x\left[-i\partial_y+eA'_y(x)\right]-
\alpha_y\left[-i\partial_x+eA'_x(y)\right]\right\}
\right]\Xi'_{\jmath}=4\sigma k_{\jmath}\Xi'_{\jmath}.
\end{equation}
I observe that the energy eigenvalue $w_{\jmath}$ does not appear in this equation then, 
looking also at Eqs. (\ref{Bess}) and (\ref{K}), any continuous 
value $w_{\jmath}\equiv E\ge 0$ is acceptable. \emph{The fact that in 
the presence of the gravitational field described by the Rindler metric tensor (\ref{g_R}) the energy of the electron has  continuous eigenvalues from zero to infinity 
that do not depend on the other quantum numbers 
is the most relevant difference with the case in which no gravitational field 
is present}. The physical origin of this difference 
lies on the fact that in the present case the linear momentum 
along the gravitational field is not a constant of motion. In other words, 
the ``longitudinal'' energy of the electron contains not only, 
as in absence of the gravitational field, the rest energy 
and the kinetic energy but also a \emph{negative} gravitational potential energy.

In order to solve Eq. (\ref{Eig_E_T_l}) I square it, then
\begin{equation}
\label{Eig_E_T}
\left[-\partial_x^2-\partial_y^2+
\left(\frac{eB}{2}\right)^2\left(x^2+y^2\right)+
eB\left(\mathcal{L}_z+\sigma_z\right)\right]
\Xi'_{\jmath}=16\left(k_{\jmath}^2-m^2\right)\Xi'_{\jmath}
\end{equation}
with $\mathcal{L}_z$ the $z$ component of the electron orbital angular momentum. The solutions of this equation are well known \cite{Cohen,Bagrov}. Two 
nonnegative integer quantum numbers $n_d$ and $n_g$ have to be introduced 
and
\begin{equation}
\label{k}
k_{\jmath}=k_{n_d}=\sqrt{m^2+\frac{eBn_d}{8}}
\end{equation}
can be interpreted as a sort of ``transverse'' energy of the electron 
in the spacetime with the metric (\ref{g_R}). 
With this definition the spinor $\Xi'_{\jmath}(x,y)$ is given by
\begin{equation}
\label{phi}
\Xi'_{\jmath}(x,y)=\Xi'_{n_d,n_g,\sigma}(x,y)=\frac{1}{2\sqrt{k_{n_d}}}
\begin{pmatrix}
\sqrt{k_{n_d}+m\sigma}\theta'_{n_d-1,n_g}(x,y)\\
\sqrt{k_{n_d}-m\sigma}\theta'_{n_d,n_g}(x,y)\\
i\sigma\sqrt{k_{n_d}+m\sigma}\theta'_{n_d-1,n_g}(x,y)\\
i\sigma\sqrt{k_{n_d}-m\sigma}\theta'_{n_d,n_g}(x,y)
\end{pmatrix}
\end{equation}
where the functions $\theta'_{n_d,n_g}(x,y)$ are given in Eq. (\ref{theta_p}). 
The functions $\theta'_{-1,n_g}(x,y)$ are not defined but this does not cause 
any problem because, by solving step by step Eq. (\ref{Eig_E_T}), one finds that if $n_d=0$ then $\sigma$ must be equal to $-1$ and the corresponding coefficient $\sqrt{k_{n_d}+m\sigma}$ vanishes. Finally, I quote that the coefficients in Eq. (\ref{phi}) have been chosen in such a way 
the spinors $\Xi'_{n_d,n_g,\sigma}(x,y)$ result normalized as
\begin{equation}
\label{ort_tr_e}
\int dxdy\Xi^{\prime\dag}_{n_d,n_g,\sigma}(x,y)\Xi'_{n'_d,n'_g,\sigma'}(x,y)=
\delta_{n_d,n'_d}\delta_{n_g,n'_g}\delta_{\sigma,\sigma'}.
\end{equation}

At this point I have to determine only the normalization factor in 
Eq. (\ref{u_L}). As I have said, $E$ is a continuous eigenvalue. 
For this reason if I require that the functions 
$U'_{\jmath}(\mathbf{r})=U'_{n_d,n_g,\sigma}(E;\mathbf{r})$ are normalized 
as [see Eq. (\ref{s_p^R_f})]
\begin{equation}
(U'_{n_d,n_g,\sigma}(E),U'_{n'_d,n'_g,\sigma'}(E'))^{(R)}=
\delta(E-E')\delta_{n_d,n'_d}\delta_{n_g,n'_g}
\delta_{\sigma,\sigma'}
\end{equation}
then, it is easy to show that the final form of the electron modes of 
the one-particle Hamiltonian is [see \cite{Torres} for a more detailed derivation 
of the normalization factor]
\begin{equation}
\label{U^R}
\begin{split}
&U'_{n_d,n_g,\sigma}(E;\mathbf{r})\\
&\quad=\sqrt{\frac{k_{n_d}r_G\cosh(2\pi Er_G)}{4\pi^2}}\\
&\qquad\quad\times[P_-K_{1/2+2iEr_G}(4k_{n_d}z)+P_+K_{1/2-2iEr_G}(4k_{n_d}z)]
\Xi'_{n_d,n_g,\sigma}(x,y).
\end{split}
\end{equation}

The physical meaning of the quantum numbers $E$ and $\sigma$ is clear from 
Eqs. (\ref{Eig_E_p}) and (\ref{Eig_S}). In order to understand the physical meaning 
of the remaining quantum numbers $n_d$ and $n_g$ I introduce the $z$ component $\mathcal{J}^{(1/2)}_z$ of the total angular momentum operator and the operator
\begin{equation}
\label{R_perp_d}
R_{xy}^2=
16\left[\left(\frac{x}{2}-\frac{\mathcal{P}_y}{eB}\right)^2+
\left(\frac{y}{2}+\frac{\mathcal{P}_x}{eB}\right)^2\right]
\end{equation}
corresponding to the operator (\ref{R_xy^2}) in Minkowski spacetime. Now, it can easily be shown that the two previous operators commute between them, with the time-independent form of the one-particle Hamiltonian (\ref{H_1p_R}) and with $\mathcal{S}_z$. Also, it can be shown that 
\begin{subequations}
\begin{align}
\label{J_u}
\mathcal{J}^{(1/2)}_z U'_{n_d,n_g,\sigma}(E) &=
\left(n_d-n_g-\frac{1}{2}\right)U'_{n_d,n_g,\sigma}(E),\\
\label{R_perp}
R_{xy}^2 U'_{n_d,n_g,\sigma}(E) &=\frac{16(2n_g+1)}{eB}
U'_{n_d,n_g,\sigma}(E).
\end{align}
\end{subequations}
Moreover, in the following I will use the operator $\rho_{xy}^2$ that is defined as \cite{Cohen}
\begin{equation}
\label{r_perp_d}
\rho_{xy}^2=16\left[\left(\frac{x}{2}+\frac{\mathcal{P}_y}{eB}\right)^2+
\left(\frac{y}{2}-\frac{\mathcal{P}_x}{eB}\right)^2\right]
\end{equation}
and that corresponds to the square of the radius of the helix 
along which a classical electron moves in the presence of the magnetic field $\mathbf{B}'$ given in Eq. (\ref{B_p}) and in the spacetime with metric tensor (\ref{g_R}) [see \cite{Cohen}]. It can be shown that the states (\ref{U^R}) are not eigenstates of $\rho_{xy}^2$, but that if $n_d\gg 1$ then 
\begin{align}
\label{r_perp}
\rho_{xy}^2 U'_{n_d,n_g,\sigma}(E)\simeq\frac{32n_d}{eB}U'_{n_d,n_g,\sigma}(E) && n_d\gg 1.
\end{align}
In conclusion, the previous eigenvalue equations allow me to conclude that, as in Minkowski spacetime, the quantum numbers $n_d$ and $n_g$ are connected with the motion of the electron and of the positron in the plane perpendicular to the magnetic field.

The positron modes can be built in an analogous way and the final result is
\begin{equation}
\label{V^R}
\begin{split}
&V'_{n_d,n_g,\sigma}(E;\mathbf{r})\\
&\quad=\sqrt{\frac{k_{n_g}r_G\cosh(2\pi Er_G)}{4\pi^2}}\\
&\qquad\quad\times[P_-K_{1/2-2iEr_G}(4k_{n_g}z)+P_+K_{1/2+2iEr_G}(4k_{n_g}z)]
\tilde{\Xi}'_{n_d,n_g,\sigma}(x,y)
\end{split}
\end{equation}
with
\begin{equation}
\label{tr_p}
\tilde{\Xi}'_{n_d,n_g,\sigma}(x,y)=\frac{1}{2\sqrt{k_{n_g}}}\begin{pmatrix}
i\sigma\sqrt{k_{n_g}-m\sigma}\theta'_{n_g-1,n_d}(x,y)\\
i\sigma\sqrt{k_{n_g}+m\sigma}\theta'_{n_g,n_d}(x,y)\\
\sqrt{k_{n_g}-m\sigma}\theta'_{n_g-1,n_d}(x,y)\\
\sqrt{k_{n_g}+m\sigma}\theta'_{n_g,n_d}(x,y)
\end{pmatrix}.
\end{equation}
These modes satisfy the eigenvalue equations
\begin{subequations}
\begin{align}
\mathcal{H}^{(R)}V'_{n_d,n_g,\sigma}(E) &=-EV'_{n_d,n_g,\sigma}(E),\\
\mathcal{S}_zV'_{n_d,n_g,\sigma}(E) 
&=-\sigma\frac{k_{n_g}}{2m}V'_{n_d,n_g,\sigma}(E),\\
\label{J_v}
\mathcal{J}^{(1/2)}_z V'_{n_d,n_g,\sigma}(E) &=
-\left(n_d-n_g+\frac{1}{2}\right)V'_{n_d,n_g,\sigma}(E),\\
\label{R_xy_2_p_g}
R_{xy}^2 V'_{n_d,n_g,\sigma}(E) &=\frac{16(2n_d+1)}{eB}
V'_{n_d,n_g,\sigma}(E),\\
\label{r_xy_2_p_g}
\rho_{xy}^2 V'_{n_d,n_g,\sigma}(E) &\simeq\frac{32n_g}{eB}
V'_{n_d,n_g,\sigma}(E) && n_g\gg 1
\end{align}
\end{subequations}
with the constraint that if $n_g=0$ then $\sigma =+1$ and they are such that
\begin{subequations}
\begin{align}
(V'_{n_d,n_g,\sigma}(E),V'_{n'_d,n'_g,\sigma'}(E'))^{(R)} 
&=\delta(E-E')\delta_{n_d,n'_d}\delta_{n_g,n'_g}
\delta_{\sigma,\sigma'},\\
(U'_{n_d,n_g,\sigma}(E),V'_{n'_d,n'_g,\sigma'}(E'))^{(R)} &=0.
\end{align}
\end{subequations}
Finally, in Appendix E I show that the set of spinors 
$U'_{n_d,n_g,\sigma}(E;\mathbf{r})$ and 
$V'_{n_d,n_g,\sigma}(E;\mathbf{r})$ is complete.

As usual, it is preferable to deal with normalizable 
modes then I have to 
find a convenient boundary condition at a given surface $z=b$ that 
discretizes the energies $E$. Since 
the procedure is identical for the electron and the positron modes, I will 
consider only the electron modes. Now, as I have said, the functions $K_{1/2\pm 2iEr_G}(4k_{n_d}z)$ go exponentially to zero for large values of $k_{n_d}z$ and go to infinity as 
$\left(k_{n_d}z\right)^{-1/2}$ for small values of $k_{n_d}z$ \cite{Abramowitz}. For this reason, it is clear that 
\begin{enumerate}
\item the modes $U'_{n_d,n_g,\sigma}(E;\mathbf{r})$ can not satisfy a ``zero'' condition 
at a given $k_{n_d}b\ll 1$ or a canonical periodicity 
condition between two points $k_{n_d}b_1\ll 1$ and $k_{n_d}b_2\gg 1$;
\item if we want to build modes with a finite normalization integral 
we have to modify 
the functions $K_{1/2\pm 2iEr_G}(4k_{n_d}z)$ in the region with 
$k_{n_d}z\ll 1$.
\end{enumerate}
On the other hand, we already know that the modified Bessel functions $I_{1/2\pm 2iEr_G}(4k_{n_d}z)$ have a regular behaviour in the region
$k_{n_d}z\ll 1$ and that they also satisfy the longitudinal equation (\ref{Bess}) [see Eq. (\ref{I_K})]. For this reason, I consider an arbitrary fixed 
surface $z=b$ such that $k_{n_d}b\ll 1$ and assume that the electron 
modes of the one-particle Hamiltonian and of the spin operator 
(\ref{S}) are the spinors $U'_J(\mathbf{r})$ defined as
\begin{equation}
\label{u_d}
U'_J(\mathbf{r})=
\begin{cases}
\begin{aligned}
&N^{(<)}_J\sqrt{\frac{k_{n_d}r_G
\cosh(2\pi E_{n,n_d}r_G)}{4\pi^2}}\\
&\qquad\times\left[P_-I_{1/2+2iE_{n,n_d}r_G}(4k_{n_d}z)\right.\\
&\qquad\quad\;\left.+P_+I_{1/2-2iE_{n,n_d}r_G}(4k_{n_d}z)\right]
\Xi'_{n_d,n_g,\sigma}(x,y) && \text{if $z\le b$}
\end{aligned}\\
\begin{aligned}
&N^{(>)}_J\sqrt{\frac{k_{n_d}r_G
\cosh(2\pi E_{n,n_d}r_G)}{4\pi^2}}\\
&\qquad\times\left[P_-K_{1/2+2iE_{n,n_d}r_G}(4k_{n_d}z)\right.\\
&\qquad\quad\;\left.+P_+K_{1/2-2iE_{n,n_d}r_G}(4k_{n_d}z)\right]
\Xi'_{n_d,n_g,\sigma}(x,y) && \text{if $z> b$}
\end{aligned}
\end{cases}
\end{equation}
where $J\equiv\{n,n_d,n_g,\sigma\}$ with $n$ a new integer quantum number characterizing the discrete energies (as I will see the discrete energies will also depend on the quantum number 
$n_d$) and where $N^{(<)}_J$ and $N^{(>)}_J$ are two real 
normalization factors to be determined. It is evident that the spinor $U'_J(\mathbf{r})$ satisfies Eqs. (\ref{Eig_E}) and (\ref{Eig_S}) in the regions $z<b$ and $z>b$ 
[see also Eq. (\ref{I_K})]. Also, since Eq. (\ref{Eig_E}) is a first-order 
equation in the variable $z$ I only require that the 
spinor $U'_J(\mathbf{r})$ is continuous at $z=b$. 
By means of this condition and by requiring that the norm of 
$U'_J(\mathbf{r})$ is unit, I make the energies discrete and determine 
the normalization factors $N^{(<)}_J$ and 
$N^{(>)}_J$. The details of the 
calculations are given in Appendix F and here I only quote the final expression of the 
discrete energies [see Eq. (\ref{E_n_n_d})]
\begin{align}
E_{n,n_d}=\frac{n\pi}{2r_G}\log^{-1} \left(k_{n_d}b\right) && k_{n_d}b\to 0
\end{align}
and of the coefficients $N^{(<)}_J$ and 
$N^{(>)}_J$ [see Eqs. (\ref{N_m}) and (\ref{N_M})]:
\begin{subequations}
\label{N_m_M_t}
\begin{align}
\label{N_m_t}
N^{(<)}_J &=N^{(<)}_{n,n_d}=
\frac{\pi}{8k_{n_d}b}\frac{\sqrt{1+(4E_{n,n_d}r_G)^2}}
{\cosh(2\pi E_{n,n_d}r_G)}\frac{1}{\sqrt{\varrho_{n_d}}} && k_{n_d}b\to 0,\\
\label{N_M_t}
N^{(>)}_J &=N^{(>)}_{n_d}=\frac{1}{\sqrt{\varrho_{n_d}}} 
 && k_{n_d}b\to 0
\end{align}
\end{subequations}
with [see Eq. (\ref{rho})]
\begin{align}
\label{rho_t}
\varrho_{n_d}=-\frac{2r_G}{\pi}\log(k_{n_d}b) && k_{n_d}b\to 0
\end{align}
the density of the energy levels. Obviously, all these 
quantities will be used in the intermediate calculations but at the end I have to perform 
the limit $b\to 0$ and the physically relevant results must be independent of $b$.

%
%
\subsection{Calculation of the presence probability}
We already know that, in order to obtain in the framework of the adiabatic perturbation theory the presence probability of a pair in the presence of the slowly-varying magnetic field (\ref{B_exp}), I have to calculate the corresponding presence matrix element. If the pair is present at time $t$ with the electron 
in the state $U'_J(t,\mathbf{r})$ and the positron in the state 
$V'_{J'}(t,\mathbf{r})$ with $J\equiv\{n,n_d,n_g,\sigma\}$ and $J'\equiv\{n',n'_d,n'_g,\sigma'\}$, the presence matrix element is given by
\begin{equation}
\label{M_E_i}
\begin{split}
\dot{H}^{\prime(R)}_{JJ'}(t)&\equiv\langle JJ'(t)|\dot{H}^{\prime(R)}(t)|0(t)\rangle\\
&=\frac{16e\dot{B}^{\text{exp}}_{\shortuparrow}(t)}{r_G}\int d\mathbf{r}U_J^{\prime\dag}(t,\mathbf{r})
z\left(x\alpha_y-y\alpha_x\right)V'_{J'}(t,\mathbf{r})
\end{split}
\end{equation}
[see Eqs. (\ref{H_tot_R}), (\ref{H_R}) and (\ref{H_1p_R})]. A more useful form of the previous 
matrix element can be given by using the matrices $\alpha_{\pm}$ and the expressions (\ref{x}) and (\ref{y}) of the operators $x$ and $y$:
\begin{equation}
\label{M_E}
\begin{split}
&\dot{H}^{\prime(R)}_{JJ'}(t)\\
&\quad=\frac{32ie\dot{B}^{\text{exp}}_{\shortuparrow}(t)}{r_G\sqrt{2eB^{\text{exp}}_{\shortuparrow}(t)}}\int d\mathbf{r}U_J^{\prime\dag}(t,\mathbf{r})
z[\alpha_-(a_d+a_g^{\dag})-
\alpha_+(a_g+a_d^{\dag})]V'_{J'}(t,\mathbf{r}).
\end{split}
\end{equation}

Now, as I have said at the end of the previous Section, 
in order to calculate these matrix elements I should 
use the expression (\ref{u_d}) of $U'_J(t,\mathbf{r})$ with 
$N^{(<)}_{n,n_d}(t)$ and $N^{(>)}_{n_d}(t)$ given 
by Eqs. (\ref{N_m_M_t}) and an analogous expression of $V_{J'}^{\prime}(t,\mathbf{r})$.\footnote{I remind that these quantities are now time-dependent because the magnetic field depends on time.} Actually, an easy power 
counting shows that the contribution of the integral on the variable $z$ from 
$0$ to $b$ goes to $0$ in the limit $b\to 0$. In fact, each spinor contains a factor 
$[k_{n_d}(t)b\sqrt{\log(k_{n_d}(t)b)}]^{-1}$ 
coming from $N^{(<)}_{n,n_d}(t)$ 
[see Eqs. (\ref{N_m_t}) and (\ref{rho_t})]. Also, from Eq. (\ref{I_a}) we see that 
the modified Bessel functions $I_{1/2+2iE_{n,n_d}(t)r_G}(4k_{n_d}(t)z)$ 
behave as $\sqrt{k_{n_d}(t)z}$ in the integration 
domain $0\le z\le b$. Finally, because of the presence of the $z$ factor in 
the matrix element (\ref{M_E}) the result of the integral on $z$ depends on $b$ as 
$k_{n_d}(t)b\log^{-1}(k_{n_d}(t)b)$ and then it goes to zero in the limit $b\to 0$. In this way, since at the end of the calculations the limit $b\to 0$ has to be performed, the matrix element 
(\ref{M_E}) can be calculated by using in the whole region $z\ge 0$ the expressions of the spinors $U'_J(t,\mathbf{r})$ and $V'_{J'}(t,\mathbf{r})$ valid in the region $z>b$. Actually, I can use directly the spinors $U'_{n_d,n_g,\sigma}(E;\mathbf{r})$ and 
$V'_{n'_d,n'_g,\sigma'}(E';\mathbf{r})$ multiplied by $N^{(>)}_{n_d}(t)$ and 
$N^{(>)}_{n'_g}(t)$ respectively because the presence of the factor $z$ in 
the matrix element (\ref{M_E}) makes finite the resulting integral 
from $0$ to $\infty$ [see also the general formula (\ref{int})]:
\begin{equation}
\label{M_E_1}
\begin{split}
&\dot{H}^{\prime(R)}_{n_d,n_g,\sigma;n'_d,n'_g,\sigma'}(E,E';t)=\frac{32ie\dot{B}^{\text{exp}}_{\shortuparrow}(t)}{r_G\sqrt{2eB^{\text{exp}}_{\shortuparrow}(t)\varrho_{n_d}(t)\varrho_{n'_g}(t)}}\\
&\quad\times\int d\mathbf{r}U_{n_d,n_g,\sigma}^{\prime\dag}(E;t,\mathbf{r})z[\alpha_-(a_d+a_g^{\dag})-\alpha_+(a_g+a_d^{\dag})]V'_{n'_d,n'_g,\sigma'}(E';t,\mathbf{r}).
\end{split}
\end{equation}
At this point, I have to substitute Eqs. (\ref{U^R}) and (\ref{V^R}) [with the time-varying magnetic field $B^{\text{exp}}_{\shortuparrow}(t)$ instead of $B$] in place 
of $U_{n_d,n_g,\sigma}^{\prime\dag}(E;t,\mathbf{r})$ and 
$V'_{n'_d,n'_g,\sigma'}(E';t,\mathbf{r})$ and apply the various operators. By using the intermediate matrix elements
\begin{subequations}
\begin{align}
\begin{split}
&\int dxdy \Xi^{\prime\dag}_{n_d,n_g,\sigma}(t,x,y)
P_{\pm}\alpha_-\tilde{\Xi}'_{n'_d,n'_g,\sigma'}(t,x,y)\\
&\qquad =(1\pm i\sigma)(1\mp i\sigma')
\sqrt{\frac{\left[k_{n_d}(t)-m\sigma\right][k_{n'_g}(t)-m\sigma']}
{64k_{n_d}(t)k_{n'_g}(t)}}\delta_{n_d,n'_g-1}\delta_{n_g,n'_d},
\end{split}\\
\begin{split}
&\int dxdy \Xi^{\prime\dag}_{n_d,n_g,\sigma}(t,x,y)
P_{\pm}\alpha_+\tilde{\Xi}'_{n'_d,n'_g,\sigma'}(t,x,y)\\
&\qquad =(1\mp i\sigma)(1\pm i\sigma')
\sqrt{\frac{\left[k_{n_d}(t)+m\sigma\right][k_{n'_g}(t)+m\sigma']}
{64k_{n_d}(t)k_{n'_g}(t)}}\delta_{n_d-1,n'_g}\delta_{n_g,n'_d}
\end{split}
\end{align}
\end{subequations}
and
\begin{subequations}
\begin{align}
\begin{split}
&\int dxdy \Xi^{\prime(-)\dag}_{n_d,n_g,\sigma}(t,x,y)
P_{\pm}\alpha_-\tilde{\Xi}'_{n'_d,n'_g,\sigma'}(t,x,y)\\
&\qquad =(1\pm i\sigma)(1\mp i\sigma')
\sqrt{\frac{n_d\left[k_{n_d}(t)-m\sigma\right][k_{n'_g}(t)-m\sigma']}
{64k_{n_d}(t)k_{n'_g}(t)}}\delta_{n_d,n'_g}\delta_{n_g,n'_d},
\end{split}\\
\begin{split}
&\int dxdy \Xi^{\prime(+)\dag}_{n_d,n_g,\sigma}(t,x,y)
P_{\pm}\alpha_+\tilde{\Xi}'_{n'_d,n'_g,\sigma'}(t,x,y)\\
&\qquad =(1\mp i\sigma)(1\pm i\sigma')
\sqrt{\frac{n_d\left[k_{n_d}(t)+m\sigma\right][k_{n'_g}(t)+m\sigma']}
{64k_{n_d}(t)k_{n'_g}(t)}}\delta_{n_d,n'_g}\delta_{n_g,n'_d}
\end{split}
\end{align}
\end{subequations}
that can be easily checked, the matrix element (\ref{M_E_1}) can be written as
\begin{equation}
\label{M_E_2}
\begin{split}
&\dot{H}^{\prime(R)}_{n_d,n_g,\sigma;n'_d,n'_g,\sigma'}(E,E';t) =\frac{ie\dot{B}^{\text{exp}}_{\shortuparrow}(t)}{\pi^2}
\sqrt{\frac{\cosh(2\pi Er_G)\cosh(2\pi E'r_G)}{2[eB^{\text{exp}}_{\shortuparrow}(t)]^3}}\\
&\qquad\times\left\{\sqrt{\frac{(n_g+1)\left[k_{n_d}(t)-m\sigma\right]
\left[k_{n_d+1}(t)-m\sigma'\right]}{\varrho_{n_d}(t)\varrho_{n_d+1}(t)}}\right.\\
&\qquad\qquad\times\mathrm{Re}\left[(1-i\sigma)(1+i\sigma')I_{n_d,n_d+1}(E,E';t)\right]
\delta_{n_d,n'_g-1}\delta_{n_g+1,n'_d}\\
&\qquad\quad+\sqrt{\frac{n_d\left[k_{n_d}(t)-m\sigma\right]
\left[k_{n_d}(t)-m\sigma'\right]}{\varrho^2_{n_d}(t)}}\\
&\qquad\qquad\times\mathrm{Re}\left[(1-i\sigma)(1+i\sigma')I_{n_d,n_d}(E,E';t)\right]
\delta_{n_d,n'_g}\delta_{n_g,n'_d}\\
&\qquad\quad-\sqrt{\frac{n_g\left[k_{n_d}(t)+m\sigma\right]
\left[k_{n_d-1}(t)+m\sigma'\right]}{\varrho_{n_d}(t)\varrho_{n_d-1}(t)}}\\
&\qquad\qquad\times\mathrm{Re}\left[(1+i\sigma)(1-i\sigma')I_{n_d,n_d-1}(E,E';t)\right]
\delta_{n_d-1,n'_g}\delta_{n_g-1,n'_d}\\
&\qquad\quad-\sqrt{\frac{n_d\left[k_{n_d}(t)+m\sigma\right]
\left[k_{n_d}(t)+m\sigma'\right]}{\varrho^2_{n_d}(t)}}\\
&\qquad\qquad\times\mathrm{Re}\left[(1+i\sigma)(1-i\sigma')I_{n_d,n_d}(E,E';t)\right]
\delta_{n_d,n'_g}\delta_{n_g,n'_d}\Biggr\}
\end{split}
\end{equation}
where the adimensional function
\begin{equation}
\label{inte}
\begin{split}
&I_{l,l'}(E,E';t)\\
&\quad\equiv\int_0^{\infty}ds s
K_{1/2-2iEr_G}\left[\frac{4k_l(t)}{\sqrt{2eB^{\text{exp}}_{\shortuparrow}(t)}}s\right]
K_{1/2+2iE'r_G}\left[\frac{4k_{l'}(t)}{\sqrt{2eB^{\text{exp}}_{\shortuparrow}(t)}}s\right]
\end{split}
\end{equation}
has been introduced.

Before continuing, I want to point out that from Eq. (\ref{M_E_2}) it can be 
seen that the total angular momentum of the field $\Psi'(t,\mathbf{r})$ is conserved in the 
transition. In fact, in any case [see Eqs. (\ref{J_u}) and (\ref{J_v})]
\begin{equation}
n_d-n_g-\frac{1}{2}+n'_d-n'_g+\frac{1}{2}=n_d-n_g+n'_d-n'_g=0.
\end{equation}
Of course, this selection rule is a consequence of the fact that 
the time evolution of the magnetic field does not break the rotational symmetry of the 
system around the $z$ axis or, in other words, of the fact that 
$\mathcal{J}^{(1/2)}_z$ and $\dot{\mathcal{H}}^{\prime(R)}(t)$ commute.

Now, since I am interested only in the strong magnetic field regime 
in which $B^{\text{exp}}_{\shortuparrow}(t)\gg B_{cr}$, I can simplify the expression of the 
transition matrix element (\ref{M_E_2}) by taking into account only those 
transitions whose probabilities are proportional to the lowest power of $B_{cr}/B^{\text{exp}}_{\shortuparrow}(t)$. In the framework of the 
adiabatic perturbation theory the first-order transition amplitude in $\dot{B}^{\text{exp}}_{\shortuparrow}(t)$ of the presence of a pair at time $t$ in the state with quantum numbers $\{E,n_d,n_g,\sigma;E',n'_d,n'_g,\sigma'\}$ is given by
\begin{equation}
\label{ampl}
\begin{split}
&\gamma^{(R)}_{n_d,n_g,\sigma;n'_d,n'_g,\sigma'}(E,E';t)\\
&\qquad=\frac{1}{E+E'}\int_0^tdt'
\dot{H}^{\prime(R)}_{n_d,n_g,\sigma;n'_d,n'_g,\sigma'}(E,E';t')\exp[i(E+E')t']
\end{split}
\end{equation}
and the corresponding probability is the square modulus of this quantity. It 
is evident that, since the energies $E$ do not depend on $B^{\text{exp}}_{\shortuparrow}(t)$, we can perform the $(B_{cr}/B^{\text{exp}}_{\shortuparrow}(t))$-power counting directly on the matrix element (\ref{M_E_2}). To this end I need the general behaviour of two particular classes of the integral (\ref{inte}) that is of $I_{0,n'_d}(E,E';t)$ with $n'_d>0$ and of $I_{n_d,n_d}(E,E';t)$ with $n_d>0$. By reminding the expression (\ref{k}) with the time-dependent 
magnetic field for $k_{n_d}(t)$ and by using the general formula (\ref{int}), it 
can easily be seen that
\begin{subequations}
\begin{align}
I_{0,n'_d}(E,E';t) &\sim \sqrt[4]{\frac{1}{n^{\prime 6}_d}\frac{B_{cr}}{B^{\text{exp}}_{\shortuparrow}(t)}} 
&& \text{if $n'_d>0$ and $B^{\text{exp}}_{\shortuparrow}(t)\gg B_{cr}$},\\
\label{I_nn}
I_{n_d,n_d}(E,E';t) &\sim \frac{1}{n_d} 
&& \text{if $n_d>0$ and $B^{\text{exp}}_{\shortuparrow}(t)\gg B_{cr}$}
\end{align}
\end{subequations}
where, for later convenience, I also pointed out the dependence on the quantum 
numbers $n_d$ and $n'_d$. Obviously, the integrals 
$I_{n_d,n_d\pm 1}(E,E';t)$ behave as the integral (\ref{I_nn}) and then 
this criterion allows me to neglect the transitions 
in which the electron is in a $(n_d=0,\sigma=-1)$-state 
or the positron in a $(n_g=0,\sigma=+1)$-state [see Eq. (\ref{M_E_2})].

Another criterion I will use to select only the most probable transitions 
is the dependence of the corresponding probabilities 
on the quantum numbers $n_d$ and $n_g$. As previously, I can work directly on 
the matrix element (\ref{M_E_2}) by keeping in mind that at the end 
I will sum the probabilities with different values of $n_d$ and $n_g$. Now, 
according to what I have said at the end of Chap. 3, the internal consistency of the model 
requires here that the sum on $n_g$ (and on $n_d$) 
cannot be extended up to infinity but that they must be stopped 
up to a certain $N^{\text{exp}}_{\shortuparrow}(t)$ corresponding through the 
relation
\begin{equation}
\label{N_max}
N^{\text{exp}}_{\shortuparrow}(t)\equiv \frac{eB^{\text{exp}}_{\shortuparrow}(t)}{32}R^2_{\perp M}
\end{equation}
to a fixed $R_{\perp M}$ [see Eqs. (\ref{R_perp}), (\ref{r_perp}), (\ref{R_xy_2_p_g}) and (\ref{r_xy_2_p_g})]. By reminding the physical meaning of the operators 
$R_{xy}^2$ and $\rho_{xy}^2$, it is clear that the 
transverse motion of a classical electron (positron) 
is confined in a circle with radius $2R_{\perp M}$, in such a way this quantity 
can be assumed as the radius of the quantization cylinder whose axis is parallel to the $z$ axis. 

Coming back to the matrix element (\ref{M_E_2}), we see that it contains 
two kinds of terms, the first one being proportional 
essentially to $\sqrt{n_g}$ and the second one to $\sqrt{n_d}$. By taking into account 
Eq. (\ref{I_nn}) it is easy to see that the first kind of terms gives rise 
to final probabilities proportional to $(N^{\text{exp}}_{\shortuparrow}(t))^2\log N^{\text{exp}}_{\shortuparrow}(t)$, while the 
second one to final probabilities proportional to $(N^{\text{exp}}_{\shortuparrow}(t))^2$. For all these reasons I can consider only the transitions to states with 
$n_d> 0$ and $n'_g> 0$ in such a way only the following four kinds of transitions amplitudes result different from zero:
\begin{subequations}
\begin{align}
\begin{split}
&\gamma^{(R)}_{n_d,n_g,\sigma;n_d+1,n_g+1,\sigma}(E,E';t)\\
&\qquad=\frac{i}{2\pi^2}\sqrt{\frac{(n_g+1)\sqrt{n_d(n_d+1)}\cosh(2\pi Er_G)\cosh(2\pi E'r_G)}
{\varrho_{n_d}(t)\varrho_{n_d+1}(t)}}\\
&\qquad\qquad\qquad\times\frac{\mathrm{Re}\left(I_{n_d,n_d+1}(E,E')\right)}{E+E'}F(E,E';t),
\end{split}\\
\begin{split}
&\gamma^{(R)}_{n_d,n_g,\sigma;n_d+1,n_g+1,-\sigma}(E,E';t)\\
&\qquad=\frac{i\sigma}{2\pi^2}\sqrt{\frac{(n_g+1)\sqrt{n_d(n_d+1)}\cosh(2\pi Er_G)\cosh(2\pi E'r_G)}{\varrho_{n_d}(t)\varrho_{n_d+1}(t)}}\\
&\qquad\qquad\qquad\times\frac{\mathrm{Im}\left(I_{n_d,n_d+1}(E,E')\right)}{E+E'}F(E,E';t),
\end{split}\\
\begin{split}
&\gamma^{(R)}_{n_d+1,n_g+1,\sigma;n_d,n_g,\sigma}(E,E';t)\\
&\qquad=-\frac{i}{2\pi^2}\sqrt{\frac{(n_g+1)\sqrt{n_d(n_d+1)}\cosh(2\pi Er_G)\cosh(2\pi E'r_G)}
{\varrho_{n_d}(t)\varrho_{n_d+1}(t)}}\\
&\qquad\qquad\qquad\times\frac{\mathrm{Re}\left(I_{n_d+1,n_d}(E,E')\right)}{E+E'}F(E,E';t),
\end{split}\\
\begin{split}
&\gamma^{(R)}_{n_d+1,n_g+1,-\sigma;n_d,n_g,\sigma}(E,E';t)\\
&\qquad=-\frac{i\sigma}{2\pi^2}\sqrt{\frac{(n_g+1)\sqrt{n_d(n_d+1)}\cosh(2\pi Er_G)\cosh(2\pi E'r_G)}{\varrho_{n_d}(t)\varrho_{n_d+1}(t)}}\\
&\qquad\qquad\qquad\times\frac{\mathrm{Im}\left(I_{n_d+1,n_d}(E,E')\right)}{E+E'}F(E,E';t)
\end{split}
\end{align}
\end{subequations}
where
\begin{equation}
\label{F}
F(E,E';t)\equiv\int_0^tdt'\frac{\dot{B}^{\text{exp}}_{\shortuparrow}(t')}{B^{\text{exp}}_{\shortuparrow}(t')}\exp[i(E+E')t']
\end{equation}
and where I pointed out that in the strong magnetic field regime, 
if $n_d> 0$ the integrals 
$I_{n_d,n_d\pm 1}(E,E')$ do not depend on time. By squaring these amplitudes, by summing on the polarization $\sigma$ and 
by multiplying by the number of electronic states $\varrho_{n_d}(t)dE$ with energies 
between $E$ and $E+dE$ and by the number of positronic states 
$\varrho_{n_d+1}(t)dE'$ with energies 
between $E'$ and $E'+dE'$, I obtain the differential probabilities
\begin{subequations}
\label{dP_1_2}
\begin{align}
\label{dP_1}
\begin{split}
&dP^{(R)}_{n_d,n_g;n_d+1,n_g+1}(E,E';t) \\
&\qquad=\frac{1}{2\pi^4}(n_g+1)\sqrt{n_d(n_d+1)}\cosh(2\pi Er_G)\cosh(2\pi E'r_G)\\
&\qquad\qquad\times\frac{\left\vert I_{n_d,n_d+1}(E,E')\right\vert^2}{(E+E')^2}
\left\vert F(E,E';t)\right\vert^2dEdE',
\end{split}\\
\label{dP_2}
\begin{split}
&dP^{(R)}_{n_d+1,n_g+1;n_d,n_g}(E,E';t)\\
&\qquad=\frac{1}{2\pi^4}(n_g+1)\sqrt{n_d(n_d+1)}\cosh(2\pi Er_G)\cosh(2\pi E'r_G)\\
&\qquad\qquad\qquad\times\frac{\left\vert I_{n_d+1,n_d}(E,E')\right\vert^2}{(E+E')^2}
\left\vert F(E,E';t)\right\vert^2dEdE'
\end{split}
\end{align}
\end{subequations}
that, as expected, do not depend on the unphysical parameter $b$. 
Now, I want to calculate the probability $dP^{(R)}(E,E';t)$ 
that a pair is present at time $t$ with the electron with 
energy between $E$ and $E+dE$ and the positron 
with energy between $E'$ and $E'+dE'$. To do this I have to sum on 
the remaining quantum numbers $n_d$ and $n_g$. As we already know, both 
the series on $n_d$ and $n_g$ are diverging, then I can perform the summations 
by assuming $n_g\simeq n_g+1$ and $n_d\simeq n_d+1$ because the most relevant 
terms are those with $n_g\gg 1$ and $n_d\gg 1$. Starting 
from Eqs. (\ref{dP_1_2}) we have
\begin{equation}
\label{dP}
\begin{split}
dP^{(R)}(E,E';t)\simeq
\frac{1}{\pi^4}&\left[\sum_{n_g=1}^{N^{\text{exp}}_{\shortuparrow}(t)}n_g\right]\cosh(2\pi Er_G)\cosh(2\pi E'r_G)\\
&\times\frac{\sum_{n_d=1}^{N^{\text{exp}}_{\shortuparrow}(t)}n_d
\left\vert I_{n_d,n_d}(E,E')\right\vert^2}{(E+E')^2}
\left\vert F(E,E';t)\right\vert^2dEdE'
\end{split}
\end{equation}
where $N^{\text{exp}}_{\shortuparrow}(t)$ has been defined in Eq. (\ref{N_max}). 

The next step is the explicit calculation of the functions 
$I_{n_d,n_d}(E,E')$ and $F(E,E';t)$ defined in Eqs. (\ref{inte}) 
and (\ref{F}). By using the general formula (\ref{int}) and the properties 
of the $\Gamma$ function (\ref{mod_gamma}) and \cite{Abramowitz} 
\begin{align}
\Gamma(1+i\xi)\Gamma(1-i\xi)=\left\vert\Gamma(1+i\xi)\right\vert^2=
\frac{\pi\xi}{\sinh(\pi\xi)} && \text{with $\xi\in \mathbb{R}$} 
\end{align}
it can easily be shown that
\begin{equation}
\label{int_f}
I_{n_d,n_d}(E,E')=\frac{\pi}{2n_d}
\frac{1-i(E-E')r_G}{\cosh \left[\pi(E-E')r_G\right]}
\frac{\pi(E+E')r_G}{\sinh \left[\pi(E+E')r_G\right]}.
\end{equation}

Instead, after some calculations the following expression of $F(E,E';t)$ can be obtained:
\begin{equation}
\label{F_1}
F(E,E';t)=\int_0^{t/\tau}ds'\frac{B_f-B_i}{B_f+(B_i-B_f)\exp{(-s')}}
\exp{\left[-s'+i(E+E')\tau s'\right]}
\end{equation}
with $s'=t'/\tau$. Now, as usual I am interested only in energetic electrons and positrons such that $E\tau\gg 1$ and $E'\tau\gg 1$ and in large times. For this 
reason, I can give the following asymptotic estimate of the integral (\ref{F_1}):
\begin{equation}
F(E,E';t)\sim\frac{B_f-B_i}{B_i}\frac{i}{(E+E')\tau}.
\end{equation}
By substituting this expression and Eq. (\ref{int_f}) in Eq. (\ref{dP}), I can write 
the asymptotic value of the probability $dP^{(R)}(E,E';t)$ as
\begin{equation}
\label{dP_f}
\begin{split}
dP^{(R)}(E,E';t) &\sim\frac{1}{\pi}\left(\frac{eB_fR^2_{\perp M}}{64}\right)^2
\log\left(\frac{eB_fR^2_{\perp M}}{32}\right)
\left(\frac{B_f-B_i}{B_i\tau}\right)^2r_G^4\\
&\quad\times\frac{1+[(E-E')r_G]^2}{r_G[(E+E')r_G]^2}\delta^{(r_G)}(E,E')dEdE'
\end{split}
\end{equation}
where I made the substitutions valid for large limes [see Eq. (\ref{N_max})]
\begin{subequations}
\begin{align}
\sum_{n_g=1}^{N^{\text{exp}}_{\shortuparrow}(t)} n_g&=\frac{1}{2}(N^{\text{exp}}_{\shortuparrow}(t))^2=
\frac{1}{2}\left(\frac{eB_fR^2_{\perp M}}{32}\right)^2,\\
\sum_{n_d=1}^{N^{\text{exp}}_{\shortuparrow}(t)} \frac{1}{n_d}&=
\log \left(N^{\text{exp}}_{\shortuparrow}(t)\right)=
\log\left(\frac{eB_fR^2_{\perp M}}{32}\right)
\end{align}
\end{subequations}
and where the function
\begin{equation}
\label{G_delta}
\delta^{(r_G)}(E,E')=\frac{\pi r_G}{2}
\frac{\cosh(2\pi Er_G)\cosh(2\pi E'r_G)}
{\cosh^2[\pi (E-E')r_G]\sinh^2[\pi (E+E')r_G]}
\end{equation}
has been introduced. I pointed out the dependence of 
$\delta^{(r_G)}(E,E')$ on the Schwarzshild radius $r_G$ because from a physical point of view I am interested in energies $E$ and $E'$ such that $Er_G\gg 1$ and $E'r_G\gg 1$. In this energy region the function $\delta^{(r_G)}(E,E')$ strongly depends even on small 
changes of $E$ and $E'$ through the hyperbolic functions. This can be 
seen more clearly by writing Eq. (\ref{G_delta}) as
\begin{equation}
\begin{split}
\delta^{(r_G)}(E,E') &=\frac{\pi r_G}{4}
\frac{\cosh[2\pi (E-E')r_G]+\cosh[2\pi (E+E')r_G]}
{\cosh^2[\pi (E-E')r_G]\sinh^2[\pi (E+E')r_G]}\\
&=\frac{\pi r_G}{2}
\left\{\frac{1}{\sinh^2[\pi (E+E')r_G]}+
\frac{1}{\cosh^2[\pi (E-E')r_G]}\right\}.
\end{split}
\end{equation}
From this expression and by reminding that $E,E'\ge 0$, I obtain
\begin{equation}
\begin{split}
&\lim_{r_G\to\infty}\delta^{(r_G)}(E,E')\\
&\qquad=\lim_{r_G\to\infty}\frac{\pi r_G}{2}\left\{\frac{1}{\sinh^2[\pi (E+E')r_G]}+
\frac{1}{\cosh^2[\pi (E-E')r_G]}\right\}\\
&\qquad=\lim_{r_G\to\infty}
\frac{\pi r_G}{2 \cosh^2[\pi (E-E')r_G]}=\begin{cases}
0 & \text{if $E\neq E'$}\\
\infty & \text{if $E=E'$}.
\end{cases}
\end{split}
\end{equation}
Finally, by observing that
\begin{equation}
\frac{\pi}{2}\int_{-\infty}^{\infty}\frac{r_GdE'}{\cosh^2[\pi (E-E')r_G]}=
\int_0^{\infty}\frac{d\eta}{\cosh^2\eta}=1,
\end{equation}
I can conclude that
\begin{equation}
\lim_{r_G\to\infty}\delta^{(r_G)}(E,E')=\delta (E-E')
\end{equation}
and then that
\begin{align}
\delta^{(r_G)}(E,E')\sim\delta(E-E') && 
\text{if $E,E'\ge 0$ and $Er_G,E'r_G\gg 1$}.
\end{align}
With this result and by integrating Eq. (\ref{dP_f}) 
with respect to the positron energy $E'$, I finally obtain 
the probability that an electron is present at large times with 
an energy between $E$ and $E+dE$ such that $E\tau\gg 1$ and $Er_G\gg 1$ in the form
\begin{equation}
dP^{(R)}(E)\sim \frac{1}{4\pi}
\left(\frac{eB_fR^2_{\perp M}}{64}\right)^2
\log\left(\frac{eB_fR^2_{\perp M}}{32}\right)
\left(\frac{B_f-B_i}{\tau B_i}\right)^2\frac{r_GdE}{E^2}.
\end{equation}
In order to obtain a probability per unit volume I have to give an estimate 
of the height of the quantization cylinder. Now, I have said that 
the modified Bessel functions $K_{1/2\pm 2iEr_G}(4k_{n_d}(t)z)$ 
(I refer to the electron wave functions but an identical 
conclusion can be drawn for the positron ones) are 
exponentially decreasing as $k_{n_d}(t)z\gg 1$. In particular, it can easily be 
shown that if $Er_G\gg 1$ the exponential behaviour of the function $K_{1/2\pm 2iEr_G}(4k_{n_d}(t)z)$ for large times starts at
\begin{equation}
z_0=\frac{2Er_G}{4k_{n_d}(t)}\le\frac{2Er_G}{\sqrt{2eB_f}}.
\end{equation}
For this reason I can assume $L_z=2Er_G/\sqrt{2eB_f}$ and
\begin{equation} 
V=\pi (2R_{\perp M})^2\frac{2Er_G}{\sqrt{2eB_f}}=
\frac{8\pi Er_GR_{\perp M}^2}{\sqrt{2eB_f}}
\end{equation}
as the volume of the quantization cylinder in such a way
\begin{equation}
\label{dP_f_f}
\frac{dP^{(R)}(E;t)}{dVdE}\sim \frac{1}{\pi}
\left(\frac{eB_f}{128}\right)^{5/2}
\log\left(\frac{eB_fR^2_{\perp M}}{32}\right)
\left(\frac{B_f-B_i}{\tau B_i}\right)^2\frac{R^2_{\perp M}}{E^3}.
\end{equation}
This probability does not depend on the electron mass nor on the gravitational 
radius of the black hole but this is due only to the fact that I am working in the 
strong magnetic field regime and in the high-energy region. Instead, it is not so obvious how to explain the dependence of the final presence probability (\ref{dP_f_f}) on the logarithm of $R_{\perp M}$. Now, before comparing Eq. (\ref{dP_f_f}) to the analogous result Eq. (\ref{dP_lin}) where no gravitational field effects were taken into account I have to 
integrate Eq. (\ref{dP_f_f}) with respect to the electron energy. Nevertheless, I stress that also in this case the comparison can be only qualitative because a linear dependence on time of the magnetic field was used to obtain Eq. (\ref{dP_lin}). Before doing that, I note 
that the presence probability scales here as $E^{-3}$ in the 
high-energy region while in the preceding case the corresponding probability per unit longitudinal momentum $k$ behaved as $k^{-4}$ [see the details in \cite{Calucci}]. This means that the production of high-energy electrons (positrons) is favoured in the 
presence of the gravitational field. Now, to be coherent with the approximations I have made, I perform the integration of Eq. (\ref{dP_f_f}) from $E_m=100\;r_G^{-1}$ to infinity in fact, by assuming $\tau=1\;\text{s}$ and $r_G=3.0\times 10^6\;\text{cm}$ as for a $10$ solar masses black hole, it also results $E_m\tau\gg 1$. By indicating the resulting total probability per unit volume as $dP^{(R)}(t)/dV$, I obtain
\begin{equation}
\frac{dP^{(R)}(t)}{dV}\sim \frac{1}{\pi}
\left(\frac{eB_f}{64}\right)^{5/2}
\log\left(\frac{eB_fR^2_{\perp M}}{32}\right)
\left(\frac{B_f-B_i}{\tau B_i}\right)^2\frac{R^2_{\perp M}}{E_m^2}.
\end{equation}
In general, the presence probability per unit volume depends here on the $(5/2)$-power of the magnetic field strength while in Eq. (\ref{dP_lin}) it depended on its $(3/2)$-power [remind that in Eq. (\ref{dP_lin}) $b=\dot{B}_{\shortuparrow}^{\text{lin}}(t)$]. To give a more quantitative estimate I use the typical values $R_{\perp M}=10^5\;\text{cm}$, $B_i\sim B_f=10^{15}\;\text{gauss}$ and I assume the magnetic field (\ref{B_lin}) to be such that $B_{\shortuparrow}^{\text{lin}}(t)\sim B_0=B_i$ and $b=(B_f-B_i)/\tau$. In this way I obtain [see also Eq. (\ref{dP_lin})]
\begin{equation}
\label{confr_str_Min}
\frac{dP^{\text{lin}}_{\shortuparrow}(t)}{dP^{(R)}(t)}\sim \frac{E_m^2}{eB_f}=6.7\times 10^{-29}
\end{equation}
that clearly allows me to conclude that the effects of the gravitational field in 
the pair production process are really relevant and they can not be neglected at all.  
%
%
\section{Summary and conclusions}
This Chapter has been devoted to the study of the effects that the presence of a gravitational field can have on the production of electron-positron pairs in the presence of a strong, uniform and slowly-varying magnetic field. The motivation of this analysis comes from the fact that in the model I have in mind the pair production is assumed to happen near astrophysical compact objects such as magnetars or black holes. Then, it is reasonable that, especially in the last case, the gravitational effects can be relevant. Two different physical situations have been treated here: in the first one the pair is imagined to be produced far from the black hole event horizon (Sect. \ref{weak}) and in the second one it is assumed to be produced near the black hole event horizon (Sect. \ref{strong}). In both cases the presence of the gravitational field has been taken into account only in the determination of the electron and positron one-particle modes and energies while the calculation of the pair presence probabilities has been performed by using the first-order adiabatic perturbation theory.

In particular, if the pair is produced far from the black hole event horizon, the effects of the gravitational field have been treated perturbatively. I have shown how the modified one-particle modes and energies of the electron and of the positron reflect on the pair presence probabilities [see Eq. (\ref{dP_fd_f})]. In particular, I have examined the case of the production of a pair in the presence of a magnetic field varying only in strength and always perpendicular to the gravitational field. Firstly, I have found that the presence probability contains a factor $g_t^{-1}$ that makes it growing and growing as one gets closer and closer to the event horizon of the black hole. More important, I have also obtained a new new qualitative result: \emph{even in the presence of a weak gravitational field, even if only the strength of the magnetic field changes with time it is possible to create a pair with both the electron and the positron in a TGS}. Actually, this probability is a small quantity with respect to the total probability that a pair is created in Minkowski spacetime in the presence of a time-varying magnetic field with fixed direction [see Eq. (\ref{eta_g})], but this result is a consequence of the fact that the gravitational field has been treated perturbatively. From this point of view, the information we have gained is that in the presence of a gravitational field this new effect is there. 

Instead, if the pair is produced near the black hole event horizon, the situation is very different mostly because \emph{the one-particle energy of the electron (positron) is an independent continuous nonnegative quantum number}. Also in this case, I have considered a situation in which the magnetic field does not change its direction with time but, here, always remaining parallel to the gravitational field. In this case [see Eq. (\ref{dP_f_f})] we have seen that the presence probability depends on the $(5/2)$-power of the magnetic field strength while the analogous quantity in Minkowski spacetime depended only on its $(3/2)$-power. Also, the presence probability scales here as $E^{-3}$ in the high-energy region while in the flat-spacetime case the probability behaved as $k^{-4}$ with $k$ $(-k)$ the longitudinal momentum of the electron (positron). In this way, \emph{the production of high-energy electrons (positrons) is strongly favoured in the 
presence of the gravitational field}. Moreover, the ratio (\ref{confr_str_Min}) allowed 
me to safely conclude that \emph{the effects of a strong gravitational field in the pair production process are dramatically important and they can not be neglected at all}. A final observation concerns the fact that the presence of a strong gravitational field makes possible the creation of pairs that can not fly to infinity because they do not have enough energy (as I have said, the one-particle energy spectrum of the electrons and of positrons extends down to zero). \emph{These electrons (positrons) created with such energies annihilate inside the gravitational field producing low-energy photons which may fly away and, eventually, contribute to the low-energy part of the GRBs spectra}.
\backmatter
\chapter{Appendix A}
\fancyhf{}
\fancyhead[LE,RO]{\thepage}
\fancyhead[LO,RE]{Appendix A}
\setcounter{equation}{0}
\renewcommand{\theequation}{A.\arabic{equation}}
In this Appendix I study the general features of the presence amplitudes of pairs in which the electron and/or the positron is not in a TGS in the presence of the rotating magnetic field (\ref{B_rot}). Firstly, I write the corresponding presence matrix elements by using the compact notation $\dot{H}_{jj'}(t)\equiv\langle jj' (t)|\dot{H}(t)|0(t)\rangle$ [see Eq. (\ref{H_der_jjp})]. In this way, since in the presence of a purely rotating magnetic field these matrix elements and the energies $w_j(t)$ and $\tilde{w}_{j'}(t)$ are actually time-independent, the corresponding amplitudes are given by [see Eq. (\ref{gamma_jjp})]
\begin{equation}
\label{gamma_jjpB}
\gamma_{jj'}(t)=2\frac{\dot{H}_{jj'}(t)}{\left(w_j+ \tilde{w}_{j'}\right)^2}
\exp\left[\frac{i}{2}\left(w_j+\tilde{w}_{j'}\right)t\right]\sin\left[\frac{1}{2}\left(w_j+ \tilde{w}_{j'}\right)t\right].
\end{equation}
The presence matrix elements different from zero are
\begin{subequations}
\label{Oth_matr_el}
\begin{align}
\begin{split}
&\dot{H}_{n_d,k,\sigma,n_g;n_g+1,-k,-\sigma,n_d}(t)=\dot{H}_{n_d,k,\sigma,n_g+1;n_g,-k,-\sigma,n_d}(t)\\
&\qquad =\sigma \Omega N_{jj'}\sqrt{\frac{eB_{\nnearrow}}{2}(n_g+1)}
\left[1+\sigma\left(\frac{W_{jj'}^2-k^2}{eB_{\nnearrow}}+2n_d+1\right)\right],
\end{split}\\
\begin{split}
\label{div}
&\dot{H}_{n_d+1,k,\sigma,n_g;n_g,-k,-\sigma,n_d}(t)=\dot{H}_{n_d,k,\sigma,n_g;n_g,-k,-\sigma,n_d+1}(t)\\
&\qquad =\sigma \Omega N_{jj'}\sqrt{\frac{eB_{\nnearrow}}{2}(n_d+1)}\left[3+
\sigma\left(\frac{W_{jj'}^2-k^2}{eB_{\nnearrow}}+2n_d+2\right)\right],
\end{split}\\
\begin{split}
&\dot{H}_{n_d,k,\sigma,n_g;n_g,k',-\sigma,n_d+1}(t)=\dot{H}_{n_d+1,k,\sigma,n_g;n_g,k',-\sigma,n_d}(t)\\
&\qquad\qquad\qquad\qquad\qquad\;\; =-\Omega N_{jj'}\sqrt{\frac{eB_{\nnearrow}}{2}(n_d+1)}\frac{2\pi \Delta k}{L_z}\delta^{\prime(L_z)}_K,
\end{split}\\
\begin{split}
&\dot{H}_{n_d,k,\sigma,n_g;n_g,k',\sigma,n_d}(t)\\
&\qquad\qquad\qquad=\Omega N_{jj'}\left[\sigma\left(kk'-W_{jj'}^2\right)\frac{2\pi i}{L_z}\delta^{\prime(L_z)}_K-i k(2n_d+1)\delta_{k,-k'}\right],
\end{split}\\
\begin{split}
&\dot{H}_{n_d,k,-1,n_g;n_g+1,-k,-1,n_d-1}(t)=\dot{H}_{n_d,k,-1,n_g+1;n_g,-k,-1,n_d-1}(t)\\
&\qquad\qquad\qquad\qquad\qquad\qquad\;\, =-2ik\Omega N_{jj'}\sqrt{n_d(n_g+1)},
\end{split}\\
\begin{split}
&\dot{H}_{n_d,k,+1,n_g;n_g+1,-k,+1,n_d+1}(t)=\dot{H}_{n_d,k,+1,n_g+1;n_g,-k,+1,n_d+1}(t)\\
&\qquad\qquad\qquad\qquad\qquad\quad\quad\;\, =2ik\Omega N_{jj'} \sqrt{(n_d+1)(n_g+1)},
\end{split}\\
\begin{split}
&\dot{H}_{n_d,k,+1,n_g;n_g,k',+1,n_d+2}(t)=\dot{H}_{n_d+2,k,-1,n_g;n_g,k',-1,n_d}(t)\\
&\qquad\qquad\qquad=-2\Omega N_{jj'}\left(i k\delta_{k,-k'}+eB_{\nnearrow}\frac{2\pi i}{L_z}\delta^{\prime(L_z)}_K\right)\sqrt{(n_d+2)(n_d+1)}.
\end{split}
\end{align}
\end{subequations}
In order to simplify the previous formulas I defined the quantities
\begin{subequations}
\begin{align}
\Delta k &=k'-k,\\
K &=k'+k,\\
W_{jj'}^2 &=(w_j+m)(\tilde{w}_{j'}+m),\\
N_{jj'} &=\frac{eB_{\nnearrow}}{2W_{jj'}^2}\sqrt{\frac{w_j+m}{2w_j}
\frac{\tilde{w}_{j'}+m}{2\tilde{w}_{j'}}}.
\end{align}
\end{subequations}
Also, by reminding the quantization conditions (\ref{k_discr}), I note that the function $\delta^{\prime(L_z)}_k$ in Eqs. (\ref{Oth_matr_el}) defined as
\begin{equation}
\delta^{\prime(L_z)}_k\equiv \frac{1}{2\pi i}\int_{-\frac{L_z}{2}}^{\frac{L_z}{2}}dz\; z\exp(-ikz)=
\begin{cases}
0 & \text{if $k=0$}\\
\frac{L_z}{2\pi}\frac{(-1)^{\ell}}{k} & \text{if $k=\frac{2\pi \ell}{L_z}\neq 0$},
\end{cases}
\end{equation}
becomes the derivative of the $\delta$ function in the limit $L_z\to \infty$. 

All the previous matrix elements can be divided into two groups: the ones related to transitions in which the longitudinal linear 
momentum conserves and the others characterized by the presence of the function $\delta^{\prime(L_z)}_K$ in which it does not. Of course, this fact is due to the dependence of the transition operators $T'_{jj'y}(t)$ on $z$ [see Eq. (\ref{T_jjp_y})]. 

Finally, I observe that if one sums the probabilities corresponding to the previous matrix elements by means of Eq. (\ref{gamma_jjpB}) with respect 
to the quantum number $n_d$, all the series converge. Only the series corresponding to the matrix elements (\ref{div}) diverges logarithmically and it is not so obvious how to give a physical interpretation of such a kind of divergence. However, we can understand qualitatively why the probability of creating a pair with larger and larger $n_d$ and 
then with larger and larger energy decreases so slowly. In fact, the quantum number $n_d$ is also connected with the radius $\rho_{\perp}$ of the helix along which a classical electron performs its motion (see Fig. \ref{class_mot}). In particular, it can be shown that $\rho_{\perp}^2\sim n_d$ \cite{Cohen}. From this point of view, while creating, for example, an electron with larger and larger $n_d$ needs an amount of energy proportional to $\sqrt{n_d}$ [see Eq. (\ref{L_l_e})], the electron 
wave function extends over a volume that also increases with $n_d$, in such a way that the magnetic energy available for the electron creation also increases with $n_d$.

\chapter{Appendix B}
\fancyhf{}
\fancyhead[LE,RO]{\thepage}
\fancyhead[LO,RE]{Appendix B}
\setcounter{equation}{0}
\renewcommand{\theequation}{B.\arabic{equation}}
I want to show here with some detail why the function $f(k,t)dk$ given in Eq. (\ref{f}) can be interpreted as the mean number of electrons (positrons) per unit volume present at time $t$ with a longitudinal momentum between $k$ and $k+dk$.

The argument I propose is closely related to the treatment given in \cite{Thirring} in dealing with multiple soft photon production. Firstly, I remind that since I am considering the production \emph{from vacuum} of electron-positron pairs up to first order in the adiabatic perturbation theory, the \emph{vacuum} persistence probability is one [see the discussion below Eq. (\ref{dot_H_mi})]. Also, following the adiabatic perturbation theory, $f(k,t)dk$ represents the probability per unit volume that a pair is present at time $t$ with the electron with a longitudinal momentum between $k$ and $k+dk$ and the positron with a longitudinal momentum between $-k$ and $-k-dk$. Since there is a sharp correlation between the electron and the positron quantum numbers, $f(k,t)dk$ is also the probability per unit volume that an electron is present at time $t$ with a longitudinal momentum between $k$ and $k+dk$ or, symmetrically, that a positron is present at time $t$ with a longitudinal momentum between $k$ and $k+dk$ [note from Eq. (\ref{f}) that $f(k,t)=f(-k,t)$]. Then, the total probability that an electron is present at time $t$ is given by the integral $\int dk dV f(k,t)$ where $dV=L_zdA_{\perp}=L_z\pi dR_{\perp M}^2$. Since no interaction is introduced among the particles produced, I would conclude that the probability of finding at time $t$ two electrons with a longitudinal momenta between $k$ and $k+dk$ and between $k'$ and $k'+dk'$ respectively is $f(k,t)f(k',t)dkdk'dVdV'$ and so on. But, really, I must take into account the overall conservation of the probability, then, by summing all the terms and by taking into account that the particles are indistinguishable, I get the normalization factor
\begin{equation}
N(t)\left[1+\int dkdV f(k,t)+\frac{1}{2}\int dk dk'dVdV' f(k,t)f(k',t)+\dots\right]=1,
\end{equation}
that is
\begin{equation}
N(t)=\exp\left[-\int dk dV f(k,t)\right].
\end{equation}
In conclusion, the total probability that $r$ electrons are present at time $t$ is given by
\begin{equation}
\frac{1}{r!}\left[\int dkdV f(k,t)\right]^r\exp\left[-\int dkdV f(k,t)\right]
\end{equation}
which is a Poissonian distribution in the number of electrons present. This allows me to interpret the quantity $\int dk dV f(k,t)$ as the mean number of electrons present at time $t$ and then $f(k,t)dk$ as the mean number of electrons per unit volume present at time $t$ with a longitudinal momentum between $k$ and $k+dk$. 

In the previous description the pair is treated as an effective boson. This is allowed when the production rate is low, so there is a little chance of having two electrons in the same cell of the phase space, where the Pauli principle would play the dominant role (the two electrons would be necessarily correlated). In order to check the consistency of this treatment I compare the mean number of electrons produced with the number of available quantum states in the same conditions. The number of the electrons produced per unit volume and unit longitudinal momentum is given, according to the previous discussion, by the function $f(k,t)$, while the number of available quantum states per unit volume and unit longitudinal momentum is \cite{Landau3}
\begin{equation}
n(k)=\frac{1}{2\pi}\frac{eB_{\nnearrow}}{2\pi}=\frac{eB_{\nnearrow}}{4\pi^2}
\end{equation}
where I have not considered the spin factor because in the present problem the electrons created have fixed spin direction [see Eq. (\ref{states})]. In this way, the ratio $f(k,t)/n(k)$ is always less than
\begin{equation}
\frac{1}{32}\left(\Omega R_{\perp M}\frac{B_{\nnearrow}}{B_{cr}}\right)^2.
\end{equation}
But, by following the same technique used to obtain the strong inequality (\ref{lim_ad_pert_R_M}), one sees that the quantity $\Omega R_{\perp M}B_{\nnearrow}/B_{cr}$ must be much less than one in order that the first-order adiabatic perturbation theory can be safely applied, then all the previous description is coherent.

\chapter{Appendix C}
\fancyhf{}
\fancyhead[LE,RO]{\thepage}
\fancyhead[LO,RE]{Appendix C}
\setcounter{equation}{0}
\renewcommand{\theequation}{C.\arabic{equation}}
In this Appendix I want to calculate the matrix element $\bar{u}Qv$ [see Eq. (\ref{dsigma})] of the pair annihilation into two photons process by using not the electron propagator in vacuum as in the main text but the so-called \emph{Schwinger propagator} \cite{Schwinger} that is the electron propagator in the presence of a constant and uniform magnetic field that I will indicate as $\mathbf{B}'_{\nnearrow}=(0,0,B_{\nnearrow})$. The structure of the Schwinger propagator is, in general, very complicated but it simplifies in the strong field approximation ($B_{\nnearrow}\gg B_{cr}$) in which I am working. In this case the propagator is obtained as a sum only over all the TGSs of the electron and of the positron and, working in the symmetric gauge (\ref{A}), it is given by \cite{Kuznetzov}:
\begin{equation}
G'(t,\mathbf{r},t',\mathbf{r}')=\exp\left[i\frac{eB_{\nnearrow}}{2}(xy'-x'y)\right]\hat{G}'(t-t',\mathbf{r}-\mathbf{r}')
\end{equation}
where $(t,\mathbf{r})$ and $(t',\mathbf{r}')$ are two fourpoints and where 
\begin{equation}
\begin{split}
\hat{G}'(T,\mathbf{R})=&\frac{ieB_{\nnearrow}}{2\pi}\exp\left[-\frac{eB_{\nnearrow}}{4}(X^2+Y^2)\right]\\
&\qquad\quad\times\int\frac{dW dK}{(2\pi)^2}\frac{W\gamma^0-K\gamma^3+m}{W^2-K^2-m^2}\Pi_-\exp[-i(WT-KZ)]
\end{split}
\end{equation}
with $\mathbf{R}=(X,Y,Z)$ and with $\Pi_-=(1-\sigma_z)/2$ is the spin-down projector.

If I use the usual notation for the photon field [see for example \cite{Mandl}], the transition amplitude of the pair annihilation into two photons can be written as
\begin{equation}
\label{S_fi_B_0}
\begin{split}
&S'_{n,n',\lambda,\lambda'}(k,k',\mathbf{q},\mathbf{q}')=-4\pi \alpha_{em}\int dtd\mathbf{r}dt'd\mathbf{r}'\\
&\qquad\qquad\qquad\times \bar{v}'_{n',k'}(\mathbf{r})\exp(-i\varepsilon' t)\gamma^{\alpha}iG'(t,\mathbf{r},t',\mathbf{r}')\gamma^{\beta}u'_{n,k}(\mathbf{r}')\exp(-i\varepsilon t')\\
&\qquad\qquad\qquad\times \left\{\frac{(\Euler{e}_{\mathbf{q},\lambda})_{\alpha}}{\sqrt{2V\omega}}\exp[i(\omega t-\mathbf{q}\cdot\mathbf{r})]\frac{(\Euler{e}_{\mathbf{q}',\lambda'})_{\beta}}{\sqrt{2V\omega'}}\exp[i(\omega' t'-\mathbf{q}'\cdot\mathbf{r}')]\right.\\
&\qquad\qquad\qquad\quad\;\;\left.+\frac{(\Euler{e}_{\mathbf{q},\lambda})_{\alpha}}{\sqrt{2V\omega}}\exp[i(\omega t'-\mathbf{q}\cdot\mathbf{r}')]\frac{(\Euler{e}_{\mathbf{q}',\lambda'})_{\beta}}{\sqrt{2V\omega'}}\exp[i(\omega' t-\mathbf{q}'\cdot\mathbf{r})]\right\}
\end{split}
\end{equation}
where the initial electron and positron are assumed to be in the TGSs $u'_{n,k}(\mathbf{r}')$ with energy $\varepsilon=\varepsilon(k)=\sqrt{m^2+k^2}$ and $v'_{n,k}(\mathbf{r}')$ with energy $\varepsilon'=\varepsilon (k')=\sqrt{m^2+k'^2}$ respectively and where the final photon states are those described in the main text before Eq. (\ref{dsigma}) ($V$ is the quantization volume and the limit of large $V$ is understood). In the following, I will calculate only the first amplitude in Eq. (\ref{S_fi_B_0}) corresponding to the creation of the photon $(\mathbf{q},\lambda)$ in $(t,\mathbf{r})$ and of the photon $(\mathbf{q}',\lambda')$ in $(t',\mathbf{r}')$ and I will call it $S^{\prime(1)}_{n,n',\lambda,\lambda'}(k,k',\mathbf{q},\mathbf{q}')$. The second amplitude can be calculated in an analogous way.

Now, from Eqs. (\ref{u_v_p_g}) one notes that the difference of the TGSs with those of the electrons (positrons) freely propagating along the $z$ axis with spin down (up) lies only in the dependence on the transverse coordinates \cite{Mandl}. By inserting Eqs. (\ref{u_v_p_g}) and the Schwinger propagator in Eq. (\ref{S_fi_B_0}), I observe that the integrals on the time variables and on the longitudinal variables can be performed exactly. In fact, they give two $\delta$ functions that allow to calculate the integrals on $W$ and on $K$ in the Schwinger propagator and that guarantee the energy and the longitudinal momentum conservation. In this way, the amplitude $S^{\prime(1)}_{n,n',\lambda,\lambda'}(k,k',\mathbf{q},\mathbf{q}')$ can be written as
\begin{equation}
\label{S_fi_B_0_2}
\begin{split}
S^{\prime(1)}_{n,n',\lambda,\lambda'}(k,k',\mathbf{q},\mathbf{q}')&=\frac{2\pi\alpha_{em}}{L_zV\sqrt{\omega\omega'}}Q^{\prime(1)}_{n,n',\lambda,\lambda'}(k,k',\mathbf{q},\mathbf{q}')\\
&\quad \times (2\pi)^2\delta(k+k'-q_z-q'_z)\delta(\varepsilon+\varepsilon'-\omega-\omega')
\end{split}
\end{equation}
where $L_z$ is the length of the quantization volume in the $z$ direction. In this expression I defined the transition matrix element
\begin{equation}
\label{Q_B}
\begin{split}
Q^{\prime(1)}_{n,n',\lambda,\lambda'}(k,k',\mathbf{q},\mathbf{q}') &=\sqrt{\frac{(\varepsilon+m)(\varepsilon'+m)}{2\varepsilon 2\varepsilon'}}N'_{n,n'}(q_x,q_y,q'_x,q'_y)\\
&\quad \times\begin{pmatrix}
0 & -\dfrac{k'}{\varepsilon'+m} & 0 -1
\end{pmatrix}\\
&\quad\times\gamma^{\alpha}(\Euler{e}_{\mathbf{q},\lambda})_{\alpha}\frac{\gamma^0(\omega-\varepsilon')-\gamma^3(q_z-k')+m}{(\omega-\varepsilon')^2-(q_z-k')^2-m^2}\\
&\quad\times\Pi_-\gamma^{\beta}(\Euler{e}_{\mathbf{q}',\lambda'})_{\beta}\begin{pmatrix}
0\\
1\\
0\\
-\dfrac{k}{\varepsilon+m}
\end{pmatrix}
\end{split}
\end{equation}
where the function
\begin{equation}
\label{N}
\begin{split}
N'_{n,n'}(q_x,q_y,q'_x,q'_y)=\frac{1}{\pi^2}&\frac{1}{\sqrt{n!n'!}} \int d\xi d\eta d\xi' d\eta'(\xi'-i\eta')^n(\xi+i\eta)^{n'}\\
&\qquad\times\exp[-(\xi^2+\eta^2+\xi'^2+\eta'^2)]\\
&\qquad\times\exp\left[\xi(\xi'-i\eta')-i\eta(\xi'+i\eta')\right]\\
&\qquad\times\exp\left[-i\sqrt{\frac{2}{eB_{\nnearrow}}}(q_x\xi+q_y\eta+q'_x\xi'+q'_y\eta')\right]
\end{split}
\end{equation}
depends only on the transverse momentum variables $q_x$, $q_y$, $q'_x$ and $q'_y$.

The transition amplitude (\ref{S_fi_B_0_2}) is to be compared with the analogous calculated in the vacuum [see for example Eq. (25.9) in \cite{Akhiezer}]
\begin{equation}
\begin{split}
S^{(1)}_{s,s',\lambda,\lambda'}(\mathbf{k},\mathbf{k}',\mathbf{q},\mathbf{q}')&=\frac{2\pi\alpha_{em}}{V^2\sqrt{\omega\omega'}}Q^{(1)}_{s,s',\lambda,\lambda'}(\mathbf{k},\mathbf{k}',\mathbf{q},\mathbf{q}')\\
&\quad\times(2\pi)^4\delta(\mathbf{k}+\mathbf{k}'-\mathbf{q}-\mathbf{q}')\delta(\varepsilon+\varepsilon'-\omega-\omega')
\end{split}
\end{equation}
with, for an electron moving along the $z$ axis with spin down and a positron moving along the same axis with spin up,
\begin{equation}
\label{Q_B=0}
\begin{split}
&Q^{(1)}_{s,s',\lambda,\lambda'}(\mathbf{k},\mathbf{k}',\mathbf{q},\mathbf{q}')\equiv Q^{(1)}_{\lambda,\lambda'}(k,k',\mathbf{q},\mathbf{q}')\\
&\quad=\sqrt{\frac{(\varepsilon+m)(\varepsilon'+m)}{2\varepsilon 2\varepsilon'}}\begin{pmatrix}
0 & -\dfrac{k'}{\varepsilon'+m} & 0 -1
\end{pmatrix}\\
&\qquad \times\gamma^{\alpha}(\Euler{e}_{\mathbf{q},\lambda})_{\alpha}\frac{\gamma^0(\omega-\varepsilon')-\gamma^1q_x-\gamma^2q_y-\gamma^3(q_z-k')+m}{(\omega-\varepsilon')^2-q_x^2-q_y^2-(q_z-k')^2-m^2}\gamma^{\beta}(\Euler{e}_{\mathbf{q}',\lambda'})_{\beta}\\
&\qquad\times\begin{pmatrix}
0\\
1\\
0\\
-\dfrac{k}{\varepsilon+m}
\end{pmatrix}.
\end{split}
\end{equation}

Now, suppose that the transverse momenta of the outgoing photons are much smaller than the electron mass $m$. In this approximation the matrix element in the vacuum has a weak dependence on them. In the same way, since $m^2/e=B_{cr}\ll B_{\nnearrow}$ then $q_x\ll \sqrt{eB_{\nnearrow}}$, $q_y\ll \sqrt{eB_{\nnearrow}}$, $q'_x\ll \sqrt{eB_{\nnearrow}}$ and $q'_y\ll \sqrt{eB_{\nnearrow}}$ and from Eq. (\ref{N}) we see that in this case $N'_{n,n'}(q_x,q_y,q'_x,q'_y)$ [and then $S^{\prime(1)}_{n,n',\lambda,\lambda'}(k,k',\mathbf{q},\mathbf{q}')$] does not depend on the magnetic field. In particular,
\begin{equation}
N'_{n,n'}(q_x,q_y,q'_x,q'_y)\simeq \delta_{n,n'}
\end{equation}
that is, the transverse structure of the electron and positron states does not influence the matrix element $Q^{\prime(1)}_{n,n',\lambda,\lambda'}(k,k',\mathbf{q},\mathbf{q}')$. Also, by comparing Eqs. (\ref{Q_B}) and (\ref{Q_B=0}), one sees that in this approximation the matrix element in the presence of the magnetic field $\mathbf{B}'_{\nnearrow}$ and the matrix element in the vacuum have a very similar structure. These observations give me the possibility to conclude that \emph{one is allowed to use the approximated treatment leading to Eq. (\ref{dsigma}) when the photons have small transverse momenta compared with the electron mass $m$}. This situation is verified trivially when the electron and positron momenta are small compared with $m$ but also when the incoming particles are in the ultrarelativistic regime: in fact, in this case, the photon production shows a pronounced peak in forward and backward directions \cite{Landau4}. Instead, when the transverse momenta of the photons are not small, Eqs. (\ref{N}) and (\ref{Q_B=0}) suggest that the corrections to the matrix element in the presence of the magnetic field $\mathbf{B}'_{\nnearrow}$ are proportional to the quantities $q_x/\sqrt{eB_{\nnearrow}}$, $q_y/\sqrt{eB_{\nnearrow}}$, $q'_x/\sqrt{eB_{\nnearrow}}$ and $q'_y/\sqrt{eB_{\nnearrow}}$ while those to the matrix element in the vacuum are proportional to $q_x/m$ and $q_y/m$. In particular, by performing the Fourier transform of the Schwinger propagator one sees that it contains exponential terms in the squared transverse momenta and this implies that, by using the ``vacuum'' quantities to calculate the differential cross section $d\sigma(k,k',\omega)/d\omega$ in Eq. (\ref{dN_domega}), one overestimates the number of photons emitted with large transverse momenta.

\chapter{Appendix D}
\fancyhf{}
\fancyhead[LE,RO]{\thepage}
\fancyhead[LO,RE]{Appendix D}
\setcounter{equation}{0}
\renewcommand{\theequation}{D.\arabic{equation}}
I want to give here the explicit expression of the zero-order electron and positron TGSs and the zero-order electron and positron states corresponding to the first-excited Landau levels. I remind that I used them to calculate the transition matrix elements (\ref{M_-_+}). These states can be easily obtained by substituting Eq. (\ref{V_p}) in Eqs. (\ref{U^P0_f}) and (\ref{V^P0_f})
\begin{subequations}
\begin{align}
\label{U_gr_0}
\begin{split}
U^{\prime (0)}_{0,k,-1,x_0}(\mathbf{r}) &=\frac{1}{\sqrt[4]{g_s^3}}
\sqrt{\frac{\varepsilon^{(0)}_k+\sqrt{g_t}m}{2\varepsilon^{(0)}_k}}\\
&\qquad\times\begin{pmatrix}
0\\
\Theta'_{0,x_0}(x,y)\\
-\sqrt{\frac{g_t}{g_s}}\frac{1}{\varepsilon^{(0)}_k+\sqrt{g_t}m}\begin{pmatrix}
0\\
k\Theta'_{0,x_0}(x,y)
\end{pmatrix}
\end{pmatrix}\frac{\exp(ikz)}{\sqrt{L_z}},
\end{split}\\
\label{V_gr_0}
\begin{split}
V^{\prime (0)}_{0,k,+1,x_0}(\mathbf{r}) &=\frac{1}{\sqrt[4]{g_s^3}}
\sqrt{\frac{\varepsilon^{(0)}_k+\sqrt{g_t}m}{2\varepsilon^{(0)}_k}}\\
&\qquad\times\begin{pmatrix}
-\sqrt{\frac{g_t}{g_s}}\frac{1}{\varepsilon^{(0)}_k+\sqrt{g_t}m}
\begin{pmatrix}
0\\
k\Theta'_{0,x_0}(x,y)
\end{pmatrix}\\
0\\
\Theta'_{0,x_0}(x,y)
\end{pmatrix}\frac{\exp(-ikz)}{\sqrt{L_z}}.
\end{split}\\
\label{U_pr_0_1}
\begin{split}
U^{\prime (0)}_{0,k,+1,x_0}(\mathbf{r}) &=\frac{1}{\sqrt[4]{g_s^3}}
\sqrt{\frac{\mathcal{E}^{(0)}_k+\sqrt{g_t}m}{2\mathcal{E}^{(0)}_k}}\\
&\qquad\times
\begin{pmatrix}
\Theta'_{0,x_0}(x,y)\\
0\\
\sqrt{\frac{g_t}{g_s}}\frac{1}{\mathcal{E}^{(0)}_k+\sqrt{g_t}m}
\begin{pmatrix}
k\Theta'_{0,x_0}(x,y)\\
i\sqrt{2eB}\Theta'_{1,x_0}(x,y)
\end{pmatrix}
\\
\end{pmatrix}\frac{\exp(ikz)}{\sqrt{L_z}},
\end{split}\\
\label{U_pr_0_2}
\begin{split}
U^{\prime (0)}_{1,k,-1,x_0}(\mathbf{r}) &=\frac{1}{\sqrt[4]{g_s^3}}
\sqrt{\frac{\mathcal{E}^{(0)}_k+\sqrt{g_t}m}{2\mathcal{E}^{(0)}_k}}\\
&\qquad\times
\begin{pmatrix}
0\\
\Theta'_{1,x_0}(x,y)\\
-\sqrt{\frac{g_t}{g_s}}\frac{1}{\mathcal{E}^{(0)}_k+\sqrt{g_t}m}
\begin{pmatrix}
i\sqrt{2eB}\Theta'_{0,x_0}(x,y)\\
k\Theta'_{1,x_0}(x,y)
\end{pmatrix}
\\
\end{pmatrix}\frac{\exp(ikz)}{\sqrt{L_z}},
\end{split}\\
\label{V_pr_0_1}
\begin{split}
V^{\prime (0)}_{0,k,-1,x_0}(\mathbf{r}) &=\frac{1}{\sqrt[4]{g_s^3}}
\sqrt{\frac{\mathcal{E}^{(0)}_k+\sqrt{g_t}m}{2\mathcal{E}^{(0)}_k}}\\
&\qquad\times
\begin{pmatrix}
\sqrt{\frac{g_t}{g_s}}\frac{1}{\mathcal{E}^{(0)}_k+\sqrt{g_t}m}
\begin{pmatrix}
-k\Theta'_{0,x_0}(x,y)\\
i\sqrt{2eB}\Theta'_{1,x_0}(x,y)
\end{pmatrix}
\\
-\Theta'_{0,x_0}(x,y)\\
0
\end{pmatrix}\frac{\exp(-ikz)}{\sqrt{L_z}},
\end{split}\\
\label{V_pr_0_2}
\begin{split}
V^{\prime (0)}_{1,k,+1,x_0}(\mathbf{r}) &=\frac{1}{\sqrt[4]{g_s^3}}
\sqrt{\frac{\mathcal{E}^{(0)}_k+\sqrt{g_t}m}{2\mathcal{E}^{(0)}_k}}\\
&\qquad\times
\begin{pmatrix}
\sqrt{\frac{g_t}{g_s}}\frac{1}{\mathcal{E}^{(0)}_k+\sqrt{g_t}m}
\begin{pmatrix}
i\sqrt{2eB}\Theta'_{0,x_0}(x,y)\\
-k\Theta'_{1,x_0}(x,y)
\end{pmatrix}\\
0\\
\Theta'_{1,x_0}(x,y)
\end{pmatrix}\frac{\exp(-ikz)}{\sqrt{L_z}}.
\end{split}
\end{align}
\end{subequations}
In these expressions I have used the definitions (\ref{Phi}) and (\ref{Chi}) of
the twodimensional spinors $\Phi'_J(\mathbf{r})$ and $\mathrm{X}'_J(\mathbf{r})$
and the definitions (\ref{E_0}) and (\ref{E_1}) of the energies $\varepsilon^{(0)}_k$
and $\mathcal{E}^{(0)}_k$.

\chapter{Appendix E}
\fancyhf{}
\fancyhead[LE,RO]{\thepage}
\fancyhead[LO,RE]{Appendix E}
\setcounter{equation}{0}
\renewcommand{\theequation}{E.\arabic{equation}}
In this Appendix I want to show that the set of spinors 
$U'_{n_d,n_g,\sigma}(E;\mathbf{r})$ 
and $V'_{n_d,n_g,\sigma}(E;\mathbf{r})$ with $E\ge 0$ is complete. Since 
it is equivalent, but mathematically easier, I will show that the set of spinors 
$U'_{n_d,n_g,\sigma}(E;\mathbf{r})$ with $-\infty <E<\infty$ is complete. 
In practice I have to show that
\begin{equation}
\int_{-\infty}^{\infty}dE\sum_{n_d,n_g=0}^{\infty}\sum_{\sigma =-1}^{1}
\big[U'_{n_d,n_g,\sigma}(E;\mathbf{r})\big]_a
\big[U^{\prime*}_{n_d,n_g,\sigma}(E;\mathbf{r}')\big]_b=
\frac{1}{64}\delta_{a,b}\delta(\mathbf{r}-\mathbf{r}')
\end{equation}
where $a,b=1,\ldots,4$ are two spinor indices and where the definition 
(\ref{s_p^R_f}) of the scalar product between two spinors 
has been taken into account. By using the 
general expression (\ref{U^R}) of the spinors 
$U'_{n_d,n_g,\sigma}(E;\mathbf{r})$ and the fact that 
the projectors $P_{\pm}$ are two real and symmetric matrices 
[see Eq. (\ref{P_pm}) and remind that I work in the Dirac representation 
of the $\gamma$ matrices], I can write the previous equation as
\begin{equation}
\label{compl}
\begin{split}
&\int_{-\infty}^{\infty}d\nu\sum_{n_d,n_g,\sigma}
\frac{8k_{n_d}\cosh(\pi\nu)}{\pi^2}
\left[P_-K_{1/2+i\nu}(4k_{n_d}z)+P_+K_{1/2-i\nu}(4k_{n_d}z)\right]_{ac}
\\
&\qquad\times\big[\Xi'_{n_d,n_g,\sigma}(x,y)\big]_c\big[\Xi^{\prime*}_{n_d,n_g,\sigma}(x',y')\big]_d\\
&\qquad\times\left[P_-K_{1/2-i\nu}(4k_{n_d}z')+P_+K_{1/2+i\nu}(4k_{n_d}z')\right]_{db}=
\delta_{a,b}\delta(\mathbf{r}-\mathbf{r}')
\end{split}
\end{equation}
where $\nu=2Er_G$ and where the summation on the spinor indices is understood. Now, 
by using the integral representation \cite{Abramowitz}
\begin{align}
\label{int_repr}
K_{\lambda}(\xi)=\frac{1}{\cos(\lambda\pi/2)}\int_{0}^{\infty}ds\cos(\xi\sinh s)
\cosh(\lambda s) && \text{if $\left\vert\text{Re}(\lambda)\right\vert< 1$ and $\xi>0$}
\end{align}
of the modified Bessel functions, it can easily be shown that
\begin{equation}
\int_{-\infty}^{\infty}d\nu\frac{8k_{n_d}\cosh(\pi\nu)}{\pi^2}
K_{1/2\mp i\nu}(4k_{n_d}z)K_{1/2\pm i\nu}(4k_{n_d}z')=2\delta(z-z').
\end{equation}

In the same way it can be seen that the integrals
\begin{equation}
\int_{-\infty}^{\infty}d\nu\frac{8k_{n_d}\cosh(\pi\nu)}{\pi^2}
K_{1/2\pm i\nu}(4k_{n_d}z)K_{1/2\pm i\nu}(4k_{n_d}z')
\end{equation}
vanish by using the integral representation (\ref{int_repr}) for one of the 
Bessel functions and the following
\begin{align}
K_{\lambda}(\xi)=\frac{1}{\sin(\lambda\pi/2)}\int_{0}^{\infty}ds\sin(\xi\sinh s)
\sinh(\lambda s) && \text{if $\left\vert\text{Re}(\lambda)\right\vert< 1$ and $\xi>0$}
\end{align}
for the other. With these results Eq. (\ref{compl}) is true if
\begin{equation}
\label{compl_2}
\begin{split}
\sum_{n_d,n_g,\sigma}
&\left[P_-\right]_{ac}\bigl[\Xi'_{n_d,n_g,\sigma}(x,y)\bigr]_c
\bigl[\Xi^{\prime*}_{n_d,n_g,\sigma}(x',y')\bigr]_d\left[P_-\right]_{db}\\
&+\left[P_+\right]_{ac}\bigl[\Xi'_{n_d,n_g,\sigma}(x,y)\bigr]_c
\bigl[\Xi^{\prime*}_{n_d,n_g,\sigma}(x',y')\bigr]_d\left[P_+\right]_{db}\\
&=\frac{1}{2}\delta_{a,b}\delta(x-x')\delta(y-y').
\end{split}
\end{equation}
Now, by using the expression (\ref{phi}) of the spinors 
$\Xi'_{n_d,n_g,\sigma}(x,y)$ it can be seen that the matrix 
$M(x,y,x',y')\equiv\sum_{n_d,n_g,\sigma}\Xi'_{n_d,n_g,\sigma}(x,y)
\Xi^{\prime\dag}_{n_d,n_g,\sigma}(x',y')$ has the general structure
\begin{equation}
\begin{split}
&M(x,y,x',y')\\
&\qquad=\begin{pmatrix}
E(x,y,x',y') & A(x,y,x',y') & -B(x,y,x',y') & 0\\
D(x,y,x',y') & E(x,y,x',y') & 0 & C(x,y,x',y')\\
B(x,y,x',y') & 0 & E(x,y,x',y') & A(x,y,x',y')\\
0 & -C(x,y,x',y') & D(x,y,x',y') & E(x,y,x',y')
\end{pmatrix}
\end{split}
\end{equation}
with
\begin{subequations}
\begin{align}
\begin{split}
A(x,y,x',y') &=\sum_{n_d,n_g,\sigma}\big[\Xi'_{n_d,n_g,\sigma}(x,y)\big]_1
\big[\Xi^{\prime*}_{n_d,n_g,\sigma}(x',y')\big]_2\\
&=\sum_{n_d,n_g,\sigma}\big[\Xi'_{n_d,n_g,\sigma}(x,y)\big]_3
\big[\Xi^{\prime*}_{n_d,n_g,\sigma}(x',y')\big]_4,
\end{split}\\
\begin{split}
B(x,y,x',y') &=\sum_{n_d,n_g,\sigma}\big[\Xi'_{n_d,n_g,\sigma}(x,y)\big]_3
\big[\Xi^{\prime*}_{n_d,n_g,\sigma}(x',y')\big]_1\\
&=-\sum_{n_d,n_g,\sigma}\big[\Xi'_{n_d,n_g,\sigma}(x,y)\big]_1
\big[\Xi^{\prime*}_{n_d,n_g,\sigma}(x',y')\big]_3,
\end{split}\\
\begin{split}
C(x,y,x',y') &=\sum_{n_d,n_g,\sigma}\big[\Xi'_{n_d,n_g,\sigma}(x,y)\big]_2
\big[\Xi^{\prime*}_{n_d,n_g,\sigma}(x',y')\big]_4\\
&=-\sum_{n_d,n_g,\sigma}\big[\Xi'_{n_d,n_g,\sigma}(x,y)\big]_4
\big[\Xi^{\prime*}_{n_d,n_g,\sigma}(x',y')\big]_2,
\end{split}\\
\begin{split}
D(x,y,x',y') &=\sum_{n_d,n_g,\sigma}\big[\Xi'_{n_d,n_g,\sigma}(x,y)\big]_2
\big[\Xi^{\prime*}_{n_d,n_g,\sigma}(x',y')\big]_1\\
&=\sum_{n_d,n_g,\sigma}\big[\Xi'_{n_d,n_g,\sigma}(x,y)\big]_4
\big[\Xi^{\prime*}_{n_d,n_g,\sigma}(x',y')\big]_3,
\end{split}\\
E(x,y,x',y') &=\sum_{n_d,n_g,\sigma}\big[\Xi'_{n_d,n_g,\sigma}(x,y)\big]_a
\big[\Xi^{\prime*}_{n_d,n_g,\sigma}(x',y')\big]_a && a=1,\ldots,4.
\end{align}
\end{subequations}
If one performs the products among the matrices 
$P_{\pm}$ and $M(x,y,x',y')$ in Eq. (\ref{compl_2}), 
it can be shown that
\begin{equation}
\begin{split}
&\left[P_-\right]_{ac}\big[M(x,y,x',y')\big]_{cd}\left[P_-\right]_{db}\\
&\quad+\left[P_+\right]_{ac}\big[M(x,y,x',y')\big]_{cd}\left[P_+\right]_{db}=
\delta_{a,b}E_a(x,y,x',y').
\end{split}
\end{equation}
In this way, it is evident that the exact expressions of the 
four functions $A(x,y,x',y'),\ldots,D(x,y,x',y')$ are not needed. Finally, by using the 
completeness of the set of spinors $\Xi'_{n_d,n_g,\sigma}(x,y)$ one sees that
\begin{align}
E_a(x,y,x',y')=\frac{1}{2}\delta(x-x')\delta(y-y') && a=1,\ldots,4
\end{align}
and then that Eq. (\ref{compl_2}) is, actually, an identity.

\chapter{Appendix F}
\fancyhf{}
\fancyhead[LE,RO]{\thepage}
\fancyhead[LO,RE]{Appendix F}
\setcounter{equation}{0}
\renewcommand{\theequation}{F.\arabic{equation}}
In this Appendix I will impose that the spinor (\ref{u_d}) is continuous at 
$z=b$ and that its norm is unit. As a result, I will discretize 
the energies $E$ and determine the two factors $N^{(<)}_J$ and 
$N^{(>)}_J$ with $J\equiv\{n,n_d,n_g,\sigma\}$ appearing 
in Eq. (\ref{u_d}). The continuity condition is satisfied if
\begin{equation}
\label{B_C}
N^{(<)}_{n,n_d}I_{1/2+2iE_{n,n_d}r_G}(4k_{n_d}b)=
N^{(>)}_{n,n_d}K_{1/2+2iE_{n,n_d}r_G}(4k_{n_d}b)
\end{equation}
where I pointed out that $N^{(<)}_{n,n_d}$ and 
$N^{(>)}_{n,n_d}$ can not depend on $n_g$ and $\sigma$ and that the energies 
depend on a new integer quantum number $n$.
Since $k_{n_d}b\ll 1$, I can use the approximated expressions 
of the modified Bessel functions near the origin \cite{Abramowitz}
\begin{subequations}
\begin{align}
\label{I_a}
I_{1/2+2iE_{n,n_d}r_G}(4k_{n_d}b) & \sim \frac{1}{\Gamma(3/2+2iE_{n,n_d}r_G)}
(2k_{n_d}b)^{1/2+2iE_{n,n_d}r_G},\\
\label{K_a}
K_{1/2+2iE_{n,n_d}r_G}(4k_{n_d}b) & \sim \frac{\Gamma(1/2+2iE_{n,n_d}r_G)}{2}
(2k_{n_d}b)^{-1/2-2iE_{n,n_d}r_G}
\end{align}
\end{subequations}
to write Eq. (\ref{B_C}) in the form
\begin{equation}
\label{B_C_2}
\frac{N^{(>)}_{n,n_d}}{N^{(<)}_{n,n_d}}
\left(\frac{1}{2}+2iE_{n,n_d}r_G\right)
\Gamma^2\left(\frac{1}{2}+2iE_{n,n_d}r_G\right)
\frac{(2k_{n_d}b)^{-1-4iEr_G}}{2}=1
\end{equation}
where the property $\Gamma (z+1)=z\Gamma (z)$ has been used. By equating 
the modulus and the phase of the left and right hand sides of Eq. (\ref{B_C_2}), I 
obtain the two real conditions
\begin{subequations}
\begin{align}
\label{N_m}
& N^{(<)}_{n,n_d}=\frac{\pi}{8k_{n_d}b}\frac{\sqrt{1+(4E_{n,n_d}r_G)^2}}
{\cosh(2\pi E_{n,n_d}r_G)}N^{(>)}_{n,n_d},\\
\label{E_n}
&\frac{\arctan\left(4E_{n,n_d}r_G\right)}{4}+\frac{\arg{\left(\Gamma
\left(\frac{1}{2}+2iE_{n,n_d}r_G\right)\right)}}{2}-E_{n,n_d}r_G\log(2k_{n_d}b)=n\frac{\pi}{2} 
\end{align}
\end{subequations}
where $n=0,\pm 1,\ldots$ and where the following property of the $\Gamma$ function 
has been used \cite{Abramowitz}: 
\begin{align}
\label{mod_gamma}
\Gamma\left(\frac{1}{2}+i\xi\right)\Gamma\left(\frac{1}{2}-i\xi\right)=
\left\vert\Gamma\left(\frac{1}{2}+i\xi\right)\right\vert^2=
\frac{\pi}{\cosh(\pi\xi)}
&& \text{with $\xi\in \mathbb{R}$}.
\end{align}
The condition (\ref{E_n}) determines the allowed discrete energies while, 
in order to determine $N^{(>)}_{n,n_d}$, I have to require that the following 
normalization condition holds [see the expression (\ref{s_p^R_f}) of the 
scalar product]:
\begin{equation}
64\int d\mathbf{r} U_{n,n_d,n_g,\sigma}^{\prime\dag}
(\mathbf{r})U'_{n,n_d,n_g,\sigma}(\mathbf{r})=1.
\end{equation}
By exploiting the orthonormalization condition (\ref{ort_tr_e}) on the 
transverse spinors $\Xi'_{n_d,n_g,\sigma}(x,y)$, 
it can be seen that the previous condition is equivalent to require that
\begin{equation}
\label{norm}
\begin{split}
64\frac{k_{n_d}r_G\cosh(2E_{n,n_d}r_G)}{4\pi^2}
&\left[\big(N^{(<)}_{n,n_d}\big)^2
\int_0^bdz\left\vert I_{1/2+2iE_{n,n_d}r_G}(4k_{n_d}z)\right\vert^2\right.\\
&\;\left.+\big(N^{(>)}_{n,n_d}\big)^2
\int_b^{\infty}dz\left\vert K_{1/2+2iE_{n,n_d}r_G}(4k_{n_d}z)\right\vert^2\right]=1.
\end{split}
\end{equation}
By using the approximated expressions (\ref{I_a}) calculated in $4k_{n_d}z$, 
the first integral gives
\begin{equation}
\label{int_0_b_I}
\int_0^bdz\left\vert I_{1/2+2iE_{n,n_d}r_G}(4k_{n_d}z)\right\vert^2
\simeq\frac{1}{k_{n_d}}
\frac{\cosh(2\pi E_{n,n_d}r_G)}{1+(4E_{n,n_d}r_G)^2}
\frac{(k_{n_d}b)^2}{\pi}.
\end{equation}
The second integral can be evaluated by using the following identity
\begin{equation}
\label{int_b_inf}
\begin{split}
&\int_b^{\infty}dz\left\vert K_{1/2+2iE_{n,n_d}r_G}(4k_{n_d}z)\right\vert^2\\
&\qquad\qquad=\lim_{\epsilon\to 0}\left[\int_0^{\infty}dz(4k_{n_d}z)^{\epsilon}
\left\vert K_{1/2+2iE_{n,n_d}r_G}(4k_{n_d}z)\right\vert^2\right.\\
&\qquad\qquad\qquad\qquad-\left.\int_0^bdz(4k_{n_d}z)^{\epsilon}
\left\vert K_{1/2+2iE_{n,n_d}r_G}(4k_{n_d}z)\right\vert^2\right].
\end{split}
\end{equation}
The first integral on the right hand side of this equation is a particular 
case of the general formula \cite{Ryzhik} 
\begin{equation}
\label{int}
\begin{split}
&\int_0^{\infty}dss^{-\rho}K_{\lambda}(as)K_{\mu}(bs)=
\frac{2^{-2-\rho}a^{-\mu+\rho-1}b^{\mu}}{\Gamma(1-\rho)}
\Gamma\left(\frac{1-\rho+\lambda+\mu}{2}\right)\\
&\qquad\qquad\times\Gamma\left(\frac{1-\rho-\lambda+\mu}{2}\right)\Gamma\left(\frac{1-\rho+\lambda-\mu}{2}\right)
\Gamma\left(\frac{1-\rho-\lambda-\mu}{2}\right)\\
&\qquad\qquad\times F\left(\frac{1-\rho+\lambda+\mu}{2},\frac{1-\rho-\lambda+\mu}{2};
1-\rho;1-\frac{b^2}{a^2}\right)
\end{split}
\end{equation}
where $a$, $b$, $\rho$, $\lambda$ and $\mu$ are complex numbers such that $\mathrm{Re}(a+b)>0$ and $\mathrm{Re}(\rho)<1-\vert\mathrm{Re}(\lambda)\vert-\vert\mathrm{Re}(\mu)\vert$ and where $F(r,s;u;z)$ is the hypergeometric function.
Instead, in the second integral on the right hand side of Eq. (\ref{int_b_inf}) 
the approximated expression (\ref{K_a}) 
of the modified Bessel function calculated in $4k_{n_d}z$ 
can be used. Obviously, even if these integrals 
are both diverging in the limit $\epsilon\to 0$, 
their divergences must cancel each other because the 
left hand side of Eq. (\ref{int_b_inf}) is finite. In fact,
\begin{equation}
\begin{split}
&\lim_{\epsilon\to 0}\left[\int_0^{\infty}dz(4k_{n_d}z)^{\epsilon}
\left\vert K_{1/2+2iE_{n,n_d}r_G}(4k_{n_d}z)\right\vert^2\right.\\
&\qquad\left.-\int_0^bdz(4k_{n_d}z)^{\epsilon}
\left\vert K_{1/2+2iE_{n,n_d}r_G}(4k_{n_d}z)\right\vert^2\right]\\
&\quad=\frac{1}{4k_{n_d}}\lim_{\epsilon\to 0}
\left[\frac{2^{-2+\epsilon}}{\Gamma(1+\epsilon)}
\Gamma\left(\frac{2+\epsilon}{2}\right)
\left\vert\Gamma\left(\frac{1+\epsilon}{2}+2iE_{n,n_d}r_G\right)\right\vert^2
\Gamma\left(\frac{\epsilon}{2}\right)\right.\\
&\left.\qquad\qquad\qquad\quad -\frac{1}{2\epsilon}
\left\vert\Gamma\left(\frac{1}{2}+2iE_{n,n_d}r_G\right)\right\vert^2
(4k_{n_d}b)^{\epsilon}\right]=\\
&\quad=\frac{1}{4k_{n_d}}
\left\vert\Gamma\left(\frac{1}{2}+2iE_{n,n_d}r_G\right)\right\vert^2
\lim_{\epsilon\to 0}
\left[\frac{1}{2\epsilon}-\frac{1}{2\epsilon}
\exp{\left(\epsilon\log{(4k_{n_d}b)}\right)}\right]\\
&\quad=-\frac{1}{8k_{n_d}}
\left\vert\Gamma\left(\frac{1}{2}+2iE_{n,n_d}r_G\right)\right\vert^2\log{(4k_{n_d}b)}
\end{split}
\end{equation}
where I used the property 
$\Gamma(\epsilon/2)=2\Gamma(1+\epsilon/2)/\epsilon$. Finally, by exploiting 
Eq. (\ref{mod_gamma}), then
\begin{equation}
\label{int_b_inf_K}
\int_b^{\infty}dz\left\vert K_{1/2+2iE_{n,n_d}r_G}(4k_{n_d}z)\right\vert^2=
-\frac{\pi}{8k_{n_d}}\frac{\log\left(4k_{n_d}b\right)}{\cosh(2E_{n,n_d}r_G)}
\end{equation}
and, by substituting Eqs. (\ref{N_m}), (\ref{int_0_b_I}) and (\ref{int_b_inf_K}) in 
Eq. (\ref{norm}), I obtain the following expression of $N^{(>)}_{n,n_d}$
\begin{equation}
N^{(>)}_{n,n_d}=N^{(>)}_{n_d}=\sqrt{\frac{4\pi}{r_G(1-8\log(4k_{n_d}b))}}.
\end{equation}
Since, at the end of the calculations the limit $b\to 0$ will be performed, I 
give the expression of $N^{(>)}_{n_d}$ in this limit:
\begin{align}
N^{(>)}_{n_d}=\sqrt{-\frac{\pi}{2r_G\log\left(k_{n_d}b\right)}} 
&& k_{n_d}b\to 0.
\end{align}
In the same limit an easy expression of the density of the energy levels 
$\varrho\left(E_{n,n_d}\right)$ can be obtained. In fact, this quantity is defined as
\begin{equation}
\varrho\left(E_{n,n_d}\right)\equiv
\left\vert\frac{dn}{dE_{n,n_d}}\right\vert.
\end{equation}
Now, if $k_{n_d}b\to 0$ then Eq. (\ref{E_n}) gives simply 
\begin{align}
\label{E_n_n_d}
E_{n,n_d}=\frac{n\pi}{2r_G}\log^{-1}\left(k_{n_d}b\right) && k_{n_d}b\to 0
\end{align}
and the density of the energy levels does not depend on the energy itself:
\begin{align}
\label{rho}
\varrho\left(E_{n,n_d}\right)=\varrho_{n_d}=
-\frac{2r_G}{\pi}\log\left(k_{n_d}b\right) && k_{n_d}b\to 0.
\end{align}
Finally, with this definition the normalization factor $N^{(>)}_{n_d}$ 
can be written in the limit $k_{n_d}b\to 0$ simply as
\begin{align}
\label{N_M}
N^{(>)}_{n_d}=\frac{1}{\sqrt{\varrho_{n_d}}} && k_{n_d}b\to 0.
\end{align}
\chapter{Acknowledgments}
\fancyhf{}
\fancyhead[LE,RO]{\thepage}
\fancyhead[LO,RE]{Acknowledgements}
I am very grateful to my advisor Prof. G. Calucci for his continuous ``scientific'' presence during these three years and for his human kindness and comprehension. I also want to thank Dott. E. Spallucci and Dott. S. Ansoldi for stimulating discussions and useful advice on quantum field theory in curved spacetimes. I can not forget about my friend Francesco for helping me in understanding some ``astrophysics'' on neutron stars and GRBs: I think we have calculated together more than one thousand orders of magnitude! Finally, thanks also to Dott. E. Pian for reading and correcting very carefully the first Chapter. 

The human support is fundamental in a three-years work such as a Phd so many thanks to Ennio, really my second advisor, for his encouragements and continuous advices just since I arrived in Trieste three years ago. Living far from home and from my family sometimes has been very hard to me. I have overcome many difficulties also thanks to my brothers, Salvo and Luca for their...brotherhood, to my mother, because it is not easy to help other people when you first need help and to my father, who has managed to ``pull out'' his great humanity. Last but not least, I would like to thank deeply ``la mia Giusi'' for her words, for her thoughts, for her...since she has entered my life, again this summer and me, ten years ago. 
\backmatter
\fancyhf{}
\fancyhead[LE,RO]{\thepage}
\fancyhead[LO,RE]{Bibliography}

\end{document}